\documentclass[pre,twocolumn,superscriptaddress]{revtex4-1}
\usepackage{epsfig,amsmath,amssymb,graphics,graphicx,color,calc}
\usepackage[colorlinks=true,linkcolor=blue,citecolor=blue,urlcolor=black]{hyperref}
\newcommand{\be}{\begin{equation}}
\newcommand{\ee}{\end{equation}}
\newcommand{\ba}{\begin{eqnarray}}
\newcommand{\ea}{\end{eqnarray}}

\newcommand{\w}{_{\mathrm{w}}}
\newcommand{\tw}{t\w}
\newcommand{\Chi}{\chi}

\begin{document}

\title{Glasses and aging: A Statistical Mechanics Perspective}

\author{Francesco Arceri}
\affiliation{Department of Physics, University of Oregon, Eugene, Oregon 97403, USA}

\author{Fran\c cois P. Landes}
\affiliation{TAU, LRI, Universit\'e Paris Sud, CNRS, INRIA, Universit\'e Paris Saclay, Orsay 91405, France}

\author{Ludovic Berthier}

\affiliation{Laboratoire Charles Coulomb (L2C), Universit\'e de Montpellier, CNRS, 34095 Montpellier, France.}

\affiliation{Department of Chemistry, University of Cambridge, Lensfield Road, Cambridge CB2 1EW, United Kingdom.}

\author{Giulio Biroli}
\affiliation{Laboratoire de Physique de l'\'Ecole normale sup\'erieure, Universit\'e PSL, CNRS, Sorbonne Universit\'e, Universit\'e Paris-Diderot, Paris, France}

\date{\today}


\maketitle

\tableofcontents

\section{Glossary}

We start with a few concise definitions
of the most important concepts
discussed in this article.

{\bf Glass transition--}For molecular liquids, the
glass transition denotes a crossover from a viscous
liquid to an amorphous solid. Experimentally, the crossover takes place
at the glass temperature, $T_g$, conventionally defined as the temperature
where the liquid's viscosity reaches the arbitrary value
of $10^{12}$~Pa.s. The glass transition more generally applies to
many different condensed matter
systems where a crossover or, less frequently, a true
phase transition, takes place between an ergodic phase and a frozen,
amorphous  glassy phase.

{\bf Aging--}In the glass phase, disordered materials
are characterized by relaxation times that exceed common
observation timescales, so that  a material quenched in its glass
phase never reaches equilibrium (neither a metastable equilibrium).
It exhibits instead an aging
behaviour during which its physical properties keep
evolving with time.

{\bf Dynamic heterogeneity--}Relaxation spectra of dynamical observables,
e.g.~the dynamical structure factor,  are very broad in
supercooled liquids.
This is associated to a spatial distribution of timescales:
at any given
time, different regions in the liquid relax at different rates. Since
the supercooled  liquid is ergodic, slow regions eventually
become fast, and vice versa. Dynamic heterogeneity refers to the
existence of these non-trivial spatio-temporal fluctuations in the local
dynamical behaviour, a phenomenon observed in virtually all
disordered systems with slow dynamics.

{\bf Effective temperature--}An aging material relaxes
very slowly, trying (in vain) to reach its equilibrium state.
During this process, the system probes
states that do not correspond to thermodynamic equilibrium,
so that its thermodynamic properties can not be
rigorously defined. Any practical measurement of its temperature
becomes a frequency-dependent operation. A `slow' thermometer
tuned to the relaxation timescale of the aging system
measures an effective temperature
corresponding to the ratio between spontaneous fluctuations (correlation)
and linear response (susceptibility). This corresponds to
a generalized form of the fluctuation-dissipation theorem
for off-equilibrium materials.

{\bf Frustration--}Impossibility of simultaneously minimizing all
the interaction terms in the energy function of
the system. Frustration might arise from quenched disorder
(as in the spin glass models), from competing
interactions (as in geometrically frustrated magnets),
or from competition between a `locally preferred order', and
global, e.g.~geometric, constraints (as in hard spheres packing problems).

{\bf Marginal Stability--}Systems are marginally stable when the number of external inputs controlling their stability is just enough to constrain all their degrees of freedom (think of a table with three legs). Marginally stable systems display an excess of zero-energy modes which makes them highly susceptible to external perturbations, and prone to extensive rearrangements.  

\section{Definition of the subject}

Glasses belong to a seemingly well-known state of matter: we easily design
glasses with desired mechanical or optical properties on an
industrial scale, they are widely
present in our daily life. Yet, a deep microscopic understanding
of the glassy state of matter remains a challenge for condensed matter
physicists~\cite{angellscience,reviewnature,berthier_theoretical_2011}.
Glasses share similarities with crystalline
solids (they are both mechanically rigid), but also with liquids
(they both have similar disordered structures at the molecular level).
It is mainly this mixed character that makes them fascinating objects, 
even to non-scientists.

A glass can be obtained by cooling the temperature of a liquid
below its glass temperature, $T_g$. The quench must be fast enough, such 
that the more standard first order phase transition towards the crystalline
phase is avoided. The glass `transition' is not a thermodynamic
transition at all, since $T_g$ is only empirically defined as the temperature
below which the material has become too viscous to flow on a `reasonable'
timescale (and it is hard to define the word `reasonable' in
any reasonable manner). Therefore, $T_g$ cannot play a very fundamental role, as
a phase transition temperature would.
It is simply the temperature below which the material looks solid on the timescale of the observer.
When quenched in the glass phase below $T_g$, liquids slowly
evolve towards an equilibrium state they will never reach on experimental
timescales. Physical properties  are then found to evolve slowly
with time in far from equilibrium states, a process known as
`aging'~\cite{struik1977physical}.

Describing theoretically and quantifying experimentally 
the physical mechanisms responsible 
for  the viscosity increase of liquids
 approaching the glass transition
and  for  aging phenomena below the glass transition
certainly stand as central
open challenges in condensed matter physics. Since
statistical mechanics aims at understanding the collective
behaviour of large assemblies of interacting objects, it comes as no surprise
that it is a central tool in the glass field. We shall therefore
summarize the understanding gained from statistical mechanics
perspectives into the problem of glasses and aging.

The subject has quite broad implications. A material is said
to be `glassy' when its typical relaxation timescale becomes of the order of,
and often much larger than, the typical duration of an experiment
or a numerical simulation. Under this generic definition, a large number
of systems can be considered as glassy~\cite{youngbook}.
One can be interested in the physics of liquids (window glasses are then
the archetype), in `hard' condensed matter (for instance
type II superconductors in the presence of disorder such as
high-$T_c$ superconductors), charge density waves or spin glasses,
`soft' condensed matter with numerous complex fluids such as
colloidal assemblies, emulsions, foams, but also granular materials, proteins,
etc. All these materials exhibit, in a part of their phase diagrams,
some sort of glassy dynamics characterized by a very rich
phenomenology with effects such as aging, hysteresis, creep, memory,
effective temperatures,
rejuvenation, dynamic heterogeneity, non-linear response, etc.

This long list explains why this research field has received
increasing attention from physicists in the last four decades.
`Glassy' topics now go much beyond the physics of simple liquids (glass
transition physics) and models and concepts developed for one
system often find applications elsewhere in physics, from
algorithmics to biophysics~\cite{complexbook}.
Motivations to study glassy materials
are numerous. Glassy materials are everywhere around us
and therefore obviously attract interest beyond academic research.
At the same time, the glass conundrum
provides theoretical physicists with
deep fundamental questions since classical tools are
sometimes  not sufficient to properly account for the glass state.
Moreover, numerically simulating the dynamics of microscopically
realistic material on timescales that are experimentally relevant remains a difficult challenge, even with modern computers.

Studies on glassy materials constitute an exciting
 research area where experiments, simulations and theoretical
calculations can meet, where both applied and fundamental problems
are considered. How can one observe, understand, and theoretically
describe the rich phenomenology of glassy materials? What are the
fundamental quantities and concepts that emerge from these descriptions?

The outline of the article is as follows. In Sec.~\ref{phenomenology}
the phenomenology of glass-forming liquids is discussed.
In Sec.~\ref{beyondglass} different types of glasses are described.
We then describe how computer simulations can provide deep insights
into the glass problem in Sec.~\ref{simu}.
The issue of dynamic heterogeneity is tackled in Sec.~\ref{dh}.
The main theoretical perspectives currently available in the field
are then summarized in Sec.~\ref{theory}. The mean-field analysis of the amorphous solid phase is reviewed in Sec.~\ref{theoryamorphous}.
In Sec.~\ref{sec:newComputMethods}, we discuss novel developments in computational studies. Aging and off-equilibrium phenomena occupy Sec.~\ref{aging}.
Finally, issues that seem important for future
research are discussed in Sec.~\ref{nofuture}.

\section{Phenomenology}

\label{phenomenology}

\subsection{Basic facts}

A vast majority of liquids (molecular liquids, polymeric liquids, etc)
form a glass if cooled fast enough in order to avoid
the crystallisation transition~\cite{angellscience}.
Typical values of cooling rate
in laboratory experiments are $0.1-100$~K/min.
The metastable phase reached in this way is called `supercooled liquid'.
In this regime the typical timescales increase in a dramatic way
and they end up to be many orders of magnitudes larger than microscopic
timescales at $T_g$, the glass transition temperature.

 For example, around the melting temperature $T_m$,
the typical timescale $\tau_\alpha$ on which density fluctuations relax,
is of the order of
$\sqrt{m a^2/K_BT}$, which corresponds to few picoseconds ($m$ is the molecular
mass, $T$ the temperature, $K_B$ the Boltzmann constant and $a$ a
typical distance between molecules).
At $T_g$, which as a rule of thumb is about $\frac{2}{3} T_m$, this timescale $\tau_\alpha$ has become of the
order of $100$~s, i.e.~$14$ orders of magnitude larger! This
phenomenon is accompanied by a concomitant increase
of the shear viscosity $\eta$. This can be understood by a
simple Maxwell model in which
$\eta$ and $\tau$ are related by $\eta=G_{\infty}\tau_\alpha$, where
$G_{\infty}$ is the instantaneous (elastic) shear modulus which does not vary
considerably in the supercooled regime. In fact,
viscosities at the glass transition temperature are of the order of
$10^{12}$~Pa.s.
In order to grasp  how viscous this is, recall
that the typical viscosity
of water (or wine) at ambient temperature is of the order of $10^{-2}$ Pa.s.
How long would one have to wait to drink a glass of
wine with a viscosity $10^{14}$ times larger?

\begin{figure}
\includegraphics[width=8.5cm]{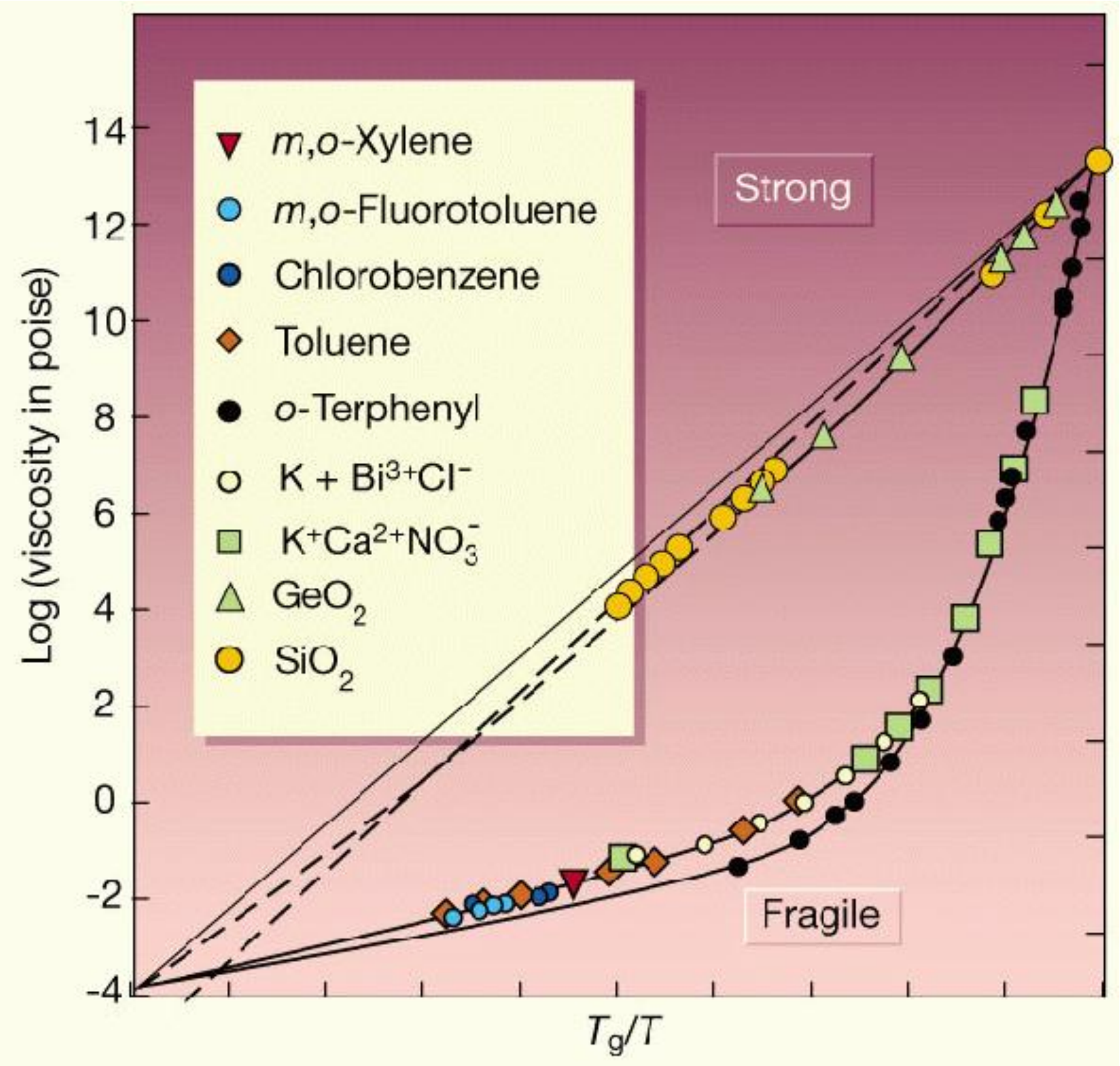}
\caption{Arrhenius plot of the viscosity of several
glass-forming liquids approaching the glass
temperature $T_g$~\cite{reviewnature}.
For `strong' glasses, the viscosity increases in an Arrhenius
manner as temperature is decreased, $\log \eta \sim E/(K_B T)$, where
$E$ is an activation energy and the plot is a straight line,
as for silica. For `fragile' liquids, the plot is bent and the effective
activation energy increases when $T$ is decreased towards $T_g$,
as for ortho-terphenyl.}
\label{angellfig}
\end{figure}

As a matter of fact,
the temperature at which the liquid does not flow anymore and becomes an
amorphous solid, called a `glass', is protocol dependent. It depends
on the cooling
rate and on the patience of the person carrying out the experiment:
solidity is
a timescale-dependent notion. Pragmatically, $T_g$ is defined as the
temperature
at which the shear viscosity is equal to $10^{13}$ Poise (also
$10^{12}$~Pa.s).

\begin{table*}
\begin{tabular}{|c|c|c|c|c|}
        \hline
        \hline
{\bf Substance}    & o-terphenyl & 2-methyltetra-hydrofuran &
n-propanol & 3-bromopentane \\
        \hline
$T_g$   & 246 & 91 & 97 & 108 \\
$T_0$  & 202.4 & 69.6 & 70.2 & 82.9   \\
$T_K$   & 204.2 & 69.3 & 72.2 & 82.5   \\
$T_K/T_0$   & 1.009 & 0.996 & 1.028 & 0.995 \\
        \hline
        \hline
\end{tabular}
\caption{\label{table} Values of glass transition temperature, VFT
singularity and Kauzmann temperatures for four supercooled
liquids~\cite{angellrichert}.}
\end{table*}

The increase of the relaxation timescale of supercooled liquids is remarkable not only because of the large number of decades involved but
also because of its
temperature dependence. This is vividly demonstrated by
plotting the logarithm of the
viscosity (or of the relaxation time) as a function
of $T_g/T$, as in Fig.~\ref{angellfig}.
This is called the `Angell' plot~\cite{angellscience}, which
is very helpful in classifying supercooled liquids.
A liquid is called strong or fragile depending on its position in the
Angell plot. Straight lines correspond
to `strong' glass-formers and to an Arrhenius behaviour. In this case, one can extract from the plot an effective activation energy, suggesting
quite a simple mechanism for relaxation, for instance by `breaking'
locally a chemical bond. The typical relaxation time is then dominated by the energy barrier to activate this process and, hence, has an Arrhenius behaviour.
Window glasses fall in this category~\footnote{The terminology `strong'
and `fragile' is not related to the mechanical properties of the glass but
to the evolution of the short-range
order close to $T_g$. Strong liquids, such as SiO$_2$, have a locally
tetrahedric  structure which persists both below
and above the glass transition contrary to fragile liquids
whose short-range amorphous structure disappears rapidly upon
heating above $T_g$.}.
If one tries to define an effective activation energy for
fragile glass-formers
using the slope of the curve in Fig.~\ref{angellfig}, then one
finds that this energy scale increases
when the temperature decreases, a `super-Arrhenius' behaviour.
This increase of energy
barriers immediately suggests that glass formation is a
collective phenomenon for fragile supercooled liquids.
Support for this interpretation is provided by the
fact that a good fit of the
relaxation time or the viscosity is given by the
Vogel-Fulcher-Tamman law (VFT):
\be
\tau_\alpha = \tau_0 \exp \left[ \frac{DT_0}{(T-T_0)}\right],
\label{vft}
\ee
which suggests a
divergence of the relaxation time, and therefore
a phase transition of some kind, at a finite temperature $T_0$.
A smaller $D$ in the VFT law corresponds to a more fragile
glass.
Note that there are other comparably good fits of these curves, such as the
B\"assler law~\cite{bassler},
\be
\tau_\alpha = \tau_0
\exp \left( K \left( \frac{T_*}{T} \right)^2 \right),
\ee
that only lead to a divergence at zero temperature.
Actually, although the relaxation time increases by $14$
orders of magnitude, the increase of its logarithm, and therefore
of the effective activation energy is very modest, and experimental
data do not allow one to unambiguously determine
the true underlying functional law without any
reasonable doubt. For this and other reasons, physical
interpretations in terms of
a finite temperature phase transition must always
be taken with a grain of salt.

However, there are other experimental facts that
shed some light and reinforce this interpretation. Among them is an empirical connection found between kinetic
and thermodynamic behaviours.
Consider the part of the entropy of liquids, $S_{\rm exc}$, which is
in excess compared to the entropy of the corresponding
crystal. Once this quantity, normalized by its value at the melting
temperature,
is plotted as a function of $T$, a remarkable connection with the
dynamics emerges. As for the
relaxation time one cannot follow this curve below $T_g$ in thermal
equilibrium. However, extrapolating the curve below $T_g$
apparently indicates
that the excess entropy vanishes at some finite
temperature, called $T_K$, which is very close to zero for strong glasses
and, generically, very close to $T_0$,
the temperature at which a VFT fit diverges.
This coincidence is quite remarkable: for materials with glass transition
temperatures that vary from $50$~K to $1000$~K
the ratio $T_K/T_0$ remains close to 1, up to a few percents.
Examples reported in Ref.~\cite{angellrichert}
are provided in Table~\ref{table}.

The chosen subscript for $T_K$ stands for
Kauzmann~\cite{kauzmann}, who recognized $T_K$ as an important
temperature scale for glasses. Kauzmann further claimed that some change of behaviour
(phase transition, crystal nucleation, etc.) must take place
above $T_K$, because below $T_K$ the entropy of the liquid, a disordered
state of matter,
becomes less than the entropy of the crystal, an ordered state of matter.
The situation that seemed paradoxical
at that time is actually not a serious conceptual problem~\cite{berthier_theoretical_2011,berthier_configurational_2019}. There is no general principle that constrains
the entropy of the liquid to be larger than that of the crystal.
As a matter of fact, the crystallisation
transition for hard spheres takes place precisely because the crystal
becomes the state with the largest entropy at sufficiently high density \cite{holyst}.

On the other hand, the importance of $T_K$ stands,
partially because it is experimentally very close
to $T_0$. Additionally,  the quantity $S_{\rm exc}$ which
vanishes at $T_K$, is thought to be a proxy
for the so-called configurational entropy, $S_c$,
which quantifies the number of metastable states. A popular
physical picture due to Goldstein~\cite{Goldstein} is that close to $T_g$
the system explores a part of the energy landscape
(or configuration space) which is full of minima separated by barriers that
increase when temperature decreases.
The dynamic evolution in the energy landscape
 would then consist in a rather short equilibration
inside a minimum followed by infrequent `jumps' between different
minima. At $T_g$ the barriers between states become so large that the system
remains trapped in one minimum, identified as one of the
possible microscopic amorphous configurations of a glass.
Following this interpretation, one can split
the entropy into two parts. A first contribution is due to the fast relaxation
inside one minimum and a second one, called the `configurational' entropy, counts
the number of metastable states: $S_c=\log N_{\rm
metastable}$.
Assuming that the contribution to the entropy due to the `vibrations'
around an amorphous glass configuration
is not very different from the entropy of the crystal, one finds that
$S_{\rm exc} \approx S_c$. In that case,
$T_K$ would correspond to a temperature at which the configurational
entropy vanishes. This in turn
would lead to a discontinuity
(a downward jump) of the specific heat and would truly correspond
to a thermodynamic phase transition.

\subsection{Static and dynamic correlation functions}

\label{subsec:Static_dynamic_corr_func}

At this point the reader might have reached the conclusion that
the glass transition may not be such a difficult problem:
there are experimental indications of a diverging timescale and
a concomitant singularity in the thermodynamics.
It simply remains to find static correlation functions displaying a
diverging correlation length related to the emergence
of `amorphous order', which would indeed classify the glass transition as
a standard second order phase transition.
Remarkably, this conclusion remains an open and debated question despite several decades of research.
Simple static correlation function are quite featureless in the supercooled
regime, notwithstanding the dramatic changes in the dynamics.
A simple static quantity is the structure factor defined by
\be
S(q)= \left\langle
\frac{1}{N} \delta \rho_{\bf q} \delta \rho_{\bf -q} \right\rangle,
\label{eq:sq}
\ee
where the Fourier component of the density reads
\be
\delta \rho_{\bf q} = \sum_{i=1}^N e^{i {\bf q} \cdot {\bf r}_i} -
\frac{N}{V} \delta_{{\bf q},0},
\ee
with $N$ is the number of particles, $V$ the volume,
and ${\bf r}_i$ is the position of particle $i$. The structure
factor measures the spatial correlations of
particle positions, but it
does not show any diverging peak in contrast to what
happens, for example, at the liquid-gas critical point where
there is a divergence at small ${\bf q}$.
More complicated static correlation functions have been
studied~\cite{debenedetti}, especially in numerical work,
but until now there are no strong indications of a diverging,
or at least substantially growing, static lengthscale~\cite{static, fernandez2006critical, cavagna}.
A snapshot of a supercooled liquid configuration in fact
just looks like a glass configuration, despite their widely different
dynamic properties~\cite{berthier_configurational_2019}.
What happens then at the glass transition? Is it a transition
or simply  a dynamic crossover?
A more refined understanding can be gained by studying dynamic
correlations or response functions.

\begin{figure}
\psfig{file=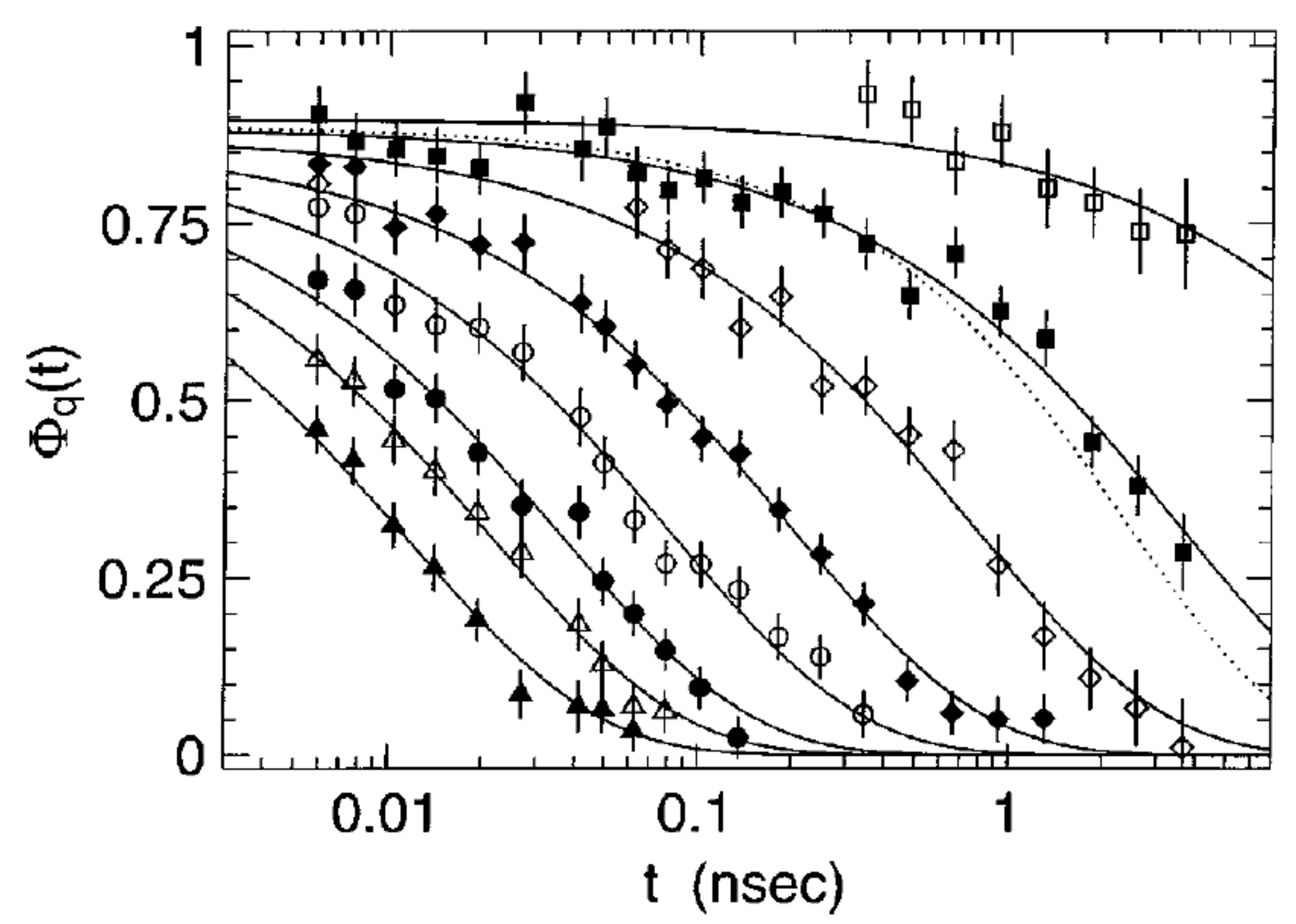,width=8.5cm}
\caption{Temperature evolution of the intermediate scattering
  function normalized by its value at time equal to zero for supercooled
  glycerol~\cite{glycerol}.
  Temperatures decrease from 413~K to 270~K from right to left.
  The solid lines are fit with a stretched exponential with
  exponent $\beta=0.7$. The dotted line represents another
  fit with $\beta=0.82$.}
\label{fqt} 
\end{figure}

A dynamic observable studied in light and neutron scattering
experiments is the
intermediate scattering function,
\begin{equation}
F({\bf q},t) = \left\langle \frac 1 N \delta \rho_{\bf q}(t)
\delta \rho_{\bf -q}(0)
\right\rangle.
\label{isf}
\end{equation}
Different $F({\bf q},t)$ measured
by neutron scattering in supercooled
glycerol~\cite{glycerol}
are shown for different temperatures in Fig.~\ref{fqt}.
These curves show a first, rather fast, relaxation to a plateau followed
by a second, much slower, relaxation.
The plateau is due to the fraction of density fluctuations that
are frozen on intermediate timescales, but
eventually relax during the second relaxation. The latter is called
`alpha-relaxation', and corresponds to the structural relaxation
of the liquid. This plateau is akin to the Edwards-Anderson order parameter
$q_{EA}$ defined for spin glasses, which measures the fraction
of frozen spin fluctuations~\cite{binderkob}.
Note that $q_{EA}$ continuously increases from zero
at the spin glass transition. Instead, for structural glasses,
a finite plateau value already appears above any transition.

\begin{figure}
\psfig{file=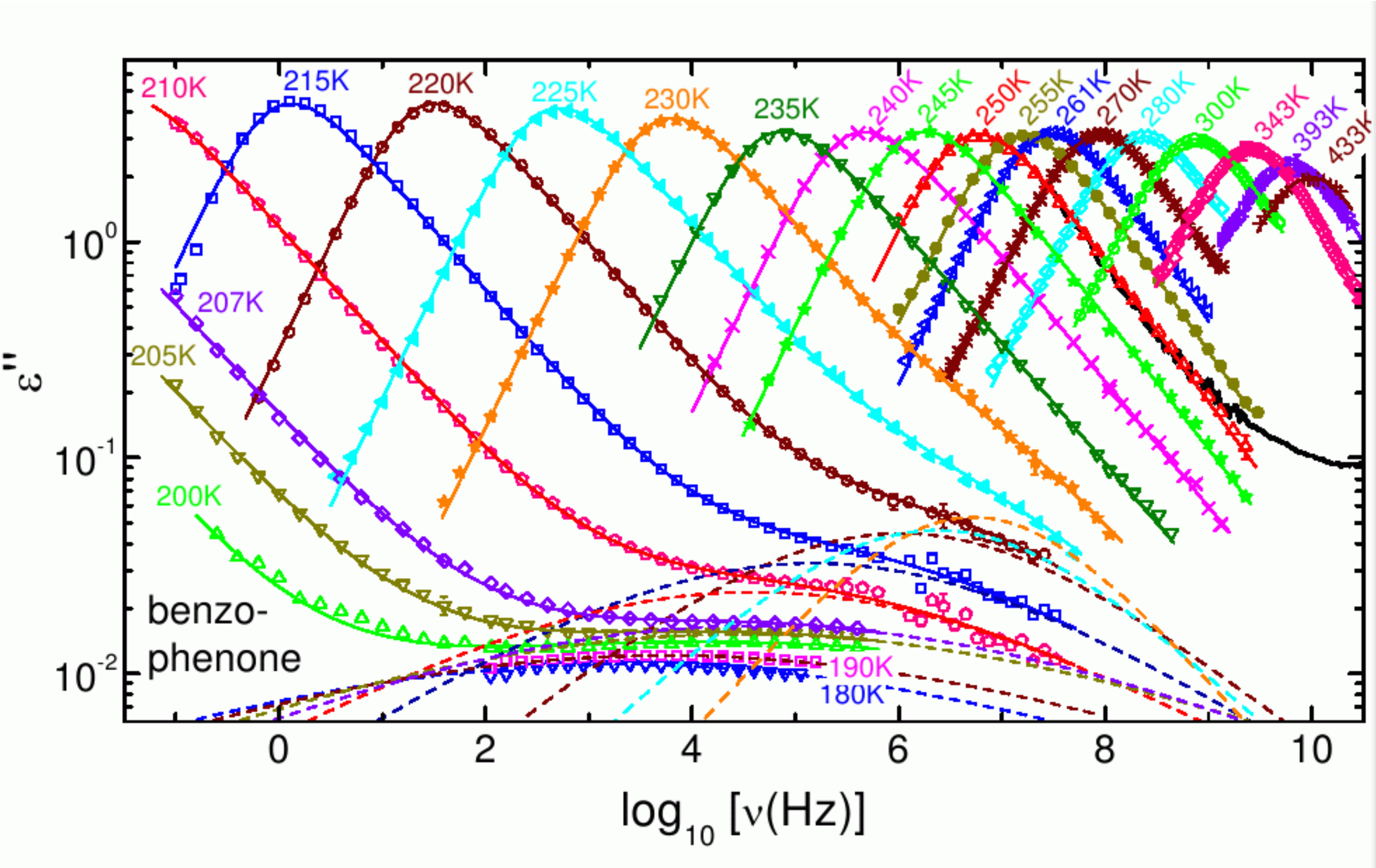,width=8.5cm}
\caption{Temperature evolution of the dielectric
susceptibility of the glass-former benzophenone measured over more than
10 decades of relaxation times~\cite{lunkenheimer}.
Dynamics slows down dramatically as
temperature is decreased and relaxation spectra become very broad
at low temperature.}
\label{lunk}
\end{figure}

The intermediate scattering function can be probed only on a
relatively small regime of temperatures.
In order to track the dynamic slowing down from microscopic
to macroscopic timescales, other correlators have been studied.
A popular one is obtained from
the dielectric susceptibility, which is related
by the fluctuation-dissipation theorem to
the time correlation of polarization fluctuations.
It is generally admitted that different dynamic probes
reveal similar temperature dependencies of the relaxation time.
The temperature
evolution of the imaginary part of the dielectric susceptibility,
$\epsilon''(\omega)$, is shown in Fig.~\ref{lunk} which covers
a very wide temperature window~\cite{lunkenheimer}. At high temperature,
a good representation of the data is given by a Debye law,
$\epsilon(\omega) = \epsilon(\infty) + \Delta \epsilon /
(1 + i \omega \tau_\alpha)$, which corresponds to an
exponential relaxation in the time domain. When temperature is decreased,
however,
the relaxation spectra become very broad and strongly non-Debye.
One particularly well-known feature of the spectra
is that they are well fitted, in the time domain,
for times corresponding to the alpha-relaxation
with a stretched exponential,
$\exp ( -( t /\tau_\alpha )^{\beta} ).$
In the Fourier domain, forms such as the Havriliak-Negami
law are used,
$\epsilon(\omega) = \epsilon(\infty) + \Delta \epsilon /
(1 + (i \omega \tau_\alpha)^\alpha)^\gamma$,
which generalizes the Debye law.
The exponents $\beta$, $\alpha$ and $\gamma$
depend in general on temperature and on the particular
dynamic probe chosen, but they capture the fact that relaxation
is increasingly non-exponential when $T$ decreases towards $T_g$.
A connection was  empirically established
between fragility and degree of non-exponentiality,
more fragile liquids being characterized by broader relaxation
spectra~\cite{reviewnature}.

To sum up, there are many remarkable phenomena that take place when
a supercooled liquid approaches the
glass transition. Striking ones have been
presented, but many others have been left out for lack of
space~\cite{angellscience,reviewnature,debenedetti,binderkob,berthier2015facets}. We have discussed physical behaviours, relationships or
empirical correlations observed in a broad class of
materials. This is quite remarkable and suggests that
there is some physics (and not only chemistry) to the
problem of the glass transition, which
we see as a collective (critical?)
phenomenon which should be relatively independent of
microscopic details. This justifies our statistical
mechanics perspective on this problem.

\section{Taxonomy of `glasses' in science}

\label{beyondglass}

We now introduce a wider range of systems whose
phenomenological behaviour is close or related to the one of glass-forming liquids, showing that glassiness is truly ubiquitous.
It does not only appear
in many different physical situations but
also in more abstract contexts,
such as computer science.

\subsection{Colloidal glass transition}

\label{colloidalglass}

Colloidal suspensions consist of large particles suspended in a
solvent~\cite{larson}.
The typical radii of the particles are in the range $R=1-500$~nm.
The solvent, which is at equilibrium at temperature $T$, renders
the short-time dynamics of the particles Brownian.
The microscopic timescale for this diffusion
is given by $\tau = R^2/D$ where $D$ is the short-time
self-diffusion coefficient. A typical value is of the order of
$\tau \sim 1$~ms, which is thus much larger than microscopic timescales for molecular liquids (in the picosecond regime).
The interaction potential between particles depends on the systems,
and this large tunability makes colloids very attractive
objects for technical applications.
A particularly relevant case, on which we will focus in the
following, is a purely hard sphere potential, which is zero when
particles do not overlap and
infinite otherwise. In this case the temperature scale becomes
an irrelevant number, apart from a trivial rescaling of the
microscopic timescale.
Colloidal hard sphere systems
have been intensively studied~\cite{larson} in experiments, simulations
and theory, varying their density $\rho$,
or their volume fraction $\varphi=\frac{4}{3} \pi R^3 \rho $.
Hard spheres display a fluid phase from $\varphi=0$ to intermediate volume fractions,
a  freezing-crystallisation transition at $\varphi \simeq 0.494$,
and a melting transition at $\varphi\simeq 0.545$.
Above this latter value the system can be compressed
until the close packing point $\varphi\simeq 0.74$,
which corresponds to the FCC crystal. Interestingly for our purposes,
a small amount of size polydispersity can suppress crystallization.
In this case, the system can be `supercompressed' above the
freezing transition without nucleating
the crystal, at least on experimental timescales.
In this regime the relaxation timescale increases
very fast~\cite{naturepusey}. At a packing fraction $\varphi_g
\simeq  0.58-0.60$ it becomes so large compared to typical
experimental timescales that the system does not relax anymore:
it is frozen.
This `colloidal glass transition' is obviously reminiscent of the
glass transition of molecular systems.
In particular, the location $\varphi_g$ of the
colloidal glass transition is as ill-defined as
the glass temperature $T_g$.

Actually, the phenomena that take place when increasing the volume
fraction are analogous to the ones seen in molecular supercooled liquid when decreasing temperature: the
relaxation timescales increases very fast and can be
fitted~\cite{chaikin,berthier2009glass} by a VFT
law in density similar to Eq.~(\ref{vft}),
dynamical correlation functions display a broad spectrum of timescales
and develop a plateau,
no static growing correlation length has been found, etc.
Also the phenomenon of dynamic heterogeneity that will be addressed
in Sec.~\ref{dh} is also observed in colloids~\cite{kegel,weeks}. However, it is important to underline a
major difference: because the microscopic
timescale for colloids is so large, experiments can only track
the first 5 decades of slowing
down. A major consequence is that
the comparison between the glass and colloidal transitions
must be performed by focusing
in both cases on the first 5 decades of the slowing down,
which corresponds to relatively high temperatures
in molecular liquids~\cite{Brambilla}.
Understanding how much and to what extent the glassiness of colloidal
suspensions is related to the one of molecular liquids
remains an active domain of research. Recently, by using colloids of smaller size and hence decreasing the microscopic timescale $\tau$, it has been possible to explore a larger range of relaxation times~\cite{hallett2018local}. This is a very promising research direction to explore the colloidal glassy regime.

\subsection{Jamming transition}

\label{jamming}

Every day life offers many examples of jammed solids.
Grains and beans poured into a container, foams and emulsions
produced by a large shear stress, sand and colloidal particles under
very high pressure. 
Depending on how compressed they are,
these materials behave as fluids or solids: a
handful of sand will flow from our open hand,
while we experience rigidity when closing the fist. 
Formally, the jamming transition is reached at infinite pressure
when all the droplets, bubbles or colloids 
are forced to come into enduring
kissing contact with one another~\cite{torquato, donev2004jamming}. 
All these systems share a fundamental feature:
they can be considered athermal in the sense
that thermal fluctuations at room-temperature are, by far, not
able to allow the system to explore its phase space.

As an immediate consequence, the glass and jamming transitions 
are of a very different nature.
The former describes a fluid-to-solid transition in a
system controlled by thermal fluctuations, 
and its location depends on the cooling or compression rate. 
The latter is a purely geometrical transition 
happening at $T=0$ in the absence of any dynamics, 
but it also corresponds to the change between 
a viscous liquid to a solid mechanical response~\cite{liunagel}. 
In recent years, the connection between the 
two phenomena has been elucidated~\cite{glassjammingreview}, 
and both transitions can be observed under 
different physical conditions in the hard spheres model.

\subsection{Granular glass transition}

\label{granularglass}

Driven granular media represent another family of systems that have recently been studied from the point of view
of their glassiness. Grains are macroscopic objects and, as a
consequence, are not affected by thermal fluctuations. A granular material is therefore frozen in a given configuration
if no energy is injected into the system~\cite{grainsbook}. However,
it can be forced in a steady state by an external drive, such as shearing
or tapping. The dynamics in this steady state shows remarkable
similarities (and differences) with simple fluids. The physics
of granular materials is a very wide subject~\cite{grainsbook}.
In the following we only address briefly what happens to a polydisperse granular fluid at
very high packing fractions.
As for colloids, the timescales for relaxation or diffusion increase very
fast when density is increased,
without any noticeable change in structural properties.
It is now established~\cite{gremaud,dauchot,durian}
that many phenomonelogical properties of the glass and jamming transitions also occur
in granular assemblies. Going beyond the mere analogy
and understanding how much colloids and granular materials are related
is a very active domain of research.

\subsection{Active glasses}

Active matter has recently emerged as a new field in physics~\cite{marchetti2013hydrodynamics,bechinger2016active}, fueled by the observation that systems such as a school fish, bacterial colonies and biological tissues display physical properties and phase transitions which can be described by statistical physics tools, and captured by simplified theoretical models. Physical systems mimicking the behaviour of natural systems have also been developed to perform controlled experiments on model systems of active materials. In particular, colloidal particles and macroscopic objects similar to the systems displaying a granular glass transition have also been developed so that these particles can become `active'~\cite{deseigne2010collective,theurkauff2012dynamic,buttinoni2013dynamical}, i.e.~self-motile objects that can move in the absence of thermal fluctuations, similar to animals, cells or bacteria. 

It is natural to expect that dense assemblies of active particles will undergo some form of dynamic arrest~\cite{henkes2011active,angelini_glass-like_2011}. As human beings, we are well aware that it becomes difficult to walk very fast in a dense crowd, as observed in the streets of large cities or in subway corridors at peak times. Indeed, there are several indications from experimental observations that a transition from a fluid-like state to an arrested glassy state can be observed in active materials~\cite{angelini_glass-like_2011,garcia2015physics,mongera2018fluid,klongvessa2019active}. 

From a conceptual viewpoint, the main difference between these observations and the glass transition observed in molecular and colloidal systems is that the driving force for single particle motion is not of thermal origin, but is instead chemical, mechanical, or biological. This means that any theoretical model for dense active materials must include some sort of non-equilibrium sources of microscopic motion and consequently  the glassy phenomena that will be described must necessarily occur far from equilibrium~\cite{berthier_non-equilibrium_2013}. In that sense, the situation is conceptually not very different from the phenomenon of the granular glass transition discussed in the previous paragraph. The difference between the two essentially lies in the details of the driving motion, which very much resembles a quasi-equilibrium thermal bath in granular glasses~\cite{durian}. 

Several numerical and theoretical studies of the glass transition in active materials have been recently published, see Refs.~\cite{berthier2019glassy} for a specific review on this topic. In particular, the glassy dynamics of so-called self-propelled particles have received considerable attention~\cite{ni2013pushing,berthier2014nonequilibrium,mandal2016active,bi_motility-driven_2016,berthier2017active,matoz2017cell}. In these models, particles interacting with simple pair interactions (similar to the ones studied to model the equilibrium glass transition) are driven by active forces that tend to displace the particles in straight lines over a finite persistence length. 

It is now understood that for dense active fluids, due to these non-equilibrium driving forces both the structure and the dynamics are very different from their equilibrium counterparts~\cite{berthier2019glassy}. It is observed that a large persistence length has a profound influence on the static correlations of the fluid since both density-density (as in Eq.~(\ref{eq:sq})) and velocity-velocity correlation functions develop non trivial non-equilibrium features~\cite{berthier2017active}. As particle crowding is increased, slow dynamics develop and is accompanied by a phenomenology similar to that of equilibrium glassy materials, with motion becoming gradually arrested at large enough density or small enough activity. Dynamic arrest in dense active materials is therefore called a non-equilibrium glass transition~\cite{berthier_non-equilibrium_2013}. Understanding this new class of glassy dynamics is an exciting new direction for research. 

\subsection{Random pinning glass transition}

\label{sec:randomPinning}

The standard control parameters used to induce a glass transition are temperature and pressure. A new way introduced in the 2000s is `random pinning'~\cite{scheidler2002growing,kim2003effects}, which consists in freezing the positions of a fraction $c$ of particles from an equilibrium configuration of a supercooled liquid. 
Theoretical arguments \cite{cammarota_ideal_2012} and numerical simulations \cite{berthier_static_2012,karmakar2011exposing} suggested that 
the dynamics of the remaining free particles slow down and undergo a glass transition by increasing $c$.  
The study of the random pinning glass transition has been the focus of several theoretical analyses \cite{cammarota2013random,krakoviack2011mode,szamel2013glassy,franz2013universality,cammarota2013general,franz2013glassy,krakoviack2014simple,phan2018theory,cammarota2016first,ikeda2017fredrickson} and numerical simulations  \cite{berthier_static_2012,kob2013probing,charbonneau2013decorrelation,karmakar2013random,chakrabarty2015dynamics,kob2014nonlinear,jack2013dynamical,fullerton2014investigating,li2015decoupling,chakrabarty2016understanding,angelani2018probing,ozawa2018ideal,niblett2018effects} in the decade 2005-2015. 
It has also been studied in experiments, in particular in colloidal glasses by optical microscopy \cite{gokhale2014growing,gokhale2016deconstructing,ganapathi2018measurements,williams2018experimental}. 

The interest in the `random-pinning glass transition' has been twofold. First, it represents a new way to test theories of the glass transition. In fact, RFOT theory and MCT (see Sec.~\ref{theory}) predict that it should have the same properties as the usual glass transition \cite{cammarota_ideal_2012,szamel2013glassy}, i.e.~increasing the pinned fraction $c$ plays the same role as lowering the temperature. This phenomenon was studied and confirmed in Ref.~\cite{kob2013probing}, where the equilibrium phase diagram of a randomly pinned glass-former was fully characterized. The dynamical behavior is not as well understood. Although it is clear that dynamics dramatically slows down when increasing $c$, hence the name `random-pinning glass transition', it has proven difficult to disentangle trivial effects due to steric constraints from collective ones. In consequence, numerical simulations could not validate or disprove the competing predictions from dynamical facilitation theories~\cite{jack2012random} and RFOT theory~\cite{cammarota_ideal_2012}. 
The other reason for the random pinning procedure was to produce configurations equilibrated very close to the glass transition (just after pinning, the remaining unpinned particles are in an equilibrium configuration for the constrained systems~\cite{krakoviack2010statistical}). This allowed to perform the first computational study of ultrastable glasses~\cite{hocky2014equilibrium}, see below. 

The random pinning procedure has interesting connections, similarities and differences with other ways to constrain the dynamics of glassy liquids introduced in recent years. Those can be grouped in three main categories: modifications of the Hamiltonian ($\epsilon$-coupling, see Sec.~\ref{subsec:epsilonCoupling_VQ}), in the dynamical rules ($s$-ensemble dynamics, see Sec.~\ref{subsec:s-ensemble}), and in the space available to the system (liquids in quenched environments~\cite{thalmann2000phase}). 

\subsection{Ultrastable glasses}

\label{sec:ultrastable}

Glassy materials are typically prepared by slowly cooling or compressing a dense fluid across the glass transition described above. The glass transition temperature or density is set by the competition between an extrinsic time scale imposed by the experimentalist (for instance the duration of the experiment, or the cooling rate), and an intrinsic timescale of the material, such as the structural relaxation time. The degree of supercooling observed in most glassy materials is then set by the typical duration of an experiment, which corresponds to an intrinsic timescale of about 100~s in molecular liquids. 

It has recently become possible to prepare `ultrastable' glasses~\cite{swallen2007organic,ediger2017perspective}, namely glassy materials which reach a degree of supercooling which is equivalent to cooling glasses at rates that are $10^5 - 10^{10}$ times slower than usual. Ultrastable glasses are not prepared by cooling bulk liquids across the glass transition, but using a completely different route called physical vapor deposition. In this procedure, the glassy material is prepared directly at the desired temperature (there is no cooling involved) by the slow deposition of individual molecules suspended in a gas phase onto a glassy film whose heigth increases slowly as more molecules are deposited. 

The degree of supercooling that can be achieved by physical vapor deposition is again set by the competition between two timescales~\cite{swallen2007organic,berthier2017origin}. The extrinsic timescale is now related to the deposition rate of the molecules, whereas the intrinsic timescale is set by the relaxation time of molecules diffusing at the free surface of the glassy film. It is known that bulk and surface dynamics may differ by many orders of magnitude in glasses~\cite{zhu2011surface}, because the molecular mobility at a free surface is much less constrained than in the bulk. Therefore, for a similar extrinsic timescale imposed by the experimental setup, a much deeper degree of supercooling is achieved by the vapor deposition process. 

Ultrastable glasses prepared using physical vapor deposition are thus expected to behave as extraordinarily-slowly cooled supercooled liquids, with a cooling rate or a preparation time that is impractically large. As such, theses glasses have physical properties that can differ rather drastically from ordinary glassy materials. 
In particular, it was shown that their mechanical, thermodynamic and kinetic properties differ quantitatively from ordinary glasses, and display specific dynamic phenomena~\cite{kearns2010one,chen2013dynamics,perez2014suppression,sepulveda2014role,rafols2018high,vila2020nucleation}. As such, they are currently the subject of intense theoretical investigations as well~\cite{wolynes2009spatiotemporal,leonard2010macroscopic,lyubimov2013model,jack2016melting,gutierrez2016front,fullerton2017density,flenner2019front,khomenko2019depletion}. The goal is to better understand the deposition process itself,  but also to better characterise the physical properties of ultrastable glasses in view of their many potential practical applications. Also, since ultrastable glasses offer a way to access much deeper supercooled states, it can be hoped that they can be used to shed new light on the glass transition phenomenon itself.  

\subsection{Other glasses in physics, and beyond}

There are many other physical contexts in which glassiness plays an
important role~\cite{youngbook}.
One of the most famous examples is the field of spin glasses.
Real spin glasses are magnetic impurities
interacting by quenched random couplings. At low
temperatures, their dynamics become extremely slow and they freeze in
amorphous spin configuration dubbed a `spin glass' by Anderson.
There are many other physical systems, often characterized by quenched
disorder, that show glassy behavior, like Coulomb glasses,
Bose glasses, etc.
In many cases, however,
one does expect quite a different physics from structural
glasses: the similarity between these systems is
therefore only qualitative.

Quite remarkably, glassiness also emerges
in other branches of science~\cite{complexbook}.
In particular, it has been discovered
recently that concepts and techniques developed for glassy systems
turn out to apply and be very useful tools in the field
of computer science.
Problems like combinatorial optimization
display phenomena completely analogous to phase transitions, and actually, to
glassy phase transitions. {\it A posteriori}, this is quite natural,
because a typical optimization problem
consists in finding a solution in a presence
of a large number of constraints.
This can be defined, for instance, as a set
of $N$ Boolean variables that satisfies $M$ constraints.
For $N$ and $M$ very large at fixed
$\alpha=M/N$, this problem very much resembles
finding a ground state in a statistical mechanics problem with quenched
disorder. Indeed one can
define an energy function (a Hamiltonian)
as the number of unsatisfied constraints, that has to be minimized,
as in a $T=0$ statistical mechanics problem.
The connection with glassy systems originates from the fact that in both cases
the energy landscape is extremely complicated, full of minima and saddles.
The fraction of constraints per degree of freedom, $\alpha$, plays a role
similar to the density in a hard sphere system.
For instance, a central problem in optimization, random k-satisfiability, has been shown to undergo a glass transition when $\alpha$ increases, analogous to the one of structural glasses~\cite{PNAS,antenucci2019glassy}.

Glassiness also plays an important role in machine learning and signal processing. In those cases, one wants to learn a specific task from many examples or retrieve a specific signal from a huge amount of data. This is done in practice by minimizing a cost function. For example, imagine that one is given a tensor $T_{i_1,i_2,i_3}=v_{i_1}v_{i_2}v_{i_3}$, constructed from a vector $v_i$ ($i=1,...,N$) of norm $\sqrt N$, and that this tensor is corrupted by noise $J_{i_1,i_2,i_3}$, which for simplicity we take independent and Gaussian for each triple  $(i_1,i_2,i_3)$. The problem called tensor PCA, which appears in image and video analysis~\cite{tensorPCAa}, consists in retrieving the signal $v_i$ from the noisy tensor $T_{i_1,i_2,i_3}+J_{i_1,i_2,i_3}$. The simplest procedure to solve this problem is to find the vector $x_i$ minimizing the following cost function: 
\[
H(\{x_i\})=\sum_{{i_1,i_2,i_3}}\left(v_{i_1}v_{i_2}v_{i_3}+ J_{i_1,i_2,i_3}-x_{i_1}x_{i_2}x_{i_3}\right)^2 .
\]  
By developing the square, one finds that the cost function $H$ is identical to the one of a 3-spin spherical glass mean-field model with quenched random couplings  $J_{i_1,i_2,i_3}$ and a term favoring configurations in the direction of 
$v_i$ \cite{tensorPCAb}.  These two contributions are competing for determining the ground-state properties, the strength of the latter with respect to the former is proportional to the signal-to-noise ratio.  
This example illustrates one way in which glassiness plays an important role in machine learning: one has to find 
a signal (the $v_i$'s) buried in a rough landscape (induced by the $J_{i_1,i_2,i_3}$'s). Practical algorithms, such as gradient descent and its stochastic version, lead to dynamics which are very similar to the ones of physical systems after a quench to low temperature. One of the main questions in this area is characterizing the algorithmic threshold, i.e.~the critical value of the signal-to-noise ratio such that the original signal can be recovered with some given accuracy. Glassy dynamics plays a central role: it is the main obstacle for recovering the signal, as the dynamics can be lost and trapped in bad minima instead of converging toward the good one correlated with the signal \cite{sarao}. 

Finally, hunting for a signal in a rough landscape~\cite{flole} is not the only context in which glassiness emerged in machine learning in recent years. In fact, more generally, there has been a lot of work aimed at characterizing the landscapes over which optimization dynamics take places in general machine learning problems (from high-dimensional statistics to deep neural networks), and at assessing to which extent glassy dynamics and rough landscapes play a relevant role~\cite{sagun2014explorations,baity}.   

\section{Numerical simulations}

\label{simu}

Studying the glass transition of molecular liquids
at a microscopic level is in principle straightforward since one
must answer a very simple question: how do particles move in
a liquid close to $T_g$? It is of course
a daunting task to attempt answering this question
experimentally because one should then
resolve the dynamics of single molecules
to be able to follow the trajectories of objects that are a few Angstroms
large on timescales of tens or hundreds of seconds, which sounds like
eternity when compared to typical molecular dynamics usually
lying in the picosecond regime.
In recent years, such direct experimental investigations have been developed
using time and space resolved techniques such as atomic
force microscopy~\cite{israeloff} or single molecule spectroscopy~\cite{single,paeng2015ideal}, but this
remains a very difficult task.

\begin{figure}
\psfig{file=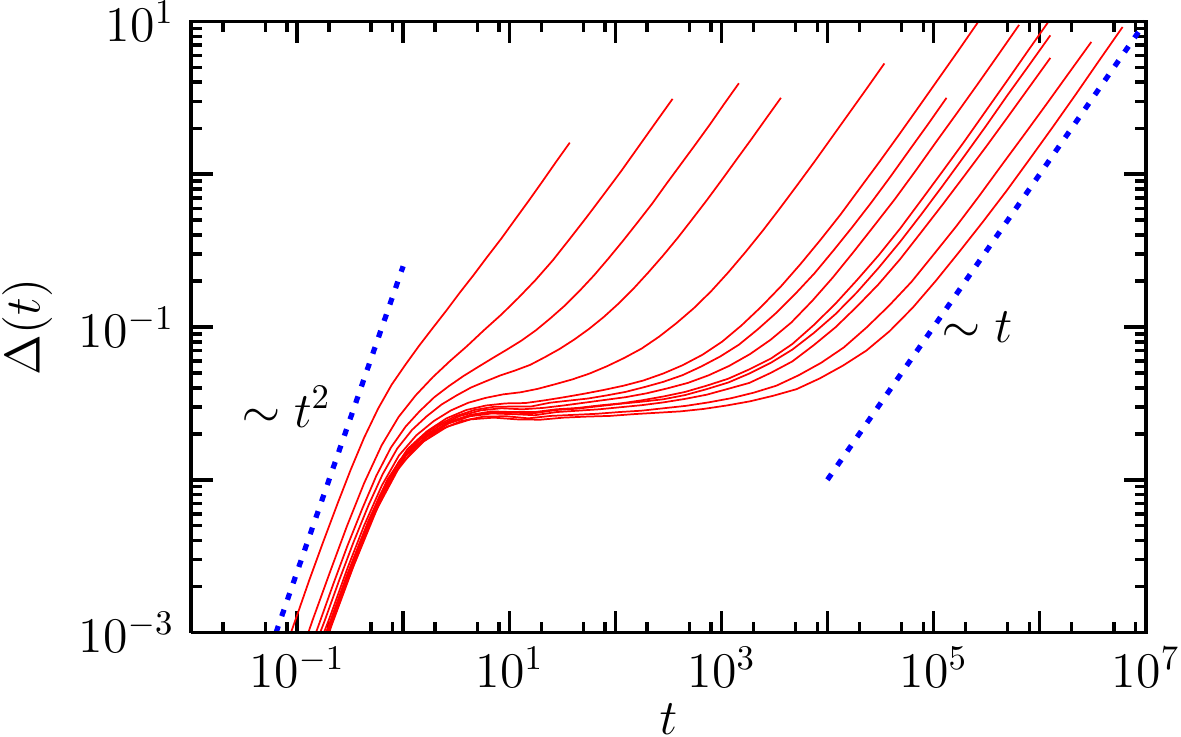,width=8.5cm}
\caption{Mean-squared displacements of individual
particles in a simple model of a glass-forming liquid composed
of Lennard-Jones particles observed
on a wide time window. When temperature decreases (from left to right),
the particle displacements become increasingly slow with several distinct
time regimes corresponding to (in this order) ballistic,
localized, and diffusive regimes.}
\label{msd}
\end{figure}

In numerical simulations, by contrast,
the trajectory of each particle in the system can,
by construction, be followed at all times.
This allows one to quantify easily single particle dynamics, as
proved  in Fig.~\ref{msd} where the averaged
mean-squared displacement $\Delta(t)$
measured in a simple Lennard-Jones glass-former
is shown and is defined as
\begin{equation}
\Delta(t) = \left\langle \frac{1}{N} \sum_{i=1}^N
| {\bf r}_i(t) -{\bf r}_i(0) |^2 \right\rangle,
\end{equation}
where ${\bf r}_i(t)$ represents the position of particle $i$ at time
$t$ in a system composed of $N$ particles and the brackets
indicate an ensemble average. The particle displacements
considerably slow down when $T$ is decreased and the self-diffusion constant
decreases by orders of magnitude, mirroring the behaviour of the
viscosity shown in Fig.~\ref{angellfig} for real systems.
Moreover, a rich dynamics is
observed, with a plateau regime at intermediate timescales,
corresponding to an extended time window during which
particles vibrate around their initial positions,
exactly as in a crystalline solid. The difference with a crystal
is of course that this localization is only transient, and all particles
eventually escape and diffuse at long times with a diffusion
constant $D_s$, so that
$\Delta(t) \sim 6 D_s t$ when $t \to \infty$.

In recent years, computer experiments have played an increasingly
important role in glass transition studies. It could almost be said
that particle trajectories in numerical work have been studied
under so many different angles that probably very little remains to be learnt
from such studies in the regime that is presently
accessible using present day computers. Unfortunately, this
does not imply complete knowledge of the physics
of supercooled liquids. As shown in Fig.~\ref{msd},
it is presently possible to follow the
dynamics of a simple glass-forming liquid
over more than eight decades of time, and over a temperature
window in which average relaxation timescales
increase by more than five decades.
This might sound impressive, but a quick
look at Fig.~\ref{angellfig}
shows, however, that at the lowest
temperatures studied in the computer,
the relaxation timescales are still orders of magnitude
faster than in experiments performed close to the glass
transition temperature.
They can be directly compared to experiments performed
in this high temperature regime, but
this also implies that simulations
focus on a relaxation regime that is about eight to ten decades
of times faster than in experiments performed close to $T_g$.
Whether numerical works are useful to understand
the kinetics of the glass transition itself at all is therefore an open,
widely debated, question. We believe that it is now possible to
numerically access temperatures which are low enough that
many features associated to the
glass transition physics can be observed:
strong decoupling phenomena, clear deviations
from fits to the mode-coupling theory (which
are experimentally known to hold only at high
temperatures), and crossovers towards truly activated dynamics. In Sec.~\ref{sec:newComputMethods}, we discuss recent developments in the field of computational studies that are able to address novel challenges regarding the static properties of supercooled liquids over a broad temperature range. 

\begin{figure}
\psfig{file=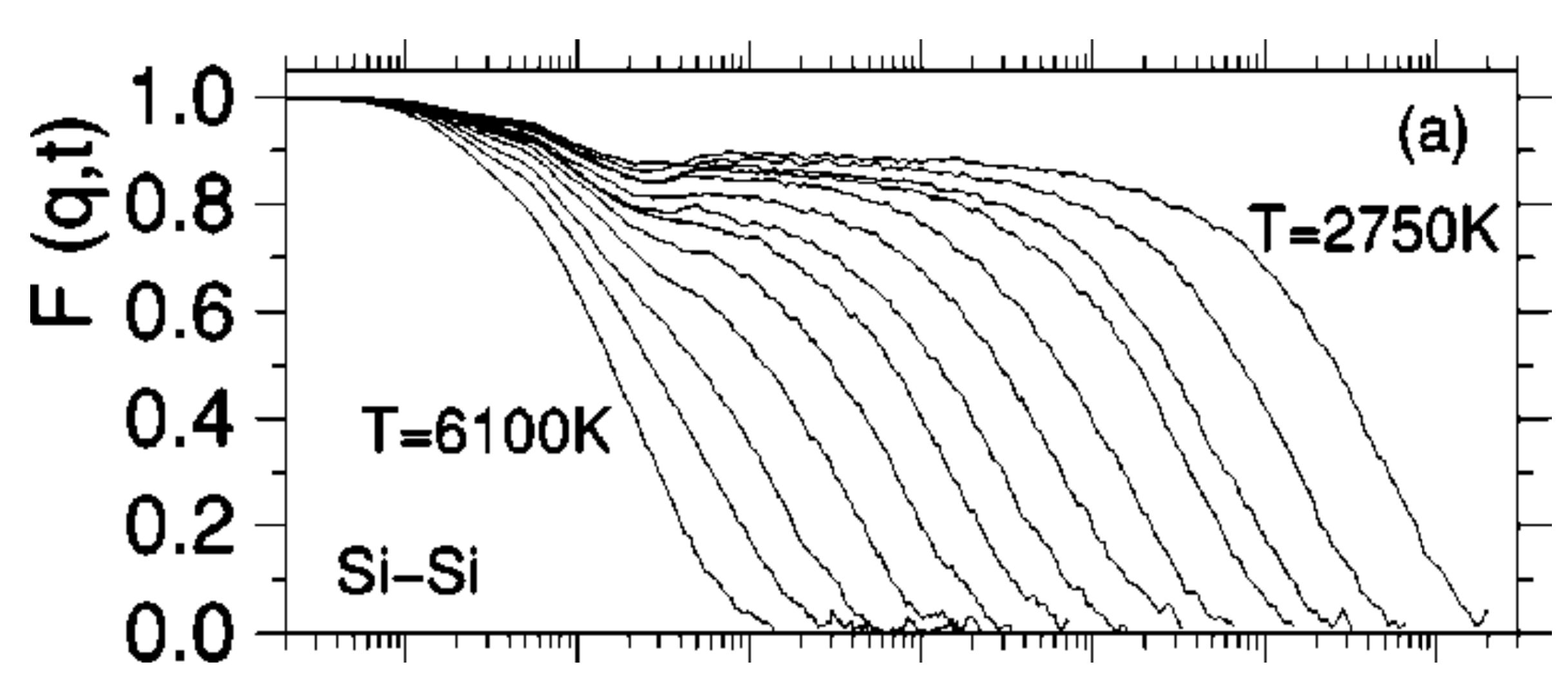,height=8.5cm,angle=-90}
\caption{Intermediate scattering function at wavevector
1.7~\AA$^{-1}$ for the $Si$ particles at $T=2750$~K obtained from molecular
dynamics simulations of a model for silica~\cite{horbachkob}.}
\label{si:fig} 
\end{figure}

Classical computer simulations of supercooled liquids
usually proceed by solving a cleverly discretized version
of Newton's equations for a given potential interaction
between particles~\cite{allen}.
If quantitative agreement with experimental data
on an existing specific material is sought, the interaction must be
carefully chosen in order to reproduce reality, for instance
by combining classical to {\it ab-initio} simulations.
From a more fundamental perspective one rather seeks the simplest
model that is still able to reproduce qualitatively the phenomenology
of real glass-formers, while being considerably simpler to study.
The implicit, but quite strong, hypothesis is that molecular details are
not needed to explain the behaviour of supercooled liquids, so that
the glass transition is indeed a topic for statistical mechanics, not for
chemistry.
A considerable amount of work has therefore been dedicated to
studying
models such as hard spheres, soft spheres, or Lennard-Jones particles.
More realistic materials are also studied focusing for instance on
the physics of network forming materials, multi-component ones,
anisotropic particles, or molecules with internal degrees of freedom.
Connections to experimental work can be made by computing
quantities that are experimentally accessible such as the
intermediate scattering function,
static structure factors, $S({\bf q})$, or
thermodynamic quantities such as specific heat or configurational
entropy, which are
directly obtained from particle trajectories
and can be measured in experiments as well.
As an example we show in Fig. \ref{si:fig} the
intermediate scattering function $F({\bf q},t)$
obtained from a molecular dynamics simulation
of a classical model for
SiO$_2$ as a function of time for different
temperatures~\cite{horbachkob}.

An important role is played by simulations also because
a large variety of dynamic and static quantities can be simultaneously
measured in a single model system. As we shall discuss below,
there are scores of different theoretical approaches to
describe the physics of glass-formers, and sometimes they
have their own set of predictions that can be readily tested
by numerical work. Indeed, quite a large amount of numerical
papers have been dedicated to testing in detail the predictions
formulated by the mode-coupling theory of the glass transition,
as reviewed in Ref.~\cite{gotze}.
Here, computer simulations are
particularly well-suited as the theory specifically
addresses
the relatively high temperature window that is studied in computer
simulations.

While Newtonian dynamics is mainly used in
numerical work on supercooled liquids,
a most appropriate choice for these materials,
it can be interesting to consider
alternative dynamics that are not deterministic, or which do
not conserve the energy. In colloidal glasses and physical
gels, for instance, particles undergo Brownian motion
arising from collisions with molecules in the solvent, and a
stochastic dynamics is more appropriate. Theoretical
considerations might also suggest the study of different sorts of
dynamics for a given interaction between particles, for instance,
to assess the role of conservation laws and
structural information. Of course, if a given dynamics
satisfies detailed balance with respect to the Boltzmann
distribution, all structural quantities remain unchanged, but the
resulting dynamical behaviour might be very different.
Several papers~\cite{gleim,szamel,berthierkob}
have studied in detail the influence of the
chosen microscopic dynamics on the dynamical behaviour in
glass-formers using either
stochastic dynamics (where a friction term and a random
noise are added to Newton's equations, the amplitude of both
terms being related by a fluctuation-dissipation theorem),
Brownian dynamics (in which
there are no momenta, and positions evolve with Langevin
dynamics), or Monte-Carlo dynamics
(where the potential energy between two configurations
is used to accept or reject a trial move). Quite surprisingly,
the equivalence between these three types of
stochastic dynamics and the originally
studied Newtonian dynamics
was established at the level of the averaged dynamical
behaviour~\cite{gleim,szamel,berthierkob},
except at very short times where obvious differences
are indeed expected.
This strongly suggests that an explanation for
the appearance of slow dynamics in these materials
originates from their amorphous structure.
However, important differences
were found when dynamic fluctuations were
considered~\cite{berthierkob,jcpI,jcpII},
even in the long-time regime comprising the structural relaxation.

\begin{figure}
\psfig{file=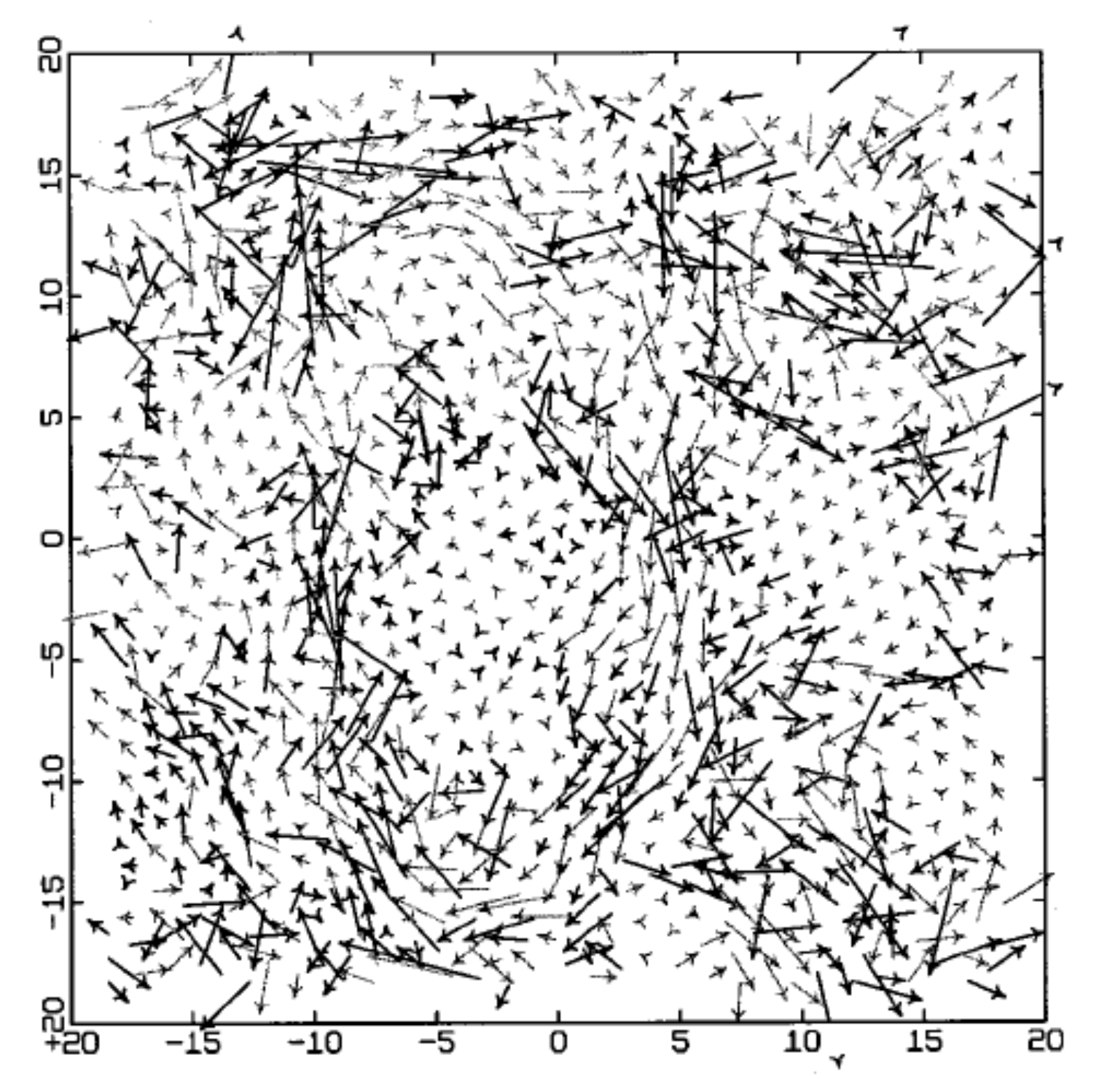,width=6.5cm,height=6.5cm}
\caption{Spatial map of single particle displacements in
the simulation of a binary mixture of soft spheres in two
dimensions~\cite{harrowell}.
Arrows show the displacement
of each particle in a trajectory of length about 10 times the structural
relaxation time. The map reveals the existence of
particles with different mobilities during relaxation, but
also the existence of spatial correlations between these
dynamic fluctuations.}
\label{peter} 
\end{figure}

Another crucial advantage of molecular simulations
is illustrated in Fig.~\ref{peter}.
This figure shows a spatial map of single particle
displacements recorded during the simulation of a binary soft sphere system
in two dimensions~\cite{harrowell}.
This type of measurement, out of reach of most
experimental techniques that study the liquid state, reveals that
dynamics might be very different from
one particle to another. More importantly, Fig.~\ref{peter} also
unambiguously reveals
the existence of spatial correlations between these dynamic
fluctuations. The presence of non-trivial spatio-temporal fluctuations
in supercooled liquids is now called `dynamic heterogeneity'~\cite{ediger}, as we now discuss. 

\section{Dynamic heterogeneity}

\label{dh}

\subsection{Existence of
spatio-temporal dynamic fluctuations}

A new facet of the relaxational behaviour of supercooled liquids
has emerged in the last two decades thanks to a considerable experimental
and theoretical effort. It is called `dynamic heterogeneity' (DH),
and now plays a central role in modern descriptions of
glassy liquids~\cite{ediger,berthier2011dynamical}.
As anticipated in the previous section, the
phenomenon of dynamic heterogeneity is related to the
spatio-temporal fluctuations of the dynamics.
Initial motivations stemmed from the search of
an explanation for the
non-exponential nature of relaxation processes in supercooled liquids,
related to the existence of a broad relaxation spectrum.
Two natural but fundamentally different explanations can be put
forward. (1) The relaxation is locally exponential,
but the typical relaxation timescale varies
spatially. Hence, global correlation
or response functions become non-exponential
upon spatial averaging over this spatial
distribution of relaxation times.
(2) The relaxation is complicated and
inherently non-exponential, even locally.
Experimental and theoretical works~\cite{ediger} suggest
that both mechanisms are likely at play, but
definitely conclude that relaxation is spatially
heterogeneous, with regions that are
faster and slower than the average.
Since supercooled liquids are ergodic materials,
a slow region will eventually become fast, and vice-versa.
A physical characterization of DH entails the determination
of the typical lifetime of the heterogeneities, as well as
their typical lengthscale.

\begin{figure}
\psfig{file=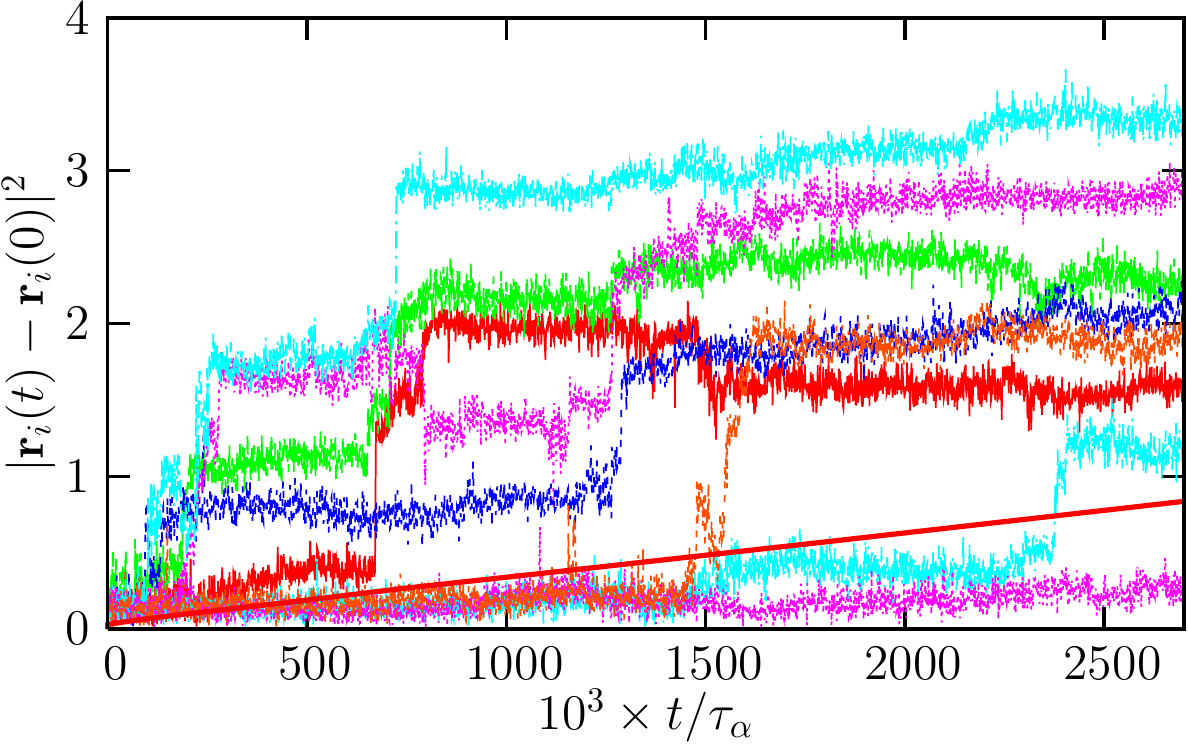,width=8cm}
\caption{Time resolved squared displacements of individual
particles in a simple model of a glass-forming liquid composed
of Lennard-Jones particles. The average is shown as a smooth full line.
Trajectories are composed of long periods of time during which particles
vibrate around well-defined positions, separated by rapid jumps that
are widely distributed in time, underlying the importance of
dynamic fluctuations.}
\label{msd2} 
\end{figure}

A clear and more direct
confirmation of the heterogenous character of the dynamics also stems
from simulation studies. For example, whereas the simulated average
mean-squared displacements are smooth
functions of time, time signals for individual
particles clearly exhibit specific features that
are not observed unless dynamics is resolved both in space and time.
These features are displayed in Fig.~\ref{msd2}. What do we see?
We mainly observe that particle trajectories are not smooth but
rather composed of a succession of long periods of time
where particles simply vibrate around
well-defined locations, separated by rapid `jumps'.
Vibrations were previously inferred from the plateau observed
at intermediate times in the mean-squared displacements
of Fig.~\ref{msd}, but the existence of jumps that are
clearly statistically widely distributed in time cannot be guessed
from averaged quantities only. The fluctuations in Fig.~\ref{msd2} suggest,
and direct measurements confirm, the importance played by
fluctuations around the averaged dynamical behaviour.

A simple type of such
fluctuations has been studied in much detail. When looking
at Fig.~\ref{msd2},
it is indeed natural to ask, for any given time, what is the distribution
of particle displacements. This is quantified by the self-part
of the van-Hove function defined as
\begin{equation}
G_s({\bf r},t) = \left\langle
\frac{1}{N} \sum_{i=1}^N \delta ({\bf r} - [{\bf r}_i(t) -
{\bf r}_i(0)] ) \right\rangle .
\end{equation}
For an isotropic Gaussian diffusive process, one gets
$G_s({\bf r},t) = \exp(-|{\bf r}|^2/(4 D_s t))/(4\pi D_s t)^{3/2}$.
Simulations reveal instead strong
deviations from Gaussian behaviour on the timescales
relevant for structural relaxation~\cite{glotzerkob}.
In particular they reveal
`fat' tails in the distributions that are much wider than expected
from the Gaussian approximation. These tails are in fact
better described by an exponential decay rather than a Gaussian one, in a wide time window comprising the
structural relaxation, such that
$G_s({\bf r},t) \sim \exp(-|{\bf r}|/\lambda(t))$~\cite{pinaki}.
Thus, they reflect the existence of a
population of particles that moves distinctively further
than the rest and appears therefore to be much more
mobile.
This observation implies that relaxation
in a viscous liquid qualitatively differs  from that of a normal liquid
where diffusion is close to Gaussian,
and that a non-trivial single
particle displacements statistics exists.

A long series of questions immediately
follows this seemingly simple observation. Answering them
has been the main occupation of many workers in this field
over the last decade. What are the particles in the tails
effectively doing? Why are they faster than the rest? Are
they located randomly in space or do they cluster? What is
the geometry, time and temperature evolution of the clusters?
Are these spatial fluctuations correlated to geometric
or thermodynamic properties of the liquids? Do similar correlations
occur in all glassy materials? Can one predict
these fluctuations theoretically? Can one understand glassy
phenomenology using fluctuation-based arguments? Can
these fluctuations be detected experimentally?

\begin{figure}
\psfig{file=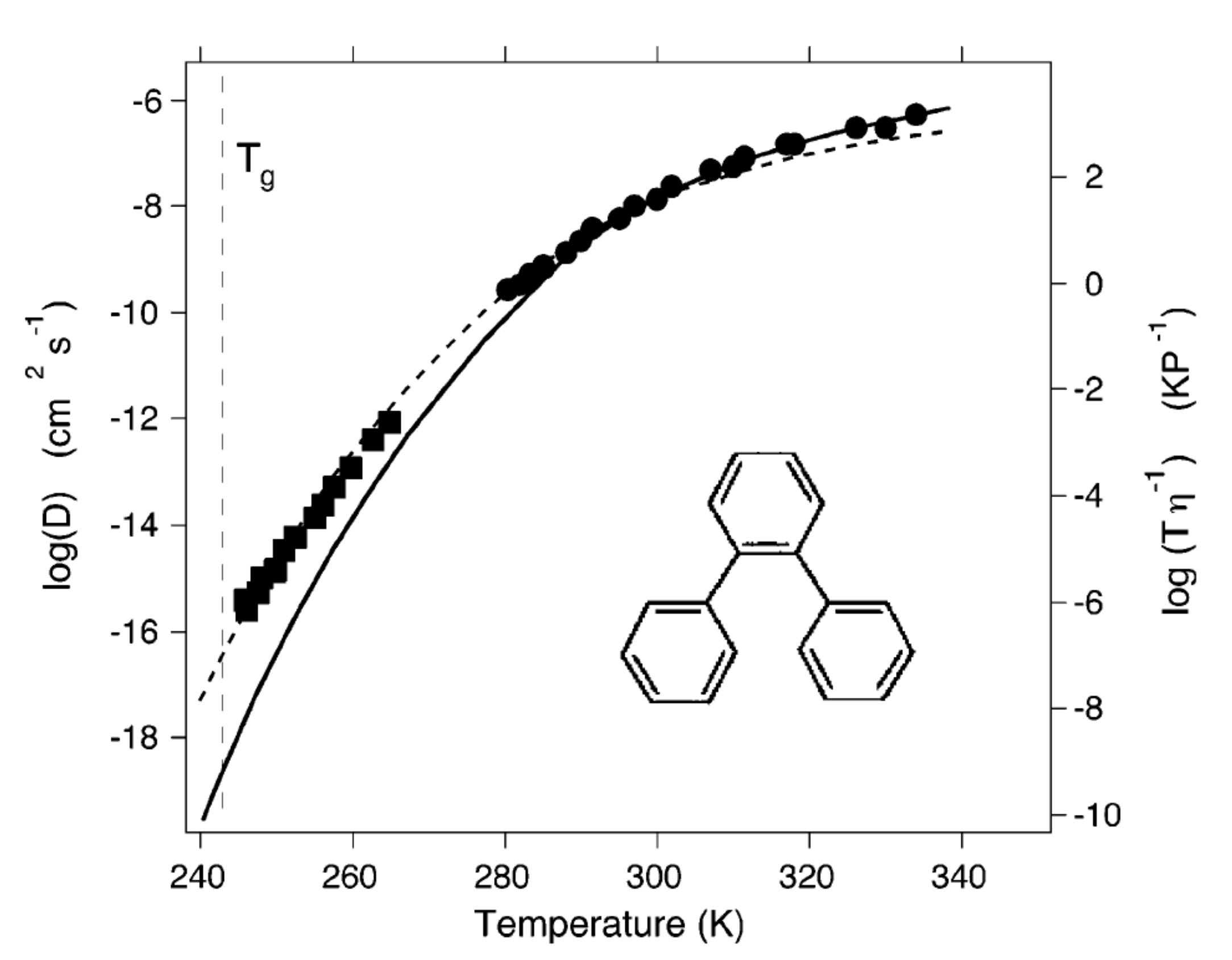,width=8.5cm}
\caption{Decoupling between viscosity (full line) and
self-diffusion coefficient (symbols) in supercooled
ortho-terphenyl~\cite{edigerotp}.
The dashed line shows a fit with a `fractional' Stokes-Einstein
relation, $D_s \sim (T / \eta)^\zeta$ with $\zeta \sim 0.82$ .}
\label{otp} 
\end{figure}

Another influential phenomenon that was related early on
to the existence of DH
is the decoupling of self-diffusion ($D_s$) and
viscosity ($\eta$). In the high temperature
liquid self-diffusion and viscosity are related by the
Stokes-Einstein relation~\cite{hansen},
$D_s \eta / T = const$.
For a large particle moving in a fluid the constant is equal to $1/(6\pi R)$
where $R$ is the particle radius. Physically, the Stokes-Einstein relation
means that two different measures of the relaxation time $R^2/D_s$ and $\eta
R^3/T$ lead to the same timescale up to a constant factor. In supercooled
liquids this phenomenological law
breaks down, as shown in Fig.~\ref{otp} for
ortho-terphenyl~\cite{edigerotp}. It is commonly found that $D_s^{-1}$
does not increase as fast as $\eta$ so that,
at $T_g$, the product $D_s \eta$ has increased by 2-3 orders
of magnitude as compared to its Stokes-Einstein value.
This phenomenon, although less spectacular than the overall change of
viscosity, is a
significative indication that different ways to measure relaxation times
lead to different answers and thus is a strong hint of the existence
of a distribution of relaxation timescales.

Indeed, a natural explanation of this effect is that different observables
probe the underlying distribution of relaxation
times in different ways~\cite{ediger}.
For example, the self-diffusion coefficient of tracer particles is dominated
by the more mobile particles whereas the viscosity or other measures of
structural relaxation probe the timescale needed for every particle
to move. An unrealistic but instructive example is a model where
there is a small, non-percolative subset of particles that are
blocked forever, coexisting with a majority of mobile
particles. In this case, the structure never relaxes but the
self-diffusion coefficient is non-zero because of the mobile particles.
Of course, in reality all particles move, eventually, but this
shows how different observables are likely to probe
different moments of the distribution of
timescales, as explicitely shown within several theoretical
frameworks~\cite{Gilles,jung}.

\begin{figure}
\psfig{file=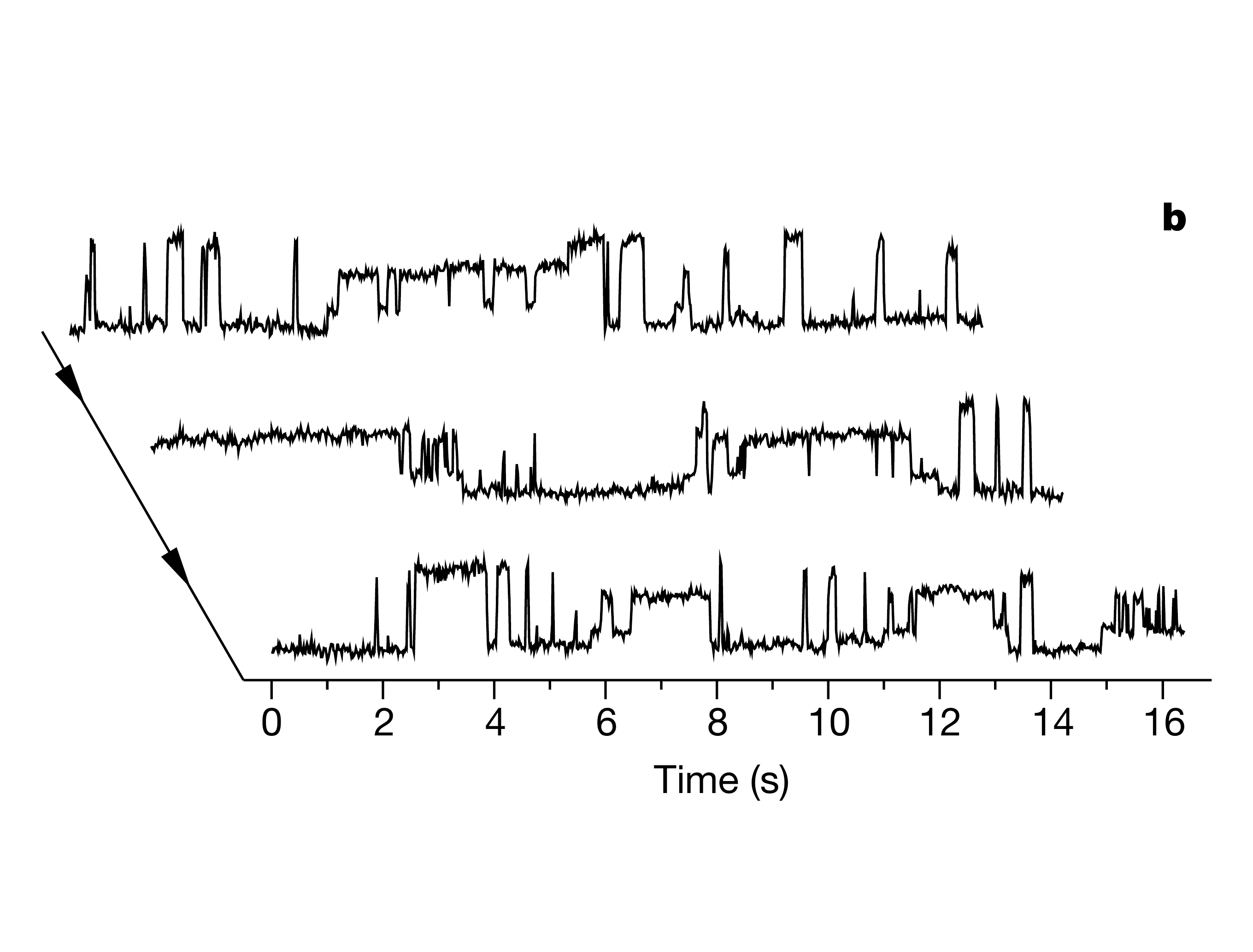,width=6.5cm,angle=-90}
\caption{Time series of polarization in the AFM experiment
performed by Vidal Russell and Israeloff~\cite{israeloff} on PVAc
at $T=300$~K. The signal intermittently switches between periods
with fast or slow dynamics, suggesting that extended regions of space
indeed transiently behave as fast and slow regions.}
\label{israeloff} 
\end{figure}

The phenomena described above, although certainly an indication
of spatio-temporal fluctuations, do not allow one to study
how these fluctuations are correlated in space.
This is however a fundamental issue both from the experimental and
theoretical points of view.
How large are the regions that are faster or
slower than the average? How does their size depend on temperature? Are these
regions compact or fractal?
These important questions were first addressed in pioneering works using
four-dimensional NMR~\cite{nmr},
or by directly
probing fluctuations at the nanoscopic scale using microscopy techniques.
In particular, Vidal Russel and
Israeloff using Atomic Force Microscopy techniques \cite{israeloff} measured
the polarization fluctuations in a volume of size of few tens of nanometers
in a supercooled polymeric liquid (PVAc) close to $T_g$.
In this spatially resolved measurement, the hope is to probe a small enough
number of dynamically correlated regions, and detect their dynamics.
Indeed, the signal shown in Fig.~\ref{israeloff} shows a dynamics
which is very intermittent in time,  switching between
periods with intense activity and other periods with no dynamics
at all, suggesting that extended regions of space
indeed transiently behave as fast or slow regions.
A much smoother signal would be measured
if such dynamically correlated `domains'
were not present.
Spatially resolved and NMR experiments are quite difficult.
They give undisputed information about the typical
lifetime of the DH, but their determination of
a dynamic correlation lengthscale is
rather indirect and/or performed on a small number of
liquids in a small temperature window.
Nevertheless, the outcome is that a
non-trivial dynamic correlation length emerges at
the glass transition, where it reaches a value of the
order of $5-10$ molecule diameters~\cite{ediger}.

\subsection{Multi-point correlation functions}

\label{subsec:multiPointCorrFunc}

More recently,
substantial progress in characterizing spatio-temporal
dynamical fluctuations was obtained
from theoretical~\cite{franzparisi,TWBBB,jcpI,jcpII} and numerical
results~\cite{harrowell,onuki,glotzerfranzparisi,glotzer,lavcevic2003spatially,berthier}.
In particular, it is now
understood that dynamical fluctuations can be measured
and characterized through the use of four-point
correlation functions. These multi-point functions can be
seen as a generalization of the spin glass
susceptibility measuring the extent of amorphous
long-range order in spin glasses. In this subsection,
we introduce these correlation functions and
summarize the main results obtained using them.

Standard experimental probes of the
averaged dynamics of liquids give access to the
time-dependent auto-correlation function of the spontaneous fluctuations of
some observable $O(t)$, $F(t)= \left\langle \delta O(0)
\delta O(t)\right\rangle $, where $\delta O(t)=O(t)-\left\langle
O\right\rangle$ represents the instantaneous value of the
deviation of $O(t)$ from its ensemble
average $\langle O \rangle$ at time $t$.
One can think of $F(t)$ as being the average of a two-point quantity,
$C(0,t)=\delta O(0) \delta O(t)$, characterizing the dynamics.
A standard example corresponds to $O$ being
equal to the Fourier transform of the
density field. In this case
$F(t)$ is the dynamical structure factor as in Eq.~(\ref{isf}).
More generally,
a correlation function $F(t)$
measures the global relaxation in the system.
Intuitively, in a system with important  dynamic correlations,
the fluctuations of
$C(0,t)$ are stronger. Quantitative information on the
amplitude of those fluctuations is provided by the variance
\begin{equation}
\chi_4(t)=
N\left\langle \delta C(0,t)^{2}\right\rangle,
\label{chi4def}
\end{equation}
where $\delta
C(0,t)=C(0,t)-F(t)$, and $N$ is the total number of particles in the system.
The associated spatial correlations show up more clearly when considering a
`local' probe of the dynamics,
like for instance an orientational correlation function
measured by dielectric or light scattering experiments, which can
be expressed as
\begin{equation}
\label{eq3:equation}
C(0,t) = \frac{1}{V}\int d^{3}r \, c(\mathbf{r};0,t),
\end{equation}
where $V$ is the volume of the sample and $c(\mathbf{r};0,t)$
characterizes the dynamics between times $0$ and $t$ around point
$\mathbf{r}$. For example, in the above mentioned case of orientational
correlations, $c(\mathbf{r};0,t)\propto \frac{V}{N}\sum_{i,j=1}^{N}
\delta(\mathbf{r}-\mathbf{r}_{i})Y(\Omega_{i}(0))Y(\Omega_{j}(t))$,
where $\Omega_{i}$ denotes the angles describing the orientation of
molecule $i$, $\mathbf{r}_{i}(0)$  is the position of that molecule
at time $0$, and
$Y(\Omega)$ is some appropriate rotation matrix element. 
Here, the
`locality' of the probe comes from the fact that it is dominated
by the self-term involving the same molecule at times $0$ and $t$,
or by the contribution coming from neighboring molecules.
The dynamic susceptibility $\chi_4(t)$ can thus be rewritten as
\begin{equation}
\label{eq4:equation}
\chi_{4}(t) = \rho \int d^{3}r G_4(\mathbf{r};0,t),
\end{equation}
where
\begin{equation}
G_4(\mathbf{r};0,t)= \left\langle \delta c(\mathbf{0};0,t)\delta
  c(\mathbf{r};0,t) \right \rangle,
\label{g4def}
\end{equation}
and translational invariance has been taken
into account ($\rho=N/V$ denotes the mean density). The above equations show
that $\chi_{4}(t)$ measures the extent of spatial correlation between
dynamical events at times $0$ and $t$ at different points,
\textit{i.e.}~the spatial extent of dynamically heterogeneous regions over a time span $t$.

\begin{figure}
\psfig{file=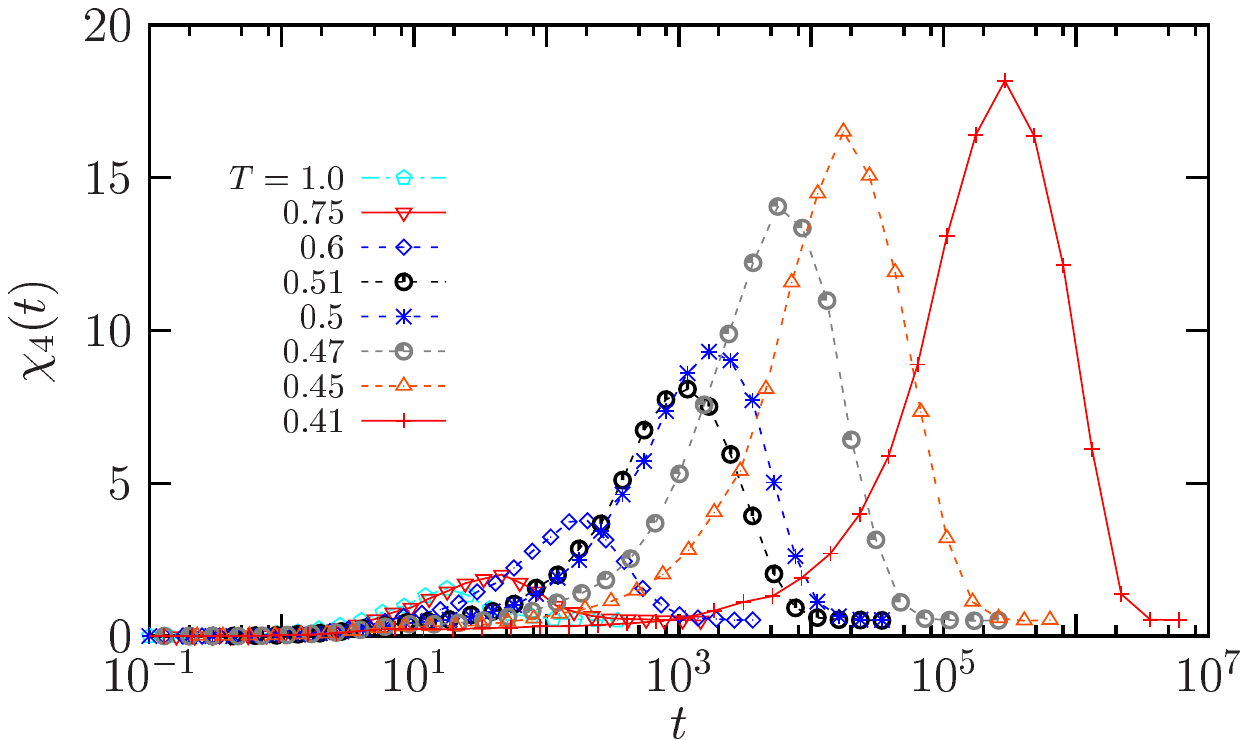,width=8.5cm}
\caption{Time dependence of $\chi_4(t)$,
quantifying the spontaneous fluctuations of the intermediate scattering
function in a Lennard-Jones supercooled liquid.
For each temperature, $\chi_4(t)$ has a maximum, which shifts to
larger times and has a larger value when $T$ is decreased,
revealing the increasing lengthscale of dynamic heterogeneity in
supercooled liquids approaching the glass transition.}
\label{chi4ludo} 
\end{figure}

The function $\chi_4(t)$ has been measured by molecular dynamics,
Brownian and Monte Carlo simulations in different
liquids~\cite{glotzerfranzparisi,glotzer,lavcevic2003spatially,berthier,glotzersilica,berthiersilica}.
An example is shown in Fig.~\ref{chi4ludo} for a Lennard-Jones liquid.
The qualitative behaviour is similar in all cases
\cite{franzparisi,TWBBB,jcpI}:
as a function of time $\chi_4(t)$ first
increases, it has a peak on a timescale that tracks the structural relaxation
timescale and then it decreases~\footnote{The decrease at long times
constitutes a   
  major difference with spin glasses. In a spin glass,
$\chi_4$ would be a
  monotonically increasing
function of time whose long-time limit coincides with the static spin
  glass susceptibility. Physically, the difference is that spin glasses
  develop long-range static amorphous order while
structural glasses do not or, at
least, in a different and more subtle way.}.
Thus the peak value measures the volume
on which the structural relaxation processes are correlated.
It is found to increase when
the temperature decreases
and the dynamics slows down.
By measuring directly $G_4(\mathbf{r};0,t)$ it has also been checked
that the increase of the peak of $\chi_4(t)$
corresponds, as expected, to a growing dynamic
lengthscale $\xi$~\cite{glotzer,lavcevic2003spatially,berthier,jcpI}, although these
measurements are much harder in computer simulations, because
very large systems need to be simulated to determine
$\xi$ unambiguously. Note that if
the dynamically correlated regions were compact, the peak
of $\chi_4$ would be proportional to $\xi^3$ in three dimensions,
directly relating $\chi_4$ measurements to that of the relevant
lengthscale of DH.

These results are also relevant because many theories of the glass
transition assume or predict,
in a way or another, that the dynamics slows down because there are
increasingly large regions on which particles have to relax in a
correlated or cooperative way. However, this lengthscale remained elusive
for a long time. Measures of the spatial extent of dynamic heterogeneity,
in particular $\chi_4(t)$ and   $G_4(\mathbf{r};0,t)$, seem
to provide the long-sought evidence of this phenomenon.
This in turn suggests that the glass transition is indeed a critical
phenomenon
characterized by growing timescales and lengthscales.
A clear and conclusive understanding of the relationship between the
lengthscale
obtained from $G_4(\mathbf{r};0,t)$ and the relaxation timescale is
still the focus of an
intense research activity.

One major issue is that obtaining information on the behaviour of
$\chi_4(t)$ and $G_4(\mathbf{r};0,t)$ from experiments is difficult.
Such measurements are necessary
because  numerical simulations can only be
performed rather far from $T_g$, see Sec.~\ref{simu}.
Up to now, direct experimental
measurements of $\chi_4(t)$ have been restricted to
colloidal~\cite{weeks2}
and granular materials~\cite{dauchotbiroli,durian}
close to the jamming transition,
because dynamics is more easily spatially resolved in those cases.
Unfortunately, similar measurements are currently not available
in molecular liquids.

Recently, an approach based on fluctuation-dissipation relations and
rigorous inequalities has been developed in order to overcome this difficulty
\cite{science,jcpI,jcpII,cecile}. The main idea is to obtain a rigorous lower bound
on $\chi_4(t)$ using the Cauchy-Schwarz inequality
$\left\langle \delta H(0)\delta C(0,t)\right\rangle^{2}\leqslant \left\langle
  \delta H(0)^{2}\right\rangle \left\langle \delta C(0,t)^{2}\right\rangle$,
where $H(t)$ denotes the enthalpy at time $t$.
By using fluctuation-dissipation relations
the previous inequality can be rewritten as~\cite{science}
\begin{equation}
\label{eq7:equation}
\chi_{4}(t) \geq \frac{k_{B}T^{2}}{c_{P}} \left[ \chi_{T}(t)
\right]^{2},
\end{equation}
where the multi-point response function $\chi_T(t)$ is defined by
\begin{equation}
\chi_T(t)= \frac{\partial F(t)}{\partial T}
\bigg\vert_{N,P}=\frac{N}{k_{B}T^{2}}\left\langle \delta H(0)\delta
  C(0,t)\right\rangle.
\end{equation}
In this way, the experimentally accessible response
$\chi_{T}(t)$ which quantifies the sensitivity of average
correlation functions $F(t)$ to an infinitesimal temperature change,
can be used in Eq.~(\ref{eq7:equation}) to yield
a lower bound on $\chi_{4}(t)$.
Moreover, detailed
numerical simulations and theoretical arguments~\cite{jcpI,jcpII}
strongly suggest that the right hand side of (\ref{eq7:equation})
actually provides a good estimation of $\chi_4(t)$, not just a lower bound.

\begin{figure}
\psfig{file=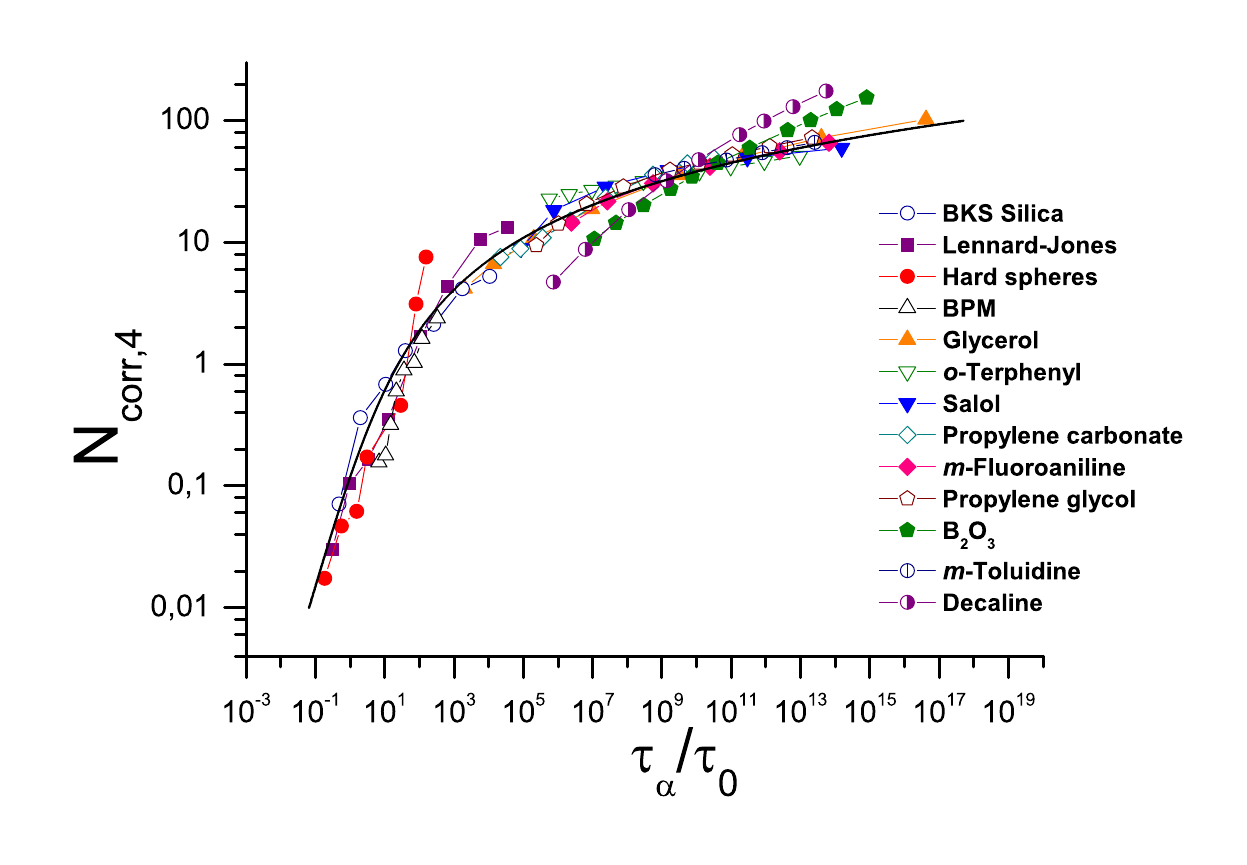,width=9.5cm}
\caption{Universal dynamic scaling relation
between number of dynamically correlated particles, $N_{{\rm corr},4}$,
and relaxation timescale, $\tau_\alpha$, for a number of
glass-formers~\cite{cecile}, determined using Eq.~(\ref{eq7:equation}).}
\label{cecile} 
\end{figure}

Using this method, Dalle-Ferrier {\it et al.}~\cite{cecile} have been able to
obtain the evolution of the peak value of $\chi_4$
for many different glass-formers in the entire supercooled
regime. In Fig. \ref{cecile} we show some of these results as a
function of the relaxation
timescale. The value on the
$y$-axis, the peak of $\chi_4$, is a proxy for the number of
molecules, $N_{{\rm corr},4}$
that have to evolve in a correlated way in order to
relax the structure of the liquid. Note that
$\chi_4$ is expected to be equal to $N_{{\rm corr},4}$ only up to a
proportionality constant that is not known
from experiments, which probably explains why
the high temperature values of $N_{{\rm corr},4}$ are smaller than one.
Figure~\ref{cecile} also indicates that $N_{{\rm corr},4}$ grows faster
when $\tau_{\alpha}$ is not very large,
close to the onset of slow dynamics, and
a power law relationship between $N_{{\rm corr},4}$ and $\tau_\alpha$
fits this regime well ($\tau_\alpha /\tau_0 < 10^4$).
The growth of $N_{{\rm corr},4}$
becomes much slower closer to $T_{g}$.
A change of 6 decades in time corresponds to a mere increase of a
factor about 4 of $N_{{\rm corr},4}$, suggesting logarithmic
rather than power law growth of dynamic correlations.
This is in agreement with several theories
of the glass transition which are based on
activated dynamic scaling~\cite{rfotwolynes,gcpnas,gillesreview}.

Understanding quantitatively this relation between
timescales and lengthscales
is one of the main recent topics addressed in theories of
the glass transition, see Sec.~\ref{theory}. Furthermore, numerical works
are also devoted to characterizing better the geometry of
the dynamically heterogeneous
regions~\cite{glotzerstring,kobdemos,kob2012non}.

\subsection{Non-linear response function}

Diverging responses are characteristic signatures of  phase transitions. Linear static responses measuring the change in the order parameter due to external fields diverge at second order phase transitions \cite{chaikin1995principles}.
By using fluctuation-dissipation relations one can show that such divergence is intimately related to the divergence of the correlation length emerging in two point-functions. 
Spin-glasses, the archetypal example of disordered systems, display a diverging static magnetic non-linear cubic  response \cite{binder1986spin,baity2013critical}. In Ref.~\cite{bouchaud2005nonlinear} it was argued that the counterpart of these phenomena for supercooled liquids can be found in non-linear dynamical susceptibilities, which 
should grow approaching the glass transition, thus providing a complementary way (compared to $\chi_4$) to reveal its collective nature. In experiments on molecular liquids, non-linear dielectric susceptibility are a natural probe to unveil this phenomenon. 

\begin{figure*}
\includegraphics[width=18cm]{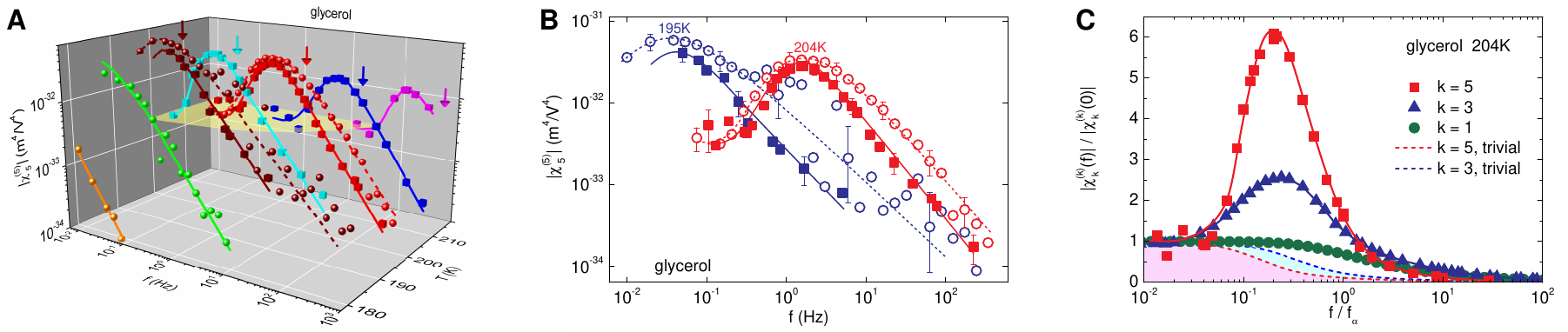}
\caption{Modulus of the fifth-order susceptibility in super-cooled glycerol as a function of frequency (from \cite{albert_fifth-order_2016}). {\bf(A)} The susceptibilities $\chi_5^{(5)}$ reported are obtained directly by monitoring the response of the sample at $5 \omega$, when applying an electric field $E$ at angular frequency $\omega$. Two independent setups were used. Lines are guides for the eyes. {\bf{(B)}} Projection onto the susceptibility-frequency plane of the data of panel A at $204$K and at $195$K. {\bf{(C)}} Comparison of the fifth-order, cubic, and linear susceptibilities. Symbols, with line to guide the eyes. The higher the order $k$, the stronger the hump of $\vert \chi_k^{(k)} \vert$.} 
\label{fig:fifths}
\end{figure*}

The simplest explanation \cite{albert_fifth-order_2016} for this scenario is based on the assumption 
that $N_{\text{corr}} = (\ell/a)^{d_f}$ molecules are amorphously ordered over the lengthscale $\ell$, where $a$ is the molecular size 
and $d_f$ is the fractal dimension of the ordered clusters. In consequence, their dipoles, which are oriented in apparently random positions, are essentially locked 
together during a time $\tau_\alpha$. In the presence of an external electric field $E$ oscillating at frequency $\omega \ge \tau_\alpha^{-1}$, the dipolar degrees of freedom of these molecules contribute to the polarisation per unit volume as 
\begin{equation} \label{scaling}
p = \mu_{dip} \frac{\sqrt{(\ell /a)^{d_f}}}{(\ell /a)^d} F\left(\frac{\mu_{dip} E \sqrt{(\ell /a)^{d_f}}}{kT}\right)
\end{equation}
where $\mu_{dip}$ is an elementary dipole moment, $F$ is an odd scaling function, and $d=3$ the dimension of space.
This states that randomly locked dipoles have an overall moment $\sim \sqrt{N_{\text{corr}}}$, and that we should compare the thermal energy with the energy of this `super-dipole' in a field. 

Expanding Eq.~\ref{scaling} in powers of $E$, one finds the `glassy' contribution to
$p$:
\begin{eqnarray} \label{eq2}
\frac{p}{\mu_{dip}} & = & F'(0) \left(\frac{\ell}{a} \right)^{d_f - d} \left(\frac{\mu_{dip} E}{kT}\right) +\nonumber  \\
\ \ &\ & + 
\frac{1}{3!} F^{(3)}(0) \left(\frac{\ell}{a} \right)^{2d_f - d} \left(\frac{\mu_{dip} E}{kT}\right)^3 + \nonumber  \\
\ \ &\ & + \frac{1}{5!}F^{(5)}(0) \left(\frac{\ell}{a} \right)^{3d_f - d} \left(\frac{\mu_{dip} E}{kT}\right)^5 + \dots
\end{eqnarray}

Because $d_f \leq d$, the first term, contributing to the usual linear dielectric susceptibility, $\chi_1(\omega)$,
cannot grow as $\ell$ increases. This simple theoretical argument explains why spatial glassy correlations does not show up in $\chi_1(\omega)$, in experiments. The second term, contributing to the third-order dielectric susceptibility $\chi_3(\omega)$, does grow with $\ell$ provided $d_f > d/2$. Several theories indeed suggest that ordered domains are compact, $d_f=d$ \cite{wolynes2012structural,gillesreview}. 
The third term leads to the fifth-order susceptibility $\chi_5(\omega)$, which should diverge as $\ell^{3d_f - d}$ (higher $n$-order susceptibilities diverge as $\ell^{(n+1)d_f/2 - d}$).
This line of arguments shows that measuring non-linear susceptibilities is a way to probe and characterize the collective dynamical behavior associated to glassy dynamics. 

This challenge was taken up in the series of works \cite{crauste2010evidence,bauer2013cooperativity,brun2012evidence,albert_fifth-order_2016}. The main outcomes of these experiments have been: (i) to show that indeed a growth of the non-linear responses goes along with the glass transition, (ii) to measure in a new way the number of correlated molecules close to $T_g$, (iii) to estimate that $d_f\simeq d$ (for $d=3$). As an example, we reproduce the results of \cite{albert_fifth-order_2016} in Fig.~\ref{fig:fifths}, which shows the increase of the fifth-order susceptibility with temperature (panel A), its humped shape in frequency (panel B), and the stronger singularity of the 
fifth-order susceptibility compared to the third-order one (panel C), as expected for a collective phenomenon.

Another set of experiments on non-linear responses was performed in colloids: non-linear mechanical susceptibilities were probed approaching the colloidal glass transition~\cite{seyboldt2016divergence} and shown to grow approaching it. Differently from the dielectric case, third-harmonic shear responses have a peak at a frequency associated to the $\beta$-relaxation, and not to the $\alpha$-relaxation. The reason is related to the fact that a very slow shear-strain does not affect the relaxation time-scale, whereas an external electric field (even a static one) does, see Ref.~\cite{seyboldt2016divergence}. 

Finally, a word about theories. Given that the growth of non-linear dynamical susceptibility is a relatively newly established fact in the glass-physics arena, one can wonder how the different theoretical framework developed to explain the glass transitions cope with it. Thermodynamic theories based on the increases of some kind of medium range order naturally do, as explained above. Mode-Coupling-Theory also predicts diverging dynamical non-linear susceptibility at the MCT transition \cite{tarzia2010anomalous,seyboldt2016divergence}. Purely local theory are instead at odds. Dynamical facilitation theory was argued to be compatible with such findings in Ref.~\cite{speck2019dynamic}, even though the general arguments put forward in Ref.~\cite{albert_fifth-order_2016} indicate the opposite conclusion. 

\section{Theory of the glass transition}

\label{theory}

We now present some theoretical approaches
to the glass transition. It is impossible to cover all of them
in a brief review, simply
because there are way too many of them, perhaps the clearest
indication that the glass transition
remains an open problem. We choose to present approaches that
are keystones and have a solid statistical
mechanics basis. Loosely speaking, they
have an Hamiltonian, can be simulated numerically,
or studied analytically with statistical mechanics tools.
Of course, the choice of Hamiltonians
is crucial and contains very important assumptions about the nature
of the glass transition. All these
approaches have given rise to
unexpected results. One finds more in them than what was supposed
at the beginning, which
leads to new, testable predictions.
Furthermore, with models that are precise enough, one
can test (and hopefully falsify!) these approaches
by working out all their predictions in great detail,
and comparing the outcome to actual data. This is not possible
with `physical pictures', or simpler approaches of the problem
which we therefore do not discuss.

Before going into the models, we would like to state
the few important questions that theoreticians face.

\begin{itemize}

\item Why do
the relaxation time and the viscosity
increase when $T_g$ is approached?
Why is this growth super-Arrhenius?

\item Can one understand and describe quantitatively the average
dynamical behaviour of
supercooled liquids, in particular broad relaxation
spectra, non-exponential behaviour, and their
evolution with fragility?

\item Is there a relation between kinetics and
thermodynamics (like $T_0 \simeq T_K$),
and why?

\item Can one understand and describe quantitatively
the spatio-temporal fluctuations of
the dynamics? How and why are these fluctuations related to the
dynamic slowing down?

\item Is the glass transition a collective phenomenon? If yes,
of which kind?
Is there a finite temperature or zero temperature ideal glass transition?

\item Is the slowing down of the dynamics driven by the growth of amorphous order and a static length? Or is its origin purely dynamic?

\item Is there a geometric, real space explanation for the dynamic
slowing down that takes into account molecular degrees of freedom?

\end{itemize}

The glass transition
appears as a kind of `intermediate coupling' problem, since
for instance typical growing lengthscales are
found to be at most a few tens of particle large close to $T_g$.
It would therefore be difficult to recognize the correct
theory even if one bumped into it.
To obtain quantitative, testable predictions, one must therefore
be able to work out also preasymptotic effects.
This is particularly difficult,
especially in cases where the asymptotic theory itself
has not satisfactorily been worked out.
As a consequence, at this time, theories can only be judged
by their overall predictive power and their theoretical consistency.

\subsection{Random First Order Transition Theory}

\label{giuliotheory}

\subsubsection{Mean-field models and a zest of replica theory}

In the last three decades, three independent
lines of research,
Adam-Gibbs theory~\cite{ag}, mode-coupling theory~\cite{gotze}
and spin glass theory~\cite{beyond}, have merged to produce
a theoretical ensemble that now goes under the name of
Random First Order Transition theory (RFOT), a terminology introduced by
Kirkpatrick, Thirumalai and Wolynes~\cite{KTW,KTWbis}
who played a major role in this unification. Instead of following the rambling development of history, we
summarize it in a more modern and unified way.

A key ingredient of RFOT theory is the existence of a chaotic or complex
free energy landscape with a specific evolution with temperature and/or
density. Analysing it in a controlled way
for three dimensional interacting particles is an impossible task.
This can be achieved, however, in simplified models or using
mean-field approximations, that have therefore
played a crucial role in the development of RFOT theory.

A first concrete example is given by
`lattice glass models'~\cite{BiroliMezard}.
These are models of hard particles sitting
on a lattice. The Hamiltonian is infinite
either if there is more than one particle on a site or if the number of
occupied neighbors of an occupied site is larger than a parameter $m$, 
but is zero otherwise. 
Tuning the parameter $m$, or changing
the type of lattice, in particular its connectivity,
yields different models. Lattice glasses are constructed as simple
statmech models to study the
glassiness of hard sphere systems. The constraint on the number of
occupied neighbors mimicks the
geometric frustration~\cite{Nelson} encountered when trying to pack
hard spheres in three dimensions. Numerical simulations 
show that their phenomenological glassy behavior is indeed analogous to the one of supercooled liquids \cite{darst2010dynamical,seif2016structure,nishikawa2020lattice}.
Other models with a finite energy are closer to
molecular glass-formers, and can also be
constructed \cite{mccullagh2005finite}. 
These models can be solved exactly on a
Bethe lattice~\footnote{In order to have a
well-defined thermodynamics, Bethe lattices are generated as random
graphs with fixed connectivity, also
called random regular graphs.}, which reveals a rich
physical behaviour~\cite{bethe}. In particular their
free energy landscape can be analyzed in full details and turns
out to have the properties that are also found
in several `generalized spin glasses'.
Probably the most studied example of such spin glasses
is the $p$-spin model, defined by the Hamiltonian~\cite{MezardGross}
\begin{equation}
\label{pspin}
H = - \sum_{i_1,...,i_p}J_{i_1,...,i_p}S_{i_1}...S_{i_{p}},
\end{equation}
where the $S_i$'s are $N$ Ising or spherical spins,
$p>2$ is the number of interacting spins in a single term of the sum, 
and $J_{i_1,...,i_p}$ quenched random couplings extracted from a distribution 
which, with no loss of generality,
can be taken as the Gaussian distribution
with zero mean and variance $p!/(2N^{p-1})$.
In this model, the couplings $J_{i_1,...,i_p}$ play the role
of self-induced disorder in glasses, and promotes a glass phase at low temperature. 

All these models can be analyzed using the so-called replica theory~\cite{beyond}. 
Given its importance in setting the foundations 
of the theory of glasses at the mean-field level, 
we now present its main technical steps. 
To keep the discussion as simple as possible, 
we focus on $p$-spin models. Note that the theory holds 
for more complex models but it is technically more involved. 
The starting point is the computation of the free-energy 
which is obtained as an average
over the distribution of couplings:
\begin{equation}
F = \lim_{N\rightarrow \infty} -\frac{1}{\beta N} \overline{\log{Z_J}},
\end{equation}
where $\overline{\cdots}$ represents the average over the disorder.
Performing this average is possible thanks to the replica trick
\begin{equation}
\overline{\log{Z_J}} = \lim_{n\rightarrow 0} \frac{1}{n} \log{\overline{Z^n}},
\end{equation}
where $n$ is the index of replicas, i.e.~clones of the same system with different
couplings $J_{i_1,...,i_p}$ extracted from the same distribution.
The use of the replica trick may seem purely
mathematical, yet it has a profound physical sense.
If the system is ergodic, averages of thermodynamical observables 
for two replicas of the same system (with identical disorder) coincide, 
whereas they differ if ergodicity is broken. 
We can define the overlaps between two replicas $a,b$ as $Q_{ab}$,
which defines the $n \times n$ overlap matrix:
\begin{equation}
Q_{ab} = \frac{1}{N} \sum_{i=1}^N S_i^a S_i^b,
\end{equation}
where the product between spins represents
a dot product for spherical spins~\cite{hessian}.
After some computations, the free energy can be
expressed as a function of $Q_{ab}$, 
which therefore plays the role of the order parameter. 
In the ergodic phase one expects symmetry 
between replicas~\footnote{If additional symmetries are broken then one can have ergodicity breaking also in the RS phase.}, 
and the so-calledreplica-symmetic (RS) 
parametrization of $Q_{ab}$ is adopted:
all the off-diagonal elements of $Q_{ab}$ are equal to $q_0 < 1$
and the diagonal elements are $Q_{aa}=1$.
The parametrization that corresponds to the glass phase, 
when ergodicity is broken, is the so-called 
one-step replica symmetry breaking (1RSB) solution.
Here, the overlap matrix is divided into blocks of dimension $m \times m$;
elements belonging to blocks far from the diagonal are equal to $q_0$,
while off-diagonal elements of blocks along the diagonal are
equal to $q_1$ with $1> q_1> q_0$. On the diagonal $Q_{aa}=1$. 
This parametrization encodes the existence of 
many thermodynamically equivalent basins, 
hence two replicas can either fall in the same basin 
and have overlap $q_1$, or fall in two different basins and have overlap $q_0$. 
The crucial simplification introduced by the mean-field approximation 
is that barriers between basins have a 
free energy cost which grows exponentially with $N$, so that truly metastable states can be defined in the thermodynamic limit~\cite{supercavagna}.
At high temperature (or low density) 
the RS solution has a lower free energy. 
Below the ideal glass transition temperature 
the 1RSB solution instead becomes dominant. 

\subsubsection{Liquids and glasses in infinite dimensions}

A major theoretical breakthrough of the last years is the analysis of the glass transition for interacting particle systems in the limit of infinite dimensions~\cite{KurParZam,charbonneau_fractal_2014,charbonneau_exact_2014,KurParUrbZam,parisi2020theory}. The starting point approach is the definition of a pair interaction potential with a proper scaling with dimension $d$ to ensure a non trivial thermodynamic limit:
\begin{equation}
    v(r) =\tilde v [d(r/\ell-1)] 
\end{equation}
where $\ell$ defines the range of the interaction. 
Many different potentials used to model glasses can be written in this way by using a suitable function $\tilde v(x)$, such as hard spheres, Lennard-Jones, Yukawa, square-well, harmonic, and Weeks-Chandler-Andersen potentials~\cite{parisi2020theory}.   
In the limit of infinite space dimension, $d \rightarrow \infty$, and using the scaling above, the thermodynamics and the dynamics of liquids and glasses can be analyzed exactly~\footnote{For large $d$ the crystalline phase does not intervene. In fact, the amorphous and crystalline solid phases are well separated in configuration space
and issues related to finite dimensions, such as the crystallization of monodisperse particles, are suppressed \cite{Skoge, vanmeelhardsphere2009}.}. The resulting theory is qualitatively very similar to the one obtained from the simple models discussed in the previous section (both for the statics, in terms of replica formalism, and for the dynamics, in terms of self-consistent Langevin equations). 

\begin{figure}
\psfig{file=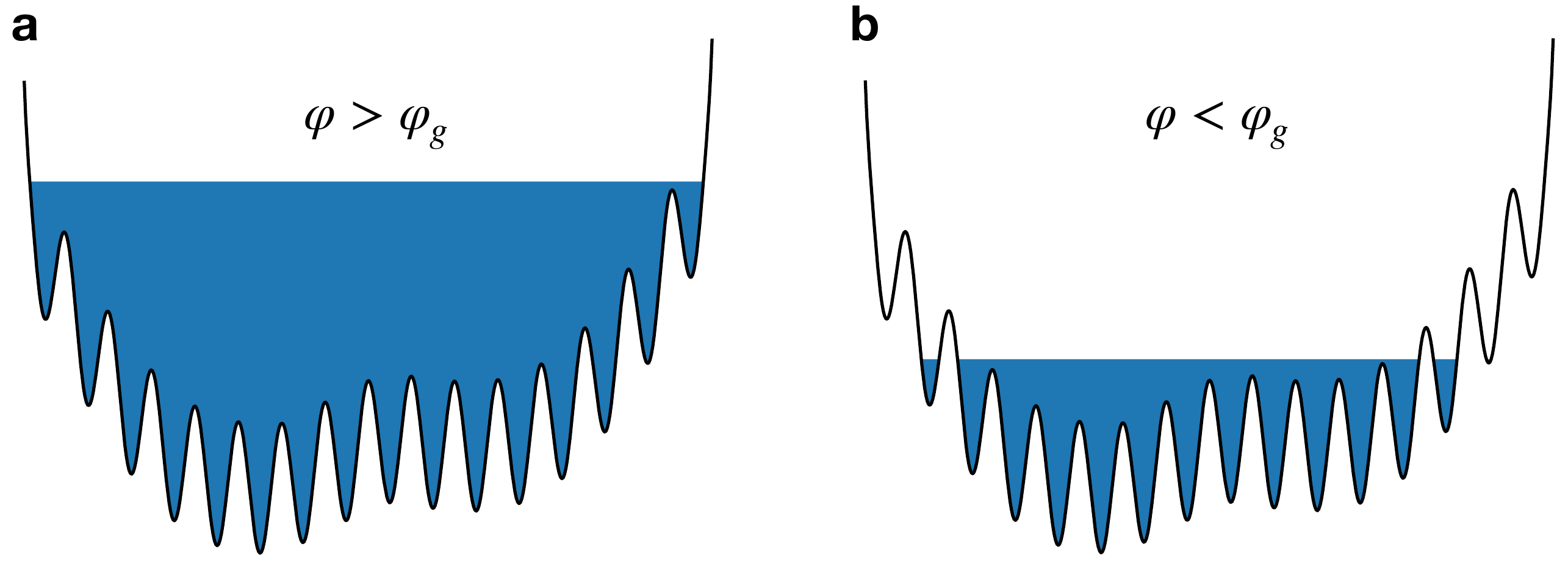,width=8.5cm}
\caption{Sketch of the evolution of free-energy
landscape of hard spheres across the glass transition. In the liquid phase
({\bf a}) at low packing fractions $\varphi < \varphi_g$,
every portion of the phase space is accessible. For $\varphi > \varphi_g$ the system is in the glass phase ({\bf b}) and remains trapped in one of the many equivalent basins.}
\label{landscape} 
\end{figure}

In fact, all these models belong
to the universality class of 1RSB systems \cite{charbonneau_exact_2014},
with a free-energy landscape evolving as in the sketch in Fig.~\ref{landscape}.
At low densities or high enough temperatures,
they all describe an ergodic liquid phase,
analogous to the paramagnetic phase of a spin glass.
Under cooling or application of an external pressure,
the free energy breaks up into many different minima
which eventually trap the dynamics, and the system enters the glass phase, as described further below.

The merit of the infinite dimensional theory is that it offers quantitative results and applies directly to microscopic models of liquids and glasses. Moreover, it directly reveals the nature of `mean-field' theories and approximations, such as the diagrammatic liquid theory and Mode-Coupling Theory. Last but not least, it establishes once and for all that the 1RSB phase and associated physics and phase transition is the correct 
and universal mean-field theory of glass-forming models. 

\subsubsection{Random first order transitions}

\label{RFOT-i}

\begin{figure}
\psfig{file=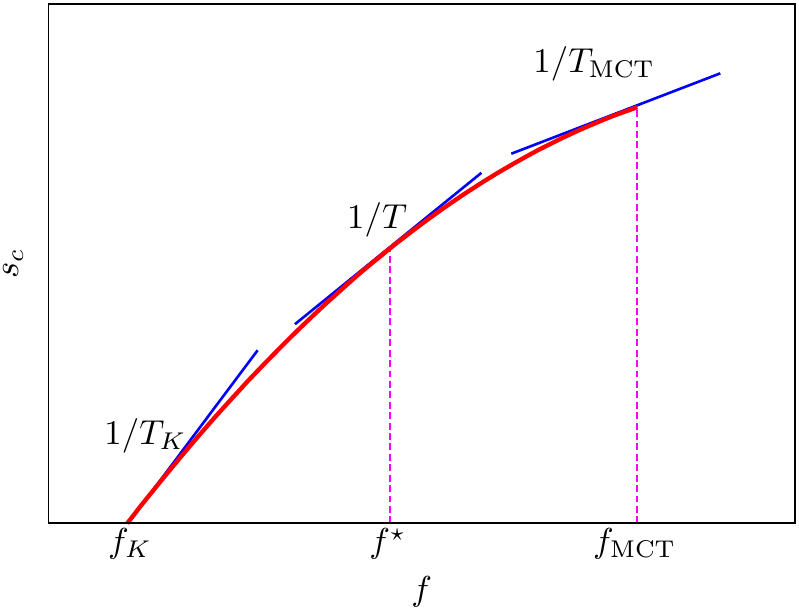,width=8.5cm}
\caption{Typical shape of the configurational entropy,
$s_c$, as a function of free energy density, $f$ in the range
$T_k < T <T_{MCT}$ for random first order landscapes.
A graphical solution of Eq.~(\ref{betaf}) is obtained by
finding the value of $f$ at which the
slope of the curve is $1/T$. Note that $s_c$
is also a function of temperature,
so this curve changes with $T$.}
\label{sc:fig} 
\end{figure}

We now discuss the physics associated to the 1RSB phase transition, and more generally to RFOT. The free energy landscape of glassy systems is `rugged', as shown in Fig.~\ref{landscape}. It is 
characterized by many minima and saddle points of various order.
Actually, the number of stationary points is so large that in
order to count them one has
to introduce an entropy, called configurational entropy or complexity,
$s_c = \frac{1}{N} \log {\cal N}(f)$,
where ${\cal N}(f)$ is the
number of stationary points with a given free energy density
$f$. The (real space) density profile
corresponding to one given minimum is amorphous and lacks any type of
periodic long-range order, and different minima are
very different from one another. Defining a similarity measure
between them, an `overlap' $Q$ (see Eq.~(\ref{eq:defoverlap}) below for a precise definition), one typically finds that two minima with the same free energy $f$ have zero overlap. The typical shape of the
configurational entropy as a function of $f$ is shown in Fig~\ref{sc:fig}.

At high temperature,
there is typically a single minimum, the high temperature liquid state.
There is a temperature below which
an exponentially large (in the system size) number of minima appears.
Within  mean-field models,
corresponding to Bethe lattices, completely connected lattices, and interacting particles for $d\rightarrow \infty$ these minima
correspond to macroscopic
physical states analogous to the periodic minimum corresponding to the
crystal~\footnote{There is of course no crystal state in disordered
systems such as in Eq.~(\ref{pspin}). In the case of lattice glass models,
there is a crystal phase but it can disappear depending whether the Bethe
lattice is a Cayley tree or a random regular graph.}.
Once the system is in one of these states it remains
trapped there forever, since the barriers separating states diverge with the system size. However, when transposed to finite dimensional systems, these states
become metastable and have a finite lifetime. As a consequence, in order
to compute thermodynamic properties, one has to sum over all of them
using the Boltzmann weight $\exp(-\beta N f_\alpha)$ for each
state $\alpha$~\cite{monasson}:
\begin{equation}
\label{zsum}
Z=\sum_{\alpha}e^{-\beta N f_\alpha} = \int df
\exp [ N s_c(f;T) ] e^{-\beta N f},
\end{equation}
where $\beta=1/(K_B T)$.
Evaluating this sum by saddle point method yields three regimes.
At high temperature, $T>T_{MCT}$,
the liquid corresponding to a flat density profile dominates
the sum. The landscape is simple and has a single minimum.
This is followed by
an intermediate temperature regime, $T_K<T<T_{MCT}$,
where the sum is dominated
by all terms with free energy density satisfying
\begin{equation}\label{betaf}
\left.\frac{\partial s_c(f,T)}{\partial f} \right|_{f=f^*}= \beta .
\end{equation}
There are many of them, the logarithm of their number being given by
$N s_c(f^*,T)$, see Fig.~\ref{sc:fig} for a graphical solution of
Eq.~(\ref{betaf}).
Upon decreasing the temperature, $s_c(f^*,T)$
decreases until a temperature, $T_K$, below
which the sum in Eq.~(\ref{zsum}) becomes
dominated by only few terms corresponding to states with
free energy density $f_K$ given by $s_c(f_K,T)=0$, see Fig.~\ref{sc:fig}.
The entropy in the intermediate temperature range above
$T_K$ has two contributions: the one counting the number of minima,
given by $s_c$, and the intra-state entropy, $s_{in}$,
counting the number of configurations inside each state.
At $T_K$, the configurational entropy vanishes, $s_c(T_K)=0$,
As a consequence the specific heat undergoes a jump
towards a smaller value across $T_K$, an exact realization
of the `entropy vanishing' mechanism
conjectured by Kauzmann~\cite{kauzmann}.

Let us discuss the dynamical behaviour which results from the
above analysis.
We have already mentioned that
relaxation processes do not occur below $T_{MCT}$ because
states have an infinite lifetime. The stability
of these states can be analyzed  by computing the free energy Hessian
in the minima~\cite{hessian}. One finds that states become
more fragile when $T \to T_{MCT}^-$, are marginally stable
at $T=T_{MCT}$, unstable for $T>T_{MCT}$.
The relaxation dynamics of these models can be
analyzed exactly~\cite{barrat2004slow,MaimbourgKurchanZamponi}. Coming from
high temperature, the dynamics slows down and the relaxation
time diverges at $T_{MCT}$ in a power law manner,
\be
\tau_{\alpha} \sim \frac{1}{(T-T_{MCT})^\gamma},
\label{gamma}
\ee
where $\gamma$ is a critical exponent.
The physical reason is the incipient stable states
that appear close to $T_{MCT}$.
The closer the temperature is to $T_{MCT}$, the longer it takes
to find an unstable direction to relax.

Amazingly, the dynamical transition
that appears upon approaching $T_{MCT}$ in random first
order landscapes is completely analogous
to the one predicted for supercooled liquids
by the Mode-Coupling Theory (MCT) of the glass transition, and developed
independently  by Leuthesser, Bengtzelius,
G\"otze, Sj\"olander and coworkers~\cite{gotze}. Actually, MCT can be
considered as an approximation which becomes controlled and exact
for these mean-field models.
Originally, MCT was developed using
projector operator formalism~\cite{Leuthesser,BGS}
and field-theory methods~\cite{DM} to yield
closed integro-differential equations for the dynamical structure factor
in supercooled liquids.
These approaches were recently generalized~\cite{BB,BBMR}
to deal with dynamic heterogeneity and make predictions
for the multi-point susceptibilities and correlation
functions discussed in Sec.~\ref{dh}.
Within MCT, the relaxation timescale diverges in a power law
fashion at $T_{MCT}$, as in Eq.~(\ref{gamma}). This divergence
is accompanied by critical behaviour that
appears both in space (long range spatial dynamic correlations),
and in time (time-dependent power laws).

Comparing Eqs.~(\ref{vft}) and (\ref{gamma}) makes it clear
that MCT cannot be used to describe viscosity data close
to $T_g$ since it does not predict activated behaviour.
It is recognized that an MCT transition
at $T_{MCT}$ does not occur in real materials, so that
$T_{MCT}$ is, at best, a dynamical crossover.
A central advantage of MCT, compared to many other theories is that it can yield
quantitative predictions from microscopic input obtained for a particular material.
As such it has been applied to scores of different systems, with predictions that can be
directly confronted to experimental or numerical
measurements. A major drawback is the freedom offered
by the `crossover' nature of the MCT transition,
so that `negative' results can often be attributed
to corrections to asymptotic predictions rather than deficiencies of
the theory itself.
Nevertheless, MCT has proven to be useful
and continues to be developed,
applied and generalized to study many different
physical situations~\cite{gotze},
including aging systems and
non-linear rheology of
glassy materials~\cite{BBK,dave,fuchs}, see also Sec.~\ref{aging}.

What happens below $T_{MCT}$ in a finite dimensional
system if the relaxation time does not diverge
as predicted in Eq.~(\ref{gamma})? Why is the transition avoided?
In fact, the plethora of states that one finds in mean-field
are expected to become (at best) metastable in finite dimension,
with a finite lifetime, even below $T_{MCT}$.
What is their typical lifetime, and how these metastable states
are related to the structural relaxation are issues that
still await for a complete microscopic analysis.

There exist, however, phenomenological arguments~\cite{rfotwolynes,KTW2,BB2},
backed by microscopic computations~\cite{Schmalian,Franz} that yield
a possible solution dubbed `mosaic state' by Kirkpatrick,
Thirumalai and Wolynes~\cite{KTW2}.
Schematically, the mosaic picture states that, in the regime $T_K<T<
T_{MCT}$, the liquid is
composed of domains of linear size $\xi$. Inside
each domain, the system is in one of the mean-field states.
The length of the domains is fixed by a competition
between energy and configurational entropy. A state in a finite but large
region of linear size $l$ can be selected by
appropriate boundary conditions that decrease
its free energy by an amount which scales
as $\Upsilon l^{\theta}$ with $\theta \le 2$.
On the other hand,
the system can gain entropy,
which scales as $s_c l^3$,  if it visits
the other numerous states.
Entropy obviously gains on large lengthscales,
the crossover length $\xi$ being obtained
by balancing the two terms,
\begin{equation}
\label{eqxi}
\xi = \left(\frac{\Upsilon}
{Ts_c(T)}\right)^{1/(3-\theta)}.
\end{equation}
In this scenario,
the configurational entropy
on scales smaller than $\xi$ is too small to stir the
configurations efficiently and win over the dynamically generated
pinning field due to the environment, while ergodicity
is restored at larger scale.
Hence, the relaxation time of the system is the relaxation time, $\tau(\xi)$, of finite size
regions. Barriers are finite, unlike in the mean-field
treatment. Smaller length scales are faster
but unable to decorrelate, whereas larger scales are orders of magnitude
slower. Assuming thermal activation over
energy barriers which are supposed to grow
with size as $\xi^\psi$, one finally predicts, using
Eq.~(\ref{eqxi}), that~\cite{BB2}
\be
\log \left( \frac{\tau_\alpha}{\tau_0} \right) =
c \frac{\Upsilon}{k_B T}
\left(\frac{\Upsilon}
{Ts_c(T)}\right)^{\psi/ (3-\theta)},
\label{ag}
\ee
where $c$ is a constant.

The above argument is rather generic and therefore not very predictive.
There exist microscopic computations~\cite{Schmalian,Franz,biroli2017fluctuations} aimed at putting these phenomenological arguments on a firmer
basis and computing the exponents $\theta$ and $\psi$.
The results are not yet fully conclusive
because they involve replica calculations with some
assumptions, but they do confirm
the phenomenological scenario presented above and suggest
that $\theta = 2$. Some other phenomenological
arguments suggest the  value of $\theta=3/2$~\cite{KTW2}.
There are no computation available
for $\psi$, only the suggestion
that $\psi=\theta$ \cite{KTW2}.

Note that using the value $\theta=3/2$ with $\theta=\psi$ simplifies
Eq.~(\ref{ag}) into a form that is well-known experimentally
and relates $\log \tau_\alpha$ directly
to $1/S_c$, which is the
celebrated Adam-Gibbs relation~\cite{ag} between
relaxation time and configurational entropy that
is in rather good quantitative agreement with many
experimental results~\cite{Angell,Hodge,Johari,ozawa2019does}.
The Random First Order Transition theory can be considered, therefore,
as a microscopic theory that reformulates and generalizes
the Adam-Gibbs mechanism. Furthermore, using the fact  that
the configurational entropy vanishes linearly at $T_K$,
a VFT divergence of the relaxation
time as in Eq.~(\ref{vft}) is predicted, with the identification that
\be
T_0 = T_K.
\label{totk}
\ee

The equality (\ref{totk}) between two temperatures that are
commonly used in the description of experimental data
certainly constitutes a central achievement of RFOT theory since
it accounts for the empirical
relation found between the kinetics and the
thermodynamics of supercooled liquids.
Furthermore RFOT theory naturally contains MCT, which
can be used to describe the first decades of the dynamical slowing down,
while the spin glass side of RFOT theory qualitatively explains
the dynamics in terms of the peculiar
features of the free energy landscape that have been detailed above.
Dynamics first slows down because there appears incipient metastable states, and once these metastable states are formed, the dynamics becomes dominated by the thermally activated barrier crossing
from one metastable state to another, 
which is consistent with
the relation between dynamical correlation length and timescale
discussed in Sec.~\ref{dh}. Quite importantly, microscopic
computations of $T_{MCT}$ and $T_0$ for realistic models of liquids are
possible~\cite{MezardParisi,parisi2020theory}.

Probably the most serious weakness of the RFOT theory construction
is that the theory, although worked out in full details within
mean-field models or the large dimensionality limit, is based for finite dimensions on asymptotic results valid, e.g., for $T\rightarrow T_K$. The application of RFOT theory to temperatures accessible in experiments (hence not very close to $T_K$) requires additional phenomenological assumptions.
Moreover, the dynamical processes leading to the VFT law are not understood completely.
Although the ultimate consequences of the theory
are sometimes in very good agreement with experiments, as
Eq.~(\ref{totk}), direct tests of the mosaic
state picture are rare and difficult~\cite{cavagna,ozawa_does_2017}. 

\subsubsection{Heterogeneous disorder and mapping to the Random Field Ising Model}

The study of second-order phase transitions shows that field theory provides 
a natural framework to go beyond mean-field theory \cite{chaikin1995principles}. 
With this in mind, researchers in glass physics have also gone down this route in recent years. 
One of their main achievements has been to identify fluctuations that are neglected by mean-field theory and 
play a very important role in shaping the physical behavior of supercooled liquids. These fluctuations have been related to the notion of `self-induced disorder' in Ref.~\cite{franz2011field}. 
In the following, we may prefer the name `self-induced heterogeneity'~\footnote{This terminology was suggested to us by Jean-Philippe Bouchaud.} to make the distinction with the `self-induced disorder' discussed in Sec.~\ref{subsec:epsilonCoupling_VQ}. The main idea is that when observing equilibrium relaxation from time $t$ to time $t+\tau_\alpha$, the state of the system 
at time $t$ is spatially heterogeneous: for instance it can have higher density in one region and lower density in another. Actually, theoretical analysis shows \cite{stevenson2008constructing,biroli2018randoma,biroli2018randomb} that even key mean-field quantities, such as the configurational entropy, are heterogeneously distributed in space. Since the amount of slowing down is directly linked to those quantities (at least within RFOT theory), these static fluctuations  induce strong dynamical fluctuations, leading in particular to dynamical heterogeneities~\cite{franz2011field}. From the theoretical point of view, 
they play a key role in changing properties of the MCT transition and the ideal glass transition. In fact, even though the MCT transition is like a spinodal instability within mean-field theory~\cite{kirkpatrick1987stable}, one finds that once
these fluctuations are included, MCT enters the universality class of a {disordered} spinodal, like, e.g.~the spinodal of the Random Field-Ising  Model \cite{franz2011field}. Using results obtained on this problem, this connection implies that even in the absence of activated hopping, the MCT transition changes nature in any finite dimension: it is either wiped out by non-perturbative fluctuations \cite{rizzo2016dynamical} or it becomes dominated by rare and non-perturbative events, as happens for the spinodal of the RFIM~\cite{nandi2016spinodals}. 
One nevertheless expects that the higher the spatial dimension, the more obvious an echo of the MCT mean-field transition should persist~\cite{biroli2012random,berthier2019finite,MaimbourgKurchanZamponi}.  

Finally, the role of heterogeneous disorder on the 
ideal glass transition has been investigated in Refs.~\cite{stevenson2008constructing,biroli2018randoma,biroli2018randomb}. The main outcome of these studies is an effective model for the glass transition that takes the form of an RFIM with extra long-range anti-ferromagnetic and multi-body interactions. These new couplings depress the ideal glass transition temperature but do not lead to qualitative changes. 
The strength of the disorder is, however, crucial: a strong enough disorder (a system-dependent feature) can destroy the ideal glass transition, as may happen for the RFIM. Another relation with the RFIM was also found in Ref.~\cite{biroli2012random}, where it was shown that amorphous interfaces between rearranging regions behave statistically as the ones of domain walls in the RFIM.

Let us conclude with a word of caution: not everything is understood about self-induced heterogeneity. Since the disorder is strongly linked to the state of the system, it is  also to a large extent renewed after a time $\tau_\alpha$ , thus it {evolves} and at the same time it {affects} the dynamics, i.e.~it is not truly quenched. This is a first difficulty in assessing precisely its role for glassy relaxation~\cite{berthier_can_2019}. A second one is that although for static properties, such as configurational entropy or the Franz-Parisi potential, one can establish a mapping to the RFIM, for the dynamics the situation is more intricate and no mapping has been found up to now~\footnote{The difficulty is that the mapping to the RFIM proceeds by relating the overlap (for glasses) to the magnetization (for the RFIM). There are no natural dynamical equations for the overlap, and in the only cases where those have been established--the $\beta$-regime of MCT--these proved to be quite complex and different from the corresponding equations for the magnetization of the RFIM.}.

\subsubsection{Renormalization group for the glass transition}

In parallel with the efforts described in the previous section, developing a renormalization group (RG) analysis of the glass transition has been a new important theoretical activity in the last decade. Different methods have been used, and were applied on lattice disordered models and replica lattice field theories which display a glass transition. Given that the ideal glass transition has a mixed character, intermediate between first and second-order phase transition, usual perturbative RG techniques developed 
for continuous phase transitions do not work. 
Therefore researchers had to focus on non-perturbative methods. In Ref.~\cite{castellana2010hierarchical} the hierarchical Dyson RG method was employed to analyse the Random Energy Model in finite dimension. The authors found an ideal glass phase transition similar to the one taking place within mean-field but with a non-analytical behavior of the free-energy at $T_K$, leading in particular to a specific heat exponent different from one, as assumed within RFOT theory. The real-space properties, correlation length and energy barrier, and the nature of the fixed-point were first studied in \cite{yeo2012origin,cammarota2011renormalization} by Migdal-Kadanoff RG. A complete and more advanced analysis was performed in \cite{angelini2017real}, in which it was shown that 
the ideal glass transition is associated to a so-called {zero temperature fixed point} \cite{fisher1986scaling}. The main implication is that the correlation length and the typical energy barrier have power law divergences with non-trivial exponents, hence implying a super-Arrhenius behavior. 
However, such a fixed point was found only in dimensions higher than three. 
In consequence, this RG treatment predicts that in three dimensions the glass transition is actually an avoided phase transition \cite{gillesphysica}: glassy behavior is still driven by the RG fixed point present in higher dimension, but the correlation length and the timescale do not truly diverge (arguably an irrelevant fact since one 
cannot approach the transition close enough, but an important conceptual one).

The RG approaches we reviewed above offer a new perspective on the nature of the glass transition. They provide important guidelines for more controlled non-perturbative RG treatments. Ideally, one would like to tackle 
directly interacting particle systems in the continuum and use 
methods that have been proved to be precise and reliable in previous studies, such as the one developed by Wetterich \cite{berges2002non}. This is a formidable challenge 
as those techniques do not seem to be able to handle 
the kind of rare and localized non-perturbative events that are relevant for glassy dynamics \cite{rulquin2016nonperturbative}. 

\subsection{Free volume, defects, and facilitated models}

\label{theorykcm}

\subsubsection{Lattice gases}

In this subsection we motivate and briefly  summarize
studies of a different family of statistical mechanics models
that turns out to yield a rich variety of physical behaviors.
Their starting point are physical
assumptions that might seem similar to the models described in
Sec.~\ref{giuliotheory}, but the outcome yields a different
physical explanation of the glass transition. Although the two theoretical
approaches cannot be simultaneously
correct, they both
have been influential and very instructive in order to develop
a theoretical understanding of glassy phenomena.

As in Sec.~\ref{giuliotheory}, we first consider hard sphere systems. 
We follow the lattice gas description introduced by Kob and Andersen~\cite{KAgas}, 
and work on a three dimensional cubic lattice.
As in a hard sphere system, we assume no interaction
between particles beyond the hard-core constraint
that the occupation number $n_i$ at site $i$ is at most
equal to 1,
\begin{equation}
H[\{ n_i \}] = 0, \quad n_i = 0, \, 1.
\label{hamka}
\end{equation}
In contrast to the lattice glass model, all configurations
respecting the hard-core constraint are allowed and are equally
probable. Geometric frustration is instead introduced at the level
of the kinetic rules, that are defined as constrained
local moves. Namely, a particle can jump to a nearest neighbor site
only if that site is empty (to satisfy the hard-core constraint),
but, additionally, only if the sites occupied before and after the move
have less than $m$ neighbors, $m$ being an adjustable parameter,
which Kob and Andersen choose as $m=4$ for $d=3$ ($m=6$ corresponds
to the unconstrained lattice gas). The model captures
the idea that if the liquid if locally very dense,
no movement is possible while regions with low density
move more easily.

Of course, such kinetically constrained lattice gases
have been studied in various spatial dimensions, for different values
of $m$, for different constraints, or even different
lattice geometries~\cite{solrit}.
These models capture the idea of a `cage' effect
in a strict sense, meaning that a particle with a
dense neighbor shell cannot diffuse.
Although the cage seems a purely local concept,
it turns out that diffusion in constrained lattice gases
arises from cooperative rearrangements, so that
slow dynamics can be directly shown to be driven
by the growth of dynamic lengthscales for these cooperative
moves~\cite{silvioKA,TBF,panetal}.
This strongly suggests that such cooperative moves most
probably have a role in the dynamics of real liquids.

\subsubsection{Free volume, dynamic criticality}

In the lattice gas picture, the connection with the liquid
is not obvious because it is the density (`free volume') rather than the temperature that
controls the dynamics. Thermal models with similar features
can in fact be defined along the following lines. In a liquid,
low temperature implies a very small
probability to find a location with enough free volume to move.
The idea of a small concentration of `hot spots' is in fact
reminiscent of another picture of the glass transition
based on the idea of `defects' which is captured by
the defect model proposed by Glarum~\cite{glarum} in the 60's,
where relaxation proceeds via the diffusion
of a low concentration of independent defects. In the mid-80's,
using both ideas of kinetic constraints
and rare defects, Fredrickson and Andersen
defined a family of kinetic Ising models for the glass
transition~\cite{FA}.
They studied an assembly of non-interacting spins,
\begin{equation}
H[\{ n_i \}] = \sum_{i=1}^N n_i , \quad n_i = 0, \, 1,
\label{hamfa}
\end{equation}
where $n_i=1$ represents the defects, whose concentration
becomes exponentially small at low temperature,
$\langle n_i \rangle \approx \exp(-1/T)$. As for the Kob-Andersen
lattice gas, the non-trivial ingredient lies in the chosen rates
for the kinetic transitions between states. The kinetic rules stipulate
that a transition at site $i$ can happen with a
usual Glauber rate, but only if site $i$ is surrounded by
at least $k$ defects ($k=0$ corresponds to the unconstrained
limit). Again, one can easily imagine studying such models
in different spatial dimensions, on different lattices, and with
slightly different kinetic rules, yielding a large number of possible
behaviors~\cite{solrit,leonard}.
The similarity between those spin facilitated models
and the kinetically constrained lattice gases is striking.
Altogether, they form a large family of models generically called
kinetically constrained models (KCMs)~\cite{solrit}.

The connection between KCMs and the much older
concept of free volume is obvious from our presentation.
Free volume models are among the most widely used models to
analyze experimental data, especially in polymeric systems.
They have been thoroughly reviewed before~\cite{debenedetti,freevol}, and the
main prediction is that dynamic slowing down occurs because
the free volume available to each particle, $v_f$, vanishes at some
temperature $T_0$ as $v_f \approx \alpha(T-T_0)$. 
Statistical arguments then relate relaxation timescales to free volume
assuming that a movement is possible if locally there is `enough'
available free volume, more than a typical value
$v_0$. This is clearly reminiscent of
the above idea of a kinetic constraint for local moves in
lattice gases.
An appealing VFT divergence is then predicted:
\be
\frac{\tau_\alpha}{\tau_0} \sim \exp\left(\gamma \frac{v_0}{v_f}\right) \sim
\exp \left( \frac{\gamma v_0/\alpha}{[T-T_0]^{\mu}} \right),
\label{vftfree}
\ee
where $\gamma$ is a numerical factor and $\mu=1$.
Predictions such as Eq.~(\ref{vftfree}) justify the wide use
of free volume approaches, despite the many (justified) criticisms
that have been raised.

Initially it was suggested that KCMs would
similarly display finite temperature
or finite density dynamic transitions
similar to the one predicted by the
mode-coupling theory of supercooled liquids~\cite{FA},
but it was soon realized~\cite{FA2,harro} that most KCMs do not display such
singularity, and timescales in fact only diverge in the
limit of zero temperature ($T=0$) or maximal density ($\rho=1$).
Models displaying a $T_c>0$ or $\rho_c<1$
transition have also been introduced and analyzed \cite{TBF2}. They provide
a microscopic realization, based on well-defined statistical mechanics models,
of the glass transition predicted by free volume arguments. Their relaxation
timescale diverges with a VFT-like form but with an exponent $\mu \simeq 0.64$.
Understanding their universality classes or the degree of generality of the mechanism leading to the transition is still an open problem
\cite{Elmatad2009,elmatad_finite-temperature_2010,Elmatad2012}; the most recent results on this front come from mathematical physicists who have been able to classify many of the different possible behaviors on the basis of the microscopic dynamical rules \cite{hartarsky2019universality,martinelli2019universality,hartarsky2019universality2,martinelli2019towards}. 

Extensive studies have shown that
KCMs have a macroscopic behavior which resembles
the phenomenology of supercooled liquids, displaying in particular
an Arrhenius or super-Arrhenius increase of
relaxation timescales upon decreasing temperature, 
and non-exponential relaxation functions at
equilibrium~\cite{solrit}.
Early studies also demonstrated that,
when suddenly quenched to very low temperatures,
the subsequent non-equilibrium aging dynamics of these models
compares well with experimental observations on the aging
of liquids~\cite{FA2}.
The diverse definitions of such models
suggest a broad variety of different behaviors. 
This feature is both positive and negative:
on the one hand one can explore various scenarios to describe
glass transition phenomena, but on the other hand,
one would like to be able to decide what particular model
should be used to get a quantitative description for a particular liquid.
It is not straightforward to perform microscopic predictions using the framework of KCMs, since there is no direct observable parameter to use as input of the theory (unlike $g(r)$, for MCT). An operative definition of KCMs' defects was provided (see next section and \cite{keys_calorimetric_2013,Keys2011}), however this does not allow to directly choose which KCM (which rules) are appropriate for a given liquid.

Despite this caveat, it is quite useful to use KCMs as
theoretical tools to define concepts and obtain new ideas.
It is precisely in this perspective that
interest in KCMs has increased, in large part
since it was realized that their dynamics is spatially
heterogeneous~\cite{harro,silvioKA,gc},
a central feature of supercooled liquids dynamics.
In particular, virtually all the aspects
related to dynamic heterogeneity mentioned in Sec.~\ref{dh}
can be investigated and rationalized, at least qualitatively, in terms of KCMs. The dynamics of these systems can be understood by considering where `relaxation' happens and then propagate, which is dictated by the underlying defect motion~\cite{solrit}.
Depending on the particular model, defects
can diffuse or have a more complicated motion. 
Furthermore, they can be point-like or `cooperative'
(formed by point-like defects moving in a cooperative way).
A site can relax only when it is visited by a defect. 
As a consequence, the heterogeneous character of the dynamics is entirely
encoded in the defect configuration and defect motion~\cite{gc}.
For instance, a snapshot similar to Fig.~\ref{peter} in a
KCM shows clusters which have
relaxed within the time interval $t$~\cite{steve,nef}.
These are formed by all sites visited by
a defect between $0$ and $t$. 
The other sites are instead frozen in their initial state.
In these models the dynamics slows down
because the defect concentration decreases. 
As a consequence, in the regime
of slow dynamics there are few defects and strong dynamic heterogeneity.
Detailed numerical and analytical studies 
have indeed shown that in these systems,
non-exponential relaxation patterns do stem
from a spatial, heterogeneous distribution of timescales,
directly connected to a distribution of dynamic
lengthscales~\cite{gc,steve,mayerjack,TBF,TBF2,panetal}.
Decoupling phenomena also appear naturally in KCMs and can be shown
to be very direct, quantifiable, consequences of
the dynamic heterogeneity~\cite{jung}, which also deeply
affects the process of self-diffusion in a system close to
its glass transition~\cite{berthierepl}. More fundamentally,
multi-point susceptibilities and multi-point spatial
correlation functions such as the ones defined
in Eqs.~(\ref{chi4def}) and (\ref{g4def}), can be studied
in much greater detail than in molecular systems, 
relating their evolution to time and length scales~\cite{TWBBB,panetal,jcpII,chandler,steve2}.
This type of scaling behavior
has been observed close to $T=0$ and $\rho=1$ in spin models
and lattice gases without a transition~\footnote{A critical (different) behaviour is expected and predicted for models having a transition \cite{TBF2}.}. 
Different theoretical approaches have shown that 
these particular points of the phase diagram correspond to genuine critical
points where timescales and dynamic lengthscales diverge with well-defined
critical laws~\cite{steve2,mayerjack}. Such `dynamic criticality' implies the existence of universal scaling behavior in the physics of supercooled liquids,
of the type reported for instance in Fig.~\ref{cecile}.

\subsubsection{Defects: connection with Hamiltonian dynamics models}

A central criticism about the free volume approach,
that is equally relevant for KCMs, concerns the identification,
at the molecular level, of the vacancies (in lattice gases),
mobility defects (in spin facilitated models), or of the free volume itself.
The attempts to provide reasonable coarse-graining
from molecular models with continuous degrees
of freedom to lattice models with kinetic rules have been, for a very long time,
quite limited and not fully convincing~\cite{kennett,glotzersilica}.

For molecular models, this issue was first attacked in 2010, 
with a definition of the cumulated dynamical activity $K$, 
an extensive quantity characterizing the frequency of state changes 
(from excited to non-excited and vice versa) 
\cite{hedges_dynamic_2009, elmatad_finite-temperature_2010}. 
This definition was made more concrete and studied in models of supercooled liquids in \cite{Keys2011}, where an appropriate functional is explicitly designed as a recorder of excitations (or defects). Defects or excitations are not to be mistaken with displacements. Indeed, some locations providing opportunities for structural reorganization do coincide with defects, and their presence can be inferred by observing nontrivial particle displacements associated with transitions between relatively long-lived configurations. Displacements instead refer to dynamical moves in short segments of a trajectory, while defects refer to underlying configurations. The explicit definition of defects has since been used to estimate the role of facilitation in glassy dynamics, to provide a microscopic validation of KCMs~\cite{Keys2015,Isobe2016}, to explain dynamical heterogeneities in glassy materials, or to compute the dynamical facilitation volume~\cite{Elmatad2012}.

The conceptual proof that kinetic rules
emerge effectively and induce a slow dynamics
has been obtained for simple lattice spin models~\cite{juanpe},
with a dynamics that directly maps
onto constrained models. A deeper study of this kind of model was performed more recently~\cite{turner_overlap_2015}. 
Several examples are available
but here we only mention the simple case of the bidimensional
plaquette model defined by
a Hamiltonian of a $p$-spin type on a square lattice
of linear size $L$,
\be
H = -J \sum_{i=1}^{L-1} \sum_{j=1}^{L-1} S_{i,j} S_{i+1,j}
S_{i,j+1} S_{i+1,j+1},
\label{spm}
\ee
where $S_{i,j}=\pm 1$ is an Ising variable lying at node $(i,j)$ of the lattice.
Contrary to KCMs, the Hamiltonian in Eq.~(\ref{spm})
contains genuine interactions, 
which are no less (or no more) physical than $p$-spin models
discussed in Sec.~\ref{giuliotheory}. Interestingly
the dynamics of this system is (trivially) mapped onto
that of a KCM by analyzing its behavior in terms of plaquette
variables, $p_{i,j} \equiv S_{i,j} S_{i+1,j}
S_{i,j+1} S_{i+1,j+1}$, such that the Hamiltonian
becomes a non-interacting one, $H = -J \sum_{i,j} p_{i,j}$,
as in Eq.~(\ref{hamfa}).
More interestingly, the analogy also applies to the dynamics~\cite{juanpe}.
The fundamental moves are spin-flips, but when a single spin is
flipped the states of the four plaquettes surrounding that spin
change. Considering the different types of moves, one quickly realizes
that excited plaquettes, $p_{i,j} = + 1$, act as
sources of mobility, since the energetic barriers to spin flips
are smaller in those regions. This observation allows to
identify the excited plaquettes as defects,
by analogy with KCMs. Spatially heterogeneous dynamics,
diverging lengthscales accompanying diverging timescales
and scaling behavior sufficiently close to
$T=0$ can be established by further analysis~\cite{spm},
providing a simple but concrete example of how an interacting
many body system might effectively behave as a model with
kinetic constraints~\footnote{This type of plaquette models, and
other spin models, were introduced originally~\cite{Lipowsky,Sethna}
to show how ultra-slow glassy dynamics can emerge because
of growing free energy barriers.}.

\subsubsection{Connection with other perspectives}

An essential drawback of facilitated models is that among the microscopic
`details' thrown away to arrive at
simple statmech models such as the ones in Eqs.~(\ref{hamka}) and
(\ref{hamfa}), information on the thermodynamical behavior
of the liquids has totally disappeared. In particular,
a possible coincidence between
VFT and Kauzmann temperatures, $T_0$ and $T_K$ is not expected,
nor can the dynamics be deeply connected to thermodynamics,
as in Adam-Gibbs relations.
The thermodynamical behavior of KCMs appears
different from the one of real glass-formers close to $T_g$~\cite{BBT}.
This is probably the point where
KCMs and RFOT approaches differ most obviously.
Even though the dynamics of KCMs shares similarities
with systems characterized with a complex energy
landscape~\cite{nontopo,steve3},
thermodynamical behaviors are widely different
in both cases, as has been recently highlighted in Ref.~\cite{rob2}
by focusing on the concrete examples of
plaquette models such as in Eq.~(\ref{spm}).

Finally, when KCMs were first defined, they were argued to display
a dynamical transition of a very similar nature
to the one predicted by MCT~\cite{FA}. Although the claim
has been proven wrong~\footnote{Most KCMs do not have a finite
temperature dynamical transition and the ones displaying
a transition have critical properties different
from MCT.}, it bears some truth: both approaches
basically focus on the kinetic aspects of the glass transition
and they both predict the existence of some dynamic criticality with
diverging lengthscales and timescales. This
similarity is even deeper, since a mode-coupling singularity is
present when (some) KCMs are studied on the
Bethe lattice~\cite{TBF}, but is `avoided' when more realistic
lattice geometries are considered~\cite{berthier2012finite}.

\subsection{Geometric frustration, avoided criticality, and locally preferred structures}

\label{subsec:frustration}

In all of the above models,
`real space' was present in the sense that special attention was paid to
different lengthscales characterizing the physics of the models
that were discussed. However, apart from the `packing models' with
hard-core interactions,
no or very little attention was paid to the geometric structure
of local arrangements in molecular liquids close to a glass transition.
This slight oversight is generally justified using concepts such as
`universality' or `simplicity', meaning that
one studies complex phenomena using simple models, a typically
statistical mechanics perspective.
However, important questions remain: what is the liquid structure
within mosaic states? How do different states differ? What is the
geometric origin of the defects invoked in KCMs? Are they similar
to defects found in crystalline materials (disclinations, dislocations, vacancies, etc.)?
Some lines of research attempt to provide answers to these questions, making heavy use of the concept of geometric frustration.

\subsubsection{Geometric frustration}

Broadly speaking, frustration refers to the
impossibility of simultaneously minimizing all
the interaction terms in the energy function of
a system. Frustration might arise from quenched disorder
(as in the spin glass models described above), but liquids have
no quenched randomness.
In liquids, instead, frustration has a purely geometrical origin.
It is attributed to a competition between a short-range
tendency for the extension of a `locally preferred order', and
global constraints that prevent the periodic tiling of space
with this local structure.

This can be illustrated
by considering once more the packing problem of spheres
in three dimensions. In that case, locally the preferred cluster of
spheres is an icosahedron. However, the 5-fold rotational symmetry
characteristic of icosahedral order is not compatible
with translational symmetry, and formation of a periodic
icosahedral crystal is impossible~\cite{frank}.
The geometric frustration that
affects spheres in three dimensional Euclidean space
can be relieved in curved space~\cite{Nelson}.
In Euclidian space,
the system possesses topological
defects (disclination lines), as the
result of forcing the ideal icosahedral
ordering into a `flat' space.
Nelson and coworkers developed a solid theoretical framework based
on this picture to suggest that the slowing down of supercooled liquids
is due to the slow wandering of these topological defects~\cite{Nelson},
but their treatment remains too abstract to obtain quantitative, explicit results.
Further theoretical work extended the integral-equation approach to calculate the pair correlation function in hyperbolic geometry, making it easier to compare predictions and simulation data~\cite{sausset_thermodynamics_2009}. 
MCT was also re-derived in curved space (on a sphere), showing that it still cannot capture quantitatively the transition
\cite{vest_dynamics_2014,vest_mode-coupling_2015}.
Explicit numerical results were also obtained recently~\cite{Turci2017},
showing a clear first-order like transition from liquid to ordered solid in an appropriately curved space, becoming avoided as Euclidean space is retrieved. 

\subsubsection{Coulomb frustrated theories}

The picture of sphere packing disrupted by frustration has been
further developed in simple statistical
models characterized by geometric frustration,
in a pure statistical mechanics
approach~\cite{gillesreview}.
To build such a model, one must be able to identify, then
capture, the physics of geometric frustration. Considering
a locally ordered domain of linear size $L$,
Kivelson {\it et al.}~\cite{gillesphysica}
suggest that the corresponding free energy scales as
\be
F(L,T) = \sigma(T) L^2 - \phi(T) L^3 +s(T) L^5.
\label{fgilles}
\ee
The first two terms express the tendency of growing locally preferred order
and represent respectively the energy cost of having an interface
between two phases and a bulk free energy gain inside the domain.
Geometric frustration is encoded in the third term which represents
the strain free energy resulting from frustration.
The remarkable feature of Eq.~(\ref{fgilles}) is the super-extensive
scaling of the energy cost due to frustration which opposes
the growth of local order.
The elements in Eq.~(\ref{fgilles}) can then be directly
incorporated into ferromagnetic models where `magnetization'
represents the local order, ferromagnetic interactions represent the
tendency to local ordering, and Coulombic antiferromagnetic
interactions represent the opposite effect, coming from the frustration. The
following Hamiltonian
possesses these minimal ingredients:
\be
H = -J \sum_{\langle i,j \rangle} {\bf S}_i \cdot {\bf S}_j
+ K \sum_{i \neq j} \frac{{\bf S}_i \cdot {\bf S}_j }{|{\bf x}_i
- {\bf x}_j|},
\label{coulomb}
\ee
where the spin ${\bf S}_i$ occupies the site $i$ at position
${\bf x}_i$.
Such Coulomb frustrated models have been studied in great detail, using
various approximations to study models for various space and spin
dimensions~\cite{gillesreview}.

The general picture is that the ferromagnetic
transition occurring at $T=T_c^0$ in the pure model with no frustration ($K=0$)
is either severely displaced to lower temperatures for $K>0$,
sometimes with a genuine discontinuity at $K\to0$, yielding the
concept of `avoided criticality'. For the simple case of
Ising spins in $d=3$, the situation
is different since the second order transition becomes
first-order between a paramagnetic phase and a spatially modulated phase
(stripes). For $K>0$ and $T < T_c^0$ the system is described
as a `mosaic' of domains corresponding to some local order, 
the size of which increases (but does not diverge!) when $T$ decreases.
Tarjus, Kivelson and co-workers clearly
demonstrated  that such a structuration
into mesoscopic domains allows one to understand most of the fundamental
phenomena occurring in supercooled liquids~\cite{gillesreview}.
Their picture as a whole
is very appealing because it directly addresses the
physics in terms of the `real space', and the presence of domains
of course connects to ideas such as cooperativity, dynamic heterogeneity
and spatial fluctuations, that directly explains, at
least qualitatively, non-exponential
relaxation, decoupling phenomena or super-Arrhenius increase of the
viscosity. However, as for the RFOT mosaic picture, direct
confirmations of this scenario are rare~\cite{coslovich}, or difficult to obtain.

\subsubsection{Locally preferred structures}

Going back to the geometric structure of local arrangements in real space, 
a line of research has emerged that is based upon some kind of local order~\cite{Coslovich2011,royall2015role,malins2013identification,malins2013identificationb}. Icosahedral order was initially shown to be linked to the dynamics of some binary mixtures. But more generally, for simple enough glass-formers, a broader variety of locally ordered structures  can be defined~\cite{royall2015role}, such as, e.g., defective icosahedron~\cite{Royall2017} A vast body of literature has shown that these locally preferred structures (LPS) seem to correlate with structural relaxation.
Using trajectory path sampling (see Sec.~\ref{subsec:s-ensemble}), it was found that LPS and dynamic activity play equivalent roles, and are therefore strongly correlated~\cite{turci_nonequilibrium_2017,turci_structural-dynamical_2018}. The increased number of LPS has also been linked to hindered crystallization~\cite{turci_devitrification_2019}. These multiple effects suggest the LPS impacts the dynamics in various ways. 

For several systems in $d=2,3$, simple local order parameters 
measuring the distance from sterically favored structures 
(sixfold symmetry, angles and closeness to local tetrahedron) 
have been spatially coarse-grained \cite{tong_revealing_2018}, thus revealing a clear structure-dynamics relationship. The interpretation in Ref.~\cite{tong_revealing_2018} is that these LPS represent an indirect measurement of the true amorphous order. 
The very simple and widespread tetrahedral order has been observed 
in a variety of materials \cite{shi_distinct_2019}. 
The LPS-dynamics correlation has also been confirmed recently in experiments on colloids~\cite{pinchaipat_experimental_2017,hallett2018local},  
providing further evidence of the important role of LPS in glassy behavior.

Despite these important advances, the concrete application of LPS-based techniques 
on any particular realistic glass-forming material is hindered 
by the lack of a universal operative definition of the LPS. For relatively simple systems, an operative scheme was designed to automatically find these LPS \cite{mossa_operational_2006,royall2015role}.

A natural alternative to this tedious exercise is the use of modern machine learning techniques. In particular, one may consider the unsupervised learning task 
of grouping together similar local structures (this is called clustering). Once clusters (corresponding to automatically-found LPS) are found, any local environment may be assigned to its most similar group (LPS), but without the burden of defining the LPS's one by one~\cite{ronhovde2011detecting,ronhovde2012detection,paret2020assessing,boattini2020autonomously}.

\section{Mean-field theory of the amorphous phase}

\label{theoryamorphous}

The phase transition between liquid and glass 
is not the only interesting phenomenon 
characterising the phase diagram of glassy materials. 
Since the transition occurs at finite pressure and temperature, 
glasses can be further compressed or cooled within the glass phase itself~\cite{KurParUrbZam,KurParZam,BiroliUrbani,RainoneUrbani,RainoneUrbaniYoshinoZamponi,softmeanfield}. 
How do physical properties of glasses change in this context? 
In mean-field theory, this question has been widely 
investigated by using the hard spheres glass model~\cite{ParisiSlanina, parisi_mean-field_2010}, a favorite canonical example of a glass-former system because of its analytical simplicity. Eventually, by compressing a hard sphere glass, 
the system undergoes the jamming transition in the limit of infinite pressure~\cite{donev2004jamming}. 
In this section, we briefly survey recent progress 
in the development of an analytic theory of 
the glass phase in the large $d$ limit, with a particular emphasis on 
hard spheres~\cite{parisi2020theory}. 

\subsection{Mean-field glassy phase diagrams}

\label{subsec:gardnerJamming}

When a glass-forming liquid undergoes the glass transition, 
it becomes confined into a single free energy minimum 
and the timescale to explore different minima becomes infinite. 
It is formally possible to define thermodynamic properties 
by restricting the available statistical configurations 
to a single free energy minimum.
This can be enforced in the replica formalism 
by considering two copies of the system and constraining 
the distance between them~\cite{glassjammingreview}. 
First, an equilibrium reference configuration
$\underline{Y}$ at $(T_g, \hat{\varphi}_g)$ is introduced, 
where $\hat{\varphi}$ is the scaled packing fraction $\hat{\varphi} = 2^d\varphi /d$. 
Second, a copy of the equilibrium configuration $\underline{X}(t)$ 
is created and evolved in time. Let us define now the mean-squared displacement (MSD) 
between the two copies as $\overline{\langle \Delta(\underline{X}, \underline{Y})\rangle} = \Delta_r$. 
The properties of $\underline{X}(t)$ are sampled 
in a restricted region of phase space close to the equilibrium configuration. 
Within this state following construction, 
the system at $(T_g, \hat{\varphi}_g)$ with initial configuration $\underline{Y}$ 
can be adiabatically followed anywhere in the glass phase diagram.

Concretely, for the glass state selected by $\underline{Y}$ and followed until $(T,\hat{\varphi})$, we can write the restricted partition function as:
\begin{equation}
Z[T, \hat{\varphi}|\underline{Y}, \Delta_r] = \int d\underline{X} e^{-\beta V(\underline{X})} \delta(\Delta_r - \Delta(\underline{X},\underline{Y})) ,
\end{equation}
where $V(\underline{X})$ is the potential energy 
of the configuration $\underline{X}$, 
and the delta function enforces the restricted average. 
In order to obtain the glass free energy, 
we need to compute its average over 
the chosen reference configuration $\underline{Y}$, 
which acts as a source of quenched disorder:
\begin{equation}
\begin{aligned}
f_g(T, \hat{\varphi}|T_g, \hat{\varphi}_g, \Delta_r) = & -\frac{T}{N} \int \frac{d\underline{Y}}{Z[T_g, \hat{\varphi}_g]} e^{-\beta_g V(\underline{Y})} \\
& \times \ln{Z[T, \hat{\varphi}|\underline{Y}, \Delta_r]}
\end{aligned}
\end{equation}
where $Z[T_g, \hat{\varphi}_g]=\int dY \exp^{-\beta_g V(\underline{Y})}$
is the partition function at $(T_g, \hat{\varphi}_g)$.
Mathematically, the quenched disorder is handled using the replica method. 
We then introduce $(n+1)$ replicas of the original system, 
with the initial glass at $(T_g, \hat{\varphi}_g)$
being the master replica, while
the $n$ other slave replicas describe the
glass at $(T, \hat{\varphi})$.
The glass free energy is finally 
expressed in terms of the average MSD between 
the slave replicas and the master replica $\Delta_r$, 
and the average distance between the slave replicas $\Delta$. 
At this step, we assume that the symmetry between
slave replicas is not broken, which corresponds to the 1RSB ansatz described in Sec.~\ref{giuliotheory}.

By choosing the state point at $(T, \hat{\varphi})=(T_g, \hat{\varphi}_g)$, 
the recursive equations for $\Delta$ and $\Delta_r$ 
have to satisfy $1/{\hat{\varphi}} = \mathcal{F}_\beta(\Delta)$,
where $\mathcal{F}_\beta(\Delta)$ is a positive function
which vanishes for both $\Delta \rightarrow \infty$ and $\Delta \rightarrow 0$,
with an absolute maximum in between.
This equation can then be satisfied only if
\begin{equation}
\frac{1}{\hat{\varphi_d}} \leq \max_\Delta \mathcal{F}_\beta(\Delta).
\end{equation}
This condition occurs for volume fractions larger than a critical value $\hat{\varphi}_d(\beta_g)$, which corresponds to the dynamical glass transition.

\begin{figure}
\psfig{file=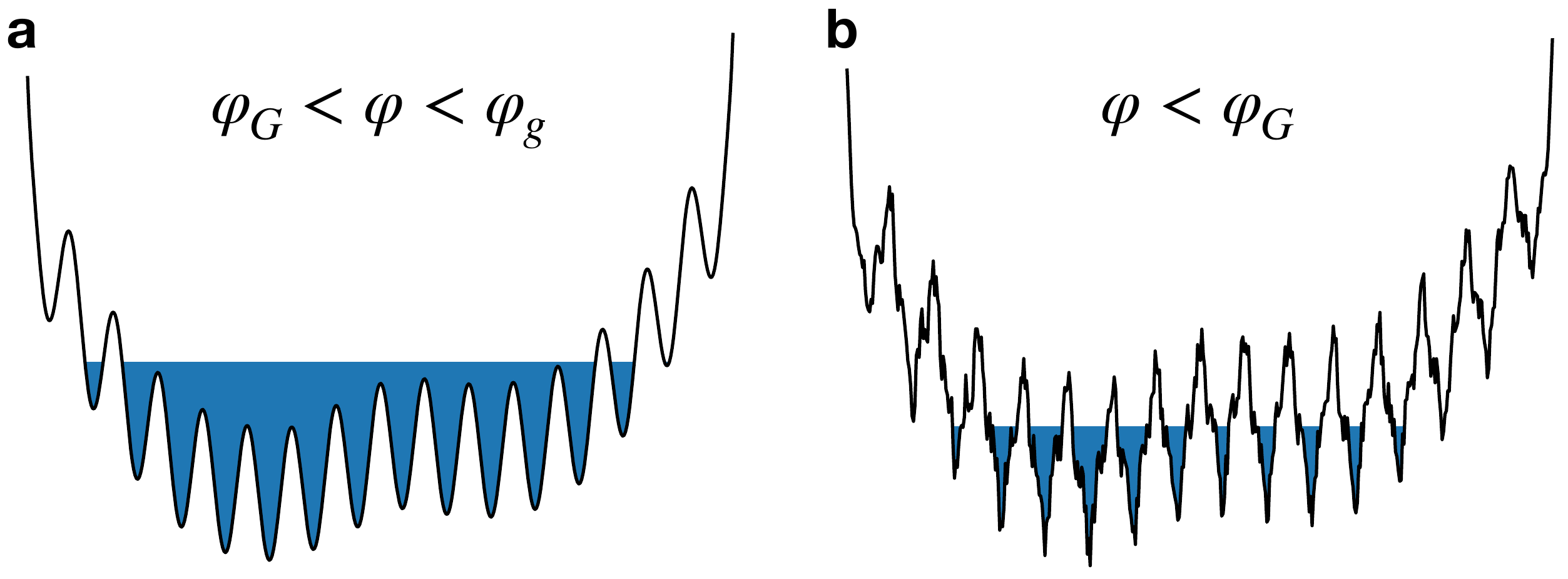,width=8.5cm}
\caption{(a) Sketch of the free energy structure deep in the hard sphere glass phase, where each basin breaks down in sub-basins corresponding to secondary relaxations. At the Gardner transition in (b), the sub-basins become fractal and ergodicity is broken.}
\label{landscape-marginal} 
\end{figure}

We can explore the glass phase following 
the glass prepared at the glass transition $(T_g, \hat{\varphi}_g)$ 
at different temperatures and packing fractions. At low $T$ and high $\hat{\varphi}$
one eventually meets another phase transition~\cite{KurParUrbZam}, 
where the 1RSB assumption fails~\cite{beyond} 
and the more complex full-replica symmetry breaking (fullRSB) 
solution is necessary to compute the glass free energy, 
the so-called Gardner phase transition~\cite{Gardner,berthier2019gardner}. 
Here, the fullRSB solution corresponds to a 
hierarchical organisation of the distances 
between the slave replicas and 
the glass becomes marginally stable~\cite{RainoneUrbani,universality_franz_2017}. 
The emergence of a complex free energy landscape 
gives rise to non-trivial dynamical processes~\cite{CamilleLudo,liao2019hierarchical,scalliet2019nature}. 
A pictorial representation of the Gardner transition is shown in Fig.~\ref{landscape-marginal}. 

\begin{figure}
\psfig{file=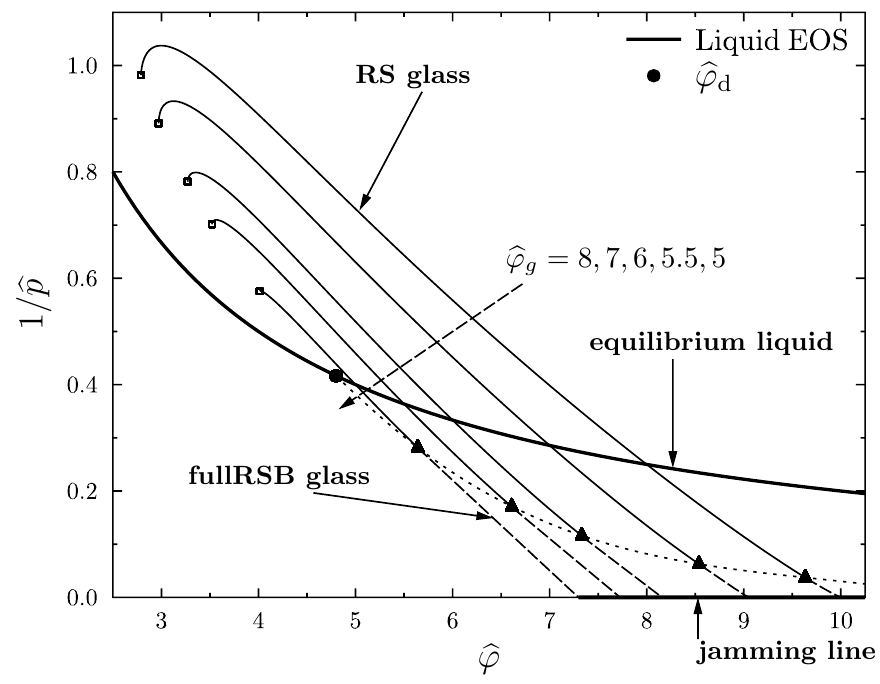,width=8.5cm}
\caption{Phase diagram of hard spheres
in the inverse reduced pressure -- reduced packing fraction  $(1/\hat{p}, \hat{\varphi})$ plane. The glass transition is marked by a full circle. The glass equations of state are reported as full lines in the region where the replica symmetric solution is stable.
The Gardner transition is marked by triangles, beyond which the fullRSB solution is stable (dashed lines). The glass equations of state end at the jamming transition.
Upon decompression, glasses are stable until a spinodal instability arises (open squares).}
\label{hardphasediagram} 
\end{figure}

It is worth noting that the derivation sketched above 
is completely general and can be used for 
any glassy pair potential mentioned in Sec.~\ref{giuliotheory}. 
In the following we will apply this formalism 
to the hard spheres model, for which several implications 
from the mean field picture have been 
successfully tested numerically~\cite{PNASgardner,charbonneau_exact_2014}. 
Here, the relevant state parameter 
is the scaled reduced pressure $\hat{p} \equiv \beta P/\rho d$.
We refer to Refs.~\cite{berthier2019gardner,biroli2016breakdown,scalliet2017absence,scalliet2019nature,softmeanfield,BiroliUrbani} for more results regarding systems made of soft potentials. 

Starting from an equilibrated hard sphere liquid configuration at $\hat{\varphi}_g$, 
we can apply the state following formalism to explore the hard sphere phase diagram in Fig.~\ref{hardphasediagram}.
The reduced pressure can be computed from the equation of state of an infinite dimensional hard sphere liquid $\hat{p} \sim \hat{\varphi}/2$, derived from a Virial expansion of the free energy~\cite{MaimbourgKurchanZamponi}.
Starting from $\hat{\varphi}_g$ and decompressing the system,
the glass eventually undergoes a melting transition: 
the 1RSB solution becomes unstable and 
the glass melts into the liquid via a spinodal instability~\cite{KurParUrbZam}.
Upon compression instead, the glass enters deeper 
into the glass phase and remains dynamically arrested. 
Numerically, this has been proven by measuring $\Delta$ as the long-time limit of the MSD $\Delta(t)$  between the system at time $t$ and the initial configuration at $t=0$. 
The order parameter of the transition $\Delta_r$ 
is instead computed as the long-time limit of the distance $\Delta_{AB}(t)$ 
between two copies $A$ and $B$ of the same initial system 
evolved with different initial velocities:
\begin{equation}
\Delta_{AB} = \left\langle \frac{1}{N} \sum_{i=1}^N
| {\bf r}_i^A -{\bf r}_i^B |^2 \right\rangle .
\end{equation}

Upon further compression, the glass eventually undergoes 
the Gardner transition at a finite pressure $\hat{p}_G$.
Here, the relation between $\Delta_{r}$ and $\Delta$ breaks down 
and $\Delta(t)$ is characterized by a logarithmic growth in time, suggesting
the emergence of a complex free energy landscape~\cite{PNASgardner}. 
The copies $A$ and $B$ cannot occupy the same sub-basin 
and are no longer able to explore the entire metabasin. 
Due to the fractal nature of the free energy landscape, 
the excitations required to move around the fractal states 
correspond to soft modes~\cite{scalliet2019nature}. The correlation length of these modes can be estimated by measuring the dynamical susceptibility, 
computed as the variance of $\Delta_{AB}$, which indeed shows a divergence at the Gardner transition~\cite{PNASgardner}.

Compressing further within the Gardner phase, 
the pressure eventually diverges as 
the system reaches its jamming density $\hat{\varphi}_J$,
which depends explicitly on the selected initial condition $(T_g, \hat{\varphi}_g)$. 
In particular, there exists a range of jamming points, or 
a `jamming line'~\cite{chaudhuri2010jamming}, whose extension increases with $d$~\cite{KurParZam}. 

\subsection{Jamming}

\label{jammingscaling}

During the last two decades, a large research effort has shed light on the critical behavior characterizing the jamming transition~\cite{liu2010jamming}. Jamming can be seen from two different perspectives. An assembly of Brownian hard spheres under compression becomes rigid at a finite density, at which point the pressure diverges. On the other hand, athermal packings of soft repulsive spheres reach the jamming point under decompression when the pressure vanishes. In both situations, each particle is constrained by enduring contacts with the neighbor particles and the system is rigid.
In particular, at jamming the average number of contacts per particle $Z$ reaches the critical value $Z_c=2d$, 
which represents the lower limit for mechanical stability~\cite{ohern_random_2002} (Maxwell's criterion for rigidity). From the hard spheres side,
$Z$ jumps from zero to $Z_c$ at the transition,
while from the soft spheres side,
as the pressure decreases toward zero the excess number of contacts scales as~\cite{ohern_random_2003, durian95}:
\begin{equation}
\Delta Z \equiv Z - Z_c \sim \Delta\varphi^{1/2},
\end{equation}
where $\Delta\varphi = \varphi - \varphi_J$
is the amount of compression above the jamming threshold.
A connection between hard and soft spheres at jamming 
is observed in the pair correlation function~\cite{ohern_random_2003,torquato},
confirming that allowed configurations of hard and soft spheres are identical at jamming.

When $\Delta Z = 0$ the system is isostatic, i.e.
there are just enough contacts to ensure mechanical stability
and the system is marginally stable:
breaking a bond between contacts can lead to an excitation that causes
a collective motion throughout the whole system~\cite{wyart2012marginal}.
Not surprisingly, this critical behavior fits well into the free energy picture
of marginal glasses reported above.

Marginality in athermal jammed solids can be explained in real space by the so-called {cutting argument}~\cite{cutting-argument}.
Imagine removing the contacts between a subsystem of linear size $l$ and the rest of the system. If we slightly compress the system, this cutting will lead to a competition between the overall excess contacts $\Delta Z$ created by the compression, and the missing contacts at the boundary of the subsystem.
If the total number of contacts
is below the isostatic value $N_{iso} = NZ/2$,
then there are modes with no energetic cost, i.e.~soft modes.
The number of soft modes $N_{soft}$
then corresponds to the difference between
the number of contacts at the boundary,
proportional to $l^{d-1}$, and the number of
extra contacts created by the compression,
which scales as $\Delta Z l^d$.
There is then a critical length $l^\ast \sim \Delta \varphi^{-1/2}$
for which the system looks isostatic and for $l = l^\ast$, soft modes correlate over the whole subsystem. These extended anomalous modes correspond to random excitations 
over all the system, profoundly different from acoustic modes proper of crystalline solids.

Other anomalies of jammed solids are observed in the scaling of the elastic moduli near the transition.
These critical behaviors have been successfully
described within a force network picture,
for which en effective medium theory has been developed~\cite{wyart_scaling_2010, degiuli_lerner_wyart_theory_2015}. In particular, a jammed soft sphere configuration can be mapped onto a network of springs
with elastic contacts $k_{eff}$, computed
as second derivatives of the pairwise interaction between particles.
The resulting scaling behaviors for the bulk modulus
$B \sim k_{eff}$ and the shear modulus $G \sim k_{eff} \Delta \varphi^{1/2}$
suggest that the Poisson ratio $G/B \sim \Delta \varphi^{1/2}$
vanishes at the jamming transition~\cite{ohern_random_2003}.
This criticality reflects on the frequency of normal modes which is directly related to the elastic moduli ($B(\omega), G(\omega)$) by the dispersion relation $\omega^\ast = ck^\ast$, where $k^\ast \sim 1/l^\ast$ and $c$ is the speed of sound.
Since sound propagates either longitudinally (B) or transversely (G), two different length scales can be defined: 
the longitudinal length scale $l^\ast \sim \Delta \varphi^{-1/2}$, which matches the cutting length scaling behavior and is indeed attributed to extended soft modes, and 
the transverse length scale which follows the scaling $l_t \sim \Delta \varphi ^{-1/4}$.

Other critical scaling laws have been predicted
both by replica mean field calculations and
effective medium theory for a spring network,
with good consistency with numerical results in finite dimensions.
In particular, the distributions of interparticle voids
and interparticle forces follow universal
power-laws~\cite{charbonneau_universal_2012, charbonneau_jamming_2015, degiuli_force_2014, lerner13}. Contact forces can be either extended or localized, with distributions defined by power law exponents $\theta_e$ and $\theta_l$ respectively. Extended forces are predicted from the infinite dimensional exact solution, whereas the localized forces likely result from the presence of localised defects, such as rattling particles, which only exist in finite dimensions. Remarkably, the numerical value of the critical exponents associated to scaling laws near jamming can be predicted analytically in the mean-field approach~\cite{charbonneau_fractal_2014,charbonneau_exact_2014,parisi2020theory}, and their value is confirmed by numerical simulations in dimensions $d \geq 2$. 

The influence of temperature on the jamming criticality has also been studied~\cite{ikeda_berthier_biroli,degiuli_lerner_wyart_theory_2015}.
These works show that above jamming there exists a region in the plane $T - \varphi$
where the harmonic approximation of the soft sphere potential holds,
and the vibrational spectrum converges
to its zero temperature limit, provided that $T<T^\ast(\varphi)$.
The value of $T^\ast(\varphi)$ decreases with $\Delta \varphi \rightarrow 0$ with a trivial scaling exponent. A similar result holds below jamming for hard sphere glasses~\cite{BritoWyart}. For $T>T^\ast(\varphi)$, the harmonic approximation
breaks down, defining an anharmonic critical regime,
controlled by non-analyticities in the interparticle potential. Physically, strong anharmonicites stem from the constant breaking and reformation of particle contacts in the presence of thermal fluctuations~\cite{schreck2011repulsive}.

\subsection{Vibrational properties}

The anomalous thermal properties of low temperature glasses can be related
to the structure of the free energy landscape of glassy states. Amorphous solids behave very differently from crystalline solids.
In terms of heat capacity and thermal conductivity,
crystals are dominated by phononic excitations with a low-frequency
density of states (DOS) $D(\omega)$ given by the Debye scaling
law $D(\omega) \sim \omega^{d-1}$.
Instead, the thermal properties of glasses are dominated
by an excess of vibrational modes referred to as the boson peak and by an anomalous low-frequency scaling of $D(\omega)$. This excess of anomalous vibrations reflects, within mean-field theory, the existence of multiple free energy barriers in glassy states. In fact, when the glass enters the Gardner phase, the system becomes marginal and even infinitesimal perturbations lead to excitations that can bring the system to a different glassy state. 

\label{vibrations}
\begin{figure}
\includegraphics[width=8.5cm]{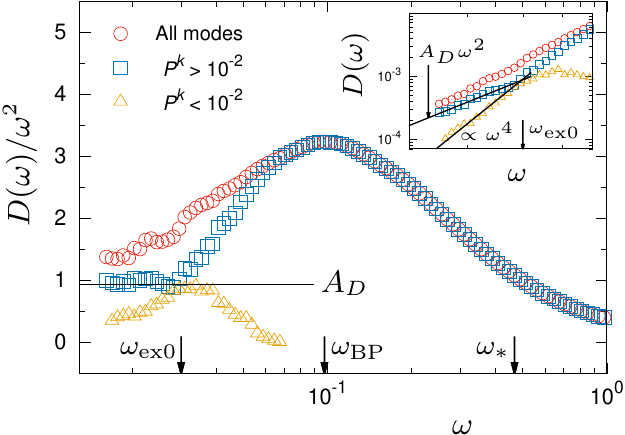}
\caption{Adapted from \cite{mizuno_continuum_2017}. Vibrational density of states of jammed harmonic soft spheres scaled by the Debye law $\omega^{d-1}$ in $d = 3$. Modes with index $k$ are classified as extended (blue) or localized (yellow) by their participation ratio $P^k$. Below the boson peak frequency $\omega_{BP}$, the density of states is the superposition of anomalous extended modes eventually obeying Debye scaling, and a population of quasi-localised modes scaling as $\omega^4$, as confirmed in the inset.}  
\label{fig:vdos} 
\end{figure}

The mean-field theory of glasses
has been explored using soft spheres in the jamming limit~\cite{franz_universal_2015}.
The theory predicts the low-frequency scaling of the vibrational density of states (vDOS) to be $D(\omega) \sim \omega^2$ in any
dimension~\cite{parisi_mean-field_2010, wyart_scaling_2010, jacquin_microscopic_2011}, quite differently from the Debye scaling.
The same result was previously obtained within the effective medium theory~\cite{degiuli_force_2014}.

Numerically, the nature of the low-frequency
vibrational spectrum has been widely studied
using soft spheres packings close to jamming.
Early studies suggested the existene of the $D(\omega) \sim \omega^2$
scaling~\cite{charbonneau_exact_2014, franz_universal_2015}
for a wide range of dimensions $d$, reinforcing the relevance of the mean-field description for finite dimensional systems~\cite{charbonneau_universal_2016}. The modes giving rise to this scaling form have been found to be extended anomalous modes. A more recent study established that the $\omega^2$ scaling is only observed over a finite frequency range, which seems to increase systematically with the space dimension $d$, which is consistent with a pure quadratic scaling when $d=\infty$. However, for any finite $d$, the density of states eventually obeys Debye scaling for sufficiently low frequencies. 

Finally, recent numerical works show that
for frequencies lower than the boson peak,
an additional family of soft modes due to marginal instabilities can be observed~\cite{lerner_statistics_2016,mizuno_continuum_2017}.
As Fig.~\ref{fig:vdos} shows, the vibrational density
of these additional modes scales as $\omega^4$. A spatial analysis of such modes shows that they correspond to quasi-localised modes, which are again absent from the large $d$ analytic description. 

\subsection{Rheology}

\label{sec:rheology}

Once the glass is created, it can be adiabatically cooled or compressed, but it can also be deformed by applying an external mechanical constraint. The rheology of amorphous solids is a very broad research field. Here, we present recent results in this field obtained using the mean field glass theory, including implications regarding elasticity, yielding and shear jamming~\cite{RainoneUrbani,RainoneUrbaniYoshinoZamponi,yoshino_shear_2014,biroli2016breakdown}. 

We report results obtained from
the same state following formalism applied
to study the amorphous phase along a compression
in the $d \rightarrow \infty$ limit.
If the master replica $\underline{Y}$ is in the dynamically arrested region,
the system reacts elastically to a small applied strain $\gamma$.
We can then obtain the stress-strain curve as a
function of the state point $(T, \hat{\varphi})$ of the slave replica $\underline{X}$.
The stress for an elastic medium increases linearly with strain, which defines the shear modulus
$\hat{\mu} = \frac{d\hat{\sigma}}{d\gamma}$ computed at zero strain,
where stress and shear modulus are scaled such that the
$d \rightarrow \infty$ limit remains finite.
In the small strain limit one finds
\begin{equation}
\hat{\mu} = \frac{1}{\Delta}
\end{equation}
where $\Delta$ is the long time limit of the MSD.
The MSD $\Delta(\underline{X}, \underline{Y})$ is the superposition of an affine component due to the strain,
and of a non-affine contribution defined by the particular shear protocol.
At the glass transition, the shear modulus jumps from a zero value (liquid state) to a finite value at $\hat{\varphi}_d$ (glass state). In finite dimensions, this sharp discontinuity becomes a crossover~\cite{parisi2020theory}.  

When the system is confined within a glass state,
it is able to sustain a shear strain on a time scale
which corresponds to the diverging time scale for which
the dynamics becomes diffusive.
One can then follow adiabatically the slave replica
until a state point $(T, \hat{\varphi})$ and
study the linear response to shear for the
different phases of the glass. 
This corresponds to exploring the strain vs volume fraction phase diagram of the system.  
Upon decompression, the shear modulus
decreases and displays a square root singularity
at the melting spinodal point~\cite{RainoneUrbaniYoshinoZamponi, parisi2020theory}.

Increasing the strain and/or the volume fraction, the glass phase may undergo a Gardner transition and transform into a marginal glass, for which 
 all non-linear elastic modulii diverge and standard elasticity theory does not hold anymore \cite{biroli2016breakdown}.
As for a simple compression without shear, the boundary of the Gardner phase transition explicitly depends on the selected glass state. 

Once the Gardner phase is entered, upon further compression or strain, two kinds of transition may occur in hard sphere glasses. First, the shear modulus may increase and eventually diverge when a jamming point is reached. At zero strain, this is the ordinary jamming transition. In that case, the power law scaling of the MSD directly implies a similar behaviour for the shear modulus. In the presence of a finite strain, this corresponds to the phenomenon of shear jamming, observed in the context of granular materials~\cite{urbani2017shear,peters2016direct}.  

A second type of instability can occur when increasing the strain of a hard sphere glass. Here, the shear stress reaches a maximum followed by a spinodal instability where the fullRSB solution for $\Delta$ and $\Delta_r$ is no longer stable. The spinodal point $\gamma_Y(\hat{\varphi}_g)$ corresponds to the glass yielding transition~\cite{urbani2017shear,parisi_shear_2017}. 
The yielding transition in glasses has been
studied for a variety of models and under different
physical conditions~\cite{lin_scaling_2014,ozawa_random_2018}. In particular, it has been suggested that the yielding transition belongs to the same universality class as the RFIM, i.e.~a spinodal transition with disorder. 

\section{New computational methods}

\label{sec:newComputMethods}

In the last decade, a number of new numerical techniques have allowed to attack the challenges presented in the above theoretical sections. These techniques typically make use of tools that go beyond the realm of standard computer simulations to either sample phase space more efficiently, or access information and observables that are not directly stored in particle trajectories.  

\subsection{The Swap Monte-Carlo method}

\label{subsec:swap}

To sample deeply supercooled states in equilibrium, one needs to run computer simulations over a duration that scales with the equilibrium structural relaxation time $\tau_\alpha$. Because $\tau_\alpha$ grows rapidly as the system approaches the glass transition, ordinary computer simulations are limited to a rather high temperature regime much above the experimental glass transition $T_g$. 

\begin{figure}
\psfig{file=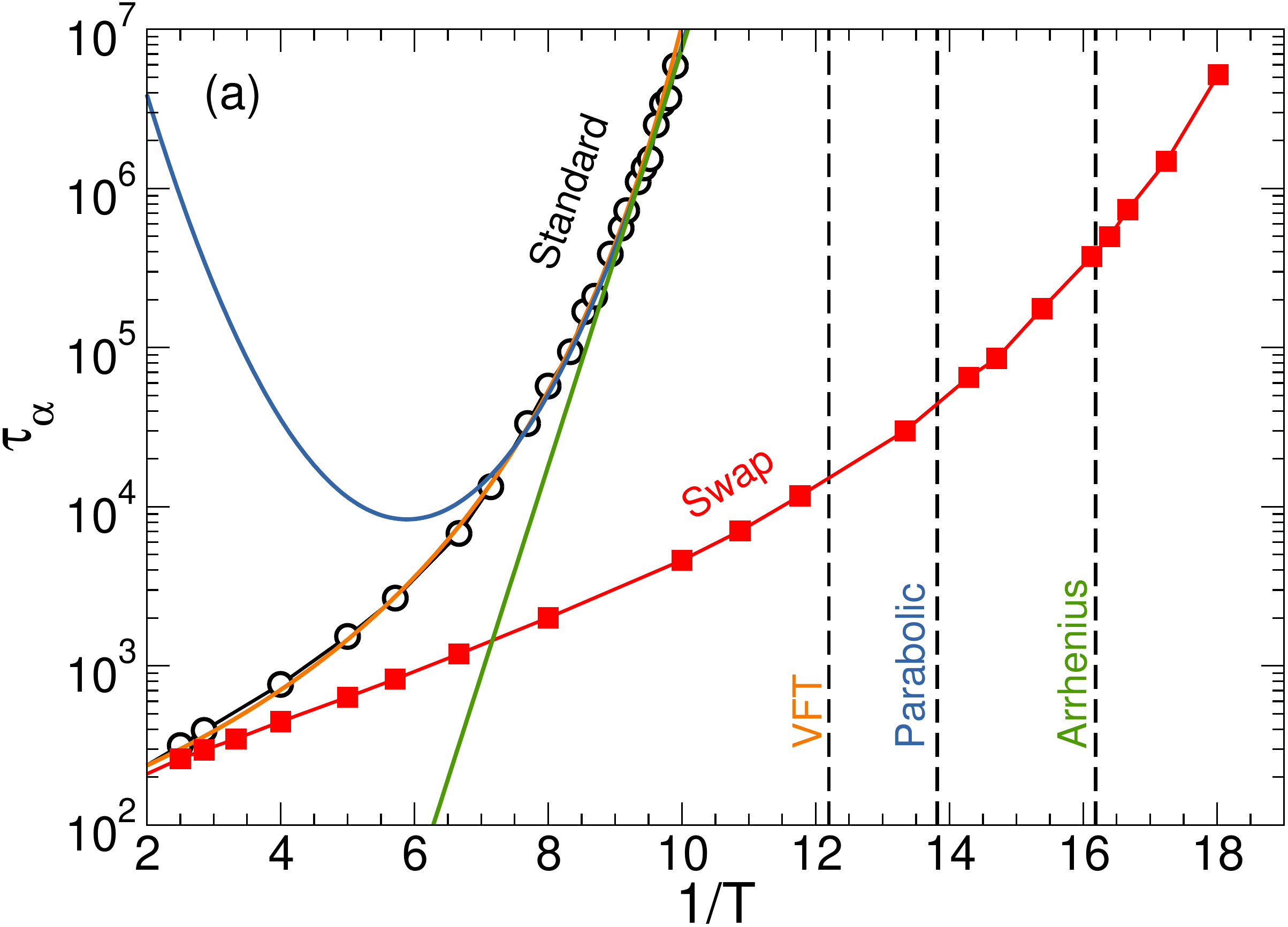,width=8.5cm}
\caption{From \cite{ninarello_models_2017}.
Relaxation times for standard (open symbols) and swap (closed symbols) Monte Carlo dynamics. The standard dynamics is fitted with the VFT, parabolic, and Arrhenius laws, which are then used to estimate the location of the experimental glass temperature $T_g$ (vertical dashed lines). The swap dynamics efficiently equilibrates the system at temperatures below $T_g$.}
\label{swapTimesfigure}
\end{figure}

The swap Monte Carlo algorithm is an efficient way to produce equilibrium configurations of a supercooled liquid on the computer~\cite{berthier_equilibrium_2016,ninarello_models_2017}. Here, `efficient' means that the equilibration time of the swap Monte Carlo algorithm increases much less than $\tau_\alpha$ as temperature decreases. Actually, it was shown that for several three dimensional model systems, the equilibration speedup can be larger than a factor $10^{11}$, as illustrated in Fig.~\ref{swapTimesfigure}. For this particular model of soft spheres, the swap algorithm continues to reach thermal equilibrium below the experimental glass transition $T_g$.      

The swap algorithm was introduced long ago~\cite{gazzillo_equation_1989}, and it was first used in the context of the glass transition by Grigera and Parisi~\cite{grigera_fast_2001}. Its ability to reach equilibrium at extremely low temperatures was established more recently~\cite{ninarello_models_2017, berthier_equilibrium_2016, ninarello_computer_nodate}. By comparison with ordinary computer simulations, the swap Monte Carlo algorithm introduces unphysical particle moves, where the identity of a pair of randomly chosen particles is exchanged. Provided the swap moves are constructed to satisfy detailed balance, the algorithm is guaranteed to reach thermal equilibrium. The swap Monte Carlo algorithm can be implemented in a molecular dynamics setting, replacing the swap moves by an exchange between the system and a reservoir of particles~\cite{brito_theory_2018,kapteijns_fast_2019,berthier2019efficient}. Typically, increasing the frequency of swap moves speeds up the dynamics but there are practical limits to the maximal allowed frequency~\cite{berthier2019efficient}. The swap algorithm was shown to work well from $d=2$ up to at least $d=8$ dimensions~\cite{berthier_bypassing_2019}.

A decisive step to increase the efficiency of the swap Monte Carlo algorithm was the development of models for supercooled liquids that were tailored to increase the swap efficiency. However, and perhaps more importantly, many models that were previously thought to be good glass-formers were in fact crystallizing very easily when swap was employed. Thus, a key step was also the development of more robust glass-forming models, using size polydispersity and non-additive interaction parameters to prevent structural ordering~\cite{ninarello_models_2017,parmar2020ultrastable}.  

In summary, the swap Monte Carlo algorithm easily and rapidly produces a large number of independent equilibrium configurations of a glass-former over a very broad range of temperatures, from the high temperature liquid, down to the mode-coupling crossover, and even below the experimental glass transition temperature. The latter regime can be explored experimentally only using physical vapor deposition, see Sec.~\ref{sec:ultrastable}. Therefore, many physical properties of glassy materials can now be measured over a temperature regime that is extremely broad, and may be compared directly to experiments with no extrapolation. This has led to an important activity 
 to address several problems related to glassy materials: the Gardner transition in finite dimensional glass-formers~\cite{PNASgardner,liao2019hierarchical,scalliet2019nature}, the link between glass and jamming transitions~\cite{berthier_equilibrium_2016,coslovich2018local}, the measurement of the configurational entropy in deeply supercooled liquids~\cite{ozawa_configurational_2018,berthier_zero-temperature_2019}, the analysis of point-to-set lengthscales~\cite{berthier_configurational_2017,yaida_point--set_2016} and of the Franz-Parisi potential~\cite{berthier_configurational_2017,guiselin2020random}, the evolution of several important properties of glasses with the glass preparation~\cite{khomenko2019depletion,wang2019low,wang2019sound}, and the physics of ultrastable glasses~\cite{flenner2019front,berthier2017origin}.

In addition, the demonstration that a very simple algorithm can speed up the equilibration dynamics of supercooled liquids can be seen as an interesting physical result in itself. If such a result appears quite natural in the context of kinetic facilitation~\cite{wyart_does_2017,gutierrez2019accelerated}, it is more challenging (but possible) to interpret in the context of the random first order transition theory~\cite{ikeda_mean_2017,szamel_theory_2018,berthier_can_2019}, where dynamics becomes highly collective at very low temperatures. 

\subsection{Franz-Parisi potential}

\label{subsec:epsilonCoupling_VQ}

In seminal work~\cite{franz_phase_1997,franz_e_1998}, Franz and Parisi introduced a quantity now called the Franz-Parisi potential, $V(Q)$. This quantity plays the role of a Landau free energy for mean-field phase transitions, in the sense that it expresses the free energy cost of the order parameter fluctuations at the mean-field level. 

For spin glass models, where the approach was first introduced, the overlap $Q$ represents indeed the spin glass order parameter. It quantifies the degree of similarity between two configurations $\mathcal{C}_0$ and $\mathcal{C}_1$. For liquids with continuous degrees of freedom, a practical definition of the overlap reads:    
\begin{align}
Q = \frac{1}{N}\sum^N_{i=1} \sum^N_{j=1}  \Theta( a - |{\bf r}_i^{\mathcal{C}_0}-{\bf r}_j^{\mathcal{C}_1}|),
\label{eq:defoverlap}
\end{align}
where ${\bf r}_i^{\mathcal{C}_0}$ represents the position of particle $i$ in configuration $\mathcal{C}_0$. The overlap is close to unity when the two configurations $\mathcal{C}_0$ and $\mathcal{C}_1$ have similar density profiles, up to thermal vibrations of spatial extension $a$ (typically, one takes $a$ as a small fraction of the particle diameter,  $a \sim 0.2 \sigma$).

The Franz-Parisi potential $V(Q)$ efficiently captures all the features associated to a random first order transition. It can be defined from the probability distribution function $P(Q)$ of the overlap as $V(Q)= -(T/N) \log P(Q)$. In particular, $V(Q)$ is characterised by a single minimum near $Q=0$ at high temperature, but develops within mean-field theory a secondary minimum at a finite $Q>0$ when $T$ decreases towards the Kauzmann transition, indicating that the glass phase is metastable with respect to the liquid. 
The free energy difference between the glass and the liquid phases in this regime is given by $T S_c(T)$, where $S_c(T)$ is the configurational entropy. Physically, this means that localising the system in a single free energy minimum in the liquid phase comes with a free energy cost of entropic nature. 

In addition, Franz and Parisi introduced a field $\varepsilon$ conjugate to the overlap $Q$: 
\begin{align}
H_{FP} [\mathcal{C}_1] = H [\mathcal{C}_1] - \varepsilon Q[\mathcal{C}_0,\mathcal{C}_1].
\end{align}
This allowed them to explore an extended phase diagram by changing both $T$ and $\varepsilon$~\cite{cardenas_constrained_1999,donati_theory_2002}. In this plane, the Kauzmann transition at $(T=T_K, \varepsilon=0)$ extends as a first order transition line, which ends at a second order critical point at a position $(T_c>T_K, \varepsilon_c)$. More recent work taking into account finite dimensional fluctuations suggest that this critical point should exist also in finite dimensions, and should be in the same universality class of the random field Ising model~\cite{biroli2014random}. 

The Franz-Parisi potential and the extended phase diagram have been studied numerically in finite dimensional models in recent years~\cite{cardenas_constrained_1999,berthier_overlap_2013,berthier_evidence_2015,turner_overlap_2015,jack_phase_2016,guiselin2020random,parisi2014liquid,berthier_configurational_2017,biroli_role_2016}. Taken together, these studies confirm the existence of both a first order transition line and a second order RFIM critical point in finite dimensional glass-formers. 

A second important outcome of the Franz-Parisi potential is the possibility to directly estimate a configurational entropy for equilibrium glass-formers using the free energy difference between the liquid and metastable glass phase~\cite{berthier_novel_2014,berthier_configurational_2017}. This definition of the configurational entropy is conceptually closer to its mean-field definition, and does not rely on an explicit definition of metastable states for a finite dimensional system~\cite{berthier_configurational_2019}. 

\subsection{Point-to-set lengthscale}

As mentioned in Sec.~\ref{subsec:Static_dynamic_corr_func},
assessing and measuring a growing static length scale is crucial to the glass problem. Standard probes used for second order phase transitions, such as 2-point and 4-point correlation functions, do not seem to provide any useful evidence. 
A possible explanation is that these correlation functions 
do not carry enough information to capture the relevant structural order,
also termed {amorphous order},
and that one has to use higher-order correlation functions
(see also sec.~\ref{subsec:multiPointCorrFunc}). 
The point-to-set correlation function $C(R)$ is effectively an $n$-point correlation function, where $n$ is the number of particles comprised in a sphere of radius $R$ (the `set'), which can indeed be a large number. 

The justification is that in order to probe amorphous order one has to proceed as in standard phase transitions: fix a suitable 
boundary condition and study whether it enforces a given arrangement 
of particles in the bulk of the system. For simple cases, such as the ferromagnetic Ising model, it is clear what type of boundary conditions are needed (all spins up or all spins down). However, for supercooled liquids this is a much harder task. The problem can be circumvented by using equilibrated configurations and freezing all particles outside a cavity of radius $R$. This provides the boundary conditions sought for: if the system is indeed ordering, then using cavities drawn from different equilibrium configurations will give access to different sets of appropriate boundary conditions. 
This method was first proposed theoretically in the context of RFOT theory~\cite{BB2,montanari2006rigorous} and signal processing~\cite{mezard2006reconstruction}, and was transformed into a concrete numerical procedure in Refs.~\cite{cavagna,biroli_thermodynamic_2008}. 
In the last decades, measurements of the point-to-set length were progressively refined~\cite{cavagna_dynamic_2012,berthier_static_2012,berthier_efficient_2016,charbonneau_linking_2016,yaida_point--set_2016,berthier_configurational_2017}. 

Let us describe the main steps and mention some important obstacles that had to be overcome. Firstly, one needs a collection of well-equilibrated configurations $\mathcal{C}_{0}$ to start with. This is not supposed to be the hardest part, but for very low temperatures, methods such as the swap Monte Carlo algorithm are necessary.

Secondly, for each sample $\mathcal{C}_{0}$
one freezes all particles outside a cavity of radius $R$,
then let the particles inside ergodically visit the remaining phase space, and record the configurations $\mathcal{C}_{1}$ that are sampled.
For small cavities, large activation barriers make conventional molecular dynamics simulations ineffective, but this problem can be solved using parallel tempering techniques~\cite{berthier_efficient_2016}, in addition to swap Monte Carlo~\cite{cavagna_dynamic_2012}. It is crucial to check that a complete sampling of the restricted configurational space has been reached inside the cavity, and careful tests have been devised to this end~\cite{cavagna_dynamic_2012}. 

Thirdly, one needs to measure the overlap distribution $P(Q)$
between the quenched reference configuration $\mathcal{C}_{0}$
and the equilibrium samples $\mathcal{C}_{1}$. This is very similar to the Franz-Parisi construction, except for the local nature of the constraint. There are two possibilities.
One is that the cavity is so small that only
one state (one configuration, up to vibrations) can be visited,
so that the peak of $P(Q)$ is in the high-overlap range.
The other is that the cavity is sufficiently large for many different states to be accessible. In that case, $P(Q)$ has a peak at low $Q$.
In between the two cases, there is a critical value of the radius, $R \sim \xi_{PTS}$, which corresponds to the crossover, with a bimodal $P(Q)$. Since finite size cavities cannot be self-averaging, 
one needs to repeat the overlap measurements for many independent quenched configurations $\mathcal{C}_{0}$, and then perform 
an average over these realizations of the disorder. Indeed, for $R \sim \xi_{PTS}$, large sample-to-sample variations of $P(Q)$ are observed.

\begin{figure}
\psfig{file=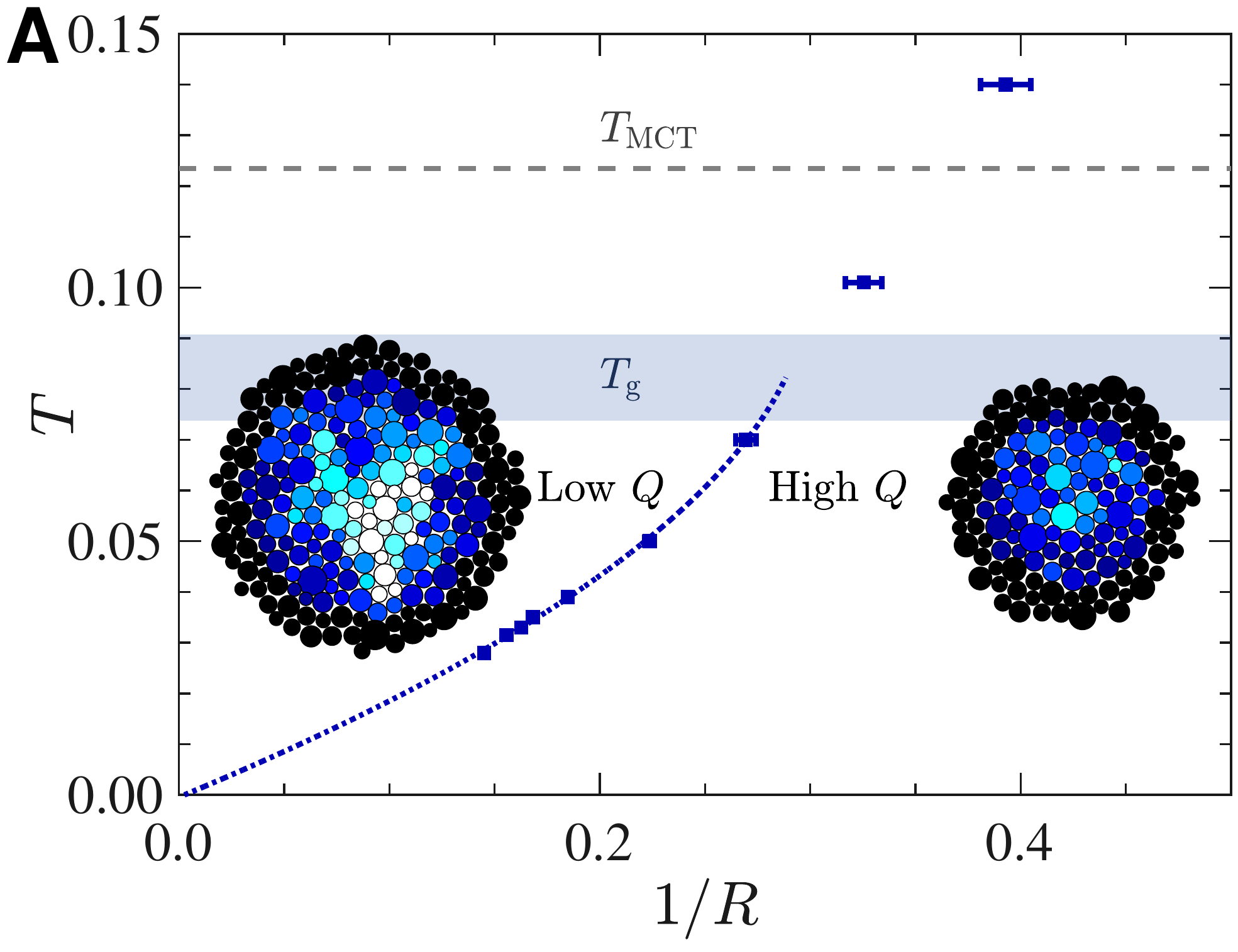,width=8.5cm}
\caption{From \cite{berthier_zero-temperature_2019}. Evolution of the inverse point to set length, denoted here as $1/R$ with the temperature $T$. A clear growth of the point to set length up to $\xi_{PTS} \approx 6.5$ is observed. Cavities smaller (larger) than $\xi_{PTS}$ have a large (small) overlap with the reference configuration, as illustrated in the snapshots, where the particle shade encodes the overlap value.}
\label{fig:pointToSet}
\end{figure}

There are a number of subtleties and extensions related to the overlap function that we now describe. A first subtlety is that an appropriate measure of the overlap needs to focus on the center of the spherical cavity, since boundaries always have a high overlap, and this makes the results difficult to interpret~\cite{biroli_thermodynamic_2008}.
Moreover, the overlap needs to be a smooth function
of the distance between configurations, and step functions are too singular to average over. A second subtlety is that for some liquids,
the simple positional information does not adequately
cover all the structural information. In that case, the conventional overlap needs to be completed with additional coordinates overlap, e.g.~bond angles. This was used in Ref.~\cite{yaida_point--set_2016} to show that hexatic order could be captured by the point-to-set correlations (see also \cite{russo2015assessing}).
 
Let us also mention a few extensions of the method that have been proposed. In Ref.~\cite{cavagna_dynamic_2012},
the authors proposed to relax the frozen configuration constraint
by letting the outside atoms vibrate, so long as they maintain a large overlap with the initial configuration: they called this reference state `frozen state' (as opposed to frozen configuration). A second type of extension is to study various confinement geometries, as in Refs.~\cite{cavagna_dynamic_2012,li_growing_2014,berthier_static_2012,kob2012non}, where it was shown that the geometry of confinement needs to be carefully considered. 

Thanks to the computational progress outlined above, several important results were established in the last decade. Among them: (i) clear evidence that the slowing down of the dynamics is accompanied by the growth of the point-to-set length (even though a mild one), (ii) established relation between point-to-set and configurational entropy: the former appears to be inversely proportional to the other \cite{berthier_configurational_2019},  thus directly linking the growth of amorphous order to the decrease in number of metastable states---a tenet of RFOT theory, (iii) clear difference between two dimensional and three dimensional behavior: two-dimensional glass-formers
display a point-to-set that appears to diverges at 
zero temperature, as shown in Fig.~\ref{fig:pointToSet}, thus indicating a $T_K=0$ Kauzman transition temperature~\cite{berthier_zero-temperature_2019}. This is in sharp contrast with the results in three dimensions, where the extrapolated $T_K$ is larger than zero. 

\subsection{$s$-ensemble and large deviations}

\label{subsec:s-ensemble}

The Franz-Parisi potential is an example of a large deviation analysis. The idea behind it is that fluctuations in the overlap field play an important role for the glass transition, analog to the magnetization field in a ferromagnet. In order to investigate such fluctuations, one can then study the large deviation function (the free-energy) associated to the spatial average of the field (the order parameter). This is the usual route followed in thermodynamic analyses of second-order phase transitions. The $s$-ensemble is the dynamical counterpart of such a procedure. 
One modifies the dynamical rules, in order to be in a particular subset of the `dynamical states', to probe the tendency of the original system to explore these  states. This idea originates both from theoretical considerations
on the large deviations of activity (rare events) predicted in KCMs
\cite{garrahan_dynamical_2007,
garrahan_first-order_2009}
and from considerations on efficient simulation schemes
(for rare events as well) that apply more generally~\cite{Noe2009}.
In practice, the $s$-ensemble dynamics
is a particular instance of transition path sampling,
where the quantity of interest is
 a measure of the {activity} of a multi-particles trajectory $\{x_i(t)\}_{i=1,\ldots,N}$, cumulated over time
\begin{align}
K[x(t)] = \frac{1}{N t_{obs}} \sum_{t\in [t_0,t_0+t_{obs}]} \sum_{i\in[1,N]} (x_i(t+\Delta t) - x_i(t))^2
\end{align}
where $\Delta t$ (resp. $t_{obs}$) is chosen to be of the order of a few ballistic times (resp. a few  relaxation times).
In the $s$-ensemble, the actual relaxation time then becomes larger than $t_{obs}$.
Here we follow the notations used in \cite{hedges_dynamic_2009},
where this biasing method was first applied to structural glasses
(for its initial introduction in KCMs, see \cite{garrahan_dynamical_2007,garrahan_first-order_2009,elmatad_finite-temperature_2010}).
The method is called $s$-ensemble because
the probability of a trajectory $x(t)$ is $P_0[x(t)]e^{-sK[x(t)]}$,
where $P_0[x(t)]$ is the probability for the unperturbed system. 
This amounts to defining a `thermodynamics of trajectories' which are biased 
(and classified) depending on their activity; $s$ is the biasing field. 
In order to numerically simulate and probe such a measure, one can perform Monte-Carlo sampling in trajectory space. This method is actually so efficient at finding low-energy states that special attention must be paid to crystallisation~\cite{hedges_dynamic_2009}.

\begin{figure}
\psfig{file=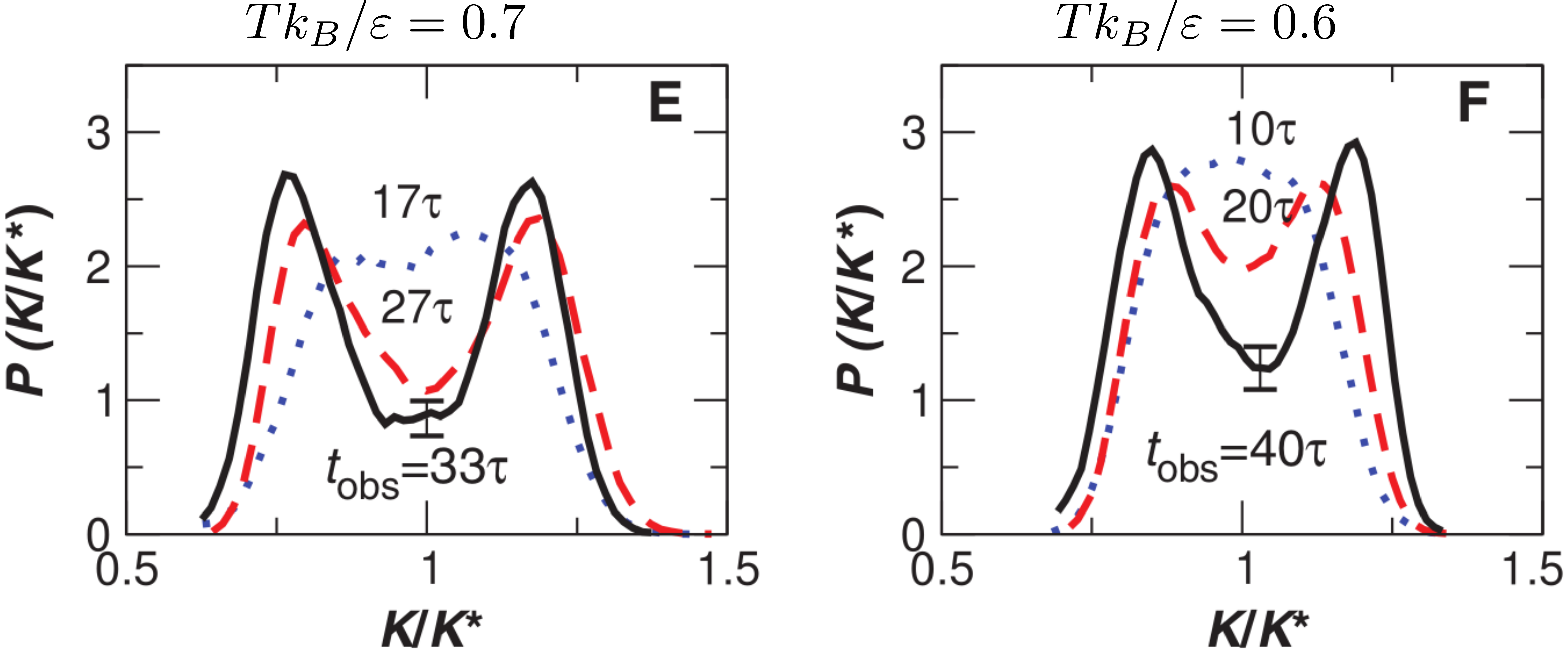,width=8.5cm}
\caption{From \cite{hedges_dynamic_2009}.
Coexistence of the active (large $K$) and inactive (small $K$) phases, as evidenced from $s$-ensemble biased simulations. As temperature is reduced, the distribution becomes increasingly bimodal, as expected when approaching a first order phase transition.}
\label{fig:sEnesemble-melting2}
\end{figure}

One of the most important results obtained by the $s$-ensemble method is a direct evidence of a first order phase transition between an active phase (for low $s$) and an inactive phase (for large $s$) in supercooled liquids: see Fig.~\ref{fig:sEnesemble-melting2}. In the $s-T$ plane this corresponds to a first order transition line $s^*(T)$ which ends in a critical point at high temperature
\cite{hedges_dynamic_2009, elmatad_finite-temperature_2010}.
Although the method focuses on the dynamical properties, inactive states do exhibit interesting structural features. 
They are more stable than equilibrium samples \cite{jack_preparation_2011}:
mechanically, in terms of their lower number of low-energy modes (depletion of $D(\omega)$),
and thermodynamically, in terms of a longer lived `melting' transient from the glass to the liquid state. This suggests that inactive states correspond to very stable glassy state, i.e. non-equilibrium glassy states with low fictive temperatures.  

The foundation of the $s$-ensemble relies on the dynamical behavior of KCMs. For
the East model scenario it was shown~\cite{Keys2015} that configurations obtained by 
(i) the $s$-ensemble, (ii) finite-rate cooling, and (iii) quenching and long time aging, are all equivalent. This provides a concrete example where glassiness is directly related to the phase transition unveiled by the  $s$-ensemble. Let us conclude by noticing that even systems having a RFOT display such a phase transition, the difference being that the line $s^*(T)$ reaches $s^*=0$ for $T=T_K$ and not for $T=0$ as for KCMs~\cite{jack2010metastable}. This shows that the $s$-ensemble is actually a general construction, which transforms the physics of metastability observed in supercooled liquids in well-defined phase transitions in the extended ensemble comprising space and time.     

\subsection{Machine Learning developments}

We have mentioned the possible use of (unsupervised) Machine Learning (ML) in sec.~\ref{subsec:frustration} for automatic identification of LPS in supercooled liquids (see also Refs.~\cite{ronhovde2011detecting,ronhovde2012detection,boattini2020autonomously,paret2020assessing}). Another set of ML techniques to be used is supervised learning:
automatically defining fluid- or solid-like structures ({features} in ML language) by labeling them for a training set of local environments which are observed to be locally fluid- or solid-like.
The general idea developed in \cite{Schoenholz2017}
is to train a neural network (or other fitting model)
to `predict' the current degree of mobility or `instantaneous activity' from the knowledge of structure only.
This may be cast as finding the function $f$
such that $f(\{\vec{r}\})(i,t) =y_{pred} \approx y_{true}(i,t)$,
where $\{\vec{r}\}(i,t)$ is the structure of the neighborhood of particle $i$ at time $t$,
and $y_{true}(i,t)$ (the label) is a measure of the dynamical activity for particle $i$ at time $t$.
Concretely, the target label $y_{true}$ has to encode
some notion of local, instantaneous mobility
\cite{Candelier2009,Candelier2010a,Candelier2010c,candelier_avalanches_2010},
while the local structure may be defined
e.g.~by the (local, instantaneous) density pair correlation function $g(r)$,
with the possible addition of angular variables~\cite{Cubuk2016,Schoenholz2016}.
This technique has been applied with success,
showing how much structure correlates with dynamics in 
supercooled liquids \cite{Schoenholz2016,landes_attractive_2019},
disordered solids
\cite{Cubuk2016,Cubuk2015,Schoenholz2016a, Schoenholz2016,landes_attractive_2019},
for the plasticity of amorphous materials \cite{Mechanics,Definiujemy2012}
and in polycrystalline materials \cite{sharp2018machine}.
%

The problem of the interpretation,
which is one of the major issues with statistical learning in general, remains open:
how to make use of predictions emerging from hundred-parameters models?
This issue is transverse to most ML applications
and is currently under active scrutiny.
In terms of basic science, the conclusion of Liu and collaborators is that the predicted $y_{pred}$ provides a new observable, called the softness field~\cite{Schoenholz2016a}, which plays a key role for the dynamics. 

We conclude mentioning a very recent work which introduced a new ML technique in the glass physics arena. The authors of  \cite{bapst2020unveiling} focused on the problem of predicting {long-time} dynamics (more precisely, the propensity field \cite{widmer2006predicting}) from knowledge of structure alone, by leveraging on recent progress in graph neural networks. The results are impressive: the ability to predict the propensity map is substantially better than existing numerical physics-based methods and the ML techniques described above, thus establishing a promising new way to study glassy dynamics.  

\section{Aging and off-equilibrium dynamics}

\label{aging}

\subsection{Why aging?}

We have dedicated most of the above discussion to properties
of materials approaching the glass transition at thermal equilibrium.
We discussed a rich phenomenology and serious
challenges for both our numerical and analytical capabilities to account
for these phenomena. For most people, however, glasses are interesting
below the glass transition, so deep in the glass phase that the material
seems to be frozen forever in a seemingly arrested amorphous state,
endowed with enough mechanical stability for a glass to retain, say, the
liquid it contains (preferably a nice red wine).
Does this mean that there is no interesting physics in the glass state?

The answer is clearly `no'. There is still life (and physics) below the glass
transition. We recall that for molecular glasses, $T_g$ is defined
as the temperature below which relaxation is too slow to
occur within an experimental timescale. Much below $T_g$, therefore,
the equilibrium relaxation timescale is so astronomically large
that thermal equilibrium is out of reach. One enters therefore
the realm of off-equilibrium dynamics.
A full physical understanding of the non-equilibrium
glassy state remains a central challenge~\cite{youngbook,barrat2004slow}.

\begin{figure}
\psfig{file=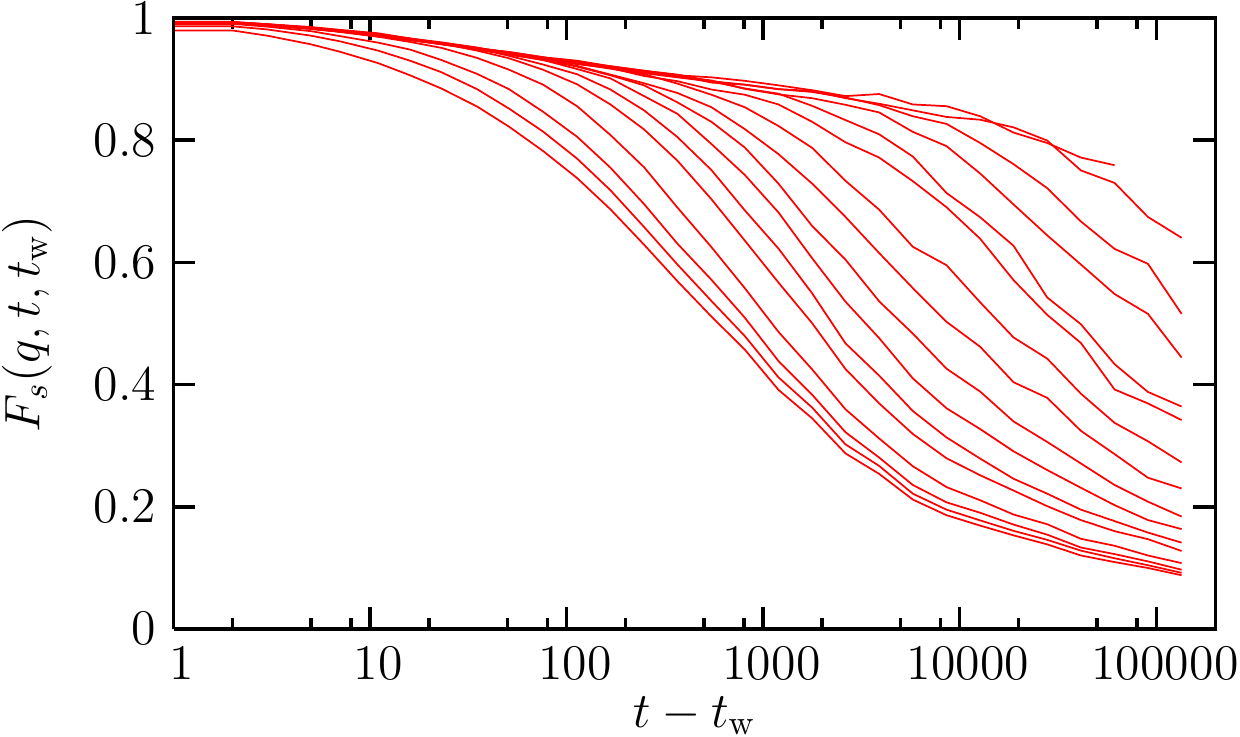,width=8.5cm}
\caption{Aging dynamics in a Lennard-Jones
glass-forming liquid at low temperature. The system is quenched at
time $\tw=0$ at low $T$, where the temperature is kept constant.
Two-time self-intermediate scattering functions
are then measured for 20 logarithmically spaced waiting times $\tw$
from $\tw=1$ to $\tw=10^5$ (from left to right). The relaxation
becomes slower when $\tw$ increases: the system ages.}
\label{agingfig} 
\end{figure}

A first consequence of studying materials in a time window
smaller than equilibrium relaxation timescales is that the system
can, in principle, remember its complete history, a most
unwanted experimental situation since all details of the experimental
protocol may then matter. The simplest protocol to study aging phenomena
in the glass phase is quite brutal~\cite{struik1977physical}:
take a system equilibrated above the glass
transition and suddenly quench it at a low temperature
at a `waiting time' $\tw=0$
which corresponds to the beginning of the experiment.
For $\tw>0$ the system is left unperturbed at constant
temperature where it
 tries to slowly reach thermal equilibrium, even though
it has no hope to ever get there. Aging means
that the system never forgets the time $\tw$ spent
in the glass phase, its `age'. The evolution of one time quantities, e.g.
the energy, as a function of time are not a good evidence of aging. In order to show that the system never equilibrates, two time quantities such as
density-density or spin-spin correlation functions are much more
useful. A typical example is presented
in Fig.~\ref{agingfig} where the self-part of the intermediate function
in Eq.~(\ref{isf}) is shown for a Lennard-Jones molecular liquid
at low temperature. Immediately after the quench, the system
exhibits a relatively fast relaxation: particles still move
substantially. However, when the age increases,
dynamics slow down and relaxation becomes much slower. When
$\tw$ becomes very large, relaxation becomes too slow to be followed
in the considered time window and the system seems frozen on that
particular timescale: it has become a glass.
A striking feature conveyed by these data is that an aging system
not only remains out-of-equilibrium for all practical purposes,
but its typical relaxation time is in fact
set by its age $\tw$. In simple cases, the effective
relaxation time after waiting a time $\tw$ scales at
$\tw$ itself, which means that since equilibration timescales
have diverged, $\tw$ is the only remaining
relevant timescale in the problem.

A popular interpretation of this phenomenon is given
by considering trap models~\cite{bouchaud1992weak}.
In this picture, reminiscent of the Goldstein
view of the glass transition
mentioned above~\cite{Goldstein},
the system is described as a single particle evolving
in a complex energy landscape with a broad distribution
of trap depths--a paradigmatic mean-field approach.
Aging in this perspective arises because the system visits traps
that become deeper when $\tw$ increases, corresponding
to more and more stable states. It therefore takes
more and more time for the system to escape, and the dynamics slow
down with time, as observed in Fig.~\ref{agingfig}.
This implies that any physical property of the glass becomes
an age-dependent quantity in aging protocols, and more
generally becomes dependent on how the glass was prepared. One can
easily imagine using this property to tune mechanical or optical
characteristics of a material by simply changing the way it
is prepared, like how fast it is cooled to the glassy state (recall our discussion of ultrastable glasses in Sec.~\ref{sec:ultrastable}).

A real space alternative picture was promoted in particular
in the context of spin glass studies, based on the ideas of
scaling and renormalization~\cite{braymoore,braymoore2,FH}. The physical picture
is that of a coarsening process, where the system
develops long-range order by growing extended domains
of lengthscale $\ell(\tw)$. On lengthscales less than $\ell(\tw)$,
the system has had the time to order since the quench at $\tw=0$.
The walls between domains evolve in a random environment. In order to move
they have to overcome free energy barriers. It is then assumed an
activated dynamic scaling which states that the typical barrier
to extend the domain from linear size $\ell(\tw)$ to, say, $2\ell(\tw)$
scales as $\ell^\psi$, where $\psi$ is some
`barrier' exponent. Using the Arrhenius law to relate
dynamics to barriers, one gets that aging corresponds to
the logarithmic growth with time of spatially correlated
domains, $\ell \sim (T \log \tw)^{1/\psi}$.
A domain growth picture of aging in spin glasses
can be directly confirmed by numerical simulations~\cite{heiko},
only indirectly by experiments.

\subsection{Memory and rejuvenation effects}

\begin{figure}
\begin{center}
\psfig{file=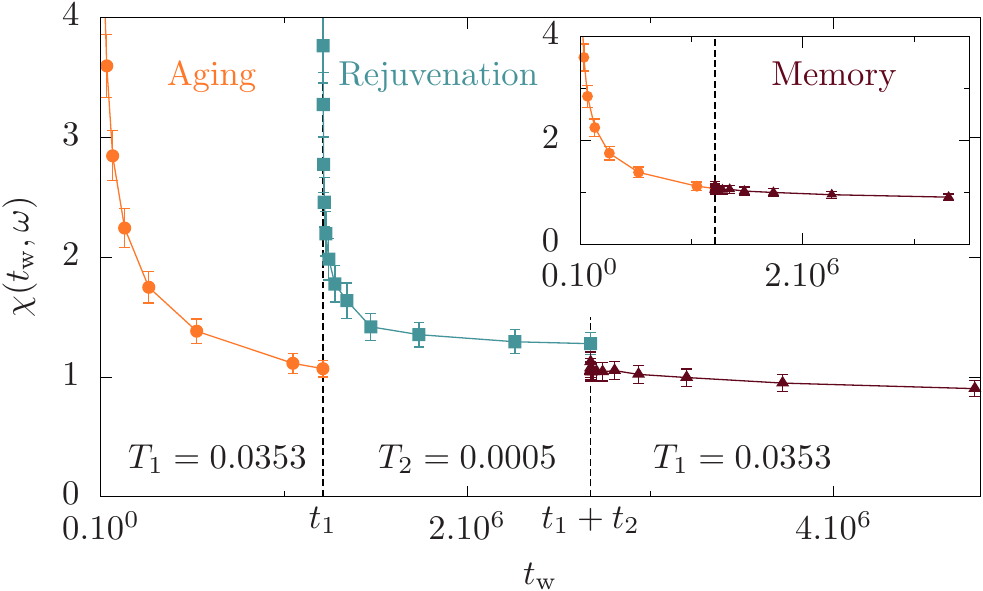,width=8.5cm}
\caption{Memory and rejuvenation effects
obtained in the numerical simulation
of a three-dimensional glass-former~\cite{CamilleLudo}.
There is a first aging step, $0< \tw < t_1$, during
which the system slowly tries to reach thermal
equilibrium at temperature $T_1$. The system `rejuvenates'
in  the second step at $T_2$, $t_1<t<t_1+t_2$, and it restart aging
(rejuvenation).
Finally in the third step, temperature is back to $T_1$, and memory
of the first step is kept intact, as shown in the inset where relaxation
during the second step of the experiment is taken away.}
\label{cycle}
\end{center}
\end{figure}

Since the complete history of a sample in the glass phase matters,
there is no reason to restrain experimental
protocols to the simple aging experiment mentioned above. Indeed,
experimentalists have investigated scores of more elaborated protocols
that have revealed an incredibly rich, and sometimes quite unexpected,
physics~\cite{youngbook}. We restrain ourselves here to a short
discussion of memory and rejuvenation effects
observed during temperature cycling experiments~\cite{refregier}
(one can imagine applying
a magnetic field or a mechanical constraint, be they constant
in time or sinusoidal, etc.).
These two effects were first observed in spin glasses, but the protocol
was then repeated in many different materials, from polymers
and organic liquids to disordered ferroelectrics. After several unsuccessful attempts,
similar effects are now observed in numerical work as well~\cite{berthieryoung,berthierbouchaud}. Recent results obtained from simulations of a three-dimensional glass-former exploring the spin-glass-like Gardner phase~\cite{CamilleLudo} are presented in Fig.~\ref{cycle}.

There are three steps in temperature cycling experiments~\cite{refregier}.
The first one is a standard aging experiment, namely a sudden
quench from high to low temperature at time $\tw=0$.
The system then ages for a duration $t_1$ at constant
temperature $T_1$. The system slowly relaxes towards equilibrium
and its dynamics slows down: 
for our spin glass example (see Fig.~\ref{cycle}) this is observed through
the measurement of some dynamic susceptibility $\chi(\tw,\omega)$.
Temperature is then suddenly shifted to $T_2 <T_1$ at time $t_1$.
There, the material restarts aging (almost) as if
the first step had not taken place. This is called `rejuvenation
effect', because the system seems to forget it is already `old'.
At total time $t_1+t_2$, temperature is then shifted back
to its initial value $T_1$. Then, aging is found to proceed
as a  quasi-perfect continuation of the first
step, as if the second step had not taken place. The system
has kept the `memory' of the first part of the experiment, despite
the rejuvenation observed in the intermediate part.
The memory effect becomes more spectacular when
relaxation during the second step is removed, as in the inset
of Fig.~\ref{cycle}. The third relaxation appears indeed as a perfect
continuation of the first one.

On top of being elegant and quite intriguing, such protocols
are relevant because they probe more deeply the dynamics
of aging materials, allowing one to ask more precise
questions beyond the simplistic observation that
`this material displays aging'.
Moreover, the observation of
similar effects in many different glassy materials
implies that these effects are intrinsic to systems
with slow dynamics. Interesting also are the subtle
differences observed from one material to the other.

Several experimental, numerical and theoretical papers have
been devoted to this type of experiments, and
these effects are not `mysterious' anymore~(see Ref.~\cite{barrat2004slow}).
A clear link between memory effects and typical
lengthscales over which the slow dynamics
takes place has been established.
Because lengthscales depend so sensitively
on timescales and on the working temperature, experiments
performed at two different temperatures typically
probe very different lengthscales, allowing the system
to store memory of its state at different temperatures over different
lengthscales~\cite{berthierbouchaud,microT}.
In return, this link has been elegantly
exploited to obtain a rather
precise experimental estimate of dynamic lengthscales
involved in the aging dynamics of spin glass materials~\cite{bert},
which seems to confirm the slow logarithmic
growth law mentioned before.

Discussion of the rejuvenation effect is slightly more subtle.
It is indeed not yet obvious that the effect as it is observed
in computer simulations and reported, e.g.~in Fig.~\ref{cycle},
is exactly similar to the one observed in experiments.
The difficulty comes from the fact that some seemingly innocuous
details of the experimental protocol, such as
the necessary use in experiments of finite cooling rates,
in fact play a crucial role and influence the physics
so that direct comparison between experiments and simulations is difficult.
In numerical work, rejuvenation can be attributed to a gradual
change with temperature of the nature of spatial correlations between
spins that develop with time~\cite{berthieryoung,berthierbouchaud}.
More drastic changes are predicted to occur
in disordered systems as a result of the chaotic evolution with temperature
of the metastable states in a spin glass (so-called `chaos
effect'~\cite{chaos1}),
that could also
be responsible for the observed rejuvenation effect~\cite{chaos2}.

\subsection{Mean-field aging and effective temperatures}

Theoretical studies of mean-field glassy models have provided important
insights into the aging dynamics of both structural and spin
glasses~\cite{CugKur1,CugKur2}.
Although such models are defined in terms of spin degrees
of freedom interacting via infinite-ranged interactions,
the deep connections between them
and the mode-coupling theory of the glass transition make them
serious candidates to investigate glassy states in general,
not only  thermodynamic properties at thermal equilibrium but also
non-equilibrium aging dynamics.
Despite their often reported
`simplicity', it took several years to derive
a proper asymptotic solution of the long-time dynamics
for a series of mean-field spin glasses (see Cugliandolo in Ref.~\cite{barrat2004slow}). These
results have then triggered an enormous activity~\cite{crisrit}
encompassing theoretical, numerical and also experimental
work trying to understand further these results, and to
check in more realistic
systems whether they have some reasonable range of
applicability beyond mean-field.

This large activity, by itself, easily demonstrates
the broad interest of these results.
More recently, the derivation of the static properties of liquids and glasses in large dimensions has renewed the interest in mean-field dynamic phenomena~\cite{parisi2020theory}. In particular, the dynamic equations governing the equilibrium properties of supercooled liquids have now been derived~\cite{MaimbourgKurchanZamponi,kurchan2016statics} and their consequences are being explored~\cite{manacorda2020numerical}. The study of the non-equilibrium (aging and sheared) dynamics is now under way~\cite{agoritsas2018out,agoritsas2019out,altieri2020dynamical}.

In these mean-field models, thermal
equilibrium is never reached, and aging proceeds by downhill motion in
an increasingly flat free energy landscape~\cite{laloux},
with subtle differences between spin glass and structural
glass models. In both cases, however, time
translational invariance is broken, and two-time correlation and
response functions depend on both their temporal arguments.
In fact, the exact dynamic solution of the equations
of motion for time correlators displays behaviours
in strikingly good agreement with the numerical results
reported in Fig.~\ref{agingfig}.

In these systems, the equations of motion in the aging regime
involve not only temporal correlations, but also time-dependent
response functions. At thermal equilibrium,
response and correlation are not independent, since the
fluctuation-dissipation theorem (FDT) relates both quantities.
In aging systems, there is no reason to expect the FDT to hold
and both quantities carry, at least in principle, distinct physical
information. Again, the asymptotic solution obtained for
mean-field models quantitatively establishes that the FDT does not
apply in the aging regime. Unexpectedly, the solution also
shows that a generalized form of the FDT holds at large
waiting times~\cite{CugKur1}.
This is defined in terms of the two-time connected correlation
function for some generic observable $A(t)$,
\be C(t,\tw) = \langle A(t) A(\tw)
\rangle - \langle A(t)\rangle \langle A(\tw) \rangle,
\label{corr}
\ee
with $t \ge
\tw$, and the corresponding two-time (impulse) response function
\be R(t,\tw) =
T \frac{\delta \langle A(t) \rangle}{\delta h (\tw)}\Bigg|_{h=0}.
\label{resp}
\ee
Here $h$ denotes the thermodynamically conjugate field to the
observable $A$ so that the perturbation to the Hamiltonian (or energy
function) is $\delta E =
-hA$, and angled
brackets indicate an average over initial conditions and any
stochasticity in the dynamics. Note that we have absorbed the
temperature $T$ in the definition of the response, for convenience.
The associated
generalized FDT reads then \be R(t,\tw) = X(t,\tw)
\frac{\partial}{\partial \tw} C(t,\tw),
\label{fdr_def}
\ee with $X(t,\tw)$ the so-called fluctuation-dissipation ratio (FDR).
At equilibrium, correlation and response functions are time
translation invariant, depending only on $\tau = t - \tw$, and
equilibrium FDT imposes that $X(t,\tw) = 1$ at all times. A parametric
fluctuation-dissipation (FD) plot of the step response or susceptibility
\be
\Chi(t,\tw)=\int_{\tw}^t dt'\,R(t,t'),
\ee
against
\be
\Delta C(t,\tw)=C(t,t)-C(t,\tw),
\ee
is then a straight line with unit slope. These simplifications do
not occur in non-equilibrium systems. But the definition of an FDR
through Eq.~(\ref{fdr_def}) becomes significant for aging
systems~\cite{CugKur1,CugKur2}.  In mean-field spin glass models the
dependence of the FDR on both temporal arguments is only through the
correlation function,
\be
X(t, \tw ) \sim X (C (t, \tw )),
\label{simple}
\ee
valid at large
waiting times, $\tw \to \infty$.
For mean-field structural glass models, the simplication
(\ref{simple}) is even more spectacular since
the FDR is shown to be characterized by only two numbers instead
of a function, namely
$X \sim 1$ at short times (large value of the correlator)
corresponding to
a quasi-equilibrium regime, with a crossover
to a non-trivial number, $X \sim X^\infty$ for
large times (small value of the correlator).
This implies that parametric FD plots are simply made of
two straight lines with slope $1$ and $X^\infty$, instead of the
single straight line of slope 1 obtained at equilibrium.

Since any kind of behaviour
is in principle allowed in non-equilibrium situations,
getting such a simple, equilibrium-like structure for the
FD relations is a remarkable result.
This immediately led to the idea that aging systems might be characterized
by an effective thermodynamic behaviour and the idea
of quasi-equilibration at different timescales~\cite{CugKurPel97}.
In particular,
generalized FD relations suggest to define an
effective temperature, as
\be
T_{\rm eff} = \frac{T}{X(t,\tw)},
\label{teffdef}
\ee
such that mean-field glasses are characterized by a unique
effective temperature, $T_{\rm eff}=T/X^\infty$.
It is thought of as the temperature at which
slow modes are quasi-equilibrated. One finds in general that
$0<X^\infty <1$, such that $T_{\rm eff} > T$,
as if the system had kept some memory of its
high temperature initial state.

The name `temperature' for the quantity defined in Eq.~(\ref{teffdef})
is not simply the result of a dimensional analysis but has a deeper,
physically appealing meaning that is revealed by asking the
following questions. How does one measure temperatures in
a many-body system whose relaxation involves well-separated
timescales? What is a thermometer (and a temperature) in a
far from equilibrium aging
material? Answers are provided in
Refs.~\cite{CugKurPel97,Kurchan} both
for mean-field models and for additional toy models
with multiple relaxation timescales. The idea is to couple
an additional degree of freedom, such as a harmonic oscillator, $x(t)$,
which plays the role of the thermometer operating at frequency
$\omega$, to an observable of interest $A(t)$ via a linear coupling,
$-\lambda x(t)A(t)$. Simple calculations show then that the
thermometer `reads' the following temperature,
\begin{equation}
\frac{1}{2} K_B T_{\rm meas}^2 \equiv \frac{1}{2} \omega^2 \langle x^2 \rangle
= \frac{\omega C'(\omega,\tw)}{2 \chi''(\omega,\tw)},
\label{meas}
\end{equation}
where $C'(\omega,\tw)$ is the real part of the
Fourier transform of Eq.~(\ref{corr}), and
$\chi(\omega,\tw)$ the imaginary part of the
Fourier transform of Eq.~(\ref{resp}), with
$h=\lambda x$. The relation (\ref{meas})
indicates that the bath temperature is measured,
$T_{\rm meas} = T$, if the frequency is high and FDT is satisfied, while
$T_{\rm meas} = T_{\rm eff} > T$ if the frequency is slow enough to be
tuned to that of the slow relaxation in the aging material.
The link between the FDR in Eq.~(\ref{fdr_def}) and the effective temperature
measured in Eq.~(\ref{meas}) was numerically confirmed in the computer
simulation of a glassy molecular liquid in Ref.~\cite{jl}.

More generally, relaxation in glassy
systems occurs in well-separated time sectors~\cite{CugKur2}; it is
then easy to imagine that each sector could be associated with an effective
temperature~\cite{Kurchan}. A thermodynamic interpretation of
effective temperatures has also been put forward, relating them to the
concept of replica symmetry breaking~\cite{FraMezParPel98}.
Interestingly, the full-step or  one-step replica symmetry
breaking schemes needed to solve the static problem in these models
have a counterpart as the FDR being a function or a number,
respectively, in the aging regime. Moreover, we note that
these modern concepts are related to, but make much more
precise, older ideas of quasi-equilibrium and fictive
temperatures in aging glasses~\cite{struik1977physical}.

\begin{figure}
\psfig{file=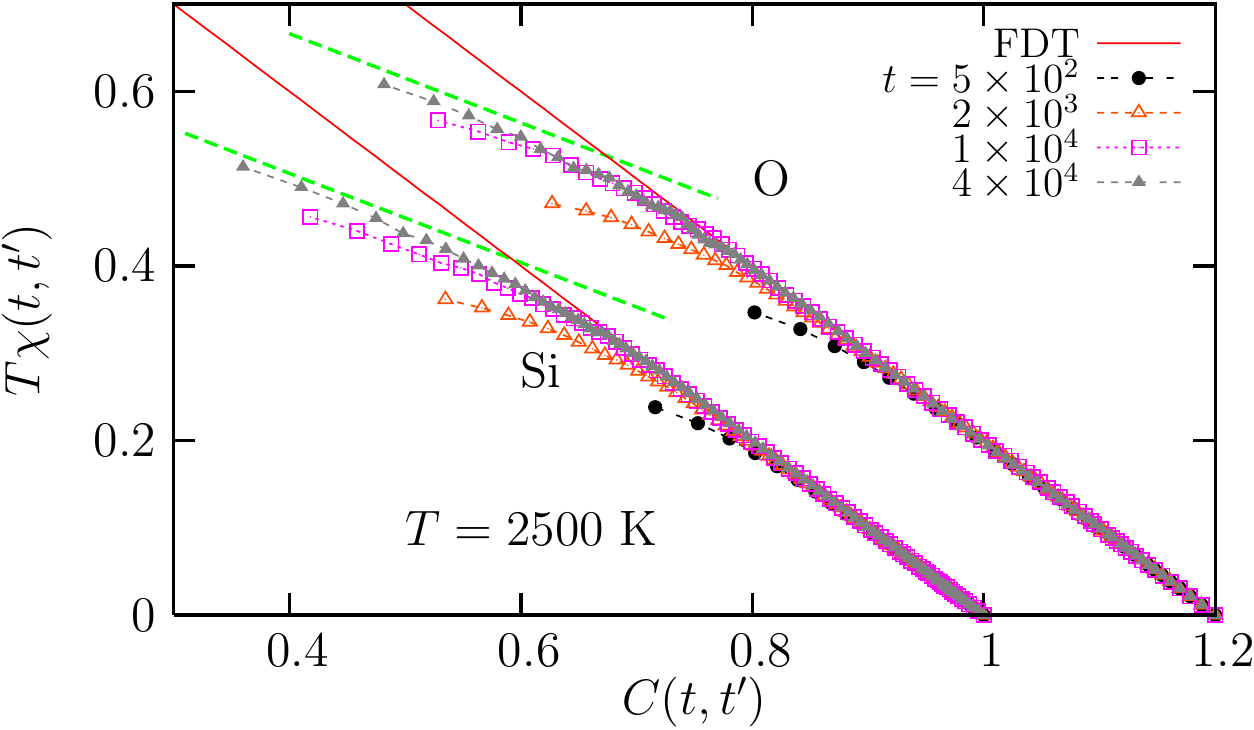,width=8.5cm}
\caption{Parametric correlation-response plots
measured in the aging regime of a numerical model for a silica
glass, SiO$_2$~\cite{prlsilica}.
The plots for both species smoothly converges
towards a two-straight line plot of slope 1 at short times (large $C$ values),
and of slope $X^\infty \approx 0.51$ at large times
(small values of $C$), yielding an effective temperature of about
$T_{\rm eff} = T/X^\infty \approx 4900$~K. Note that the glass transition
temperature of SiO$_2$ is 1446K.}
\label{fdrfig} 
\end{figure}

Taken together, these results make the mean-field description of aging
very appealing, and they nicely complement the mode-coupling/RFOT
description of the equilibrium glass transition described above.
Moreover, they
have set the agenda for a large body of numerical
and experimental work, as reviewed in~\cite{crisrit}.
In Fig.~\ref{fdrfig} we present recent numerical data obtained
in an aging silica glass~\cite{prlsilica}, presented in the form
of a parametric response-correlation plot. The measured
correlation functions are the self-part of the intermediate
scattering functions defined in Eq.~(\ref{isf}), while
the conjugated response functions quantify the response of
particle displacements to a spatially modulated
field conjugated to the density.
Plots for silicon and oxygen atoms at different
ages are presented. They seem to smoothly converge
towards a two-straight line plot, as obtained in mean-field
models (note, however, that this could be just a pre-asymptotic,
finite `$t_w$', effect).
Moreover, the second, non-trivial part of the plot
is characterized by a slope that appears to be independent
of the species, and of the wavevector chosen to quantify the dynamics,
in agreement with the idea of a unique asymptotic value of the
FDR, possibly related to a well-defined effective temperature.

\subsection{Beyond mean-field: Real Space}

Despite successful results, such as those shown in Fig.~\ref{fdrfig}, the broader
applicability of the mean-field scenario of aging dynamics remains still unclear.
While some experiments and simulations indeed seem to support the
existence of well-behaved effective
temperatures~\cite{Grigera99,Abou04,Wang06}, other studies also reveal
the limits of the mean-field scenario. Experiments have for instance
reported anomalously large FDT violations associated with intermittent
dynamics~\cite{Bellon1,Bellon2,Buisson1,Buisson2}, while theoretical
studies of model systems have also found non-monotonic or even
negative response functions~\cite{Viot03,nicodemi,kr,DepSti}, and
ill-defined or observable-dependent FDRs~\cite{FieSol02}. In
principle, these discrepancies with mean-field predictions are to be
expected, since there are many systems of physical interest in which
the dynamics are not of mean-field type, displaying both activated
processes and spatial heterogeneity.

It is thus an important task to
understand from the theoretical point of view when the mean-field
concept of an FDR-related effective temperature remains viable.
However, theoretically studying the interplay between relevant dynamic
lengthscales and thermally
activated dynamics in the non-equilibrium regime of disordered
materials is clearly a
challenging task. Nevertheless,
this problem has been approached in
different ways, as we briefly summarize in this subsection.

A first class of system that displays aging and spatial
heterogeneity is given by coarsening systems.
The paradigmatic situation  is that of an Ising ferromagnetic
model (with a transition at $T_c$)
suddenly quenched in the ferromagnetic phase at time
$\tw=0$. For $\tw > 0$, domains of positive and negative magnetizations
appear and slowly coarsen with time. The appearance of
domains that grow with time proves the presence
of both aging and heterogeneity.

The case where the quench is performed down to $T<T_c$ is
well understood. The system becomes scale invariant~\cite{reviewbray},
since the only relevant lengthscale is the growing domain size, $\ell(\tw)$.
Correlation functions display aging, and scale invariance
implies that $C(t,\tw) \sim f(\ell(t)/\ell(\tw))$. Response
functions can be decomposed into two
contributions~\cite{barrat,BBK2}: one part stems from
the bulk of the domains and behaves as the equilibrium response,
and a second one from the domain walls and becomes vanishingly small
in the long time limit, where $\ell(\tw) \to \infty$ and the density
of domain walls vanishes. This implies that for coarsening systems
in $d \geq 2$, one has $X^\infty = 0$, or equivalently an infinite
effective temperature, $T_{\rm eff} = \infty$. The case $d=1$ is special
because $T_c=0$ and the response function remains dominated by the
domain walls, which yields
the non-trivial value $X^\infty = 1/2$~\cite{ising1d,ising1dLippiello}.

Another special case has retained attention. When the quench
is performed at $T=T_c$, there is no more distinction
between walls and domains and the above argument yielding
$X^{\infty}=0$ does not hold.
Instead one studies the growth 
of critical fluctuations with time, with $\xi(\tw) \sim \tw^{1/z}$
the correlation length at time $\tw$,
where $z$ is the dynamic exponent.
Both correlation and response functions become non-trivial
at the critical point~\cite{godluck}.
It proves useful in that case to consider the dynamics
of the Fourier components of the magnetization fluctuations,
$C_q(t,\tw) = \langle m_q(t) m_{-q}(\tw) \rangle$,
and the conjugated response
$R_q(t,\tw) = \frac{\delta \langle m_q(t) \rangle}{\delta
h_{-q}(\tw)}$. From Eq.~(\ref{fdr_def}) a wavevector dependent
FDR follows, $X_q(t,\tw)$, which has
interesting properties~\cite{pre} (see \cite{pasquale} for a review).

In dimension $d=1$, it is possible to compute
$X_q(t,\tw)$ exactly in the aging regime at $T=T_c=0$.
An interesting scaling form is found, and numerical simulations
performed for $d>1$ confirm its validity:
\be
X_q(t,\tw) = {\cal X}( q^2 \tw),
\label{scaling_xq}
\ee
where the scaling function ${\cal X}(x)$ is
${\cal X}(x \to \infty) \to 1$ at small lengthscale, $q \xi \gg 1$,
and ${\cal X}(x \to 0) \to 1/2$ (in $d=1$)
at large distance,
$q \xi \ll 1$; recall that $z=2$ in that case.

Contrary to mean-field systems where geometry played no role,
here the presence of a growing correlation lengthscale
plays a crucial role in the off-equilibrium regime since
$\xi(\tw)$ allows one to discriminate between fluctuations
that satisfy the FDT at small lengthscale, $X_q \sim 1$,
and those at large lenghtscale which are still far from equilibrium,
$0< X_q \sim X^\infty < 1$. These studies suggest therefore that
generalized fluctuation-dissipation relations in fact
have a strong lengthscale dependence--a result which is
not predicted by mean-field approaches.

Another interesting result is that the FDT violation
for global observables (i.e.~those at $q=0$) takes a particularly
simple form, since the introduction of a single number is sufficient,
the FDR at zero wavevector,
$X_{q=0}(t,\tw) \equiv X^{\infty} = 1/2$ (in $d=1$).
This universal quantity takes non-trivial values
in higher dimension, e.g.~$X^\infty \approx 0.34$ is measured
in $d=2$~\cite{pre}. This shows that the study of global rather than local
quantities makes the measurement of $X^\infty$ much easier.
Finally, having a non-trivial value of $X^\infty$
for global observables suggests that the possibility to
define an effective temperature remains valid,
but it has become a more complicated object, related to global
fluctuations on large lengthscale.

Kinetically constrained spin models represent a second
class of non-mean-field
systems whose off-equilibrium has been
thoroughly studied recently~\cite{leonard}.
This is quite a natural thing to do
since these systems have
local, finite ranged
interactions, and they
combine the interesting features of being defined in
terms of (effective)
microscopic degrees of freedom, having local dynamical rules, and
displaying thermally activated and heterogeneous dynamics.

\begin{figure}
\psfig{file=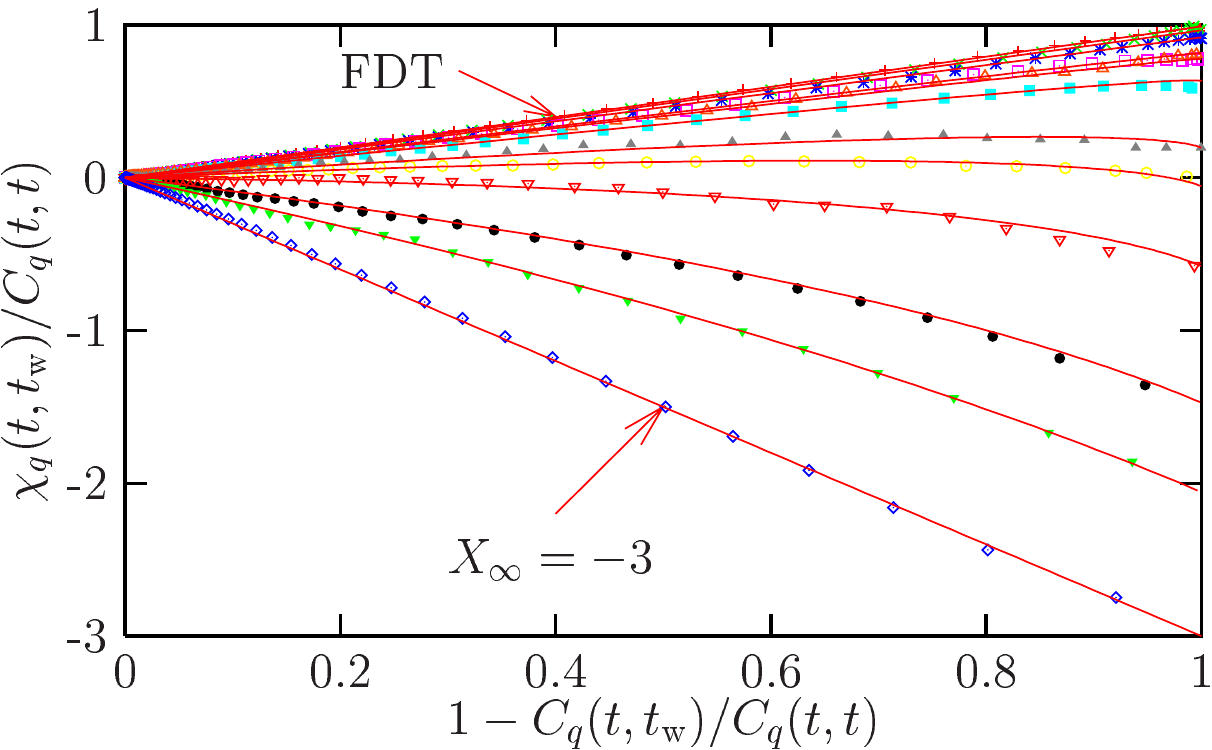,width=8.5cm}
\caption{Parametric response-correlation plots for the Fourier components
of the mobility field in the $d=3$ Fredrickson-Andersen model.
Symbols are from simulations, lines from analytic calculations,
and wavevectors decrease from top to bottom.
The FDT is close to being satisfied at large $q$
corresponding to local equilibrium. At larger distance
deviations from the FDT are seen, with an asymptotic
FDR which becomes negative. Finally, for energy fluctuations
at $q=0$ (bottom curve), the plot becomes a pure straight line
of (negative!) slope -3, as a result of thermally activated dynamics.}
\label{fdrfig2}
\end{figure}

The case of the Fredrickson-Andersen model, described in Sec.~\ref{theory},
has been studied in great detail~\cite{leonard}, and we summarize the main
results. Here,
the relevant dynamic variables are the Fourier components
of the mobility field, which also correspond in that case
to the fluctuations of the energy density.
Surprisingly, the structure of the generalized fluctuation-dissipation
relation remains once more very simple. In particular, in dimension
$d>2$, one finds a scaling form similar to (\ref{scaling_xq}),
$X_q(t,\tw) = {\cal X}( q^2 \tw)$, with a well-defined
limit at large distance $X_{q=0}(t,\tw) \equiv X^\infty$.
The deep analogy with critical Ising models stems from the fact
that mobility defects in KCMs diffuse in a way similar to
domain walls in coarsening Ising models. It is in fact
by exploiting this analogy that analytic results are
obtained in the aging regime of the
Fredrickson-Andersen model~\cite{peters}.

There is however a major qualitative difference between
the two families of model. The (big!) surprise lies in the sign
of the asymptotic FDR, since calculations show that~\cite{nveX}
\be
X^\infty = -3, \quad \,\, d>2.
\ee
In dimension $d=1$, one finds
$X_{q=0}(t,\tw) = f(t/\tw)$ with
$X_{q=0}(t\to\infty,\tw) = \frac{3\pi}{16-6\pi}
\approx -3.307$.
Numerical simulations confirm these calculations.
In Fig.~\ref{fdrfig2}, we show such a comparison between
simulations (symbols) and theory (lines) in the case
of the $d=3$ Fredrickson-Andersen model~\cite{nveX}.
Fourier components
of the mobility field yield parametric FD plots that follow
scaling with the variable $q^2 \tw$, as a direct result
of the presence of a growing lengthscale for
dynamic heterogeneity, $\xi(\tw) \sim \sqrt{\tw}$.
Again, generalized fluctuation-dissipation relations
explicitely depend on the spatial lengthscale considered,
unlike in mean-field studies.
In Fig.~\ref{fdrfig2}, the limit $q =0$
corresponding to global observables is also
very interesting since the plot is a pure straight line,
as in equilibrium. Unlike equilibrium, however, the slope is not 1 but
$-3$. A negative slope in this plot means a negative FDR, and therefore
suggests a negative effective temperature,
a very non-intuitive result at first sight.

Negative response functions in fact directly follows
from the thermally activated nature of the dynamics
of these models~\cite{nveX}. First, one should note that
the global observable shown in Fig.~\ref{fdrfig2} corresponds
to fluctuations of the energy, $e(\tw)$, whose conjugated field is
temperature.
In the aging regime the system slowly drifts towards equilibrium.
Microscopic moves result from thermally activated
processes, corresponding to the local crossing of energy barriers.
An infinitesimal change in temperature, $T \to T+\delta T$
with $\delta T >0$, accelerates these barrier crossings and
makes the relaxation dynamics faster.
The energy response to a positive temperature pulse
is therefore negative,  $\delta e < 0$, which directly
yields $\delta e / \delta T <0$, which explains
the negative sign of the FDR. This result does not hold in
mean-field glasses, where thermal activation plays no role.

Finally, another scenario holds for local observables in some KCMs when
kinetic constraints are stronger, such as the
East model~\cite{leonard}
or a bidimensional triangular plaquette model~\cite{robfdt}.
Here, relaxation is governed by a hierarchy of energy
barriers that endow the systems with specific dynamic
properties. In the aging regime following a quench,
in particular, the hierarchy yields an energy relaxation that
arises in discrete steps which take place
on very different timescales, reminiscent of the `time sectors'
encountered in mean-field spin glasses.
Surprisingly, it is found that to each of these
discrete relaxations one can associate a well-defined (positive)
value of the fluctuation-dissipation ratio, again
reminiscent of the dynamics of mean-field spin
glass models.  
Therefore, a physical picture seems to have some validity,
even in models that are very far from the mean-field limit:
in that picture, slow relaxation takes place on multiple timescales,
with each timescale characterized by a different effective temperature.

\subsection{Beyond mean-field: Energy Landscape}

\label{subsec:beyondMF-landscape}

What the mean-field approach crucially lacks is the description of activated processes that allow the system to jump over larger and larger barriers at large times. One way to introduce this key effect  is to study mean-field models at finite system size $N$. In fact, the mean field theory of aging is derived by first taking the limit $N\rightarrow \infty$ to study the dynamics on large but finite timescales. Instead, if one focuses on timescales that diverge exponentially with $N$, activated processes will occur. One can then analyze in a controlled way how jumping over barriers alters the mean-field dynamics. 

This approach was pioneered numerically by Crisanti and Ritort~\cite{crisrit,crisanti2000potential,crisanti2000activated}, and more recently pursued further in~\cite{baity2018activated,baity2018activated2,billoire2005mean,stariolo2019activated}. 
One of the most important outcomes of their study is to show the existence of an effective temperature and its relation with the energy landscape, more precisely the complexity, even when barriers are crossed. 
The understanding of the dynamical evolution in this regime was reached recently by theoretical and rigorous analysis of the Random Energy Model (REM)  in~\cite{arous2002aging,gayrard2019aging,baity2018activated,baity2018activated2}. It was shown that when observing the dynamics on exponentially large (in $N$) timescales, activated processes lead to a dynamical evolution that can be mapped on the one of the trap model~\cite{bouchaud1992weak,dyre1987master}. Again, an interesting relation with the energy landscape emerges: the exponent of the trapping time power law describing aging dynamics at a given energy is directly linked to the slope of the configurational entropy at that energy. It is not yet clear whether this scenario goes beyond the REM and applies, for instance, to the Monte Carlo dynamics of the $p$-spin model. In this case the energy landscape is more complex and one needs to understand the interplay between energy and entropic barriers~\cite{barrat1995phase,cammarota2018numerical,cammarota2015spontaneous}. 

To this end, a series of recent works focused on the organization in configuration space of the barriers that can be used to escape from a given minimum for the $p$-spin spherical model. By using the Kac-Rice formalism, these works computed the entropy of the barriers at a given energy and at a certain distance from a given minimum~\cite{ros2019complexity} and obtained the dynamical instanton representing the escape from a given minimum. 

Finally, another set of works performed a complementary study on finite size models of three dimensional supercooled liquids. The key idea was to focus on a size that is large enough to be representative of the bulk behavior and small enough to be able to study the dynamics in terms of energy landscape~\cite{buchner1999potential,doliwa2003hopping,heuer2008exploring}. This led to the introduction of the notion of {metabasins}, a set of basins that corresponds to the same metastable state. A study of the dynamics in terms of energy barriers between metabasins was then performed. It showed in a very compelling way that the dynamics start to be activated even above the MCT cross-over in 3D systems~\cite{doliwa2003hopping,denny2003trap}. Interestingly, a strong relationship with the dynamics of the trap model was also found in this case~\cite{heuer2008exploring}. 

All the results cited above provide valuable information and insights on activated dynamics. Interestingly, they show that relations between aging dynamics and energy landscape found within mean-field theory seem to hold more broadly. Many questions remain open and will hopefully be addressed in future work, such as understanding and characterizing the dynamical paths and the barrier crossings leading to relaxation during the aging regime.

\subsection{Driven glassy materials}

We have introduced aging phenomena with the argument that
in a glass phase, the timescale to equilibrate becomes
so long that the system always remembers its complete history. This is
true in general, but one can wonder whether it is possible to invent
a protocol where the material history could be erased, and the system
`rejuvenated'~\cite{mckenna}.
This concept has  been known for decades in the field of polymer
glasses, where complex thermo-mechanical histories are often
used.

Let us consider an aging protocol where the system is quenched
to low temperature at time $\tw=0$, but the system is simultaneously
forced by an external mechanical constraint. Experimentally one finds
that a stationary state can be reached, which explicitly depends
on the strength of the forcing: a system which is forced more strongly
relaxes faster than a material that is less solicited, a phenomenon
called `shear-thinning'. The material
has therefore entered a driven steady state, where
memory of its age is no longer present and the dynamics have become
stationary: aging is stopped.

Many studies of these driven glassy states have been performed
in recent years. These studies are relevant for the rheology of supercooled liquids
and glasses, and the $T \ll T_g$ limit corresponds to studies
of the plasticity of amorphous solids, a broad field in itself, see Sec.~\ref{sec:rheology}.
In the colloidal world, such studies are also relevant
for the newly-defined field of the rheology of `soft glassy
materials'. These materials are (somewhat tautologically)
defined as those for which the non-linear
rheological behaviour is believed to result precisely
from the competition between intrinsically slow relaxation processes
and an external forcing~\cite{prlsollich,ikeda2012unified}. It is believed that
the rheology of dense colloidal suspensions, foams, emulsions, binary mixtures,
or even biophysical systems are ruled by such a competition, which represents a
broad scope for applications.

From the point of view of statmech modeling,
soft glassy rheology can be naturally studied
from the very same angles as the glass transition
itself. As such trap models~\cite{prlsollich,presollich},
mean-field spin glasses~\cite{BBK} and the related
mode-coupling theory approach~\cite{dave,fuchs}
have been explicitely extended
to include an external mechanical forcing. In all these cases, one
finds that a driven steady state can be reached and aging is
indeed expected to stop at a level that depends on the strength of the
forcing. Many of the results obtained in aging systems
about the properties of an effective temperature are also shown to apply
in the driven case, as shown both theoretically~\cite{BBK} and
numerically~\cite{jllong}.
A most interesting aspect is that
the broad relaxation spectra predicted to occur
in glassy materials close to a glass transition
directly translate into `anomalous' laws both
for the linear rheological behaviour (seen experimentally
in the broad spectrum of elastic, $G'(\omega)$,
and loss, $G''(\omega)$, moduli), and the non-linear
rheological behaviour (a strong dependence of the viscosity
$\eta$ upon the shear rate $\dot\gamma$).

\section{Future directions}

\label{nofuture}

The problem of the glass transition, already very exciting in itself,
has ramifications well beyond the physics of supercooled liquids.
Glassy systems figure among the even larger class of
`complex systems'. These are formed by a set of interacting degrees of freedom
showing non-trivial emergent behaviour: as a whole they exhibit properties that are 
not already encoded in the definition of the individual parts.
As a consequence the study of glass-formers as statistical mechanics models
characterized by frustrated interactions is a fertile ground to develop new
concepts and techniques that will likely be
applied to other physical, and more generally,
scientific situations.

An example, already cited in this review, are the progress
obtained in computer science and information theory~\cite{complexbook}
using techniques originally developed for spin glasses and structural
glasses. There is no doubt that progress will steadily continue in the future along these
interdisciplinary routes.
Concerning physics, glassiness is such an ubiquitous and, yet as
we showed, rather poorly understood problem that
many developments are very likely to take place in the next decade.

Instead of guessing future
developments of the field (and then very likely be proven wrong)
we prefer to list a few problems we would like to see solved in the next years.

\begin{itemize}

\item Is the glass transition related to a true phase transition? If yes, a
  static or a dynamic one? A finite or zero temperature one?

\item Do RFOT theory, defects models, or frustration-based theory
form the correct starting points of `the' theory of the glass transition?

\item Is MCT really a useful theory for the first decades
of slowing down of the dynamics? Can one find direct evidence
that an avoided MCT transition exists and controls the dynamics?

\item What is the correct physical picture for the low temperature phase of
glass-forming liquids and spin glasses?

\item Are there general principles governing 
off-equilibrium dynamics, 
and in particular aging and sheared materials?

\item Do non-disordered, finite-dimensional, finite-range
statmech model exist that display a
thermodynamically stable amorphous
phase at low temperature?

\end{itemize}

Finally, notice that we did not discuss possible interplays
between glassiness, disorder and quantum fluctuations. This is a very fascinating topic that has boomed in recent years; new phenomena such as Many-Body Localization \cite{nandkishore2015many} and Quantum Scars \cite{turner2018weak} have been discovered, revealing new facets of slow dynamics. Models of classical glasses, such as the KCMs and the $p$-spin models found new applications in this arena \cite{pancotti2020quantum,facoetti2019classical}. 

\acknowledgments

We thank all the collaborators who worked with us on glass physics.
This work was supported by a grant from the Simons Foundation (Grant
No. 454933, L. B., Grant No. 454935, G. B.)

\bibliography{glass.bib} 

\begin{thebibliography}{481}%
\makeatletter
\providecommand \@ifxundefined [1]{%
 \@ifx{#1\undefined}
}%
\providecommand \@ifnum [1]{%
 \ifnum #1\expandafter \@firstoftwo
 \else \expandafter \@secondoftwo
 \fi
}%
\providecommand \@ifx [1]{%
 \ifx #1\expandafter \@firstoftwo
 \else \expandafter \@secondoftwo
 \fi
}%
\providecommand \natexlab [1]{#1}%
\providecommand \enquote  [1]{``#1''}%
\providecommand \bibnamefont  [1]{#1}%
\providecommand \bibfnamefont [1]{#1}%
\providecommand \citenamefont [1]{#1}%
\providecommand \href@noop [0]{\@secondoftwo}%
\providecommand \href [0]{\begingroup \@sanitize@url \@href}%
\providecommand \@href[1]{\@@startlink{#1}\@@href}%
\providecommand \@@href[1]{\endgroup#1\@@endlink}%
\providecommand \@sanitize@url [0]{\catcode `\\12\catcode `\$12\catcode
  `\&12\catcode `\#12\catcode `\^12\catcode `\_12\catcode `\%12\relax}%
\providecommand \@@startlink[1]{}%
\providecommand \@@endlink[0]{}%
\providecommand \url  [0]{\begingroup\@sanitize@url \@url }%
\providecommand \@url [1]{\endgroup\@href {#1}{\urlprefix }}%
\providecommand \urlprefix  [0]{URL }%
\providecommand \Eprint [0]{\href }%
\providecommand \doibase [0]{http://dx.doi.org/}%
\providecommand \selectlanguage [0]{\@gobble}%
\providecommand \bibinfo  [0]{\@secondoftwo}%
\providecommand \bibfield  [0]{\@secondoftwo}%
\providecommand \translation [1]{[#1]}%
\providecommand \BibitemOpen [0]{}%
\providecommand \bibitemStop [0]{}%
\providecommand \bibitemNoStop [0]{.\EOS\space}%
\providecommand \EOS [0]{\spacefactor3000\relax}%
\providecommand \BibitemShut  [1]{\csname bibitem#1\endcsname}%
\let\auto@bib@innerbib\@empty
\bibitem [{\citenamefont {Angell}(1995)}]{angellscience}%
  \BibitemOpen
  \bibfield  {author} {\bibinfo {author} {\bibfnamefont {C.~A.}\ \bibnamefont
  {Angell}},\ }\href@noop {} {\bibfield  {journal} {\bibinfo  {journal}
  {Science}\ }\textbf {\bibinfo {volume} {267}},\ \bibinfo {pages} {1924}
  (\bibinfo {year} {1995})}\BibitemShut {NoStop}%
\bibitem [{\citenamefont {Debenedetti}\ and\ \citenamefont
  {Stillinger}(2001)}]{reviewnature}%
  \BibitemOpen
  \bibfield  {author} {\bibinfo {author} {\bibfnamefont {P.~G.}\ \bibnamefont
  {Debenedetti}}\ and\ \bibinfo {author} {\bibfnamefont {F.~H.}\ \bibnamefont
  {Stillinger}},\ }\href@noop {} {\bibfield  {journal} {\bibinfo  {journal}
  {Nature}\ }\textbf {\bibinfo {volume} {410}},\ \bibinfo {pages} {259}
  (\bibinfo {year} {2001})}\BibitemShut {NoStop}%
\bibitem [{\citenamefont {Berthier}\ and\ \citenamefont
  {Biroli}(2011)}]{berthier_theoretical_2011}%
  \BibitemOpen
  \bibfield  {author} {\bibinfo {author} {\bibfnamefont {L.}~\bibnamefont
  {Berthier}}\ and\ \bibinfo {author} {\bibfnamefont {G.}~\bibnamefont
  {Biroli}},\ }\href {\doibase 10.1103/RevModPhys.83.587} {\bibfield  {journal}
  {\bibinfo  {journal} {Reviews of Modern Physics}\ }\textbf {\bibinfo {volume}
  {83}},\ \bibinfo {pages} {587} (\bibinfo {year} {2011})}\BibitemShut
  {NoStop}%
\bibitem [{\citenamefont {Struik}(1977)}]{struik1977physical}%
  \BibitemOpen
  \bibfield  {author} {\bibinfo {author} {\bibfnamefont {L.}~\bibnamefont
  {Struik}},\ }\href@noop {} {\bibfield  {journal} {\bibinfo  {journal}
  {Polymer Engineering \& Science}\ }\textbf {\bibinfo {volume} {17}},\
  \bibinfo {pages} {165} (\bibinfo {year} {1977})}\BibitemShut {NoStop}%
\bibitem [{\citenamefont {Young}(1998)}]{youngbook}%
  \BibitemOpen
  \bibfield  {author} {\bibinfo {author} {\bibfnamefont {A.~P.}\ \bibnamefont
  {Young}},\ }\href@noop {} {\emph {\bibinfo {title} {Spin glasses and random
  fields}}},\ Vol.~\bibinfo {volume} {12}\ (\bibinfo  {publisher} {World
  Scientific},\ \bibinfo {year} {1998})\BibitemShut {NoStop}%
\bibitem [{\citenamefont {Bouchaud}\ \emph {et~al.}(2011)\citenamefont
  {Bouchaud}, \citenamefont {M{\'e}zard},\ and\ \citenamefont
  {Dalibard}}]{complexbook}%
  \BibitemOpen
  \bibfield  {author} {\bibinfo {author} {\bibfnamefont {J.-P.}\ \bibnamefont
  {Bouchaud}}, \bibinfo {author} {\bibfnamefont {M.}~\bibnamefont
  {M{\'e}zard}}, \ and\ \bibinfo {author} {\bibfnamefont {J.}~\bibnamefont
  {Dalibard}},\ }\href@noop {} {\emph {\bibinfo {title} {Complex systems:
  lecture notes of the Les Houches Summer School 2006}}}\ (\bibinfo
  {publisher} {Elsevier},\ \bibinfo {year} {2011})\BibitemShut {NoStop}%
\bibitem [{\citenamefont {Richert}\ and\ \citenamefont
  {Angell}(1998)}]{angellrichert}%
  \BibitemOpen
  \bibfield  {author} {\bibinfo {author} {\bibfnamefont {R.}~\bibnamefont
  {Richert}}\ and\ \bibinfo {author} {\bibfnamefont {C.}~\bibnamefont
  {Angell}},\ }\href@noop {} {\bibfield  {journal} {\bibinfo  {journal} {The
  Journal of chemical physics}\ }\textbf {\bibinfo {volume} {108}},\ \bibinfo
  {pages} {9016} (\bibinfo {year} {1998})}\BibitemShut {NoStop}%
\bibitem [{Note1()}]{Note1}%
  \BibitemOpen
  \bibinfo {note} {The terminology `strong' and `fragile' is not related to the
  mechanical properties of the glass but to the evolution of the short-range
  order close to $T_g$. Strong liquids, such as SiO$_2$, have a locally
  tetrahedric structure which persists both below and above the glass
  transition contrary to fragile liquids whose short-range amorphous structure
  disappears rapidly upon heating above $T_g$.}\BibitemShut {Stop}%
\bibitem [{\citenamefont {B{\"a}ssler}(1987)}]{bassler}%
  \BibitemOpen
  \bibfield  {author} {\bibinfo {author} {\bibfnamefont {H.}~\bibnamefont
  {B{\"a}ssler}},\ }\href@noop {} {\bibfield  {journal} {\bibinfo  {journal}
  {Physical review letters}\ }\textbf {\bibinfo {volume} {58}},\ \bibinfo
  {pages} {767} (\bibinfo {year} {1987})}\BibitemShut {NoStop}%
\bibitem [{\citenamefont {Kauzmann}(1948)}]{kauzmann}%
  \BibitemOpen
  \bibfield  {author} {\bibinfo {author} {\bibfnamefont {W.}~\bibnamefont
  {Kauzmann}},\ }\href@noop {} {\bibfield  {journal} {\bibinfo  {journal}
  {Chemical reviews}\ }\textbf {\bibinfo {volume} {43}},\ \bibinfo {pages}
  {219} (\bibinfo {year} {1948})}\BibitemShut {NoStop}%
\bibitem [{\citenamefont {Berthier}\ \emph
  {et~al.}(2019{\natexlab{a}})\citenamefont {Berthier}, \citenamefont {Ozawa},\
  and\ \citenamefont {Scalliet}}]{berthier_configurational_2019}%
  \BibitemOpen
  \bibfield  {author} {\bibinfo {author} {\bibfnamefont {L.}~\bibnamefont
  {Berthier}}, \bibinfo {author} {\bibfnamefont {M.}~\bibnamefont {Ozawa}}, \
  and\ \bibinfo {author} {\bibfnamefont {C.}~\bibnamefont {Scalliet}},\ }\href
  {\doibase 10.1063/1.5091961} {\bibfield  {journal} {\bibinfo  {journal} {The
  Journal of Chemical Physics}\ }\textbf {\bibinfo {volume} {150}},\ \bibinfo
  {pages} {160902} (\bibinfo {year} {2019}{\natexlab{a}})}\BibitemShut
  {NoStop}%
\bibitem [{\citenamefont {Ho{\l}~yst}(2001)}]{holyst}%
  \BibitemOpen
  \bibfield  {author} {\bibinfo {author} {\bibfnamefont {R.}~\bibnamefont
  {Ho{\l}~yst}},\ }\href@noop {} {\bibfield  {journal} {\bibinfo  {journal}
  {Physica A: Statistical Mechanics and its Applications}\ }\textbf {\bibinfo
  {volume} {292}},\ \bibinfo {pages} {255} (\bibinfo {year}
  {2001})}\BibitemShut {NoStop}%
\bibitem [{\citenamefont {Goldstein}(1969)}]{Goldstein}%
  \BibitemOpen
  \bibfield  {author} {\bibinfo {author} {\bibfnamefont {M.}~\bibnamefont
  {Goldstein}},\ }\href@noop {} {\bibfield  {journal} {\bibinfo  {journal} {The
  Journal of Chemical Physics}\ }\textbf {\bibinfo {volume} {51}},\ \bibinfo
  {pages} {3728} (\bibinfo {year} {1969})}\BibitemShut {NoStop}%
\bibitem [{\citenamefont {Debenedetti}(1996)}]{debenedetti}%
  \BibitemOpen
  \bibfield  {author} {\bibinfo {author} {\bibfnamefont {P.~G.}\ \bibnamefont
  {Debenedetti}},\ }\href@noop {} {\emph {\bibinfo {title} {Metastable liquids:
  concepts and principles}}}\ (\bibinfo  {publisher} {Princeton University
  Press},\ \bibinfo {year} {1996})\BibitemShut {NoStop}%
\bibitem [{\citenamefont {Menon}\ and\ \citenamefont {Nagel}(1995)}]{static}%
  \BibitemOpen
  \bibfield  {author} {\bibinfo {author} {\bibfnamefont {N.}~\bibnamefont
  {Menon}}\ and\ \bibinfo {author} {\bibfnamefont {S.~R.}\ \bibnamefont
  {Nagel}},\ }\href@noop {} {\bibfield  {journal} {\bibinfo  {journal}
  {Physical review letters}\ }\textbf {\bibinfo {volume} {74}},\ \bibinfo
  {pages} {1230} (\bibinfo {year} {1995})}\BibitemShut {NoStop}%
\bibitem [{\citenamefont {Fern{\'a}ndez}\ \emph {et~al.}(2006)\citenamefont
  {Fern{\'a}ndez}, \citenamefont {Mart{\'\i}n-Mayor},\ and\ \citenamefont
  {Verrocchio}}]{fernandez2006critical}%
  \BibitemOpen
  \bibfield  {author} {\bibinfo {author} {\bibfnamefont {L.}~\bibnamefont
  {Fern{\'a}ndez}}, \bibinfo {author} {\bibfnamefont {V.}~\bibnamefont
  {Mart{\'\i}n-Mayor}}, \ and\ \bibinfo {author} {\bibfnamefont
  {P.}~\bibnamefont {Verrocchio}},\ }\href@noop {} {\bibfield  {journal}
  {\bibinfo  {journal} {Physical Review E}\ }\textbf {\bibinfo {volume} {73}},\
  \bibinfo {pages} {020501} (\bibinfo {year} {2006})}\BibitemShut {NoStop}%
\bibitem [{\citenamefont {Cavagna}\ \emph {et~al.}(2007)\citenamefont
  {Cavagna}, \citenamefont {Grigera},\ and\ \citenamefont
  {Verrocchio}}]{cavagna}%
  \BibitemOpen
  \bibfield  {author} {\bibinfo {author} {\bibfnamefont {A.}~\bibnamefont
  {Cavagna}}, \bibinfo {author} {\bibfnamefont {T.~S.}\ \bibnamefont
  {Grigera}}, \ and\ \bibinfo {author} {\bibfnamefont {P.}~\bibnamefont
  {Verrocchio}},\ }\href@noop {} {\bibfield  {journal} {\bibinfo  {journal}
  {Physical review letters}\ }\textbf {\bibinfo {volume} {98}},\ \bibinfo
  {pages} {187801} (\bibinfo {year} {2007})}\BibitemShut {NoStop}%
\bibitem [{\citenamefont {Wuttke}\ \emph {et~al.}(1996)\citenamefont {Wuttke},
  \citenamefont {Petry},\ and\ \citenamefont {Pouget}}]{glycerol}%
  \BibitemOpen
  \bibfield  {author} {\bibinfo {author} {\bibfnamefont {J.}~\bibnamefont
  {Wuttke}}, \bibinfo {author} {\bibfnamefont {W.}~\bibnamefont {Petry}}, \
  and\ \bibinfo {author} {\bibfnamefont {S.}~\bibnamefont {Pouget}},\
  }\href@noop {} {\bibfield  {journal} {\bibinfo  {journal} {The Journal of
  chemical physics}\ }\textbf {\bibinfo {volume} {105}},\ \bibinfo {pages}
  {5177} (\bibinfo {year} {1996})}\BibitemShut {NoStop}%
\bibitem [{\citenamefont {Binder}\ and\ \citenamefont {Kob}(2011)}]{binderkob}%
  \BibitemOpen
  \bibfield  {author} {\bibinfo {author} {\bibfnamefont {K.}~\bibnamefont
  {Binder}}\ and\ \bibinfo {author} {\bibfnamefont {W.}~\bibnamefont {Kob}},\
  }\href@noop {} {\emph {\bibinfo {title} {Glassy materials and disordered
  solids: An introduction to their statistical mechanics}}}\ (\bibinfo
  {publisher} {World scientific},\ \bibinfo {year} {2011})\BibitemShut
  {NoStop}%
\bibitem [{\citenamefont {Pardo}\ \emph {et~al.}(2007)\citenamefont {Pardo},
  \citenamefont {Lunkenheimer},\ and\ \citenamefont {Loidl}}]{lunkenheimer}%
  \BibitemOpen
  \bibfield  {author} {\bibinfo {author} {\bibfnamefont {L.}~\bibnamefont
  {Pardo}}, \bibinfo {author} {\bibfnamefont {P.}~\bibnamefont {Lunkenheimer}},
  \ and\ \bibinfo {author} {\bibfnamefont {A.}~\bibnamefont {Loidl}},\
  }\href@noop {} {\bibfield  {journal} {\bibinfo  {journal} {Physical Review
  E}\ }\textbf {\bibinfo {volume} {76}},\ \bibinfo {pages} {030502} (\bibinfo
  {year} {2007})}\BibitemShut {NoStop}%
\bibitem [{\citenamefont {Berthier}\ and\ \citenamefont
  {Ediger}(2016)}]{berthier2015facets}%
  \BibitemOpen
  \bibfield  {author} {\bibinfo {author} {\bibfnamefont {L.}~\bibnamefont
  {Berthier}}\ and\ \bibinfo {author} {\bibfnamefont {M.~D.}\ \bibnamefont
  {Ediger}},\ }\href@noop {} {\bibfield  {journal} {\bibinfo  {journal}
  {Physics Today}\ }\textbf {\bibinfo {volume} {69}},\ \bibinfo {pages} {40}
  (\bibinfo {year} {2016})}\BibitemShut {NoStop}%
\bibitem [{\citenamefont {Larson}(1999)}]{larson}%
  \BibitemOpen
  \bibfield  {author} {\bibinfo {author} {\bibfnamefont {R.~G.}\ \bibnamefont
  {Larson}},\ }\href@noop {} {\emph {\bibinfo {title} {The structure and
  rheology of complex fluids}}},\ Vol.\ \bibinfo {volume} {150}\ (\bibinfo
  {publisher} {Oxford university press New York},\ \bibinfo {year}
  {1999})\BibitemShut {NoStop}%
\bibitem [{\citenamefont {Pusey}\ and\ \citenamefont
  {Van~Megen}(1986)}]{naturepusey}%
  \BibitemOpen
  \bibfield  {author} {\bibinfo {author} {\bibfnamefont {P.~N.}\ \bibnamefont
  {Pusey}}\ and\ \bibinfo {author} {\bibfnamefont {W.}~\bibnamefont
  {Van~Megen}},\ }\href@noop {} {\bibfield  {journal} {\bibinfo  {journal}
  {Nature}\ }\textbf {\bibinfo {volume} {320}},\ \bibinfo {pages} {340}
  (\bibinfo {year} {1986})}\BibitemShut {NoStop}%
\bibitem [{\citenamefont {Cheng}\ \emph {et~al.}(2002)\citenamefont {Cheng},
  \citenamefont {Zhu}, \citenamefont {Chaikin}, \citenamefont {Phan},\ and\
  \citenamefont {Russel}}]{chaikin}%
  \BibitemOpen
  \bibfield  {author} {\bibinfo {author} {\bibfnamefont {Z.}~\bibnamefont
  {Cheng}}, \bibinfo {author} {\bibfnamefont {J.}~\bibnamefont {Zhu}}, \bibinfo
  {author} {\bibfnamefont {P.~M.}\ \bibnamefont {Chaikin}}, \bibinfo {author}
  {\bibfnamefont {S.-E.}\ \bibnamefont {Phan}}, \ and\ \bibinfo {author}
  {\bibfnamefont {W.~B.}\ \bibnamefont {Russel}},\ }\href@noop {} {\bibfield
  {journal} {\bibinfo  {journal} {Physical Review E}\ }\textbf {\bibinfo
  {volume} {65}},\ \bibinfo {pages} {041405} (\bibinfo {year}
  {2002})}\BibitemShut {NoStop}%
\bibitem [{\citenamefont {Berthier}\ and\ \citenamefont
  {Witten}(2009)}]{berthier2009glass}%
  \BibitemOpen
  \bibfield  {author} {\bibinfo {author} {\bibfnamefont {L.}~\bibnamefont
  {Berthier}}\ and\ \bibinfo {author} {\bibfnamefont {T.~A.}\ \bibnamefont
  {Witten}},\ }\href@noop {} {\bibfield  {journal} {\bibinfo  {journal}
  {Physical Review E}\ }\textbf {\bibinfo {volume} {80}},\ \bibinfo {pages}
  {021502} (\bibinfo {year} {2009})}\BibitemShut {NoStop}%
\bibitem [{\citenamefont {Kegel}\ and\ \citenamefont {van
  Blaaderen}(2000)}]{kegel}%
  \BibitemOpen
  \bibfield  {author} {\bibinfo {author} {\bibfnamefont {W.~K.}\ \bibnamefont
  {Kegel}}\ and\ \bibinfo {author} {\bibfnamefont {A.}~\bibnamefont {van
  Blaaderen}},\ }\href@noop {} {\bibfield  {journal} {\bibinfo  {journal}
  {Science}\ }\textbf {\bibinfo {volume} {287}},\ \bibinfo {pages} {290}
  (\bibinfo {year} {2000})}\BibitemShut {NoStop}%
\bibitem [{\citenamefont {Weeks}\ \emph {et~al.}(2000)\citenamefont {Weeks},
  \citenamefont {Crocker}, \citenamefont {Levitt}, \citenamefont {Schofield},\
  and\ \citenamefont {Weitz}}]{weeks}%
  \BibitemOpen
  \bibfield  {author} {\bibinfo {author} {\bibfnamefont {E.~R.}\ \bibnamefont
  {Weeks}}, \bibinfo {author} {\bibfnamefont {J.~C.}\ \bibnamefont {Crocker}},
  \bibinfo {author} {\bibfnamefont {A.~C.}\ \bibnamefont {Levitt}}, \bibinfo
  {author} {\bibfnamefont {A.}~\bibnamefont {Schofield}}, \ and\ \bibinfo
  {author} {\bibfnamefont {D.~A.}\ \bibnamefont {Weitz}},\ }\href@noop {}
  {\bibfield  {journal} {\bibinfo  {journal} {Science}\ }\textbf {\bibinfo
  {volume} {287}},\ \bibinfo {pages} {627} (\bibinfo {year}
  {2000})}\BibitemShut {NoStop}%
\bibitem [{\citenamefont {Brambilla}\ \emph {et~al.}(2009)\citenamefont
  {Brambilla}, \citenamefont {El~Masri}, \citenamefont {Pierno}, \citenamefont
  {Berthier}, \citenamefont {Cipelletti}, \citenamefont {Petekidis},\ and\
  \citenamefont {Schofield}}]{Brambilla}%
  \BibitemOpen
  \bibfield  {author} {\bibinfo {author} {\bibfnamefont {G.}~\bibnamefont
  {Brambilla}}, \bibinfo {author} {\bibfnamefont {D.}~\bibnamefont {El~Masri}},
  \bibinfo {author} {\bibfnamefont {M.}~\bibnamefont {Pierno}}, \bibinfo
  {author} {\bibfnamefont {L.}~\bibnamefont {Berthier}}, \bibinfo {author}
  {\bibfnamefont {L.}~\bibnamefont {Cipelletti}}, \bibinfo {author}
  {\bibfnamefont {G.}~\bibnamefont {Petekidis}}, \ and\ \bibinfo {author}
  {\bibfnamefont {A.~B.}\ \bibnamefont {Schofield}},\ }\href@noop {} {\bibfield
   {journal} {\bibinfo  {journal} {Physical review letters}\ }\textbf {\bibinfo
  {volume} {102}},\ \bibinfo {pages} {085703} (\bibinfo {year}
  {2009})}\BibitemShut {NoStop}%
\bibitem [{\citenamefont {Hallett}\ \emph {et~al.}(2018)\citenamefont
  {Hallett}, \citenamefont {Turci},\ and\ \citenamefont
  {Royall}}]{hallett2018local}%
  \BibitemOpen
  \bibfield  {author} {\bibinfo {author} {\bibfnamefont {J.~E.}\ \bibnamefont
  {Hallett}}, \bibinfo {author} {\bibfnamefont {F.}~\bibnamefont {Turci}}, \
  and\ \bibinfo {author} {\bibfnamefont {C.~P.}\ \bibnamefont {Royall}},\
  }\href@noop {} {\bibfield  {journal} {\bibinfo  {journal} {Nature
  communications}\ }\textbf {\bibinfo {volume} {9}},\ \bibinfo {pages} {1}
  (\bibinfo {year} {2018})}\BibitemShut {NoStop}%
\bibitem [{\citenamefont {Donev}\ \emph {et~al.}(2005)\citenamefont {Donev},
  \citenamefont {Torquato},\ and\ \citenamefont {Stillinger}}]{torquato}%
  \BibitemOpen
  \bibfield  {author} {\bibinfo {author} {\bibfnamefont {A.}~\bibnamefont
  {Donev}}, \bibinfo {author} {\bibfnamefont {S.}~\bibnamefont {Torquato}}, \
  and\ \bibinfo {author} {\bibfnamefont {F.~H.}\ \bibnamefont {Stillinger}},\
  }\href@noop {} {\bibfield  {journal} {\bibinfo  {journal} {Physical Review
  E}\ }\textbf {\bibinfo {volume} {71}},\ \bibinfo {pages} {011105} (\bibinfo
  {year} {2005})}\BibitemShut {NoStop}%
\bibitem [{\citenamefont {Donev}\ \emph {et~al.}(2004)\citenamefont {Donev},
  \citenamefont {Torquato}, \citenamefont {Stillinger},\ and\ \citenamefont
  {Connelly}}]{donev2004jamming}%
  \BibitemOpen
  \bibfield  {author} {\bibinfo {author} {\bibfnamefont {A.}~\bibnamefont
  {Donev}}, \bibinfo {author} {\bibfnamefont {S.}~\bibnamefont {Torquato}},
  \bibinfo {author} {\bibfnamefont {F.~H.}\ \bibnamefont {Stillinger}}, \ and\
  \bibinfo {author} {\bibfnamefont {R.}~\bibnamefont {Connelly}},\ }\href@noop
  {} {\bibfield  {journal} {\bibinfo  {journal} {Journal of applied physics}\
  }\textbf {\bibinfo {volume} {95}},\ \bibinfo {pages} {989} (\bibinfo {year}
  {2004})}\BibitemShut {NoStop}%
\bibitem [{\citenamefont {Liu}\ and\ \citenamefont {Nagel}(1998)}]{liunagel}%
  \BibitemOpen
  \bibfield  {author} {\bibinfo {author} {\bibfnamefont {A.~J.}\ \bibnamefont
  {Liu}}\ and\ \bibinfo {author} {\bibfnamefont {S.~R.}\ \bibnamefont
  {Nagel}},\ }\href@noop {} {\bibfield  {journal} {\bibinfo  {journal}
  {Nature}\ }\textbf {\bibinfo {volume} {396}},\ \bibinfo {pages} {21}
  (\bibinfo {year} {1998})}\BibitemShut {NoStop}%
\bibitem [{\citenamefont {Charbonneau}\ \emph {et~al.}(2017)\citenamefont
  {Charbonneau}, \citenamefont {Kurchan}, \citenamefont {Parisi}, \citenamefont
  {Urbani},\ and\ \citenamefont {Zamponi}}]{glassjammingreview}%
  \BibitemOpen
  \bibfield  {author} {\bibinfo {author} {\bibfnamefont {P.}~\bibnamefont
  {Charbonneau}}, \bibinfo {author} {\bibfnamefont {J.}~\bibnamefont
  {Kurchan}}, \bibinfo {author} {\bibfnamefont {G.}~\bibnamefont {Parisi}},
  \bibinfo {author} {\bibfnamefont {P.}~\bibnamefont {Urbani}}, \ and\ \bibinfo
  {author} {\bibfnamefont {F.}~\bibnamefont {Zamponi}},\ }\href@noop {}
  {\bibfield  {journal} {\bibinfo  {journal} {Annual Review of Condensed Matter
  Physics}\ }\textbf {\bibinfo {volume} {8}},\ \bibinfo {pages} {265} (\bibinfo
  {year} {2017})}\BibitemShut {NoStop}%
\bibitem [{\citenamefont {Jaeger}\ \emph {et~al.}(1996)\citenamefont {Jaeger},
  \citenamefont {Nagel},\ and\ \citenamefont {Behringer}}]{grainsbook}%
  \BibitemOpen
  \bibfield  {author} {\bibinfo {author} {\bibfnamefont {H.~M.}\ \bibnamefont
  {Jaeger}}, \bibinfo {author} {\bibfnamefont {S.~R.}\ \bibnamefont {Nagel}}, \
  and\ \bibinfo {author} {\bibfnamefont {R.~P.}\ \bibnamefont {Behringer}},\
  }\href@noop {} {\bibfield  {journal} {\bibinfo  {journal} {Reviews of modern
  physics}\ }\textbf {\bibinfo {volume} {68}},\ \bibinfo {pages} {1259}
  (\bibinfo {year} {1996})}\BibitemShut {NoStop}%
\bibitem [{\citenamefont {D'Anna}\ and\ \citenamefont
  {Gr{\'e}maud}(2001)}]{gremaud}%
  \BibitemOpen
  \bibfield  {author} {\bibinfo {author} {\bibfnamefont {G.}~\bibnamefont
  {D'Anna}}\ and\ \bibinfo {author} {\bibfnamefont {G.}~\bibnamefont
  {Gr{\'e}maud}},\ }\href@noop {} {\bibfield  {journal} {\bibinfo  {journal}
  {Nature}\ }\textbf {\bibinfo {volume} {413}},\ \bibinfo {pages} {407}
  (\bibinfo {year} {2001})}\BibitemShut {NoStop}%
\bibitem [{\citenamefont {Marty}\ and\ \citenamefont
  {Dauchot}(2005)}]{dauchot}%
  \BibitemOpen
  \bibfield  {author} {\bibinfo {author} {\bibfnamefont {G.}~\bibnamefont
  {Marty}}\ and\ \bibinfo {author} {\bibfnamefont {O.}~\bibnamefont
  {Dauchot}},\ }\href@noop {} {\bibfield  {journal} {\bibinfo  {journal}
  {Physical review letters}\ }\textbf {\bibinfo {volume} {94}},\ \bibinfo
  {pages} {015701} (\bibinfo {year} {2005})}\BibitemShut {NoStop}%
\bibitem [{\citenamefont {Keys}\ \emph {et~al.}(2007)\citenamefont {Keys},
  \citenamefont {Abate}, \citenamefont {Glotzer},\ and\ \citenamefont
  {Durian}}]{durian}%
  \BibitemOpen
  \bibfield  {author} {\bibinfo {author} {\bibfnamefont {A.~S.}\ \bibnamefont
  {Keys}}, \bibinfo {author} {\bibfnamefont {A.~R.}\ \bibnamefont {Abate}},
  \bibinfo {author} {\bibfnamefont {S.~C.}\ \bibnamefont {Glotzer}}, \ and\
  \bibinfo {author} {\bibfnamefont {D.~J.}\ \bibnamefont {Durian}},\
  }\href@noop {} {\bibfield  {journal} {\bibinfo  {journal} {Nature physics}\
  }\textbf {\bibinfo {volume} {3}},\ \bibinfo {pages} {260} (\bibinfo {year}
  {2007})}\BibitemShut {NoStop}%
\bibitem [{\citenamefont {Marchetti}\ \emph {et~al.}(2013)\citenamefont
  {Marchetti}, \citenamefont {Joanny}, \citenamefont {Ramaswamy}, \citenamefont
  {Liverpool}, \citenamefont {Prost}, \citenamefont {Rao},\ and\ \citenamefont
  {Simha}}]{marchetti2013hydrodynamics}%
  \BibitemOpen
  \bibfield  {author} {\bibinfo {author} {\bibfnamefont {M.~C.}\ \bibnamefont
  {Marchetti}}, \bibinfo {author} {\bibfnamefont {J.-F.}\ \bibnamefont
  {Joanny}}, \bibinfo {author} {\bibfnamefont {S.}~\bibnamefont {Ramaswamy}},
  \bibinfo {author} {\bibfnamefont {T.~B.}\ \bibnamefont {Liverpool}}, \bibinfo
  {author} {\bibfnamefont {J.}~\bibnamefont {Prost}}, \bibinfo {author}
  {\bibfnamefont {M.}~\bibnamefont {Rao}}, \ and\ \bibinfo {author}
  {\bibfnamefont {R.~A.}\ \bibnamefont {Simha}},\ }\href@noop {} {\bibfield
  {journal} {\bibinfo  {journal} {Reviews of Modern Physics}\ }\textbf
  {\bibinfo {volume} {85}},\ \bibinfo {pages} {1143} (\bibinfo {year}
  {2013})}\BibitemShut {NoStop}%
\bibitem [{\citenamefont {Bechinger}\ \emph {et~al.}(2016)\citenamefont
  {Bechinger}, \citenamefont {Di~Leonardo}, \citenamefont {L{\"o}wen},
  \citenamefont {Reichhardt}, \citenamefont {Volpe},\ and\ \citenamefont
  {Volpe}}]{bechinger2016active}%
  \BibitemOpen
  \bibfield  {author} {\bibinfo {author} {\bibfnamefont {C.}~\bibnamefont
  {Bechinger}}, \bibinfo {author} {\bibfnamefont {R.}~\bibnamefont
  {Di~Leonardo}}, \bibinfo {author} {\bibfnamefont {H.}~\bibnamefont
  {L{\"o}wen}}, \bibinfo {author} {\bibfnamefont {C.}~\bibnamefont
  {Reichhardt}}, \bibinfo {author} {\bibfnamefont {G.}~\bibnamefont {Volpe}}, \
  and\ \bibinfo {author} {\bibfnamefont {G.}~\bibnamefont {Volpe}},\
  }\href@noop {} {\bibfield  {journal} {\bibinfo  {journal} {Reviews of Modern
  Physics}\ }\textbf {\bibinfo {volume} {88}},\ \bibinfo {pages} {045006}
  (\bibinfo {year} {2016})}\BibitemShut {NoStop}%
\bibitem [{\citenamefont {Deseigne}\ \emph {et~al.}(2010)\citenamefont
  {Deseigne}, \citenamefont {Dauchot},\ and\ \citenamefont
  {Chat{\'e}}}]{deseigne2010collective}%
  \BibitemOpen
  \bibfield  {author} {\bibinfo {author} {\bibfnamefont {J.}~\bibnamefont
  {Deseigne}}, \bibinfo {author} {\bibfnamefont {O.}~\bibnamefont {Dauchot}}, \
  and\ \bibinfo {author} {\bibfnamefont {H.}~\bibnamefont {Chat{\'e}}},\
  }\href@noop {} {\bibfield  {journal} {\bibinfo  {journal} {Physical review
  letters}\ }\textbf {\bibinfo {volume} {105}},\ \bibinfo {pages} {098001}
  (\bibinfo {year} {2010})}\BibitemShut {NoStop}%
\bibitem [{\citenamefont {Theurkauff}\ \emph {et~al.}(2012)\citenamefont
  {Theurkauff}, \citenamefont {Cottin-Bizonne}, \citenamefont {Palacci},
  \citenamefont {Ybert},\ and\ \citenamefont
  {Bocquet}}]{theurkauff2012dynamic}%
  \BibitemOpen
  \bibfield  {author} {\bibinfo {author} {\bibfnamefont {I.}~\bibnamefont
  {Theurkauff}}, \bibinfo {author} {\bibfnamefont {C.}~\bibnamefont
  {Cottin-Bizonne}}, \bibinfo {author} {\bibfnamefont {J.}~\bibnamefont
  {Palacci}}, \bibinfo {author} {\bibfnamefont {C.}~\bibnamefont {Ybert}}, \
  and\ \bibinfo {author} {\bibfnamefont {L.}~\bibnamefont {Bocquet}},\
  }\href@noop {} {\bibfield  {journal} {\bibinfo  {journal} {Physical review
  letters}\ }\textbf {\bibinfo {volume} {108}},\ \bibinfo {pages} {268303}
  (\bibinfo {year} {2012})}\BibitemShut {NoStop}%
\bibitem [{\citenamefont {Buttinoni}\ \emph {et~al.}(2013)\citenamefont
  {Buttinoni}, \citenamefont {Bialk{\'e}}, \citenamefont {K{\"u}mmel},
  \citenamefont {L{\"o}wen}, \citenamefont {Bechinger},\ and\ \citenamefont
  {Speck}}]{buttinoni2013dynamical}%
  \BibitemOpen
  \bibfield  {author} {\bibinfo {author} {\bibfnamefont {I.}~\bibnamefont
  {Buttinoni}}, \bibinfo {author} {\bibfnamefont {J.}~\bibnamefont
  {Bialk{\'e}}}, \bibinfo {author} {\bibfnamefont {F.}~\bibnamefont
  {K{\"u}mmel}}, \bibinfo {author} {\bibfnamefont {H.}~\bibnamefont
  {L{\"o}wen}}, \bibinfo {author} {\bibfnamefont {C.}~\bibnamefont
  {Bechinger}}, \ and\ \bibinfo {author} {\bibfnamefont {T.}~\bibnamefont
  {Speck}},\ }\href@noop {} {\bibfield  {journal} {\bibinfo  {journal}
  {Physical review letters}\ }\textbf {\bibinfo {volume} {110}},\ \bibinfo
  {pages} {238301} (\bibinfo {year} {2013})}\BibitemShut {NoStop}%
\bibitem [{\citenamefont {Henkes}\ \emph {et~al.}(2011)\citenamefont {Henkes},
  \citenamefont {Fily},\ and\ \citenamefont {Marchetti}}]{henkes2011active}%
  \BibitemOpen
  \bibfield  {author} {\bibinfo {author} {\bibfnamefont {S.}~\bibnamefont
  {Henkes}}, \bibinfo {author} {\bibfnamefont {Y.}~\bibnamefont {Fily}}, \ and\
  \bibinfo {author} {\bibfnamefont {M.~C.}\ \bibnamefont {Marchetti}},\
  }\href@noop {} {\bibfield  {journal} {\bibinfo  {journal} {Physical Review
  E}\ }\textbf {\bibinfo {volume} {84}},\ \bibinfo {pages} {040301} (\bibinfo
  {year} {2011})}\BibitemShut {NoStop}%
\bibitem [{\citenamefont {Angelini}\ \emph {et~al.}(2011)\citenamefont
  {Angelini}, \citenamefont {Hannezo}, \citenamefont {Trepat}, \citenamefont
  {Marquez}, \citenamefont {Fredberg},\ and\ \citenamefont
  {Weitz}}]{angelini_glass-like_2011}%
  \BibitemOpen
  \bibfield  {author} {\bibinfo {author} {\bibfnamefont {T.~E.}\ \bibnamefont
  {Angelini}}, \bibinfo {author} {\bibfnamefont {E.}~\bibnamefont {Hannezo}},
  \bibinfo {author} {\bibfnamefont {X.}~\bibnamefont {Trepat}}, \bibinfo
  {author} {\bibfnamefont {M.}~\bibnamefont {Marquez}}, \bibinfo {author}
  {\bibfnamefont {J.~J.}\ \bibnamefont {Fredberg}}, \ and\ \bibinfo {author}
  {\bibfnamefont {D.~A.}\ \bibnamefont {Weitz}},\ }\href {\doibase
  10.1073/pnas.1010059108} {\bibfield  {journal} {\bibinfo  {journal}
  {Proceedings of the National Academy of Sciences}\ }\textbf {\bibinfo
  {volume} {108}},\ \bibinfo {pages} {4714} (\bibinfo {year}
  {2011})}\BibitemShut {NoStop}%
\bibitem [{\citenamefont {Garcia}\ \emph {et~al.}(2015)\citenamefont {Garcia},
  \citenamefont {Hannezo}, \citenamefont {Elgeti}, \citenamefont {Joanny},
  \citenamefont {Silberzan},\ and\ \citenamefont {Gov}}]{garcia2015physics}%
  \BibitemOpen
  \bibfield  {author} {\bibinfo {author} {\bibfnamefont {S.}~\bibnamefont
  {Garcia}}, \bibinfo {author} {\bibfnamefont {E.}~\bibnamefont {Hannezo}},
  \bibinfo {author} {\bibfnamefont {J.}~\bibnamefont {Elgeti}}, \bibinfo
  {author} {\bibfnamefont {J.-F.}\ \bibnamefont {Joanny}}, \bibinfo {author}
  {\bibfnamefont {P.}~\bibnamefont {Silberzan}}, \ and\ \bibinfo {author}
  {\bibfnamefont {N.~S.}\ \bibnamefont {Gov}},\ }\href@noop {} {\bibfield
  {journal} {\bibinfo  {journal} {Proceedings of the National Academy of
  Sciences}\ }\textbf {\bibinfo {volume} {112}},\ \bibinfo {pages} {15314}
  (\bibinfo {year} {2015})}\BibitemShut {NoStop}%
\bibitem [{\citenamefont {Mongera}\ \emph {et~al.}(2018)\citenamefont
  {Mongera}, \citenamefont {Rowghanian}, \citenamefont {Gustafson},
  \citenamefont {Shelton}, \citenamefont {Kealhofer}, \citenamefont {Carn},
  \citenamefont {Serwane}, \citenamefont {Lucio}, \citenamefont {Giammona},\
  and\ \citenamefont {Camp{\`a}s}}]{mongera2018fluid}%
  \BibitemOpen
  \bibfield  {author} {\bibinfo {author} {\bibfnamefont {A.}~\bibnamefont
  {Mongera}}, \bibinfo {author} {\bibfnamefont {P.}~\bibnamefont {Rowghanian}},
  \bibinfo {author} {\bibfnamefont {H.~J.}\ \bibnamefont {Gustafson}}, \bibinfo
  {author} {\bibfnamefont {E.}~\bibnamefont {Shelton}}, \bibinfo {author}
  {\bibfnamefont {D.~A.}\ \bibnamefont {Kealhofer}}, \bibinfo {author}
  {\bibfnamefont {E.~K.}\ \bibnamefont {Carn}}, \bibinfo {author}
  {\bibfnamefont {F.}~\bibnamefont {Serwane}}, \bibinfo {author} {\bibfnamefont
  {A.~A.}\ \bibnamefont {Lucio}}, \bibinfo {author} {\bibfnamefont
  {J.}~\bibnamefont {Giammona}}, \ and\ \bibinfo {author} {\bibfnamefont
  {O.}~\bibnamefont {Camp{\`a}s}},\ }\href@noop {} {\bibfield  {journal}
  {\bibinfo  {journal} {Nature}\ }\textbf {\bibinfo {volume} {561}},\ \bibinfo
  {pages} {401} (\bibinfo {year} {2018})}\BibitemShut {NoStop}%
\bibitem [{\citenamefont {Klongvessa}\ \emph {et~al.}(2019)\citenamefont
  {Klongvessa}, \citenamefont {Ginot}, \citenamefont {Ybert}, \citenamefont
  {Cottin-Bizonne},\ and\ \citenamefont {Leocmach}}]{klongvessa2019active}%
  \BibitemOpen
  \bibfield  {author} {\bibinfo {author} {\bibfnamefont {N.}~\bibnamefont
  {Klongvessa}}, \bibinfo {author} {\bibfnamefont {F.}~\bibnamefont {Ginot}},
  \bibinfo {author} {\bibfnamefont {C.}~\bibnamefont {Ybert}}, \bibinfo
  {author} {\bibfnamefont {C.}~\bibnamefont {Cottin-Bizonne}}, \ and\ \bibinfo
  {author} {\bibfnamefont {M.}~\bibnamefont {Leocmach}},\ }\href@noop {}
  {\bibfield  {journal} {\bibinfo  {journal} {Physical Review Letters}\
  }\textbf {\bibinfo {volume} {123}},\ \bibinfo {pages} {248004} (\bibinfo
  {year} {2019})}\BibitemShut {NoStop}%
\bibitem [{\citenamefont {Berthier}\ and\ \citenamefont
  {Kurchan}(2013)}]{berthier_non-equilibrium_2013}%
  \BibitemOpen
  \bibfield  {author} {\bibinfo {author} {\bibfnamefont {L.}~\bibnamefont
  {Berthier}}\ and\ \bibinfo {author} {\bibfnamefont {J.}~\bibnamefont
  {Kurchan}},\ }\href {\doibase 10.1038/nphys2592} {\bibfield  {journal}
  {\bibinfo  {journal} {Nature Physics}\ }\textbf {\bibinfo {volume} {9}},\
  \bibinfo {pages} {310} (\bibinfo {year} {2013})}\BibitemShut {NoStop}%
\bibitem [{\citenamefont {Berthier}\ \emph
  {et~al.}(2019{\natexlab{b}})\citenamefont {Berthier}, \citenamefont
  {Flenner},\ and\ \citenamefont {Szamel}}]{berthier2019glassy}%
  \BibitemOpen
  \bibfield  {author} {\bibinfo {author} {\bibfnamefont {L.}~\bibnamefont
  {Berthier}}, \bibinfo {author} {\bibfnamefont {E.}~\bibnamefont {Flenner}}, \
  and\ \bibinfo {author} {\bibfnamefont {G.}~\bibnamefont {Szamel}},\
  }\href@noop {} {\bibfield  {journal} {\bibinfo  {journal} {The Journal of
  chemical physics}\ }\textbf {\bibinfo {volume} {150}},\ \bibinfo {pages}
  {200901} (\bibinfo {year} {2019}{\natexlab{b}})}\BibitemShut {NoStop}%
\bibitem [{\citenamefont {Ni}\ \emph {et~al.}(2013)\citenamefont {Ni},
  \citenamefont {Stuart},\ and\ \citenamefont {Dijkstra}}]{ni2013pushing}%
  \BibitemOpen
  \bibfield  {author} {\bibinfo {author} {\bibfnamefont {R.}~\bibnamefont
  {Ni}}, \bibinfo {author} {\bibfnamefont {M.~A.~C.}\ \bibnamefont {Stuart}}, \
  and\ \bibinfo {author} {\bibfnamefont {M.}~\bibnamefont {Dijkstra}},\
  }\href@noop {} {\bibfield  {journal} {\bibinfo  {journal} {Nature
  communications}\ }\textbf {\bibinfo {volume} {4}},\ \bibinfo {pages} {1}
  (\bibinfo {year} {2013})}\BibitemShut {NoStop}%
\bibitem [{\citenamefont {Berthier}(2014)}]{berthier2014nonequilibrium}%
  \BibitemOpen
  \bibfield  {author} {\bibinfo {author} {\bibfnamefont {L.}~\bibnamefont
  {Berthier}},\ }\href@noop {} {\bibfield  {journal} {\bibinfo  {journal}
  {Physical review letters}\ }\textbf {\bibinfo {volume} {112}},\ \bibinfo
  {pages} {220602} (\bibinfo {year} {2014})}\BibitemShut {NoStop}%
\bibitem [{\citenamefont {Mandal}\ \emph {et~al.}(2016)\citenamefont {Mandal},
  \citenamefont {Bhuyan}, \citenamefont {Rao},\ and\ \citenamefont
  {Dasgupta}}]{mandal2016active}%
  \BibitemOpen
  \bibfield  {author} {\bibinfo {author} {\bibfnamefont {R.}~\bibnamefont
  {Mandal}}, \bibinfo {author} {\bibfnamefont {P.~J.}\ \bibnamefont {Bhuyan}},
  \bibinfo {author} {\bibfnamefont {M.}~\bibnamefont {Rao}}, \ and\ \bibinfo
  {author} {\bibfnamefont {C.}~\bibnamefont {Dasgupta}},\ }\href@noop {}
  {\bibfield  {journal} {\bibinfo  {journal} {Soft Matter}\ }\textbf {\bibinfo
  {volume} {12}},\ \bibinfo {pages} {6268} (\bibinfo {year}
  {2016})}\BibitemShut {NoStop}%
\bibitem [{\citenamefont {Bi}\ \emph {et~al.}(2016)\citenamefont {Bi},
  \citenamefont {Yang}, \citenamefont {Marchetti},\ and\ \citenamefont
  {Manning}}]{bi_motility-driven_2016}%
  \BibitemOpen
  \bibfield  {author} {\bibinfo {author} {\bibfnamefont {D.}~\bibnamefont
  {Bi}}, \bibinfo {author} {\bibfnamefont {X.}~\bibnamefont {Yang}}, \bibinfo
  {author} {\bibfnamefont {M.~C.}\ \bibnamefont {Marchetti}}, \ and\ \bibinfo
  {author} {\bibfnamefont {M.~L.}\ \bibnamefont {Manning}},\ }\href {\doibase
  10.1103/PhysRevX.6.021011} {\bibfield  {journal} {\bibinfo  {journal}
  {Physical Review X}\ }\textbf {\bibinfo {volume} {6}} (\bibinfo {year}
  {2016}),\ 10.1103/PhysRevX.6.021011}\BibitemShut {NoStop}%
\bibitem [{\citenamefont {Berthier}\ \emph
  {et~al.}(2017{\natexlab{a}})\citenamefont {Berthier}, \citenamefont
  {Flenner},\ and\ \citenamefont {Szamel}}]{berthier2017active}%
  \BibitemOpen
  \bibfield  {author} {\bibinfo {author} {\bibfnamefont {L.}~\bibnamefont
  {Berthier}}, \bibinfo {author} {\bibfnamefont {E.}~\bibnamefont {Flenner}}, \
  and\ \bibinfo {author} {\bibfnamefont {G.}~\bibnamefont {Szamel}},\
  }\href@noop {} {\bibfield  {journal} {\bibinfo  {journal} {New Journal of
  Physics}\ }\textbf {\bibinfo {volume} {19}},\ \bibinfo {pages} {125006}
  (\bibinfo {year} {2017}{\natexlab{a}})}\BibitemShut {NoStop}%
\bibitem [{\citenamefont {Matoz-Fernandez}\ \emph {et~al.}(2017)\citenamefont
  {Matoz-Fernandez}, \citenamefont {Martens}, \citenamefont {Sknepnek},
  \citenamefont {Barrat},\ and\ \citenamefont {Henkes}}]{matoz2017cell}%
  \BibitemOpen
  \bibfield  {author} {\bibinfo {author} {\bibfnamefont {D.}~\bibnamefont
  {Matoz-Fernandez}}, \bibinfo {author} {\bibfnamefont {K.}~\bibnamefont
  {Martens}}, \bibinfo {author} {\bibfnamefont {R.}~\bibnamefont {Sknepnek}},
  \bibinfo {author} {\bibfnamefont {J.}~\bibnamefont {Barrat}}, \ and\ \bibinfo
  {author} {\bibfnamefont {S.}~\bibnamefont {Henkes}},\ }\href@noop {}
  {\bibfield  {journal} {\bibinfo  {journal} {Soft matter}\ }\textbf {\bibinfo
  {volume} {13}},\ \bibinfo {pages} {3205} (\bibinfo {year}
  {2017})}\BibitemShut {NoStop}%
\bibitem [{\citenamefont {Scheidler}\ \emph {et~al.}(2002)\citenamefont
  {Scheidler}, \citenamefont {Kob}, \citenamefont {Binder},\ and\ \citenamefont
  {Parisi}}]{scheidler2002growing}%
  \BibitemOpen
  \bibfield  {author} {\bibinfo {author} {\bibfnamefont {P.}~\bibnamefont
  {Scheidler}}, \bibinfo {author} {\bibfnamefont {W.}~\bibnamefont {Kob}},
  \bibinfo {author} {\bibfnamefont {K.}~\bibnamefont {Binder}}, \ and\ \bibinfo
  {author} {\bibfnamefont {G.}~\bibnamefont {Parisi}},\ }\href@noop {}
  {\bibfield  {journal} {\bibinfo  {journal} {Philosophical Magazine B}\
  }\textbf {\bibinfo {volume} {82}},\ \bibinfo {pages} {283} (\bibinfo {year}
  {2002})}\BibitemShut {NoStop}%
\bibitem [{\citenamefont {Kim}(2003)}]{kim2003effects}%
  \BibitemOpen
  \bibfield  {author} {\bibinfo {author} {\bibfnamefont {K.}~\bibnamefont
  {Kim}},\ }\href@noop {} {\bibfield  {journal} {\bibinfo  {journal} {EPL
  (Europhysics Letters)}\ }\textbf {\bibinfo {volume} {61}},\ \bibinfo {pages}
  {790} (\bibinfo {year} {2003})}\BibitemShut {NoStop}%
\bibitem [{\citenamefont {Cammarota}\ and\ \citenamefont
  {Biroli}(2012)}]{cammarota_ideal_2012}%
  \BibitemOpen
  \bibfield  {author} {\bibinfo {author} {\bibfnamefont {C.}~\bibnamefont
  {Cammarota}}\ and\ \bibinfo {author} {\bibfnamefont {G.}~\bibnamefont
  {Biroli}},\ }\href {\doibase 10.1073/pnas.1111582109} {\bibfield  {journal}
  {\bibinfo  {journal} {Proceedings of the National Academy of Sciences}\
  }\textbf {\bibinfo {volume} {109}},\ \bibinfo {pages} {8850} (\bibinfo {year}
  {2012})}\BibitemShut {NoStop}%
\bibitem [{\citenamefont {Berthier}\ and\ \citenamefont
  {Kob}(2012)}]{berthier_static_2012}%
  \BibitemOpen
  \bibfield  {author} {\bibinfo {author} {\bibfnamefont {L.}~\bibnamefont
  {Berthier}}\ and\ \bibinfo {author} {\bibfnamefont {W.}~\bibnamefont {Kob}},\
  }\href {\doibase 10.1103/PhysRevE.85.011102} {\bibfield  {journal} {\bibinfo
  {journal} {Physical Review E}\ }\textbf {\bibinfo {volume} {85}} (\bibinfo
  {year} {2012}),\ 10.1103/PhysRevE.85.011102}\BibitemShut {NoStop}%
\bibitem [{\citenamefont {Karmakar}\ and\ \citenamefont
  {Procaccia}(2011)}]{karmakar2011exposing}%
  \BibitemOpen
  \bibfield  {author} {\bibinfo {author} {\bibfnamefont {S.}~\bibnamefont
  {Karmakar}}\ and\ \bibinfo {author} {\bibfnamefont {I.}~\bibnamefont
  {Procaccia}},\ }\href@noop {} {\bibfield  {journal} {\bibinfo  {journal}
  {arXiv preprint arXiv:1105.4053}\ } (\bibinfo {year} {2011})}\BibitemShut
  {NoStop}%
\bibitem [{\citenamefont {Cammarota}\ and\ \citenamefont
  {Biroli}(2013)}]{cammarota2013random}%
  \BibitemOpen
  \bibfield  {author} {\bibinfo {author} {\bibfnamefont {C.}~\bibnamefont
  {Cammarota}}\ and\ \bibinfo {author} {\bibfnamefont {G.}~\bibnamefont
  {Biroli}},\ }\href@noop {} {\bibfield  {journal} {\bibinfo  {journal} {The
  Journal of chemical physics}\ }\textbf {\bibinfo {volume} {138}},\ \bibinfo
  {pages} {12A547} (\bibinfo {year} {2013})}\BibitemShut {NoStop}%
\bibitem [{\citenamefont {Krakoviack}(2011)}]{krakoviack2011mode}%
  \BibitemOpen
  \bibfield  {author} {\bibinfo {author} {\bibfnamefont {V.}~\bibnamefont
  {Krakoviack}},\ }\href@noop {} {\bibfield  {journal} {\bibinfo  {journal}
  {Physical Review E}\ }\textbf {\bibinfo {volume} {84}},\ \bibinfo {pages}
  {050501} (\bibinfo {year} {2011})}\BibitemShut {NoStop}%
\bibitem [{\citenamefont {Szamel}\ and\ \citenamefont
  {Flenner}(2013)}]{szamel2013glassy}%
  \BibitemOpen
  \bibfield  {author} {\bibinfo {author} {\bibfnamefont {G.}~\bibnamefont
  {Szamel}}\ and\ \bibinfo {author} {\bibfnamefont {E.}~\bibnamefont
  {Flenner}},\ }\href@noop {} {\bibfield  {journal} {\bibinfo  {journal} {EPL
  (Europhysics Letters)}\ }\textbf {\bibinfo {volume} {101}},\ \bibinfo {pages}
  {66005} (\bibinfo {year} {2013})}\BibitemShut {NoStop}%
\bibitem [{\citenamefont {Franz}\ and\ \citenamefont
  {Parisi}(2013)}]{franz2013universality}%
  \BibitemOpen
  \bibfield  {author} {\bibinfo {author} {\bibfnamefont {S.}~\bibnamefont
  {Franz}}\ and\ \bibinfo {author} {\bibfnamefont {G.}~\bibnamefont {Parisi}},\
  }\href@noop {} {\bibfield  {journal} {\bibinfo  {journal} {Journal of
  Statistical Mechanics: Theory and Experiment}\ }\textbf {\bibinfo {volume}
  {2013}},\ \bibinfo {pages} {P11012} (\bibinfo {year} {2013})}\BibitemShut
  {NoStop}%
\bibitem [{\citenamefont {Cammarota}(2013)}]{cammarota2013general}%
  \BibitemOpen
  \bibfield  {author} {\bibinfo {author} {\bibfnamefont {C.}~\bibnamefont
  {Cammarota}},\ }\href@noop {} {\bibfield  {journal} {\bibinfo  {journal} {EPL
  (Europhysics Letters)}\ }\textbf {\bibinfo {volume} {101}},\ \bibinfo {pages}
  {56001} (\bibinfo {year} {2013})}\BibitemShut {NoStop}%
\bibitem [{\citenamefont {Franz}\ \emph {et~al.}(2013)\citenamefont {Franz},
  \citenamefont {Parisi},\ and\ \citenamefont
  {Ricci-Tersenghi}}]{franz2013glassy}%
  \BibitemOpen
  \bibfield  {author} {\bibinfo {author} {\bibfnamefont {S.}~\bibnamefont
  {Franz}}, \bibinfo {author} {\bibfnamefont {G.}~\bibnamefont {Parisi}}, \
  and\ \bibinfo {author} {\bibfnamefont {F.}~\bibnamefont {Ricci-Tersenghi}},\
  }\href@noop {} {\bibfield  {journal} {\bibinfo  {journal} {Journal of
  Statistical Mechanics: Theory and Experiment}\ }\textbf {\bibinfo {volume}
  {2013}},\ \bibinfo {pages} {L02001} (\bibinfo {year} {2013})}\BibitemShut
  {NoStop}%
\bibitem [{\citenamefont {Krakoviack}(2014)}]{krakoviack2014simple}%
  \BibitemOpen
  \bibfield  {author} {\bibinfo {author} {\bibfnamefont {V.}~\bibnamefont
  {Krakoviack}},\ }\href@noop {} {\bibfield  {journal} {\bibinfo  {journal}
  {The Journal of chemical physics}\ }\textbf {\bibinfo {volume} {141}},\
  \bibinfo {pages} {104504} (\bibinfo {year} {2014})}\BibitemShut {NoStop}%
\bibitem [{\citenamefont {Phan}\ and\ \citenamefont
  {Schweizer}(2018)}]{phan2018theory}%
  \BibitemOpen
  \bibfield  {author} {\bibinfo {author} {\bibfnamefont {A.~D.}\ \bibnamefont
  {Phan}}\ and\ \bibinfo {author} {\bibfnamefont {K.~S.}\ \bibnamefont
  {Schweizer}},\ }\href@noop {} {\bibfield  {journal} {\bibinfo  {journal} {The
  Journal of chemical physics}\ }\textbf {\bibinfo {volume} {148}},\ \bibinfo
  {pages} {054502} (\bibinfo {year} {2018})}\BibitemShut {NoStop}%
\bibitem [{\citenamefont {Cammarota}\ and\ \citenamefont
  {Seoane}(2016)}]{cammarota2016first}%
  \BibitemOpen
  \bibfield  {author} {\bibinfo {author} {\bibfnamefont {C.}~\bibnamefont
  {Cammarota}}\ and\ \bibinfo {author} {\bibfnamefont {B.}~\bibnamefont
  {Seoane}},\ }\href@noop {} {\bibfield  {journal} {\bibinfo  {journal}
  {Physical Review B}\ }\textbf {\bibinfo {volume} {94}},\ \bibinfo {pages}
  {180201} (\bibinfo {year} {2016})}\BibitemShut {NoStop}%
\bibitem [{\citenamefont {Ikeda}\ \emph
  {et~al.}(2017{\natexlab{a}})\citenamefont {Ikeda}, \citenamefont {Miyazaki},\
  and\ \citenamefont {Biroli}}]{ikeda2017fredrickson}%
  \BibitemOpen
  \bibfield  {author} {\bibinfo {author} {\bibfnamefont {H.}~\bibnamefont
  {Ikeda}}, \bibinfo {author} {\bibfnamefont {K.}~\bibnamefont {Miyazaki}}, \
  and\ \bibinfo {author} {\bibfnamefont {G.}~\bibnamefont {Biroli}},\
  }\href@noop {} {\bibfield  {journal} {\bibinfo  {journal} {EPL (Europhysics
  Letters)}\ }\textbf {\bibinfo {volume} {116}},\ \bibinfo {pages} {56004}
  (\bibinfo {year} {2017}{\natexlab{a}})}\BibitemShut {NoStop}%
\bibitem [{\citenamefont {Kob}\ and\ \citenamefont
  {Berthier}(2013)}]{kob2013probing}%
  \BibitemOpen
  \bibfield  {author} {\bibinfo {author} {\bibfnamefont {W.}~\bibnamefont
  {Kob}}\ and\ \bibinfo {author} {\bibfnamefont {L.}~\bibnamefont {Berthier}},\
  }\href@noop {} {\bibfield  {journal} {\bibinfo  {journal} {Physical review
  letters}\ }\textbf {\bibinfo {volume} {110}},\ \bibinfo {pages} {245702}
  (\bibinfo {year} {2013})}\BibitemShut {NoStop}%
\bibitem [{\citenamefont {Charbonneau}\ and\ \citenamefont
  {Tarjus}(2013)}]{charbonneau2013decorrelation}%
  \BibitemOpen
  \bibfield  {author} {\bibinfo {author} {\bibfnamefont {P.}~\bibnamefont
  {Charbonneau}}\ and\ \bibinfo {author} {\bibfnamefont {G.}~\bibnamefont
  {Tarjus}},\ }\href@noop {} {\bibfield  {journal} {\bibinfo  {journal}
  {Physical Review E}\ }\textbf {\bibinfo {volume} {87}},\ \bibinfo {pages}
  {042305} (\bibinfo {year} {2013})}\BibitemShut {NoStop}%
\bibitem [{\citenamefont {Karmakar}\ and\ \citenamefont
  {Parisi}(2013)}]{karmakar2013random}%
  \BibitemOpen
  \bibfield  {author} {\bibinfo {author} {\bibfnamefont {S.}~\bibnamefont
  {Karmakar}}\ and\ \bibinfo {author} {\bibfnamefont {G.}~\bibnamefont
  {Parisi}},\ }\href@noop {} {\bibfield  {journal} {\bibinfo  {journal}
  {Proceedings of the National Academy of Sciences}\ }\textbf {\bibinfo
  {volume} {110}},\ \bibinfo {pages} {2752} (\bibinfo {year}
  {2013})}\BibitemShut {NoStop}%
\bibitem [{\citenamefont {Chakrabarty}\ \emph {et~al.}(2015)\citenamefont
  {Chakrabarty}, \citenamefont {Karmakar},\ and\ \citenamefont
  {Dasgupta}}]{chakrabarty2015dynamics}%
  \BibitemOpen
  \bibfield  {author} {\bibinfo {author} {\bibfnamefont {S.}~\bibnamefont
  {Chakrabarty}}, \bibinfo {author} {\bibfnamefont {S.}~\bibnamefont
  {Karmakar}}, \ and\ \bibinfo {author} {\bibfnamefont {C.}~\bibnamefont
  {Dasgupta}},\ }\href@noop {} {\bibfield  {journal} {\bibinfo  {journal}
  {Scientific reports}\ }\textbf {\bibinfo {volume} {5}},\ \bibinfo {pages}
  {12577} (\bibinfo {year} {2015})}\BibitemShut {NoStop}%
\bibitem [{\citenamefont {Kob}\ and\ \citenamefont
  {Coslovich}(2014)}]{kob2014nonlinear}%
  \BibitemOpen
  \bibfield  {author} {\bibinfo {author} {\bibfnamefont {W.}~\bibnamefont
  {Kob}}\ and\ \bibinfo {author} {\bibfnamefont {D.}~\bibnamefont
  {Coslovich}},\ }\href@noop {} {\bibfield  {journal} {\bibinfo  {journal}
  {Physical Review E}\ }\textbf {\bibinfo {volume} {90}},\ \bibinfo {pages}
  {052305} (\bibinfo {year} {2014})}\BibitemShut {NoStop}%
\bibitem [{\citenamefont {Jack}\ and\ \citenamefont
  {Fullerton}(2013)}]{jack2013dynamical}%
  \BibitemOpen
  \bibfield  {author} {\bibinfo {author} {\bibfnamefont {R.~L.}\ \bibnamefont
  {Jack}}\ and\ \bibinfo {author} {\bibfnamefont {C.~J.}\ \bibnamefont
  {Fullerton}},\ }\href@noop {} {\bibfield  {journal} {\bibinfo  {journal}
  {Physical Review E}\ }\textbf {\bibinfo {volume} {88}},\ \bibinfo {pages}
  {042304} (\bibinfo {year} {2013})}\BibitemShut {NoStop}%
\bibitem [{\citenamefont {Fullerton}\ and\ \citenamefont
  {Jack}(2014)}]{fullerton2014investigating}%
  \BibitemOpen
  \bibfield  {author} {\bibinfo {author} {\bibfnamefont {C.~J.}\ \bibnamefont
  {Fullerton}}\ and\ \bibinfo {author} {\bibfnamefont {R.~L.}\ \bibnamefont
  {Jack}},\ }\href@noop {} {\bibfield  {journal} {\bibinfo  {journal} {Physical
  review letters}\ }\textbf {\bibinfo {volume} {112}},\ \bibinfo {pages}
  {255701} (\bibinfo {year} {2014})}\BibitemShut {NoStop}%
\bibitem [{\citenamefont {Li}\ \emph {et~al.}(2015)\citenamefont {Li},
  \citenamefont {Zhu},\ and\ \citenamefont {Sun}}]{li2015decoupling}%
  \BibitemOpen
  \bibfield  {author} {\bibinfo {author} {\bibfnamefont {Y.-W.}\ \bibnamefont
  {Li}}, \bibinfo {author} {\bibfnamefont {Y.-L.}\ \bibnamefont {Zhu}}, \ and\
  \bibinfo {author} {\bibfnamefont {Z.-Y.}\ \bibnamefont {Sun}},\ }\href@noop
  {} {\bibfield  {journal} {\bibinfo  {journal} {The Journal of chemical
  physics}\ }\textbf {\bibinfo {volume} {142}},\ \bibinfo {pages} {124507}
  (\bibinfo {year} {2015})}\BibitemShut {NoStop}%
\bibitem [{\citenamefont {Chakrabarty}\ \emph {et~al.}(2016)\citenamefont
  {Chakrabarty}, \citenamefont {Das}, \citenamefont {Karmakar},\ and\
  \citenamefont {Dasgupta}}]{chakrabarty2016understanding}%
  \BibitemOpen
  \bibfield  {author} {\bibinfo {author} {\bibfnamefont {S.}~\bibnamefont
  {Chakrabarty}}, \bibinfo {author} {\bibfnamefont {R.}~\bibnamefont {Das}},
  \bibinfo {author} {\bibfnamefont {S.}~\bibnamefont {Karmakar}}, \ and\
  \bibinfo {author} {\bibfnamefont {C.}~\bibnamefont {Dasgupta}},\ }\href@noop
  {} {\bibfield  {journal} {\bibinfo  {journal} {The Journal of chemical
  physics}\ }\textbf {\bibinfo {volume} {145}},\ \bibinfo {pages} {034507}
  (\bibinfo {year} {2016})}\BibitemShut {NoStop}%
\bibitem [{\citenamefont {Angelani}\ \emph {et~al.}(2018)\citenamefont
  {Angelani}, \citenamefont {Paoluzzi}, \citenamefont {Parisi},\ and\
  \citenamefont {Ruocco}}]{angelani2018probing}%
  \BibitemOpen
  \bibfield  {author} {\bibinfo {author} {\bibfnamefont {L.}~\bibnamefont
  {Angelani}}, \bibinfo {author} {\bibfnamefont {M.}~\bibnamefont {Paoluzzi}},
  \bibinfo {author} {\bibfnamefont {G.}~\bibnamefont {Parisi}}, \ and\ \bibinfo
  {author} {\bibfnamefont {G.}~\bibnamefont {Ruocco}},\ }\href@noop {}
  {\bibfield  {journal} {\bibinfo  {journal} {Proceedings of the National
  Academy of Sciences}\ }\textbf {\bibinfo {volume} {115}},\ \bibinfo {pages}
  {8700} (\bibinfo {year} {2018})}\BibitemShut {NoStop}%
\bibitem [{\citenamefont {Ozawa}\ \emph
  {et~al.}(2018{\natexlab{a}})\citenamefont {Ozawa}, \citenamefont {Ikeda},
  \citenamefont {Miyazaki},\ and\ \citenamefont {Kob}}]{ozawa2018ideal}%
  \BibitemOpen
  \bibfield  {author} {\bibinfo {author} {\bibfnamefont {M.}~\bibnamefont
  {Ozawa}}, \bibinfo {author} {\bibfnamefont {A.}~\bibnamefont {Ikeda}},
  \bibinfo {author} {\bibfnamefont {K.}~\bibnamefont {Miyazaki}}, \ and\
  \bibinfo {author} {\bibfnamefont {W.}~\bibnamefont {Kob}},\ }\href@noop {}
  {\bibfield  {journal} {\bibinfo  {journal} {Physical review letters}\
  }\textbf {\bibinfo {volume} {121}},\ \bibinfo {pages} {205501} (\bibinfo
  {year} {2018}{\natexlab{a}})}\BibitemShut {NoStop}%
\bibitem [{\citenamefont {Niblett}\ \emph {et~al.}(2018)\citenamefont
  {Niblett}, \citenamefont {de~Souza}, \citenamefont {Jack},\ and\
  \citenamefont {Wales}}]{niblett2018effects}%
  \BibitemOpen
  \bibfield  {author} {\bibinfo {author} {\bibfnamefont {S.}~\bibnamefont
  {Niblett}}, \bibinfo {author} {\bibfnamefont {V.~K.}\ \bibnamefont
  {de~Souza}}, \bibinfo {author} {\bibfnamefont {R.}~\bibnamefont {Jack}}, \
  and\ \bibinfo {author} {\bibfnamefont {D.}~\bibnamefont {Wales}},\
  }\href@noop {} {\bibfield  {journal} {\bibinfo  {journal} {The Journal of
  chemical physics}\ }\textbf {\bibinfo {volume} {149}},\ \bibinfo {pages}
  {114503} (\bibinfo {year} {2018})}\BibitemShut {NoStop}%
\bibitem [{\citenamefont {Gokhale}\ \emph {et~al.}(2014)\citenamefont
  {Gokhale}, \citenamefont {Nagamanasa}, \citenamefont {Ganapathy},\ and\
  \citenamefont {Sood}}]{gokhale2014growing}%
  \BibitemOpen
  \bibfield  {author} {\bibinfo {author} {\bibfnamefont {S.}~\bibnamefont
  {Gokhale}}, \bibinfo {author} {\bibfnamefont {K.~H.}\ \bibnamefont
  {Nagamanasa}}, \bibinfo {author} {\bibfnamefont {R.}~\bibnamefont
  {Ganapathy}}, \ and\ \bibinfo {author} {\bibfnamefont {A.}~\bibnamefont
  {Sood}},\ }\href@noop {} {\bibfield  {journal} {\bibinfo  {journal} {Nature
  communications}\ }\textbf {\bibinfo {volume} {5}},\ \bibinfo {pages} {4685}
  (\bibinfo {year} {2014})}\BibitemShut {NoStop}%
\bibitem [{\citenamefont {Gokhale}\ \emph {et~al.}(2016)\citenamefont
  {Gokhale}, \citenamefont {Sood},\ and\ \citenamefont
  {Ganapathy}}]{gokhale2016deconstructing}%
  \BibitemOpen
  \bibfield  {author} {\bibinfo {author} {\bibfnamefont {S.}~\bibnamefont
  {Gokhale}}, \bibinfo {author} {\bibfnamefont {A.}~\bibnamefont {Sood}}, \
  and\ \bibinfo {author} {\bibfnamefont {R.}~\bibnamefont {Ganapathy}},\
  }\href@noop {} {\bibfield  {journal} {\bibinfo  {journal} {Advances in
  Physics}\ }\textbf {\bibinfo {volume} {65}},\ \bibinfo {pages} {363}
  (\bibinfo {year} {2016})}\BibitemShut {NoStop}%
\bibitem [{\citenamefont {Ganapathi}\ \emph {et~al.}(2018)\citenamefont
  {Ganapathi}, \citenamefont {Nagamanasa}, \citenamefont {Sood},\ and\
  \citenamefont {Ganapathy}}]{ganapathi2018measurements}%
  \BibitemOpen
  \bibfield  {author} {\bibinfo {author} {\bibfnamefont {D.}~\bibnamefont
  {Ganapathi}}, \bibinfo {author} {\bibfnamefont {K.~H.}\ \bibnamefont
  {Nagamanasa}}, \bibinfo {author} {\bibfnamefont {A.}~\bibnamefont {Sood}}, \
  and\ \bibinfo {author} {\bibfnamefont {R.}~\bibnamefont {Ganapathy}},\
  }\href@noop {} {\bibfield  {journal} {\bibinfo  {journal} {Nature
  communications}\ }\textbf {\bibinfo {volume} {9}},\ \bibinfo {pages} {1}
  (\bibinfo {year} {2018})}\BibitemShut {NoStop}%
\bibitem [{\citenamefont {Williams}\ \emph {et~al.}(2018)\citenamefont
  {Williams}, \citenamefont {Turci}, \citenamefont {Hallett}, \citenamefont
  {Crowther}, \citenamefont {Cammarota}, \citenamefont {Biroli},\ and\
  \citenamefont {Royall}}]{williams2018experimental}%
  \BibitemOpen
  \bibfield  {author} {\bibinfo {author} {\bibfnamefont {I.}~\bibnamefont
  {Williams}}, \bibinfo {author} {\bibfnamefont {F.}~\bibnamefont {Turci}},
  \bibinfo {author} {\bibfnamefont {J.~E.}\ \bibnamefont {Hallett}}, \bibinfo
  {author} {\bibfnamefont {P.}~\bibnamefont {Crowther}}, \bibinfo {author}
  {\bibfnamefont {C.}~\bibnamefont {Cammarota}}, \bibinfo {author}
  {\bibfnamefont {G.}~\bibnamefont {Biroli}}, \ and\ \bibinfo {author}
  {\bibfnamefont {C.~P.}\ \bibnamefont {Royall}},\ }\href@noop {} {\bibfield
  {journal} {\bibinfo  {journal} {Journal of Physics: Condensed Matter}\
  }\textbf {\bibinfo {volume} {30}},\ \bibinfo {pages} {094003} (\bibinfo
  {year} {2018})}\BibitemShut {NoStop}%
\bibitem [{\citenamefont {Jack}\ and\ \citenamefont
  {Berthier}(2012)}]{jack2012random}%
  \BibitemOpen
  \bibfield  {author} {\bibinfo {author} {\bibfnamefont {R.~L.}\ \bibnamefont
  {Jack}}\ and\ \bibinfo {author} {\bibfnamefont {L.}~\bibnamefont
  {Berthier}},\ }\href@noop {} {\bibfield  {journal} {\bibinfo  {journal}
  {Physical Review E}\ }\textbf {\bibinfo {volume} {85}},\ \bibinfo {pages}
  {021120} (\bibinfo {year} {2012})}\BibitemShut {NoStop}%
\bibitem [{\citenamefont {Krakoviack}(2010)}]{krakoviack2010statistical}%
  \BibitemOpen
  \bibfield  {author} {\bibinfo {author} {\bibfnamefont {V.}~\bibnamefont
  {Krakoviack}},\ }\href@noop {} {\bibfield  {journal} {\bibinfo  {journal}
  {Physical Review E}\ }\textbf {\bibinfo {volume} {82}},\ \bibinfo {pages}
  {061501} (\bibinfo {year} {2010})}\BibitemShut {NoStop}%
\bibitem [{\citenamefont {Hocky}\ \emph {et~al.}(2014)\citenamefont {Hocky},
  \citenamefont {Berthier},\ and\ \citenamefont
  {Reichman}}]{hocky2014equilibrium}%
  \BibitemOpen
  \bibfield  {author} {\bibinfo {author} {\bibfnamefont {G.~M.}\ \bibnamefont
  {Hocky}}, \bibinfo {author} {\bibfnamefont {L.}~\bibnamefont {Berthier}}, \
  and\ \bibinfo {author} {\bibfnamefont {D.~R.}\ \bibnamefont {Reichman}},\
  }\href@noop {} {\bibfield  {journal} {\bibinfo  {journal} {The Journal of
  chemical physics}\ }\textbf {\bibinfo {volume} {141}},\ \bibinfo {pages}
  {224503} (\bibinfo {year} {2014})}\BibitemShut {NoStop}%
\bibitem [{\citenamefont {Thalmann}\ \emph {et~al.}(2000)\citenamefont
  {Thalmann}, \citenamefont {Dasgupta},\ and\ \citenamefont
  {Feinberg}}]{thalmann2000phase}%
  \BibitemOpen
  \bibfield  {author} {\bibinfo {author} {\bibfnamefont {F.}~\bibnamefont
  {Thalmann}}, \bibinfo {author} {\bibfnamefont {C.}~\bibnamefont {Dasgupta}},
  \ and\ \bibinfo {author} {\bibfnamefont {D.}~\bibnamefont {Feinberg}},\
  }\href@noop {} {\bibfield  {journal} {\bibinfo  {journal} {EPL (Europhysics
  Letters)}\ }\textbf {\bibinfo {volume} {50}},\ \bibinfo {pages} {54}
  (\bibinfo {year} {2000})}\BibitemShut {NoStop}%
\bibitem [{\citenamefont {Swallen}\ \emph {et~al.}(2007)\citenamefont
  {Swallen}, \citenamefont {Kearns}, \citenamefont {Mapes}, \citenamefont
  {Kim}, \citenamefont {McMahon}, \citenamefont {Ediger}, \citenamefont {Wu},
  \citenamefont {Yu},\ and\ \citenamefont {Satija}}]{swallen2007organic}%
  \BibitemOpen
  \bibfield  {author} {\bibinfo {author} {\bibfnamefont {S.~F.}\ \bibnamefont
  {Swallen}}, \bibinfo {author} {\bibfnamefont {K.~L.}\ \bibnamefont {Kearns}},
  \bibinfo {author} {\bibfnamefont {M.~K.}\ \bibnamefont {Mapes}}, \bibinfo
  {author} {\bibfnamefont {Y.~S.}\ \bibnamefont {Kim}}, \bibinfo {author}
  {\bibfnamefont {R.~J.}\ \bibnamefont {McMahon}}, \bibinfo {author}
  {\bibfnamefont {M.~D.}\ \bibnamefont {Ediger}}, \bibinfo {author}
  {\bibfnamefont {T.}~\bibnamefont {Wu}}, \bibinfo {author} {\bibfnamefont
  {L.}~\bibnamefont {Yu}}, \ and\ \bibinfo {author} {\bibfnamefont
  {S.}~\bibnamefont {Satija}},\ }\href@noop {} {\bibfield  {journal} {\bibinfo
  {journal} {Science}\ }\textbf {\bibinfo {volume} {315}},\ \bibinfo {pages}
  {353} (\bibinfo {year} {2007})}\BibitemShut {NoStop}%
\bibitem [{\citenamefont {Ediger}(2017)}]{ediger2017perspective}%
  \BibitemOpen
  \bibfield  {author} {\bibinfo {author} {\bibfnamefont {M.~D.}\ \bibnamefont
  {Ediger}},\ }\href@noop {} {\bibfield  {journal} {\bibinfo  {journal} {The
  Journal of chemical physics}\ }\textbf {\bibinfo {volume} {147}},\ \bibinfo
  {pages} {210901} (\bibinfo {year} {2017})}\BibitemShut {NoStop}%
\bibitem [{\citenamefont {Berthier}\ \emph
  {et~al.}(2017{\natexlab{b}})\citenamefont {Berthier}, \citenamefont
  {Charbonneau}, \citenamefont {Flenner},\ and\ \citenamefont
  {Zamponi}}]{berthier2017origin}%
  \BibitemOpen
  \bibfield  {author} {\bibinfo {author} {\bibfnamefont {L.}~\bibnamefont
  {Berthier}}, \bibinfo {author} {\bibfnamefont {P.}~\bibnamefont
  {Charbonneau}}, \bibinfo {author} {\bibfnamefont {E.}~\bibnamefont
  {Flenner}}, \ and\ \bibinfo {author} {\bibfnamefont {F.}~\bibnamefont
  {Zamponi}},\ }\href@noop {} {\bibfield  {journal} {\bibinfo  {journal}
  {Physical review letters}\ }\textbf {\bibinfo {volume} {119}},\ \bibinfo
  {pages} {188002} (\bibinfo {year} {2017}{\natexlab{b}})}\BibitemShut
  {NoStop}%
\bibitem [{\citenamefont {Zhu}\ \emph {et~al.}(2011)\citenamefont {Zhu},
  \citenamefont {Brian}, \citenamefont {Swallen}, \citenamefont {Straus},
  \citenamefont {Ediger},\ and\ \citenamefont {Yu}}]{zhu2011surface}%
  \BibitemOpen
  \bibfield  {author} {\bibinfo {author} {\bibfnamefont {L.}~\bibnamefont
  {Zhu}}, \bibinfo {author} {\bibfnamefont {C.}~\bibnamefont {Brian}}, \bibinfo
  {author} {\bibfnamefont {S.}~\bibnamefont {Swallen}}, \bibinfo {author}
  {\bibfnamefont {P.}~\bibnamefont {Straus}}, \bibinfo {author} {\bibfnamefont
  {M.}~\bibnamefont {Ediger}}, \ and\ \bibinfo {author} {\bibfnamefont
  {L.}~\bibnamefont {Yu}},\ }\href@noop {} {\bibfield  {journal} {\bibinfo
  {journal} {Physical Review Letters}\ }\textbf {\bibinfo {volume} {106}},\
  \bibinfo {pages} {256103} (\bibinfo {year} {2011})}\BibitemShut {NoStop}%
\bibitem [{\citenamefont {Kearns}\ \emph {et~al.}(2010)\citenamefont {Kearns},
  \citenamefont {Ediger}, \citenamefont {Huth},\ and\ \citenamefont
  {Schick}}]{kearns2010one}%
  \BibitemOpen
  \bibfield  {author} {\bibinfo {author} {\bibfnamefont {K.~L.}\ \bibnamefont
  {Kearns}}, \bibinfo {author} {\bibfnamefont {M.}~\bibnamefont {Ediger}},
  \bibinfo {author} {\bibfnamefont {H.}~\bibnamefont {Huth}}, \ and\ \bibinfo
  {author} {\bibfnamefont {C.}~\bibnamefont {Schick}},\ }\href@noop {}
  {\bibfield  {journal} {\bibinfo  {journal} {The Journal of Physical Chemistry
  Letters}\ }\textbf {\bibinfo {volume} {1}},\ \bibinfo {pages} {388} (\bibinfo
  {year} {2010})}\BibitemShut {NoStop}%
\bibitem [{\citenamefont {Chen}\ \emph {et~al.}(2013)\citenamefont {Chen},
  \citenamefont {Sep{\'u}lveda}, \citenamefont {Ediger},\ and\ \citenamefont
  {Richert}}]{chen2013dynamics}%
  \BibitemOpen
  \bibfield  {author} {\bibinfo {author} {\bibfnamefont {Z.}~\bibnamefont
  {Chen}}, \bibinfo {author} {\bibfnamefont {A.}~\bibnamefont {Sep{\'u}lveda}},
  \bibinfo {author} {\bibfnamefont {M.}~\bibnamefont {Ediger}}, \ and\ \bibinfo
  {author} {\bibfnamefont {R.}~\bibnamefont {Richert}},\ }\href@noop {}
  {\bibfield  {journal} {\bibinfo  {journal} {The Journal of chemical physics}\
  }\textbf {\bibinfo {volume} {138}},\ \bibinfo {pages} {12A519} (\bibinfo
  {year} {2013})}\BibitemShut {NoStop}%
\bibitem [{\citenamefont {P{\'e}rez-Casta{\~n}eda}\ \emph
  {et~al.}(2014)\citenamefont {P{\'e}rez-Casta{\~n}eda}, \citenamefont
  {Rodr{\'\i}guez-Tinoco}, \citenamefont {Rodr{\'\i}guez-Viejo},\ and\
  \citenamefont {Ramos}}]{perez2014suppression}%
  \BibitemOpen
  \bibfield  {author} {\bibinfo {author} {\bibfnamefont {T.}~\bibnamefont
  {P{\'e}rez-Casta{\~n}eda}}, \bibinfo {author} {\bibfnamefont
  {C.}~\bibnamefont {Rodr{\'\i}guez-Tinoco}}, \bibinfo {author} {\bibfnamefont
  {J.}~\bibnamefont {Rodr{\'\i}guez-Viejo}}, \ and\ \bibinfo {author}
  {\bibfnamefont {M.~A.}\ \bibnamefont {Ramos}},\ }\href@noop {} {\bibfield
  {journal} {\bibinfo  {journal} {Proceedings of the National Academy of
  Sciences}\ }\textbf {\bibinfo {volume} {111}},\ \bibinfo {pages} {11275}
  (\bibinfo {year} {2014})}\BibitemShut {NoStop}%
\bibitem [{\citenamefont {Sep{\'u}lveda}\ \emph {et~al.}(2014)\citenamefont
  {Sep{\'u}lveda}, \citenamefont {Tylinski}, \citenamefont {Guiseppi-Elie},
  \citenamefont {Richert},\ and\ \citenamefont {Ediger}}]{sepulveda2014role}%
  \BibitemOpen
  \bibfield  {author} {\bibinfo {author} {\bibfnamefont {A.}~\bibnamefont
  {Sep{\'u}lveda}}, \bibinfo {author} {\bibfnamefont {M.}~\bibnamefont
  {Tylinski}}, \bibinfo {author} {\bibfnamefont {A.}~\bibnamefont
  {Guiseppi-Elie}}, \bibinfo {author} {\bibfnamefont {R.}~\bibnamefont
  {Richert}}, \ and\ \bibinfo {author} {\bibfnamefont {M.}~\bibnamefont
  {Ediger}},\ }\href@noop {} {\bibfield  {journal} {\bibinfo  {journal}
  {Physical review letters}\ }\textbf {\bibinfo {volume} {113}},\ \bibinfo
  {pages} {045901} (\bibinfo {year} {2014})}\BibitemShut {NoStop}%
\bibitem [{\citenamefont {R{\`a}fols-Rib{\'e}}\ \emph
  {et~al.}(2018)\citenamefont {R{\`a}fols-Rib{\'e}}, \citenamefont {Will},
  \citenamefont {H{\"a}nisch}, \citenamefont {Gonzalez-Silveira}, \citenamefont
  {Lenk}, \citenamefont {Rodr{\'\i}guez-Viejo},\ and\ \citenamefont
  {Reineke}}]{rafols2018high}%
  \BibitemOpen
  \bibfield  {author} {\bibinfo {author} {\bibfnamefont {J.}~\bibnamefont
  {R{\`a}fols-Rib{\'e}}}, \bibinfo {author} {\bibfnamefont {P.-A.}\
  \bibnamefont {Will}}, \bibinfo {author} {\bibfnamefont {C.}~\bibnamefont
  {H{\"a}nisch}}, \bibinfo {author} {\bibfnamefont {M.}~\bibnamefont
  {Gonzalez-Silveira}}, \bibinfo {author} {\bibfnamefont {S.}~\bibnamefont
  {Lenk}}, \bibinfo {author} {\bibfnamefont {J.}~\bibnamefont
  {Rodr{\'\i}guez-Viejo}}, \ and\ \bibinfo {author} {\bibfnamefont
  {S.}~\bibnamefont {Reineke}},\ }\href@noop {} {\bibfield  {journal} {\bibinfo
   {journal} {Science advances}\ }\textbf {\bibinfo {volume} {4}},\ \bibinfo
  {pages} {eaar8332} (\bibinfo {year} {2018})}\BibitemShut {NoStop}%
\bibitem [{\citenamefont {Vila-Costa}\ \emph {et~al.}(2020)\citenamefont
  {Vila-Costa}, \citenamefont {R{\`a}fols-Rib{\'e}}, \citenamefont
  {Gonz{\'a}lez-Silveira}, \citenamefont {Lopeandia}, \citenamefont
  {Abad-Mu{\~n}oz},\ and\ \citenamefont
  {Rodr{\'\i}guez-Viejo}}]{vila2020nucleation}%
  \BibitemOpen
  \bibfield  {author} {\bibinfo {author} {\bibfnamefont {A.}~\bibnamefont
  {Vila-Costa}}, \bibinfo {author} {\bibfnamefont {J.}~\bibnamefont
  {R{\`a}fols-Rib{\'e}}}, \bibinfo {author} {\bibfnamefont {M.}~\bibnamefont
  {Gonz{\'a}lez-Silveira}}, \bibinfo {author} {\bibfnamefont {A.}~\bibnamefont
  {Lopeandia}}, \bibinfo {author} {\bibfnamefont {L.}~\bibnamefont
  {Abad-Mu{\~n}oz}}, \ and\ \bibinfo {author} {\bibfnamefont {J.}~\bibnamefont
  {Rodr{\'\i}guez-Viejo}},\ }\href@noop {} {\bibfield  {journal} {\bibinfo
  {journal} {Physical Review Letters}\ }\textbf {\bibinfo {volume} {124}},\
  \bibinfo {pages} {076002} (\bibinfo {year} {2020})}\BibitemShut {NoStop}%
\bibitem [{\citenamefont {Wolynes}(2009)}]{wolynes2009spatiotemporal}%
  \BibitemOpen
  \bibfield  {author} {\bibinfo {author} {\bibfnamefont {P.~G.}\ \bibnamefont
  {Wolynes}},\ }\href@noop {} {\bibfield  {journal} {\bibinfo  {journal}
  {Proceedings of the National Academy of Sciences}\ } (\bibinfo {year}
  {2009})}\BibitemShut {NoStop}%
\bibitem [{\citenamefont {L{\'e}onard}\ and\ \citenamefont
  {Harrowell}(2010)}]{leonard2010macroscopic}%
  \BibitemOpen
  \bibfield  {author} {\bibinfo {author} {\bibfnamefont {S.}~\bibnamefont
  {L{\'e}onard}}\ and\ \bibinfo {author} {\bibfnamefont {P.}~\bibnamefont
  {Harrowell}},\ }\href@noop {} {\bibfield  {journal} {\bibinfo  {journal} {The
  Journal of chemical physics}\ }\textbf {\bibinfo {volume} {133}},\ \bibinfo
  {pages} {244502} (\bibinfo {year} {2010})}\BibitemShut {NoStop}%
\bibitem [{\citenamefont {Lyubimov}\ \emph {et~al.}(2013)\citenamefont
  {Lyubimov}, \citenamefont {Ediger},\ and\ \citenamefont
  {de~Pablo}}]{lyubimov2013model}%
  \BibitemOpen
  \bibfield  {author} {\bibinfo {author} {\bibfnamefont {I.}~\bibnamefont
  {Lyubimov}}, \bibinfo {author} {\bibfnamefont {M.~D.}\ \bibnamefont
  {Ediger}}, \ and\ \bibinfo {author} {\bibfnamefont {J.~J.}\ \bibnamefont
  {de~Pablo}},\ }\href@noop {} {\bibfield  {journal} {\bibinfo  {journal} {The
  Journal of chemical physics}\ }\textbf {\bibinfo {volume} {139}},\ \bibinfo
  {pages} {144505} (\bibinfo {year} {2013})}\BibitemShut {NoStop}%
\bibitem [{\citenamefont {Jack}\ and\ \citenamefont
  {Berthier}(2016)}]{jack2016melting}%
  \BibitemOpen
  \bibfield  {author} {\bibinfo {author} {\bibfnamefont {R.~L.}\ \bibnamefont
  {Jack}}\ and\ \bibinfo {author} {\bibfnamefont {L.}~\bibnamefont
  {Berthier}},\ }\href@noop {} {\bibfield  {journal} {\bibinfo  {journal} {The
  Journal of chemical physics}\ }\textbf {\bibinfo {volume} {144}},\ \bibinfo
  {pages} {244506} (\bibinfo {year} {2016})}\BibitemShut {NoStop}%
\bibitem [{\citenamefont {Guti{\'e}rrez}\ and\ \citenamefont
  {Garrahan}(2016)}]{gutierrez2016front}%
  \BibitemOpen
  \bibfield  {author} {\bibinfo {author} {\bibfnamefont {R.}~\bibnamefont
  {Guti{\'e}rrez}}\ and\ \bibinfo {author} {\bibfnamefont {J.~P.}\ \bibnamefont
  {Garrahan}},\ }\href@noop {} {\bibfield  {journal} {\bibinfo  {journal}
  {Journal of Statistical Mechanics: Theory and Experiment}\ }\textbf {\bibinfo
  {volume} {2016}},\ \bibinfo {pages} {074005} (\bibinfo {year}
  {2016})}\BibitemShut {NoStop}%
\bibitem [{\citenamefont {Fullerton}\ and\ \citenamefont
  {Berthier}(2017)}]{fullerton2017density}%
  \BibitemOpen
  \bibfield  {author} {\bibinfo {author} {\bibfnamefont {C.~J.}\ \bibnamefont
  {Fullerton}}\ and\ \bibinfo {author} {\bibfnamefont {L.}~\bibnamefont
  {Berthier}},\ }\href@noop {} {\bibfield  {journal} {\bibinfo  {journal} {EPL
  (Europhysics Letters)}\ }\textbf {\bibinfo {volume} {119}},\ \bibinfo {pages}
  {36003} (\bibinfo {year} {2017})}\BibitemShut {NoStop}%
\bibitem [{\citenamefont {Flenner}\ \emph {et~al.}(2019)\citenamefont
  {Flenner}, \citenamefont {Berthier}, \citenamefont {Charbonneau},\ and\
  \citenamefont {Fullerton}}]{flenner2019front}%
  \BibitemOpen
  \bibfield  {author} {\bibinfo {author} {\bibfnamefont {E.}~\bibnamefont
  {Flenner}}, \bibinfo {author} {\bibfnamefont {L.}~\bibnamefont {Berthier}},
  \bibinfo {author} {\bibfnamefont {P.}~\bibnamefont {Charbonneau}}, \ and\
  \bibinfo {author} {\bibfnamefont {C.~J.}\ \bibnamefont {Fullerton}},\
  }\href@noop {} {\bibfield  {journal} {\bibinfo  {journal} {Physical review
  letters}\ }\textbf {\bibinfo {volume} {123}},\ \bibinfo {pages} {175501}
  (\bibinfo {year} {2019})}\BibitemShut {NoStop}%
\bibitem [{\citenamefont {Khomenko}\ \emph {et~al.}(2020)\citenamefont
  {Khomenko}, \citenamefont {Scalliet}, \citenamefont {Berthier}, \citenamefont
  {Reichman},\ and\ \citenamefont {Zamponi}}]{khomenko2019depletion}%
  \BibitemOpen
  \bibfield  {author} {\bibinfo {author} {\bibfnamefont {D.}~\bibnamefont
  {Khomenko}}, \bibinfo {author} {\bibfnamefont {C.}~\bibnamefont {Scalliet}},
  \bibinfo {author} {\bibfnamefont {L.}~\bibnamefont {Berthier}}, \bibinfo
  {author} {\bibfnamefont {D.~R.}\ \bibnamefont {Reichman}}, \ and\ \bibinfo
  {author} {\bibfnamefont {F.}~\bibnamefont {Zamponi}},\ }\href@noop {}
  {\bibfield  {journal} {\bibinfo  {journal} {Physical Review Letters}\
  }\textbf {\bibinfo {volume} {124}},\ \bibinfo {pages} {225901} (\bibinfo
  {year} {2020})}\BibitemShut {NoStop}%
\bibitem [{\citenamefont {Krzakala}\ \emph {et~al.}(2007)\citenamefont
  {Krzakala}, \citenamefont {Montanari}, \citenamefont {Ricci-Tersenghi},
  \citenamefont {Semerjian},\ and\ \citenamefont {Zdeborov{\'a}}}]{PNAS}%
  \BibitemOpen
  \bibfield  {author} {\bibinfo {author} {\bibfnamefont {F.}~\bibnamefont
  {Krzakala}}, \bibinfo {author} {\bibfnamefont {A.}~\bibnamefont {Montanari}},
  \bibinfo {author} {\bibfnamefont {F.}~\bibnamefont {Ricci-Tersenghi}},
  \bibinfo {author} {\bibfnamefont {G.}~\bibnamefont {Semerjian}}, \ and\
  \bibinfo {author} {\bibfnamefont {L.}~\bibnamefont {Zdeborov{\'a}}},\
  }\href@noop {} {\bibfield  {journal} {\bibinfo  {journal} {Proceedings of the
  National Academy of Sciences}\ }\textbf {\bibinfo {volume} {104}},\ \bibinfo
  {pages} {10318} (\bibinfo {year} {2007})}\BibitemShut {NoStop}%
\bibitem [{\citenamefont {Antenucci}\ \emph {et~al.}(2019)\citenamefont
  {Antenucci}, \citenamefont {Franz}, \citenamefont {Urbani},\ and\
  \citenamefont {Zdeborov{\'a}}}]{antenucci2019glassy}%
  \BibitemOpen
  \bibfield  {author} {\bibinfo {author} {\bibfnamefont {F.}~\bibnamefont
  {Antenucci}}, \bibinfo {author} {\bibfnamefont {S.}~\bibnamefont {Franz}},
  \bibinfo {author} {\bibfnamefont {P.}~\bibnamefont {Urbani}}, \ and\ \bibinfo
  {author} {\bibfnamefont {L.}~\bibnamefont {Zdeborov{\'a}}},\ }\href@noop {}
  {\bibfield  {journal} {\bibinfo  {journal} {Physical Review X}\ }\textbf
  {\bibinfo {volume} {9}},\ \bibinfo {pages} {011020} (\bibinfo {year}
  {2019})}\BibitemShut {NoStop}%
\bibitem [{\citenamefont {Anandkumar}\ \emph {et~al.}(2014)\citenamefont
  {Anandkumar}, \citenamefont {Ge}, \citenamefont {Hsu}, \citenamefont
  {Kakade},\ and\ \citenamefont {Telgarsky}}]{tensorPCAa}%
  \BibitemOpen
  \bibfield  {author} {\bibinfo {author} {\bibfnamefont {A.}~\bibnamefont
  {Anandkumar}}, \bibinfo {author} {\bibfnamefont {R.}~\bibnamefont {Ge}},
  \bibinfo {author} {\bibfnamefont {D.}~\bibnamefont {Hsu}}, \bibinfo {author}
  {\bibfnamefont {S.~M.}\ \bibnamefont {Kakade}}, \ and\ \bibinfo {author}
  {\bibfnamefont {M.}~\bibnamefont {Telgarsky}},\ }\href@noop {} {\bibfield
  {journal} {\bibinfo  {journal} {Journal of Machine Learning Research}\
  }\textbf {\bibinfo {volume} {15}},\ \bibinfo {pages} {2773} (\bibinfo {year}
  {2014})}\BibitemShut {NoStop}%
\bibitem [{\citenamefont {Richard}\ and\ \citenamefont
  {Montanari}(2014)}]{tensorPCAb}%
  \BibitemOpen
  \bibfield  {author} {\bibinfo {author} {\bibfnamefont {E.}~\bibnamefont
  {Richard}}\ and\ \bibinfo {author} {\bibfnamefont {A.}~\bibnamefont
  {Montanari}},\ }in\ \href@noop {} {\emph {\bibinfo {booktitle} {Advances in
  Neural Information Processing Systems}}}\ (\bibinfo {year} {2014})\ pp.\
  \bibinfo {pages} {2897--2905}\BibitemShut {NoStop}%
\bibitem [{\citenamefont {Mannelli}\ \emph {et~al.}(2020)\citenamefont
  {Mannelli}, \citenamefont {Biroli}, \citenamefont {Cammarota}, \citenamefont
  {Krzakala}, \citenamefont {Urbani},\ and\ \citenamefont
  {Zdeborov{\'a}}}]{sarao}%
  \BibitemOpen
  \bibfield  {author} {\bibinfo {author} {\bibfnamefont {S.~S.}\ \bibnamefont
  {Mannelli}}, \bibinfo {author} {\bibfnamefont {G.}~\bibnamefont {Biroli}},
  \bibinfo {author} {\bibfnamefont {C.}~\bibnamefont {Cammarota}}, \bibinfo
  {author} {\bibfnamefont {F.}~\bibnamefont {Krzakala}}, \bibinfo {author}
  {\bibfnamefont {P.}~\bibnamefont {Urbani}}, \ and\ \bibinfo {author}
  {\bibfnamefont {L.}~\bibnamefont {Zdeborov{\'a}}},\ }\href@noop {} {\bibfield
   {journal} {\bibinfo  {journal} {Physical Review X}\ }\textbf {\bibinfo
  {volume} {10}},\ \bibinfo {pages} {011057} (\bibinfo {year}
  {2020})}\BibitemShut {NoStop}%
\bibitem [{\citenamefont {Zdeborov{\'a}}\ and\ \citenamefont
  {Krzakala}(2016)}]{flole}%
  \BibitemOpen
  \bibfield  {author} {\bibinfo {author} {\bibfnamefont {L.}~\bibnamefont
  {Zdeborov{\'a}}}\ and\ \bibinfo {author} {\bibfnamefont {F.}~\bibnamefont
  {Krzakala}},\ }\href@noop {} {\bibfield  {journal} {\bibinfo  {journal}
  {Advances in Physics}\ }\textbf {\bibinfo {volume} {65}},\ \bibinfo {pages}
  {453} (\bibinfo {year} {2016})}\BibitemShut {NoStop}%
\bibitem [{\citenamefont {Sagun}\ \emph {et~al.}(2014)\citenamefont {Sagun},
  \citenamefont {Guney}, \citenamefont {Arous},\ and\ \citenamefont
  {LeCun}}]{sagun2014explorations}%
  \BibitemOpen
  \bibfield  {author} {\bibinfo {author} {\bibfnamefont {L.}~\bibnamefont
  {Sagun}}, \bibinfo {author} {\bibfnamefont {V.~U.}\ \bibnamefont {Guney}},
  \bibinfo {author} {\bibfnamefont {G.~B.}\ \bibnamefont {Arous}}, \ and\
  \bibinfo {author} {\bibfnamefont {Y.~a.}\ \bibnamefont {LeCun}},\ }\href@noop
  {} {\bibfield  {journal} {\bibinfo  {journal} {arXiv preprint
  arXiv:1412.6615}\ } (\bibinfo {year} {2014})}\BibitemShut {NoStop}%
\bibitem [{\citenamefont {Baity-Jesi}\ \emph {et~al.}(2019)\citenamefont
  {Baity-Jesi}, \citenamefont {Sagun}, \citenamefont {Geiger}, \citenamefont
  {Spigler}, \citenamefont {Arous}, \citenamefont {Cammarota}, \citenamefont
  {LeCun}, \citenamefont {Wyart},\ and\ \citenamefont {Biroli}}]{baity}%
  \BibitemOpen
  \bibfield  {author} {\bibinfo {author} {\bibfnamefont {M.}~\bibnamefont
  {Baity-Jesi}}, \bibinfo {author} {\bibfnamefont {L.}~\bibnamefont {Sagun}},
  \bibinfo {author} {\bibfnamefont {M.}~\bibnamefont {Geiger}}, \bibinfo
  {author} {\bibfnamefont {S.}~\bibnamefont {Spigler}}, \bibinfo {author}
  {\bibfnamefont {G.~B.}\ \bibnamefont {Arous}}, \bibinfo {author}
  {\bibfnamefont {C.}~\bibnamefont {Cammarota}}, \bibinfo {author}
  {\bibfnamefont {Y.}~\bibnamefont {LeCun}}, \bibinfo {author} {\bibfnamefont
  {M.}~\bibnamefont {Wyart}}, \ and\ \bibinfo {author} {\bibfnamefont
  {G.}~\bibnamefont {Biroli}},\ }\href@noop {} {\bibfield  {journal} {\bibinfo
  {journal} {Journal of Statistical Mechanics: Theory and Experiment}\ }\textbf
  {\bibinfo {volume} {2019}},\ \bibinfo {pages} {124013} (\bibinfo {year}
  {2019})}\BibitemShut {NoStop}%
\bibitem [{\citenamefont {Russell}\ and\ \citenamefont
  {Israeloff}(2000)}]{israeloff}%
  \BibitemOpen
  \bibfield  {author} {\bibinfo {author} {\bibfnamefont {E.~V.}\ \bibnamefont
  {Russell}}\ and\ \bibinfo {author} {\bibfnamefont {N.}~\bibnamefont
  {Israeloff}},\ }\href@noop {} {\bibfield  {journal} {\bibinfo  {journal}
  {Nature}\ }\textbf {\bibinfo {volume} {408}},\ \bibinfo {pages} {695}
  (\bibinfo {year} {2000})}\BibitemShut {NoStop}%
\bibitem [{\citenamefont {Adhikari}\ \emph {et~al.}(2007)\citenamefont
  {Adhikari}, \citenamefont {Capurso},\ and\ \citenamefont
  {Bingemann}}]{single}%
  \BibitemOpen
  \bibfield  {author} {\bibinfo {author} {\bibfnamefont {A.~N.}\ \bibnamefont
  {Adhikari}}, \bibinfo {author} {\bibfnamefont {N.~A.}\ \bibnamefont
  {Capurso}}, \ and\ \bibinfo {author} {\bibfnamefont {D.}~\bibnamefont
  {Bingemann}},\ }\href@noop {} {\bibfield  {journal} {\bibinfo  {journal} {The
  Journal of chemical physics}\ }\textbf {\bibinfo {volume} {127}},\ \bibinfo
  {pages} {114508} (\bibinfo {year} {2007})}\BibitemShut {NoStop}%
\bibitem [{\citenamefont {Paeng}\ \emph {et~al.}(2015)\citenamefont {Paeng},
  \citenamefont {Park}, \citenamefont {Hoang},\ and\ \citenamefont
  {Kaufman}}]{paeng2015ideal}%
  \BibitemOpen
  \bibfield  {author} {\bibinfo {author} {\bibfnamefont {K.}~\bibnamefont
  {Paeng}}, \bibinfo {author} {\bibfnamefont {H.}~\bibnamefont {Park}},
  \bibinfo {author} {\bibfnamefont {D.~T.}\ \bibnamefont {Hoang}}, \ and\
  \bibinfo {author} {\bibfnamefont {L.~J.}\ \bibnamefont {Kaufman}},\
  }\href@noop {} {\bibfield  {journal} {\bibinfo  {journal} {Proceedings of the
  National Academy of Sciences}\ }\textbf {\bibinfo {volume} {112}},\ \bibinfo
  {pages} {4952} (\bibinfo {year} {2015})}\BibitemShut {NoStop}%
\bibitem [{\citenamefont {Allen}\ and\ \citenamefont
  {Tildesley}(1989)}]{allen}%
  \BibitemOpen
  \bibfield  {author} {\bibinfo {author} {\bibfnamefont {M.}~\bibnamefont
  {Allen}}\ and\ \bibinfo {author} {\bibfnamefont {D.}~\bibnamefont
  {Tildesley}},\ }\href@noop {} {\bibfield  {journal} {\bibinfo  {journal} {New
  York: Oxford}\ }\textbf {\bibinfo {volume} {385}} (\bibinfo {year}
  {1989})}\BibitemShut {NoStop}%
\bibitem [{\citenamefont {Horbach}\ and\ \citenamefont
  {Kob}(2001)}]{horbachkob}%
  \BibitemOpen
  \bibfield  {author} {\bibinfo {author} {\bibfnamefont {J.}~\bibnamefont
  {Horbach}}\ and\ \bibinfo {author} {\bibfnamefont {W.}~\bibnamefont {Kob}},\
  }\href@noop {} {\bibfield  {journal} {\bibinfo  {journal} {Physical Review
  E}\ }\textbf {\bibinfo {volume} {64}},\ \bibinfo {pages} {041503} (\bibinfo
  {year} {2001})}\BibitemShut {NoStop}%
\bibitem [{\citenamefont {G{\"o}tze}(1999)}]{gotze}%
  \BibitemOpen
  \bibfield  {author} {\bibinfo {author} {\bibfnamefont {W.}~\bibnamefont
  {G{\"o}tze}},\ }\href@noop {} {\bibfield  {journal} {\bibinfo  {journal}
  {Journal of Physics: condensed matter}\ }\textbf {\bibinfo {volume} {11}},\
  \bibinfo {pages} {A1} (\bibinfo {year} {1999})}\BibitemShut {NoStop}%
\bibitem [{\citenamefont {Gleim}\ \emph {et~al.}(1998)\citenamefont {Gleim},
  \citenamefont {Kob},\ and\ \citenamefont {Binder}}]{gleim}%
  \BibitemOpen
  \bibfield  {author} {\bibinfo {author} {\bibfnamefont {T.}~\bibnamefont
  {Gleim}}, \bibinfo {author} {\bibfnamefont {W.}~\bibnamefont {Kob}}, \ and\
  \bibinfo {author} {\bibfnamefont {K.}~\bibnamefont {Binder}},\ }\href@noop {}
  {\bibfield  {journal} {\bibinfo  {journal} {Physical review letters}\
  }\textbf {\bibinfo {volume} {81}},\ \bibinfo {pages} {4404} (\bibinfo {year}
  {1998})}\BibitemShut {NoStop}%
\bibitem [{\citenamefont {Szamel}\ and\ \citenamefont
  {Flenner}(2004)}]{szamel}%
  \BibitemOpen
  \bibfield  {author} {\bibinfo {author} {\bibfnamefont {G.}~\bibnamefont
  {Szamel}}\ and\ \bibinfo {author} {\bibfnamefont {E.}~\bibnamefont
  {Flenner}},\ }\href@noop {} {\bibfield  {journal} {\bibinfo  {journal} {EPL
  (Europhysics Letters)}\ }\textbf {\bibinfo {volume} {67}},\ \bibinfo {pages}
  {779} (\bibinfo {year} {2004})}\BibitemShut {NoStop}%
\bibitem [{\citenamefont {Berthier}\ and\ \citenamefont
  {Kob}(2007)}]{berthierkob}%
  \BibitemOpen
  \bibfield  {author} {\bibinfo {author} {\bibfnamefont {L.}~\bibnamefont
  {Berthier}}\ and\ \bibinfo {author} {\bibfnamefont {W.}~\bibnamefont {Kob}},\
  }\href@noop {} {\bibfield  {journal} {\bibinfo  {journal} {Journal of
  Physics: Condensed Matter}\ }\textbf {\bibinfo {volume} {19}},\ \bibinfo
  {pages} {205130} (\bibinfo {year} {2007})}\BibitemShut {NoStop}%
\bibitem [{\citenamefont {Berthier}\ \emph
  {et~al.}(2007{\natexlab{a}})\citenamefont {Berthier}, \citenamefont {Biroli},
  \citenamefont {Bouchaud}, \citenamefont {Kob}, \citenamefont {Miyazaki},\
  and\ \citenamefont {Reichman}}]{jcpI}%
  \BibitemOpen
  \bibfield  {author} {\bibinfo {author} {\bibfnamefont {L.}~\bibnamefont
  {Berthier}}, \bibinfo {author} {\bibfnamefont {G.}~\bibnamefont {Biroli}},
  \bibinfo {author} {\bibfnamefont {J.-P.}\ \bibnamefont {Bouchaud}}, \bibinfo
  {author} {\bibfnamefont {W.}~\bibnamefont {Kob}}, \bibinfo {author}
  {\bibfnamefont {K.}~\bibnamefont {Miyazaki}}, \ and\ \bibinfo {author}
  {\bibfnamefont {D.}~\bibnamefont {Reichman}},\ }\href@noop {} {\bibfield
  {journal} {\bibinfo  {journal} {The Journal of chemical physics}\ }\textbf
  {\bibinfo {volume} {126}},\ \bibinfo {pages} {184503} (\bibinfo {year}
  {2007}{\natexlab{a}})}\BibitemShut {NoStop}%
\bibitem [{\citenamefont {Berthier}\ \emph
  {et~al.}(2007{\natexlab{b}})\citenamefont {Berthier}, \citenamefont {Biroli},
  \citenamefont {Bouchaud}, \citenamefont {Kob}, \citenamefont {Miyazaki},\
  and\ \citenamefont {Reichman}}]{jcpII}%
  \BibitemOpen
  \bibfield  {author} {\bibinfo {author} {\bibfnamefont {L.}~\bibnamefont
  {Berthier}}, \bibinfo {author} {\bibfnamefont {G.}~\bibnamefont {Biroli}},
  \bibinfo {author} {\bibfnamefont {J.-P.}\ \bibnamefont {Bouchaud}}, \bibinfo
  {author} {\bibfnamefont {W.}~\bibnamefont {Kob}}, \bibinfo {author}
  {\bibfnamefont {K.}~\bibnamefont {Miyazaki}}, \ and\ \bibinfo {author}
  {\bibfnamefont {D.~R.}\ \bibnamefont {Reichman}},\ }\href@noop {} {\bibfield
  {journal} {\bibinfo  {journal} {The Journal of chemical physics}\ }\textbf
  {\bibinfo {volume} {126}},\ \bibinfo {pages} {184504} (\bibinfo {year}
  {2007}{\natexlab{b}})}\BibitemShut {NoStop}%
\bibitem [{\citenamefont {Hurley}\ and\ \citenamefont
  {Harrowell}(1995)}]{harrowell}%
  \BibitemOpen
  \bibfield  {author} {\bibinfo {author} {\bibfnamefont {M.}~\bibnamefont
  {Hurley}}\ and\ \bibinfo {author} {\bibfnamefont {P.}~\bibnamefont
  {Harrowell}},\ }\href@noop {} {\bibfield  {journal} {\bibinfo  {journal}
  {Physical Review E}\ }\textbf {\bibinfo {volume} {52}},\ \bibinfo {pages}
  {1694} (\bibinfo {year} {1995})}\BibitemShut {NoStop}%
\bibitem [{\citenamefont {Ediger}(2000)}]{ediger}%
  \BibitemOpen
  \bibfield  {author} {\bibinfo {author} {\bibfnamefont {M.~D.}\ \bibnamefont
  {Ediger}},\ }\href@noop {} {\bibfield  {journal} {\bibinfo  {journal} {Annual
  review of physical chemistry}\ }\textbf {\bibinfo {volume} {51}},\ \bibinfo
  {pages} {99} (\bibinfo {year} {2000})}\BibitemShut {NoStop}%
\bibitem [{\citenamefont {Berthier}\ \emph
  {et~al.}(2011{\natexlab{a}})\citenamefont {Berthier}, \citenamefont {Biroli},
  \citenamefont {Bouchaud}, \citenamefont {Cipelletti},\ and\ \citenamefont
  {van Saarloos}}]{berthier2011dynamical}%
  \BibitemOpen
  \bibfield  {author} {\bibinfo {author} {\bibfnamefont {L.}~\bibnamefont
  {Berthier}}, \bibinfo {author} {\bibfnamefont {G.}~\bibnamefont {Biroli}},
  \bibinfo {author} {\bibfnamefont {J.-P.}\ \bibnamefont {Bouchaud}}, \bibinfo
  {author} {\bibfnamefont {L.}~\bibnamefont {Cipelletti}}, \ and\ \bibinfo
  {author} {\bibfnamefont {W.}~\bibnamefont {van Saarloos}},\ }\href@noop {}
  {\emph {\bibinfo {title} {Dynamical heterogeneities in glasses, colloids, and
  granular media}}},\ Vol.\ \bibinfo {volume} {150}\ (\bibinfo  {publisher}
  {OUP Oxford},\ \bibinfo {year} {2011})\BibitemShut {NoStop}%
\bibitem [{\citenamefont {Kob}\ \emph {et~al.}(1997)\citenamefont {Kob},
  \citenamefont {Donati}, \citenamefont {Plimpton}, \citenamefont {Poole},\
  and\ \citenamefont {Glotzer}}]{glotzerkob}%
  \BibitemOpen
  \bibfield  {author} {\bibinfo {author} {\bibfnamefont {W.}~\bibnamefont
  {Kob}}, \bibinfo {author} {\bibfnamefont {C.}~\bibnamefont {Donati}},
  \bibinfo {author} {\bibfnamefont {S.~J.}\ \bibnamefont {Plimpton}}, \bibinfo
  {author} {\bibfnamefont {P.~H.}\ \bibnamefont {Poole}}, \ and\ \bibinfo
  {author} {\bibfnamefont {S.~C.}\ \bibnamefont {Glotzer}},\ }\href@noop {}
  {\bibfield  {journal} {\bibinfo  {journal} {Physical review letters}\
  }\textbf {\bibinfo {volume} {79}},\ \bibinfo {pages} {2827} (\bibinfo {year}
  {1997})}\BibitemShut {NoStop}%
\bibitem [{\citenamefont {Chaudhuri}\ \emph {et~al.}(2007)\citenamefont
  {Chaudhuri}, \citenamefont {Berthier},\ and\ \citenamefont {Kob}}]{pinaki}%
  \BibitemOpen
  \bibfield  {author} {\bibinfo {author} {\bibfnamefont {P.}~\bibnamefont
  {Chaudhuri}}, \bibinfo {author} {\bibfnamefont {L.}~\bibnamefont {Berthier}},
  \ and\ \bibinfo {author} {\bibfnamefont {W.}~\bibnamefont {Kob}},\
  }\href@noop {} {\bibfield  {journal} {\bibinfo  {journal} {Physical review
  letters}\ }\textbf {\bibinfo {volume} {99}},\ \bibinfo {pages} {060604}
  (\bibinfo {year} {2007})}\BibitemShut {NoStop}%
\bibitem [{\citenamefont {Hansen}\ and\ \citenamefont
  {McDonald}(1990)}]{hansen}%
  \BibitemOpen
  \bibfield  {author} {\bibinfo {author} {\bibfnamefont {J.-P.}\ \bibnamefont
  {Hansen}}\ and\ \bibinfo {author} {\bibfnamefont {I.~R.}\ \bibnamefont
  {McDonald}},\ }\href@noop {} {\emph {\bibinfo {title} {Theory of simple
  liquids}}}\ (\bibinfo  {publisher} {Elsevier},\ \bibinfo {year}
  {1990})\BibitemShut {NoStop}%
\bibitem [{\citenamefont {Mapes}\ \emph {et~al.}(2006)\citenamefont {Mapes},
  \citenamefont {Swallen},\ and\ \citenamefont {Ediger}}]{edigerotp}%
  \BibitemOpen
  \bibfield  {author} {\bibinfo {author} {\bibfnamefont {M.~K.}\ \bibnamefont
  {Mapes}}, \bibinfo {author} {\bibfnamefont {S.~F.}\ \bibnamefont {Swallen}},
  \ and\ \bibinfo {author} {\bibfnamefont {M.~D.}\ \bibnamefont {Ediger}},\
  }\href {\doibase 10.1021/jp0555955} {\bibfield  {journal} {\bibinfo
  {journal} {The Journal of Physical Chemistry B}\ }\textbf {\bibinfo {volume}
  {110}},\ \bibinfo {pages} {507} (\bibinfo {year} {2006})},\ \bibinfo {note}
  {pMID: 16471562},\ \Eprint
  {http://arxiv.org/abs/https://doi.org/10.1021/jp0555955}
  {https://doi.org/10.1021/jp0555955} \BibitemShut {NoStop}%
\bibitem [{\citenamefont {Tarjus}\ and\ \citenamefont
  {Kivelson}(1995)}]{Gilles}%
  \BibitemOpen
  \bibfield  {author} {\bibinfo {author} {\bibfnamefont {G.}~\bibnamefont
  {Tarjus}}\ and\ \bibinfo {author} {\bibfnamefont {D.}~\bibnamefont
  {Kivelson}},\ }\href {\doibase 10.1063/1.470495} {\bibfield  {journal}
  {\bibinfo  {journal} {The Journal of Chemical Physics}\ }\textbf {\bibinfo
  {volume} {103}},\ \bibinfo {pages} {3071} (\bibinfo {year} {1995})},\ \Eprint
  {http://arxiv.org/abs/https://doi.org/10.1063/1.470495}
  {https://doi.org/10.1063/1.470495} \BibitemShut {NoStop}%
\bibitem [{\citenamefont {Jung}\ \emph {et~al.}(2004)\citenamefont {Jung},
  \citenamefont {Garrahan},\ and\ \citenamefont {Chandler}}]{jung}%
  \BibitemOpen
  \bibfield  {author} {\bibinfo {author} {\bibfnamefont {Y.}~\bibnamefont
  {Jung}}, \bibinfo {author} {\bibfnamefont {J.~P.}\ \bibnamefont {Garrahan}},
  \ and\ \bibinfo {author} {\bibfnamefont {D.}~\bibnamefont {Chandler}},\
  }\href {\doibase 10.1103/PhysRevE.69.061205} {\bibfield  {journal} {\bibinfo
  {journal} {Phys. Rev. E}\ }\textbf {\bibinfo {volume} {69}},\ \bibinfo
  {pages} {061205} (\bibinfo {year} {2004})}\BibitemShut {NoStop}%
\bibitem [{\citenamefont {Reinsberg}\ \emph {et~al.}(2001)\citenamefont
  {Reinsberg}, \citenamefont {Qiu}, \citenamefont {Wilhelm}, \citenamefont
  {Spiess},\ and\ \citenamefont {Ediger}}]{nmr}%
  \BibitemOpen
  \bibfield  {author} {\bibinfo {author} {\bibfnamefont {S.~A.}\ \bibnamefont
  {Reinsberg}}, \bibinfo {author} {\bibfnamefont {X.~H.}\ \bibnamefont {Qiu}},
  \bibinfo {author} {\bibfnamefont {M.}~\bibnamefont {Wilhelm}}, \bibinfo
  {author} {\bibfnamefont {H.~W.}\ \bibnamefont {Spiess}}, \ and\ \bibinfo
  {author} {\bibfnamefont {M.~D.}\ \bibnamefont {Ediger}},\ }\href {\doibase
  10.1063/1.1369160} {\bibfield  {journal} {\bibinfo  {journal} {The Journal of
  Chemical Physics}\ }\textbf {\bibinfo {volume} {114}},\ \bibinfo {pages}
  {7299} (\bibinfo {year} {2001})},\ \Eprint
  {http://arxiv.org/abs/https://doi.org/10.1063/1.1369160}
  {https://doi.org/10.1063/1.1369160} \BibitemShut {NoStop}%
\bibitem [{\citenamefont {Franz}\ and\ \citenamefont
  {Parisi}(2000)}]{franzparisi}%
  \BibitemOpen
  \bibfield  {author} {\bibinfo {author} {\bibfnamefont {S.}~\bibnamefont
  {Franz}}\ and\ \bibinfo {author} {\bibfnamefont {G.}~\bibnamefont {Parisi}},\
  }\href {\doibase 10.1088/0953-8984/12/29/305} {\bibfield  {journal} {\bibinfo
   {journal} {Journal of Physics: Condensed Matter}\ }\textbf {\bibinfo
  {volume} {12}},\ \bibinfo {pages} {6335} (\bibinfo {year}
  {2000})}\BibitemShut {NoStop}%
\bibitem [{\citenamefont {Toninelli}\ \emph {et~al.}(2005)\citenamefont
  {Toninelli}, \citenamefont {Wyart}, \citenamefont {Berthier}, \citenamefont
  {Biroli},\ and\ \citenamefont {Bouchaud}}]{TWBBB}%
  \BibitemOpen
  \bibfield  {author} {\bibinfo {author} {\bibfnamefont {C.}~\bibnamefont
  {Toninelli}}, \bibinfo {author} {\bibfnamefont {M.}~\bibnamefont {Wyart}},
  \bibinfo {author} {\bibfnamefont {L.}~\bibnamefont {Berthier}}, \bibinfo
  {author} {\bibfnamefont {G.}~\bibnamefont {Biroli}}, \ and\ \bibinfo {author}
  {\bibfnamefont {J.-P.}\ \bibnamefont {Bouchaud}},\ }\href {\doibase
  10.1103/PhysRevE.71.041505} {\bibfield  {journal} {\bibinfo  {journal} {Phys.
  Rev. E}\ }\textbf {\bibinfo {volume} {71}},\ \bibinfo {pages} {041505}
  (\bibinfo {year} {2005})}\BibitemShut {NoStop}%
\bibitem [{\citenamefont {Yamamoto}\ and\ \citenamefont {Onuki}(1998)}]{onuki}%
  \BibitemOpen
  \bibfield  {author} {\bibinfo {author} {\bibfnamefont {R.}~\bibnamefont
  {Yamamoto}}\ and\ \bibinfo {author} {\bibfnamefont {A.}~\bibnamefont
  {Onuki}},\ }\href {\doibase 10.1103/PhysRevE.58.3515} {\bibfield  {journal}
  {\bibinfo  {journal} {Phys. Rev. E}\ }\textbf {\bibinfo {volume} {58}},\
  \bibinfo {pages} {3515} (\bibinfo {year} {1998})}\BibitemShut {NoStop}%
\bibitem [{\citenamefont {Franz}\ \emph {et~al.}(1999)\citenamefont {Franz},
  \citenamefont {Donati}, \citenamefont {Parisi},\ and\ \citenamefont
  {Glotzer}}]{glotzerfranzparisi}%
  \BibitemOpen
  \bibfield  {author} {\bibinfo {author} {\bibfnamefont {S.}~\bibnamefont
  {Franz}}, \bibinfo {author} {\bibfnamefont {C.}~\bibnamefont {Donati}},
  \bibinfo {author} {\bibfnamefont {G.}~\bibnamefont {Parisi}}, \ and\ \bibinfo
  {author} {\bibfnamefont {S.~C.}\ \bibnamefont {Glotzer}},\ }\href {\doibase
  10.1080/13642819908223066} {\bibfield  {journal} {\bibinfo  {journal}
  {Philosophical Magazine B}\ }\textbf {\bibinfo {volume} {79}},\ \bibinfo
  {pages} {1827} (\bibinfo {year} {1999})},\ \Eprint
  {http://arxiv.org/abs/https://doi.org/10.1080/13642819908223066}
  {https://doi.org/10.1080/13642819908223066} \BibitemShut {NoStop}%
\bibitem [{\citenamefont {Bennemann}\ \emph {et~al.}(1999)\citenamefont
  {Bennemann}, \citenamefont {Donati}, \citenamefont {Baschnagel},\ and\
  \citenamefont {Glotzer}}]{glotzer}%
  \BibitemOpen
  \bibfield  {author} {\bibinfo {author} {\bibfnamefont {C.}~\bibnamefont
  {Bennemann}}, \bibinfo {author} {\bibfnamefont {C.}~\bibnamefont {Donati}},
  \bibinfo {author} {\bibfnamefont {J.}~\bibnamefont {Baschnagel}}, \ and\
  \bibinfo {author} {\bibfnamefont {S.~C.}\ \bibnamefont {Glotzer}},\
  }\href@noop {} {\bibfield  {journal} {\bibinfo  {journal} {Nature}\ }\textbf
  {\bibinfo {volume} {399}},\ \bibinfo {pages} {246} (\bibinfo {year}
  {1999})}\BibitemShut {NoStop}%
\bibitem [{\citenamefont {La{\v{c}}evi{\'c}}\ \emph {et~al.}(2003)\citenamefont
  {La{\v{c}}evi{\'c}}, \citenamefont {Starr}, \citenamefont {Schr{\o}der},\
  and\ \citenamefont {Glotzer}}]{lavcevic2003spatially}%
  \BibitemOpen
  \bibfield  {author} {\bibinfo {author} {\bibfnamefont {N.}~\bibnamefont
  {La{\v{c}}evi{\'c}}}, \bibinfo {author} {\bibfnamefont {F.~W.}\ \bibnamefont
  {Starr}}, \bibinfo {author} {\bibfnamefont {T.}~\bibnamefont {Schr{\o}der}},
  \ and\ \bibinfo {author} {\bibfnamefont {S.}~\bibnamefont {Glotzer}},\
  }\href@noop {} {\bibfield  {journal} {\bibinfo  {journal} {The Journal of
  chemical physics}\ }\textbf {\bibinfo {volume} {119}},\ \bibinfo {pages}
  {7372} (\bibinfo {year} {2003})}\BibitemShut {NoStop}%
\bibitem [{\citenamefont {Berthier}(2004)}]{berthier}%
  \BibitemOpen
  \bibfield  {author} {\bibinfo {author} {\bibfnamefont {L.}~\bibnamefont
  {Berthier}},\ }\href {\doibase 10.1103/PhysRevE.69.020201} {\bibfield
  {journal} {\bibinfo  {journal} {Phys. Rev. E}\ }\textbf {\bibinfo {volume}
  {69}},\ \bibinfo {pages} {020201} (\bibinfo {year} {2004})}\BibitemShut
  {NoStop}%
\bibitem [{\citenamefont {Gebremichael}\ \emph {et~al.}(2004)\citenamefont
  {Gebremichael}, \citenamefont {Vogel},\ and\ \citenamefont
  {Glotzer}}]{glotzersilica}%
  \BibitemOpen
  \bibfield  {author} {\bibinfo {author} {\bibfnamefont {Y.}~\bibnamefont
  {Gebremichael}}, \bibinfo {author} {\bibfnamefont {M.}~\bibnamefont {Vogel}},
  \ and\ \bibinfo {author} {\bibfnamefont {S.~C.}\ \bibnamefont {Glotzer}},\
  }\href {\doibase 10.1063/1.1644539} {\bibfield  {journal} {\bibinfo
  {journal} {The Journal of Chemical Physics}\ }\textbf {\bibinfo {volume}
  {120}},\ \bibinfo {pages} {4415} (\bibinfo {year} {2004})},\ \Eprint
  {http://arxiv.org/abs/https://doi.org/10.1063/1.1644539}
  {https://doi.org/10.1063/1.1644539} \BibitemShut {NoStop}%
\bibitem [{\citenamefont {Berthier}(2007{\natexlab{a}})}]{berthiersilica}%
  \BibitemOpen
  \bibfield  {author} {\bibinfo {author} {\bibfnamefont {L.}~\bibnamefont
  {Berthier}},\ }\href {\doibase 10.1103/PhysRevE.76.011507} {\bibfield
  {journal} {\bibinfo  {journal} {Phys. Rev. E}\ }\textbf {\bibinfo {volume}
  {76}},\ \bibinfo {pages} {011507} (\bibinfo {year}
  {2007}{\natexlab{a}})}\BibitemShut {NoStop}%
\bibitem [{Note2()}]{Note2}%
  \BibitemOpen
  \bibinfo {note} {The decrease at long times constitutes a major difference
  with spin glasses. In a spin glass, $\chi _4$ would be a monotonically
  increasing function of time whose long-time limit coincides with the static
  spin glass susceptibility. Physically, the difference is that spin glasses
  develop long-range static amorphous order while structural glasses do not or,
  at least, in a different and more subtle way.}\BibitemShut {Stop}%
\bibitem [{\citenamefont {Weeks}\ \emph {et~al.}(2007)\citenamefont {Weeks},
  \citenamefont {Crocker},\ and\ \citenamefont {Weitz}}]{weeks2}%
  \BibitemOpen
  \bibfield  {author} {\bibinfo {author} {\bibfnamefont {E.~R.}\ \bibnamefont
  {Weeks}}, \bibinfo {author} {\bibfnamefont {J.~C.}\ \bibnamefont {Crocker}},
  \ and\ \bibinfo {author} {\bibfnamefont {D.~A.}\ \bibnamefont {Weitz}},\
  }\href {\doibase 10.1088/0953-8984/19/20/205131} {\bibfield  {journal}
  {\bibinfo  {journal} {Journal of Physics: Condensed Matter}\ }\textbf
  {\bibinfo {volume} {19}},\ \bibinfo {pages} {205131} (\bibinfo {year}
  {2007})}\BibitemShut {NoStop}%
\bibitem [{\citenamefont {Dauchot}\ \emph {et~al.}(2005)\citenamefont
  {Dauchot}, \citenamefont {Marty},\ and\ \citenamefont
  {Biroli}}]{dauchotbiroli}%
  \BibitemOpen
  \bibfield  {author} {\bibinfo {author} {\bibfnamefont {O.}~\bibnamefont
  {Dauchot}}, \bibinfo {author} {\bibfnamefont {G.}~\bibnamefont {Marty}}, \
  and\ \bibinfo {author} {\bibfnamefont {G.}~\bibnamefont {Biroli}},\ }\href
  {\doibase 10.1103/PhysRevLett.95.265701} {\bibfield  {journal} {\bibinfo
  {journal} {Phys. Rev. Lett.}\ }\textbf {\bibinfo {volume} {95}},\ \bibinfo
  {pages} {265701} (\bibinfo {year} {2005})}\BibitemShut {NoStop}%
\bibitem [{\citenamefont {Berthier}\ \emph {et~al.}(2005)\citenamefont
  {Berthier}, \citenamefont {Biroli}, \citenamefont {Bouchaud}, \citenamefont
  {Cipelletti}, \citenamefont {El~Masri}, \citenamefont {L'H{\^o}te},
  \citenamefont {Ladieu},\ and\ \citenamefont {Pierno}}]{science}%
  \BibitemOpen
  \bibfield  {author} {\bibinfo {author} {\bibfnamefont {L.}~\bibnamefont
  {Berthier}}, \bibinfo {author} {\bibfnamefont {G.}~\bibnamefont {Biroli}},
  \bibinfo {author} {\bibfnamefont {J.-P.}\ \bibnamefont {Bouchaud}}, \bibinfo
  {author} {\bibfnamefont {L.}~\bibnamefont {Cipelletti}}, \bibinfo {author}
  {\bibfnamefont {D.}~\bibnamefont {El~Masri}}, \bibinfo {author}
  {\bibfnamefont {D.}~\bibnamefont {L'H{\^o}te}}, \bibinfo {author}
  {\bibfnamefont {F.}~\bibnamefont {Ladieu}}, \ and\ \bibinfo {author}
  {\bibfnamefont {M.}~\bibnamefont {Pierno}},\ }\href@noop {} {\bibfield
  {journal} {\bibinfo  {journal} {Science}\ }\textbf {\bibinfo {volume}
  {310}},\ \bibinfo {pages} {1797} (\bibinfo {year} {2005})}\BibitemShut
  {NoStop}%
\bibitem [{\citenamefont {Dalle-Ferrier}\ \emph {et~al.}(2007)\citenamefont
  {Dalle-Ferrier}, \citenamefont {Thibierge}, \citenamefont {Alba-Simionesco},
  \citenamefont {Berthier}, \citenamefont {Biroli}, \citenamefont {Bouchaud},
  \citenamefont {Ladieu}, \citenamefont {L'H\^ote},\ and\ \citenamefont
  {Tarjus}}]{cecile}%
  \BibitemOpen
  \bibfield  {author} {\bibinfo {author} {\bibfnamefont {C.}~\bibnamefont
  {Dalle-Ferrier}}, \bibinfo {author} {\bibfnamefont {C.}~\bibnamefont
  {Thibierge}}, \bibinfo {author} {\bibfnamefont {C.}~\bibnamefont
  {Alba-Simionesco}}, \bibinfo {author} {\bibfnamefont {L.}~\bibnamefont
  {Berthier}}, \bibinfo {author} {\bibfnamefont {G.}~\bibnamefont {Biroli}},
  \bibinfo {author} {\bibfnamefont {J.-P.}\ \bibnamefont {Bouchaud}}, \bibinfo
  {author} {\bibfnamefont {F.}~\bibnamefont {Ladieu}}, \bibinfo {author}
  {\bibfnamefont {D.}~\bibnamefont {L'H\^ote}}, \ and\ \bibinfo {author}
  {\bibfnamefont {G.}~\bibnamefont {Tarjus}},\ }\href {\doibase
  10.1103/PhysRevE.76.041510} {\bibfield  {journal} {\bibinfo  {journal} {Phys.
  Rev. E}\ }\textbf {\bibinfo {volume} {76}},\ \bibinfo {pages} {041510}
  (\bibinfo {year} {2007})}\BibitemShut {NoStop}%
\bibitem [{\citenamefont {Xia}\ and\ \citenamefont
  {Wolynes}(2000)}]{rfotwolynes}%
  \BibitemOpen
  \bibfield  {author} {\bibinfo {author} {\bibfnamefont {X.}~\bibnamefont
  {Xia}}\ and\ \bibinfo {author} {\bibfnamefont {P.~G.}\ \bibnamefont
  {Wolynes}},\ }\href {\doibase 10.1073/pnas.97.7.2990} {\bibfield  {journal}
  {\bibinfo  {journal} {Proceedings of the National Academy of Sciences}\
  }\textbf {\bibinfo {volume} {97}},\ \bibinfo {pages} {2990} (\bibinfo {year}
  {2000})},\ \bibinfo {note} {tex.eprint:
  https://www.pnas.org/content/97/7/2990.full.pdf tex.publisher: National
  Academy of Sciences}\BibitemShut {NoStop}%
\bibitem [{\citenamefont {Garrahan}\ and\ \citenamefont
  {Chandler}(2003)}]{gcpnas}%
  \BibitemOpen
  \bibfield  {author} {\bibinfo {author} {\bibfnamefont {J.~P.}\ \bibnamefont
  {Garrahan}}\ and\ \bibinfo {author} {\bibfnamefont {D.}~\bibnamefont
  {Chandler}},\ }\href {\doibase 10.1073/pnas.1233719100} {\bibfield  {journal}
  {\bibinfo  {journal} {Proceedings of the National Academy of Sciences}\
  }\textbf {\bibinfo {volume} {100}},\ \bibinfo {pages} {9710} (\bibinfo {year}
  {2003})},\ \bibinfo {note} {tex.eprint:
  https://www.pnas.org/content/100/17/9710.full.pdf tex.publisher: National
  Academy of Sciences}\BibitemShut {NoStop}%
\bibitem [{\citenamefont {Tarjus}\ \emph {et~al.}(2005)\citenamefont {Tarjus},
  \citenamefont {Kivelson}, \citenamefont {Nussinov},\ and\ \citenamefont
  {Viot}}]{gillesreview}%
  \BibitemOpen
  \bibfield  {author} {\bibinfo {author} {\bibfnamefont {G.}~\bibnamefont
  {Tarjus}}, \bibinfo {author} {\bibfnamefont {S.~A.}\ \bibnamefont
  {Kivelson}}, \bibinfo {author} {\bibfnamefont {Z.}~\bibnamefont {Nussinov}},
  \ and\ \bibinfo {author} {\bibfnamefont {P.}~\bibnamefont {Viot}},\ }\href
  {\doibase 10.1088/0953-8984/17/50/r01} {\bibfield  {journal} {\bibinfo
  {journal} {Journal of Physics: Condensed Matter}\ }\textbf {\bibinfo {volume}
  {17}},\ \bibinfo {pages} {R1143} (\bibinfo {year} {2005})}\BibitemShut
  {NoStop}%
\bibitem [{\citenamefont {Donati}\ \emph {et~al.}(1998)\citenamefont {Donati},
  \citenamefont {Douglas}, \citenamefont {Kob}, \citenamefont {Plimpton},
  \citenamefont {Poole},\ and\ \citenamefont {Glotzer}}]{glotzerstring}%
  \BibitemOpen
  \bibfield  {author} {\bibinfo {author} {\bibfnamefont {C.}~\bibnamefont
  {Donati}}, \bibinfo {author} {\bibfnamefont {J.~F.}\ \bibnamefont {Douglas}},
  \bibinfo {author} {\bibfnamefont {W.}~\bibnamefont {Kob}}, \bibinfo {author}
  {\bibfnamefont {S.~J.}\ \bibnamefont {Plimpton}}, \bibinfo {author}
  {\bibfnamefont {P.~H.}\ \bibnamefont {Poole}}, \ and\ \bibinfo {author}
  {\bibfnamefont {S.~C.}\ \bibnamefont {Glotzer}},\ }\href {\doibase
  10.1103/PhysRevLett.80.2338} {\bibfield  {journal} {\bibinfo  {journal}
  {Phys. Rev. Lett.}\ }\textbf {\bibinfo {volume} {80}},\ \bibinfo {pages}
  {2338} (\bibinfo {year} {1998})}\BibitemShut {NoStop}%
\bibitem [{\citenamefont {Appignanesi}\ \emph {et~al.}(2006)\citenamefont
  {Appignanesi}, \citenamefont {Rodr\'{\i}guez~Fris}, \citenamefont {Montani},\
  and\ \citenamefont {Kob}}]{kobdemos}%
  \BibitemOpen
  \bibfield  {author} {\bibinfo {author} {\bibfnamefont {G.~A.}\ \bibnamefont
  {Appignanesi}}, \bibinfo {author} {\bibfnamefont {J.~A.}\ \bibnamefont
  {Rodr\'{\i}guez~Fris}}, \bibinfo {author} {\bibfnamefont {R.~A.}\
  \bibnamefont {Montani}}, \ and\ \bibinfo {author} {\bibfnamefont
  {W.}~\bibnamefont {Kob}},\ }\href {\doibase 10.1103/PhysRevLett.96.057801}
  {\bibfield  {journal} {\bibinfo  {journal} {Phys. Rev. Lett.}\ }\textbf
  {\bibinfo {volume} {96}},\ \bibinfo {pages} {057801} (\bibinfo {year}
  {2006})}\BibitemShut {NoStop}%
\bibitem [{\citenamefont {Kob}\ \emph {et~al.}(2012)\citenamefont {Kob},
  \citenamefont {Rold{\'a}n-Vargas},\ and\ \citenamefont
  {Berthier}}]{kob2012non}%
  \BibitemOpen
  \bibfield  {author} {\bibinfo {author} {\bibfnamefont {W.}~\bibnamefont
  {Kob}}, \bibinfo {author} {\bibfnamefont {S.}~\bibnamefont
  {Rold{\'a}n-Vargas}}, \ and\ \bibinfo {author} {\bibfnamefont
  {L.}~\bibnamefont {Berthier}},\ }\href@noop {} {\bibfield  {journal}
  {\bibinfo  {journal} {Nature Physics}\ }\textbf {\bibinfo {volume} {8}},\
  \bibinfo {pages} {164} (\bibinfo {year} {2012})}\BibitemShut {NoStop}%
\bibitem [{\citenamefont {Chaikin}\ \emph {et~al.}(1995)\citenamefont
  {Chaikin}, \citenamefont {Lubensky},\ and\ \citenamefont
  {Witten}}]{chaikin1995principles}%
  \BibitemOpen
  \bibfield  {author} {\bibinfo {author} {\bibfnamefont {P.~M.}\ \bibnamefont
  {Chaikin}}, \bibinfo {author} {\bibfnamefont {T.~C.}\ \bibnamefont
  {Lubensky}}, \ and\ \bibinfo {author} {\bibfnamefont {T.~A.}\ \bibnamefont
  {Witten}},\ }\href@noop {} {\emph {\bibinfo {title} {Principles of condensed
  matter physics}}},\ Vol.~\bibinfo {volume} {10}\ (\bibinfo  {publisher}
  {Cambridge university press Cambridge},\ \bibinfo {year} {1995})\BibitemShut
  {NoStop}%
\bibitem [{\citenamefont {Binder}\ and\ \citenamefont
  {Young}(1986)}]{binder1986spin}%
  \BibitemOpen
  \bibfield  {author} {\bibinfo {author} {\bibfnamefont {K.}~\bibnamefont
  {Binder}}\ and\ \bibinfo {author} {\bibfnamefont {A.~P.}\ \bibnamefont
  {Young}},\ }\href@noop {} {\bibfield  {journal} {\bibinfo  {journal} {Reviews
  of Modern physics}\ }\textbf {\bibinfo {volume} {58}},\ \bibinfo {pages}
  {801} (\bibinfo {year} {1986})}\BibitemShut {NoStop}%
\bibitem [{\citenamefont {Baity-Jesi}\ \emph {et~al.}(2013)\citenamefont
  {Baity-Jesi}, \citenamefont {Ba{\~n}os}, \citenamefont {Cruz}, \citenamefont
  {Fernandez}, \citenamefont {Gil-Narvion}, \citenamefont {Gordillo-Guerrero},
  \citenamefont {Iniguez}, \citenamefont {Maiorano}, \citenamefont {Mantovani},
  \citenamefont {Marinari} \emph {et~al.}}]{baity2013critical}%
  \BibitemOpen
  \bibfield  {author} {\bibinfo {author} {\bibfnamefont {M.}~\bibnamefont
  {Baity-Jesi}}, \bibinfo {author} {\bibfnamefont {R.}~\bibnamefont
  {Ba{\~n}os}}, \bibinfo {author} {\bibfnamefont {A.}~\bibnamefont {Cruz}},
  \bibinfo {author} {\bibfnamefont {L.~A.}\ \bibnamefont {Fernandez}}, \bibinfo
  {author} {\bibfnamefont {J.~M.}\ \bibnamefont {Gil-Narvion}}, \bibinfo
  {author} {\bibfnamefont {A.}~\bibnamefont {Gordillo-Guerrero}}, \bibinfo
  {author} {\bibfnamefont {D.}~\bibnamefont {Iniguez}}, \bibinfo {author}
  {\bibfnamefont {A.}~\bibnamefont {Maiorano}}, \bibinfo {author}
  {\bibfnamefont {F.}~\bibnamefont {Mantovani}}, \bibinfo {author}
  {\bibfnamefont {E.}~\bibnamefont {Marinari}},  \emph {et~al.},\ }\href@noop
  {} {\bibfield  {journal} {\bibinfo  {journal} {Physical Review B}\ }\textbf
  {\bibinfo {volume} {88}},\ \bibinfo {pages} {224416} (\bibinfo {year}
  {2013})}\BibitemShut {NoStop}%
\bibitem [{\citenamefont {Bouchaud}\ and\ \citenamefont
  {Biroli}(2005)}]{bouchaud2005nonlinear}%
  \BibitemOpen
  \bibfield  {author} {\bibinfo {author} {\bibfnamefont {J.-P.}\ \bibnamefont
  {Bouchaud}}\ and\ \bibinfo {author} {\bibfnamefont {G.}~\bibnamefont
  {Biroli}},\ }\href@noop {} {\bibfield  {journal} {\bibinfo  {journal}
  {Physical Review B}\ }\textbf {\bibinfo {volume} {72}},\ \bibinfo {pages}
  {064204} (\bibinfo {year} {2005})}\BibitemShut {NoStop}%
\bibitem [{\citenamefont {Albert}\ \emph {et~al.}(2016)\citenamefont {Albert},
  \citenamefont {Bauer}, \citenamefont {Michl}, \citenamefont {Biroli},
  \citenamefont {Bouchaud}, \citenamefont {Loidl}, \citenamefont
  {Lunkenheimer}, \citenamefont {Tourbot}, \citenamefont {Wiertel-Gasquet},\
  and\ \citenamefont {Ladieu}}]{albert_fifth-order_2016}%
  \BibitemOpen
  \bibfield  {author} {\bibinfo {author} {\bibfnamefont {S.}~\bibnamefont
  {Albert}}, \bibinfo {author} {\bibfnamefont {T.}~\bibnamefont {Bauer}},
  \bibinfo {author} {\bibfnamefont {M.}~\bibnamefont {Michl}}, \bibinfo
  {author} {\bibfnamefont {G.}~\bibnamefont {Biroli}}, \bibinfo {author}
  {\bibfnamefont {J.-P.}\ \bibnamefont {Bouchaud}}, \bibinfo {author}
  {\bibfnamefont {A.}~\bibnamefont {Loidl}}, \bibinfo {author} {\bibfnamefont
  {P.}~\bibnamefont {Lunkenheimer}}, \bibinfo {author} {\bibfnamefont
  {R.}~\bibnamefont {Tourbot}}, \bibinfo {author} {\bibfnamefont
  {C.}~\bibnamefont {Wiertel-Gasquet}}, \ and\ \bibinfo {author} {\bibfnamefont
  {F.}~\bibnamefont {Ladieu}},\ }\href@noop {} {\bibfield  {journal} {\bibinfo
  {journal} {Science}\ }\textbf {\bibinfo {volume} {352}},\ \bibinfo {pages}
  {1308} (\bibinfo {year} {2016})}\BibitemShut {NoStop}%
\bibitem [{\citenamefont {Wolynes}\ and\ \citenamefont
  {Lubchenko}(2012)}]{wolynes2012structural}%
  \BibitemOpen
  \bibfield  {author} {\bibinfo {author} {\bibfnamefont {P.~G.}\ \bibnamefont
  {Wolynes}}\ and\ \bibinfo {author} {\bibfnamefont {V.}~\bibnamefont
  {Lubchenko}},\ }\href@noop {} {\emph {\bibinfo {title} {Structural glasses
  and supercooled liquids: Theory, experiment, and applications}}}\ (\bibinfo
  {publisher} {John Wiley \& Sons},\ \bibinfo {year} {2012})\BibitemShut
  {NoStop}%
\bibitem [{\citenamefont {Crauste-Thibierge}\ \emph {et~al.}(2010)\citenamefont
  {Crauste-Thibierge}, \citenamefont {Brun}, \citenamefont {Ladieu},
  \citenamefont {L’h{\^o}te}, \citenamefont {Biroli},\ and\ \citenamefont
  {Bouchaud}}]{crauste2010evidence}%
  \BibitemOpen
  \bibfield  {author} {\bibinfo {author} {\bibfnamefont {C.}~\bibnamefont
  {Crauste-Thibierge}}, \bibinfo {author} {\bibfnamefont {C.}~\bibnamefont
  {Brun}}, \bibinfo {author} {\bibfnamefont {F.}~\bibnamefont {Ladieu}},
  \bibinfo {author} {\bibfnamefont {D.}~\bibnamefont {L’h{\^o}te}}, \bibinfo
  {author} {\bibfnamefont {G.}~\bibnamefont {Biroli}}, \ and\ \bibinfo {author}
  {\bibfnamefont {J.-P.}\ \bibnamefont {Bouchaud}},\ }\href@noop {} {\bibfield
  {journal} {\bibinfo  {journal} {Physical review letters}\ }\textbf {\bibinfo
  {volume} {104}},\ \bibinfo {pages} {165703} (\bibinfo {year}
  {2010})}\BibitemShut {NoStop}%
\bibitem [{\citenamefont {Bauer}\ \emph {et~al.}(2013)\citenamefont {Bauer},
  \citenamefont {Lunkenheimer},\ and\ \citenamefont
  {Loidl}}]{bauer2013cooperativity}%
  \BibitemOpen
  \bibfield  {author} {\bibinfo {author} {\bibfnamefont {T.}~\bibnamefont
  {Bauer}}, \bibinfo {author} {\bibfnamefont {P.}~\bibnamefont {Lunkenheimer}},
  \ and\ \bibinfo {author} {\bibfnamefont {A.}~\bibnamefont {Loidl}},\
  }\href@noop {} {\bibfield  {journal} {\bibinfo  {journal} {Physical review
  letters}\ }\textbf {\bibinfo {volume} {111}},\ \bibinfo {pages} {225702}
  (\bibinfo {year} {2013})}\BibitemShut {NoStop}%
\bibitem [{\citenamefont {Brun}\ \emph {et~al.}(2012)\citenamefont {Brun},
  \citenamefont {Ladieu}, \citenamefont {L’H{\^o}te}, \citenamefont
  {Biroli},\ and\ \citenamefont {Bouchaud}}]{brun2012evidence}%
  \BibitemOpen
  \bibfield  {author} {\bibinfo {author} {\bibfnamefont {C.}~\bibnamefont
  {Brun}}, \bibinfo {author} {\bibfnamefont {F.}~\bibnamefont {Ladieu}},
  \bibinfo {author} {\bibfnamefont {D.}~\bibnamefont {L’H{\^o}te}}, \bibinfo
  {author} {\bibfnamefont {G.}~\bibnamefont {Biroli}}, \ and\ \bibinfo {author}
  {\bibfnamefont {J.}~\bibnamefont {Bouchaud}},\ }\href@noop {} {\bibfield
  {journal} {\bibinfo  {journal} {Physical review letters}\ }\textbf {\bibinfo
  {volume} {109}},\ \bibinfo {pages} {175702} (\bibinfo {year}
  {2012})}\BibitemShut {NoStop}%
\bibitem [{\citenamefont {Seyboldt}\ \emph {et~al.}(2016)\citenamefont
  {Seyboldt}, \citenamefont {Merger}, \citenamefont {Coupette}, \citenamefont
  {Siebenb{\"u}rger}, \citenamefont {Ballauff}, \citenamefont {Wilhelm},\ and\
  \citenamefont {Fuchs}}]{seyboldt2016divergence}%
  \BibitemOpen
  \bibfield  {author} {\bibinfo {author} {\bibfnamefont {R.}~\bibnamefont
  {Seyboldt}}, \bibinfo {author} {\bibfnamefont {D.}~\bibnamefont {Merger}},
  \bibinfo {author} {\bibfnamefont {F.}~\bibnamefont {Coupette}}, \bibinfo
  {author} {\bibfnamefont {M.}~\bibnamefont {Siebenb{\"u}rger}}, \bibinfo
  {author} {\bibfnamefont {M.}~\bibnamefont {Ballauff}}, \bibinfo {author}
  {\bibfnamefont {M.}~\bibnamefont {Wilhelm}}, \ and\ \bibinfo {author}
  {\bibfnamefont {M.}~\bibnamefont {Fuchs}},\ }\href@noop {} {\bibfield
  {journal} {\bibinfo  {journal} {Soft Matter}\ }\textbf {\bibinfo {volume}
  {12}},\ \bibinfo {pages} {8825} (\bibinfo {year} {2016})}\BibitemShut
  {NoStop}%
\bibitem [{\citenamefont {Tarzia}\ \emph {et~al.}(2010)\citenamefont {Tarzia},
  \citenamefont {Biroli}, \citenamefont {Lef{\`e}vre},\ and\ \citenamefont
  {Bouchaud}}]{tarzia2010anomalous}%
  \BibitemOpen
  \bibfield  {author} {\bibinfo {author} {\bibfnamefont {M.}~\bibnamefont
  {Tarzia}}, \bibinfo {author} {\bibfnamefont {G.}~\bibnamefont {Biroli}},
  \bibinfo {author} {\bibfnamefont {A.}~\bibnamefont {Lef{\`e}vre}}, \ and\
  \bibinfo {author} {\bibfnamefont {J.-P.}\ \bibnamefont {Bouchaud}},\
  }\href@noop {} {\bibfield  {journal} {\bibinfo  {journal} {The Journal of
  chemical physics}\ }\textbf {\bibinfo {volume} {132}},\ \bibinfo {pages}
  {054501} (\bibinfo {year} {2010})}\BibitemShut {NoStop}%
\bibitem [{\citenamefont {Speck}(2019)}]{speck2019dynamic}%
  \BibitemOpen
  \bibfield  {author} {\bibinfo {author} {\bibfnamefont {T.}~\bibnamefont
  {Speck}},\ }\href@noop {} {\bibfield  {journal} {\bibinfo  {journal} {Journal
  of Statistical Mechanics: Theory and Experiment}\ }\textbf {\bibinfo {volume}
  {2019}},\ \bibinfo {pages} {084015} (\bibinfo {year} {2019})}\BibitemShut
  {NoStop}%
\bibitem [{\citenamefont {Adam}\ and\ \citenamefont {Gibbs}(1965)}]{ag}%
  \BibitemOpen
  \bibfield  {author} {\bibinfo {author} {\bibfnamefont {G.}~\bibnamefont
  {Adam}}\ and\ \bibinfo {author} {\bibfnamefont {J.~H.}\ \bibnamefont
  {Gibbs}},\ }\href@noop {} {\bibfield  {journal} {\bibinfo  {journal} {The
  journal of chemical physics}\ }\textbf {\bibinfo {volume} {43}},\ \bibinfo
  {pages} {139} (\bibinfo {year} {1965})}\BibitemShut {NoStop}%
\bibitem [{\citenamefont {M{\'e}zard}\ \emph {et~al.}(1987)\citenamefont
  {M{\'e}zard}, \citenamefont {Parisi},\ and\ \citenamefont
  {Virasoro}}]{beyond}%
  \BibitemOpen
  \bibfield  {author} {\bibinfo {author} {\bibfnamefont {M.}~\bibnamefont
  {M{\'e}zard}}, \bibinfo {author} {\bibfnamefont {G.}~\bibnamefont {Parisi}},
  \ and\ \bibinfo {author} {\bibfnamefont {M.}~\bibnamefont {Virasoro}},\
  }\href@noop {} {\emph {\bibinfo {title} {Spin glass theory and beyond: An
  Introduction to the Replica Method and Its Applications}}},\ Vol.~\bibinfo
  {volume} {9}\ (\bibinfo  {publisher} {World Scientific Publishing Company},\
  \bibinfo {year} {1987})\BibitemShut {NoStop}%
\bibitem [{\citenamefont {Kirkpatrick}\ and\ \citenamefont
  {Thirumalai}(1987)}]{KTW}%
  \BibitemOpen
  \bibfield  {author} {\bibinfo {author} {\bibfnamefont {T.~R.}\ \bibnamefont
  {Kirkpatrick}}\ and\ \bibinfo {author} {\bibfnamefont {D.}~\bibnamefont
  {Thirumalai}},\ }\href@noop {} {\bibfield  {journal} {\bibinfo  {journal}
  {Physical review letters}\ }\textbf {\bibinfo {volume} {58}},\ \bibinfo
  {pages} {2091} (\bibinfo {year} {1987})}\BibitemShut {NoStop}%
\bibitem [{\citenamefont {Kirkpatrick}\ and\ \citenamefont
  {Wolynes}(1987{\natexlab{a}})}]{KTWbis}%
  \BibitemOpen
  \bibfield  {author} {\bibinfo {author} {\bibfnamefont {T.}~\bibnamefont
  {Kirkpatrick}}\ and\ \bibinfo {author} {\bibfnamefont {P.}~\bibnamefont
  {Wolynes}},\ }\href@noop {} {\bibfield  {journal} {\bibinfo  {journal}
  {Physical Review A}\ }\textbf {\bibinfo {volume} {35}},\ \bibinfo {pages}
  {3072} (\bibinfo {year} {1987}{\natexlab{a}})}\BibitemShut {NoStop}%
\bibitem [{\citenamefont {Biroli}\ and\ \citenamefont
  {M{\'e}zard}(2001)}]{BiroliMezard}%
  \BibitemOpen
  \bibfield  {author} {\bibinfo {author} {\bibfnamefont {G.}~\bibnamefont
  {Biroli}}\ and\ \bibinfo {author} {\bibfnamefont {M.}~\bibnamefont
  {M{\'e}zard}},\ }\href@noop {} {\bibfield  {journal} {\bibinfo  {journal}
  {Physical review letters}\ }\textbf {\bibinfo {volume} {88}},\ \bibinfo
  {pages} {025501} (\bibinfo {year} {2001})}\BibitemShut {NoStop}%
\bibitem [{\citenamefont {Nelson}(2002)}]{Nelson}%
  \BibitemOpen
  \bibfield  {author} {\bibinfo {author} {\bibfnamefont {D.~R.}\ \bibnamefont
  {Nelson}},\ }\href@noop {} {\emph {\bibinfo {title} {Defects and geometry in
  condensed matter physics}}}\ (\bibinfo  {publisher} {Cambridge University
  Press},\ \bibinfo {year} {2002})\BibitemShut {NoStop}%
\bibitem [{\citenamefont {Darst}\ \emph {et~al.}(2010)\citenamefont {Darst},
  \citenamefont {Reichman},\ and\ \citenamefont {Biroli}}]{darst2010dynamical}%
  \BibitemOpen
  \bibfield  {author} {\bibinfo {author} {\bibfnamefont {R.~K.}\ \bibnamefont
  {Darst}}, \bibinfo {author} {\bibfnamefont {D.~R.}\ \bibnamefont {Reichman}},
  \ and\ \bibinfo {author} {\bibfnamefont {G.}~\bibnamefont {Biroli}},\
  }\href@noop {} {\bibfield  {journal} {\bibinfo  {journal} {The Journal of
  chemical physics}\ }\textbf {\bibinfo {volume} {132}},\ \bibinfo {pages}
  {044510} (\bibinfo {year} {2010})}\BibitemShut {NoStop}%
\bibitem [{\citenamefont {Seif}\ and\ \citenamefont
  {Grigera}(2016)}]{seif2016structure}%
  \BibitemOpen
  \bibfield  {author} {\bibinfo {author} {\bibfnamefont {A.}~\bibnamefont
  {Seif}}\ and\ \bibinfo {author} {\bibfnamefont {T.~S.}\ \bibnamefont
  {Grigera}},\ }\href@noop {} {\bibfield  {journal} {\bibinfo  {journal} {arXiv
  preprint arXiv:1611.06754}\ } (\bibinfo {year} {2016})}\BibitemShut {NoStop}%
\bibitem [{\citenamefont {Nishikawa}\ and\ \citenamefont
  {Hukushima}(2020)}]{nishikawa2020lattice}%
  \BibitemOpen
  \bibfield  {author} {\bibinfo {author} {\bibfnamefont {Y.}~\bibnamefont
  {Nishikawa}}\ and\ \bibinfo {author} {\bibfnamefont {K.}~\bibnamefont
  {Hukushima}},\ }\href {\doibase 10.1103/PhysRevLett.125.065501} {\bibfield
  {journal} {\bibinfo  {journal} {Phys. Rev. Lett.}\ }\textbf {\bibinfo
  {volume} {125}},\ \bibinfo {pages} {065501} (\bibinfo {year}
  {2020})}\BibitemShut {NoStop}%
\bibitem [{\citenamefont {McCullagh}\ \emph {et~al.}(2005)\citenamefont
  {McCullagh}, \citenamefont {Cellai}, \citenamefont {Lawlor},\ and\
  \citenamefont {Dawson}}]{mccullagh2005finite}%
  \BibitemOpen
  \bibfield  {author} {\bibinfo {author} {\bibfnamefont {G.~D.}\ \bibnamefont
  {McCullagh}}, \bibinfo {author} {\bibfnamefont {D.}~\bibnamefont {Cellai}},
  \bibinfo {author} {\bibfnamefont {A.}~\bibnamefont {Lawlor}}, \ and\ \bibinfo
  {author} {\bibfnamefont {K.~A.}\ \bibnamefont {Dawson}},\ }\href@noop {}
  {\bibfield  {journal} {\bibinfo  {journal} {Physical Review E}\ }\textbf
  {\bibinfo {volume} {71}},\ \bibinfo {pages} {030102} (\bibinfo {year}
  {2005})}\BibitemShut {NoStop}%
\bibitem [{Note3()}]{Note3}%
  \BibitemOpen
  \bibinfo {note} {In order to have a well-defined thermodynamics, Bethe
  lattices are generated as random graphs with fixed connectivity, also called
  random regular graphs.}\BibitemShut {Stop}%
\bibitem [{\citenamefont {Rivoire}\ \emph {et~al.}(2004)\citenamefont
  {Rivoire}, \citenamefont {Biroli}, \citenamefont {Martin},\ and\
  \citenamefont {M{\'e}zard}}]{bethe}%
  \BibitemOpen
  \bibfield  {author} {\bibinfo {author} {\bibfnamefont {O.}~\bibnamefont
  {Rivoire}}, \bibinfo {author} {\bibfnamefont {G.}~\bibnamefont {Biroli}},
  \bibinfo {author} {\bibfnamefont {O.~C.}\ \bibnamefont {Martin}}, \ and\
  \bibinfo {author} {\bibfnamefont {M.}~\bibnamefont {M{\'e}zard}},\
  }\href@noop {} {\bibfield  {journal} {\bibinfo  {journal} {The European
  Physical Journal B-Condensed Matter and Complex Systems}\ }\textbf {\bibinfo
  {volume} {37}},\ \bibinfo {pages} {55} (\bibinfo {year} {2004})}\BibitemShut
  {NoStop}%
\bibitem [{\citenamefont {Gross}\ and\ \citenamefont
  {M{\'e}zard}(1984)}]{MezardGross}%
  \BibitemOpen
  \bibfield  {author} {\bibinfo {author} {\bibfnamefont {D.~J.}\ \bibnamefont
  {Gross}}\ and\ \bibinfo {author} {\bibfnamefont {M.}~\bibnamefont
  {M{\'e}zard}},\ }\href@noop {} {\bibfield  {journal} {\bibinfo  {journal}
  {Nuclear Physics B}\ }\textbf {\bibinfo {volume} {240}},\ \bibinfo {pages}
  {431} (\bibinfo {year} {1984})}\BibitemShut {NoStop}%
\bibitem [{\citenamefont {Castellani}\ and\ \citenamefont
  {Cavagna}(2005)}]{hessian}%
  \BibitemOpen
  \bibfield  {author} {\bibinfo {author} {\bibfnamefont {T.}~\bibnamefont
  {Castellani}}\ and\ \bibinfo {author} {\bibfnamefont {A.}~\bibnamefont
  {Cavagna}},\ }\href@noop {} {\bibfield  {journal} {\bibinfo  {journal}
  {Journal of Statistical Mechanics: Theory and Experiment}\ }\textbf {\bibinfo
  {volume} {2005}},\ \bibinfo {pages} {P05012} (\bibinfo {year}
  {2005})}\BibitemShut {NoStop}%
\bibitem [{Note4()}]{Note4}%
  \BibitemOpen
  \bibinfo {note} {If additional symmetries are broken then one can have
  ergodicity breaking also in the RS phase.}\BibitemShut {Stop}%
\bibitem [{\citenamefont {Cavagna}(2009)}]{supercavagna}%
  \BibitemOpen
  \bibfield  {author} {\bibinfo {author} {\bibfnamefont {A.}~\bibnamefont
  {Cavagna}},\ }\href@noop {} {\bibfield  {journal} {\bibinfo  {journal}
  {Physics Reports}\ }\textbf {\bibinfo {volume} {476}},\ \bibinfo {pages} {51}
  (\bibinfo {year} {2009})}\BibitemShut {NoStop}%
\bibitem [{\citenamefont {Kurchan}\ \emph {et~al.}(2012)\citenamefont
  {Kurchan}, \citenamefont {Parisi},\ and\ \citenamefont
  {Zamponi}}]{KurParZam}%
  \BibitemOpen
  \bibfield  {author} {\bibinfo {author} {\bibfnamefont {J.}~\bibnamefont
  {Kurchan}}, \bibinfo {author} {\bibfnamefont {G.}~\bibnamefont {Parisi}}, \
  and\ \bibinfo {author} {\bibfnamefont {F.}~\bibnamefont {Zamponi}},\
  }\href@noop {} {\bibfield  {journal} {\bibinfo  {journal} {Journal of
  Statistical Mechanics: Theory and Experiment}\ }\textbf {\bibinfo {volume}
  {2012}},\ \bibinfo {pages} {P10012} (\bibinfo {year} {2012})}\BibitemShut
  {NoStop}%
\bibitem [{\citenamefont {Charbonneau}\ \emph
  {et~al.}(2014{\natexlab{a}})\citenamefont {Charbonneau}, \citenamefont
  {Kurchan}, \citenamefont {Parisi}, \citenamefont {Urbani},\ and\
  \citenamefont {Zamponi}}]{charbonneau_fractal_2014}%
  \BibitemOpen
  \bibfield  {author} {\bibinfo {author} {\bibfnamefont {P.}~\bibnamefont
  {Charbonneau}}, \bibinfo {author} {\bibfnamefont {J.}~\bibnamefont
  {Kurchan}}, \bibinfo {author} {\bibfnamefont {G.}~\bibnamefont {Parisi}},
  \bibinfo {author} {\bibfnamefont {P.}~\bibnamefont {Urbani}}, \ and\ \bibinfo
  {author} {\bibfnamefont {F.}~\bibnamefont {Zamponi}},\ }\href@noop {}
  {\bibfield  {journal} {\bibinfo  {journal} {Nature communications}\ }\textbf
  {\bibinfo {volume} {5}},\ \bibinfo {pages} {1} (\bibinfo {year}
  {2014}{\natexlab{a}})}\BibitemShut {NoStop}%
\bibitem [{\citenamefont {Charbonneau}\ \emph
  {et~al.}(2014{\natexlab{b}})\citenamefont {Charbonneau}, \citenamefont
  {Kurchan}, \citenamefont {Parisi}, \citenamefont {Urbani},\ and\
  \citenamefont {Zamponi}}]{charbonneau_exact_2014}%
  \BibitemOpen
  \bibfield  {author} {\bibinfo {author} {\bibfnamefont {P.}~\bibnamefont
  {Charbonneau}}, \bibinfo {author} {\bibfnamefont {J.}~\bibnamefont
  {Kurchan}}, \bibinfo {author} {\bibfnamefont {G.}~\bibnamefont {Parisi}},
  \bibinfo {author} {\bibfnamefont {P.}~\bibnamefont {Urbani}}, \ and\ \bibinfo
  {author} {\bibfnamefont {F.}~\bibnamefont {Zamponi}},\ }\href@noop {}
  {\bibfield  {journal} {\bibinfo  {journal} {Journal of Statistical Mechanics:
  Theory and Experiment}\ }\textbf {\bibinfo {volume} {2014}},\ \bibinfo
  {pages} {P10009} (\bibinfo {year} {2014}{\natexlab{b}})}\BibitemShut
  {NoStop}%
\bibitem [{\citenamefont {Kurchan}\ \emph {et~al.}(2013)\citenamefont
  {Kurchan}, \citenamefont {Parisi}, \citenamefont {Urbani},\ and\
  \citenamefont {Zamponi}}]{KurParUrbZam}%
  \BibitemOpen
  \bibfield  {author} {\bibinfo {author} {\bibfnamefont {J.}~\bibnamefont
  {Kurchan}}, \bibinfo {author} {\bibfnamefont {G.}~\bibnamefont {Parisi}},
  \bibinfo {author} {\bibfnamefont {P.}~\bibnamefont {Urbani}}, \ and\ \bibinfo
  {author} {\bibfnamefont {F.}~\bibnamefont {Zamponi}},\ }\href@noop {}
  {\bibfield  {journal} {\bibinfo  {journal} {The Journal of Physical Chemistry
  B}\ }\textbf {\bibinfo {volume} {117}},\ \bibinfo {pages} {12979} (\bibinfo
  {year} {2013})}\BibitemShut {NoStop}%
\bibitem [{\citenamefont {Parisi}\ \emph {et~al.}(2020)\citenamefont {Parisi},
  \citenamefont {Urbani},\ and\ \citenamefont {Zamponi}}]{parisi2020theory}%
  \BibitemOpen
  \bibfield  {author} {\bibinfo {author} {\bibfnamefont {G.}~\bibnamefont
  {Parisi}}, \bibinfo {author} {\bibfnamefont {P.}~\bibnamefont {Urbani}}, \
  and\ \bibinfo {author} {\bibfnamefont {F.}~\bibnamefont {Zamponi}},\
  }\href@noop {} {\emph {\bibinfo {title} {Theory of Simple Glasses: Exact
  Solutions in Infinite Dimensions}}}\ (\bibinfo  {publisher} {Cambridge
  University Press},\ \bibinfo {year} {2020})\BibitemShut {NoStop}%
\bibitem [{Note5()}]{Note5}%
  \BibitemOpen
  \bibinfo {note} {For large $d$ the crystalline phase does not intervene. In
  fact, the amorphous and crystalline solid phases are well separated in
  configuration space and issues related to finite dimensions, such as the
  crystallization of monodisperse particles, are suppressed \cite {Skoge,
  vanmeelhardsphere2009}.}\BibitemShut {Stop}%
\bibitem [{Note6()}]{Note6}%
  \BibitemOpen
  \bibinfo {note} {There is of course no crystal state in disordered systems
  such as in Eq.~(\ref {pspin}). In the case of lattice glass models, there is
  a crystal phase but it can disappear depending whether the Bethe lattice is a
  Cayley tree or a random regular graph.}\BibitemShut {Stop}%
\bibitem [{\citenamefont {Monasson}(1995)}]{monasson}%
  \BibitemOpen
  \bibfield  {author} {\bibinfo {author} {\bibfnamefont {R.}~\bibnamefont
  {Monasson}},\ }\href@noop {} {\bibfield  {journal} {\bibinfo  {journal}
  {Physical review letters}\ }\textbf {\bibinfo {volume} {75}},\ \bibinfo
  {pages} {2847} (\bibinfo {year} {1995})}\BibitemShut {NoStop}%
\bibitem [{\citenamefont {Barrat}\ \emph {et~al.}(2004)\citenamefont {Barrat},
  \citenamefont {Feigelman}, \citenamefont {Kurchan} \emph
  {et~al.}}]{barrat2004slow}%
  \BibitemOpen
  \bibfield  {author} {\bibinfo {author} {\bibfnamefont {J.-L.}\ \bibnamefont
  {Barrat}}, \bibinfo {author} {\bibfnamefont {M.}~\bibnamefont {Feigelman}},
  \bibinfo {author} {\bibfnamefont {J.}~\bibnamefont {Kurchan}},  \emph
  {et~al.},\ }in\ \href@noop {} {\emph {\bibinfo {booktitle} {Slow Relaxations
  and Nonequilibrium Dynamics in Condensed Matter}}}\ (\bibinfo {year}
  {2004})\BibitemShut {NoStop}%
\bibitem [{\citenamefont {Maimbourg}\ \emph {et~al.}(2016)\citenamefont
  {Maimbourg}, \citenamefont {Kurchan},\ and\ \citenamefont
  {Zamponi}}]{MaimbourgKurchanZamponi}%
  \BibitemOpen
  \bibfield  {author} {\bibinfo {author} {\bibfnamefont {T.}~\bibnamefont
  {Maimbourg}}, \bibinfo {author} {\bibfnamefont {J.}~\bibnamefont {Kurchan}},
  \ and\ \bibinfo {author} {\bibfnamefont {F.}~\bibnamefont {Zamponi}},\
  }\href@noop {} {\bibfield  {journal} {\bibinfo  {journal} {Physical review
  letters}\ }\textbf {\bibinfo {volume} {116}},\ \bibinfo {pages} {015902}
  (\bibinfo {year} {2016})}\BibitemShut {NoStop}%
\bibitem [{\citenamefont {Leutheusser}(1984)}]{Leuthesser}%
  \BibitemOpen
  \bibfield  {author} {\bibinfo {author} {\bibfnamefont {E.}~\bibnamefont
  {Leutheusser}},\ }\href@noop {} {\bibfield  {journal} {\bibinfo  {journal}
  {Physical Review A}\ }\textbf {\bibinfo {volume} {29}},\ \bibinfo {pages}
  {2765} (\bibinfo {year} {1984})}\BibitemShut {NoStop}%
\bibitem [{\citenamefont {Bengtzelius}\ \emph {et~al.}(1984)\citenamefont
  {Bengtzelius}, \citenamefont {Gotze},\ and\ \citenamefont {Sjolander}}]{BGS}%
  \BibitemOpen
  \bibfield  {author} {\bibinfo {author} {\bibfnamefont {U.}~\bibnamefont
  {Bengtzelius}}, \bibinfo {author} {\bibfnamefont {W.}~\bibnamefont {Gotze}},
  \ and\ \bibinfo {author} {\bibfnamefont {A.}~\bibnamefont {Sjolander}},\
  }\href@noop {} {\bibfield  {journal} {\bibinfo  {journal} {Journal of Physics
  C: solid state Physics}\ }\textbf {\bibinfo {volume} {17}},\ \bibinfo {pages}
  {5915} (\bibinfo {year} {1984})}\BibitemShut {NoStop}%
\bibitem [{\citenamefont {Das}\ and\ \citenamefont {Mazenko}(1986)}]{DM}%
  \BibitemOpen
  \bibfield  {author} {\bibinfo {author} {\bibfnamefont {S.~P.}\ \bibnamefont
  {Das}}\ and\ \bibinfo {author} {\bibfnamefont {G.~F.}\ \bibnamefont
  {Mazenko}},\ }\href@noop {} {\bibfield  {journal} {\bibinfo  {journal}
  {Physical Review A}\ }\textbf {\bibinfo {volume} {34}},\ \bibinfo {pages}
  {2265} (\bibinfo {year} {1986})}\BibitemShut {NoStop}%
\bibitem [{\citenamefont {Biroli}\ and\ \citenamefont {Bouchaud}(2004)}]{BB}%
  \BibitemOpen
  \bibfield  {author} {\bibinfo {author} {\bibfnamefont {G.}~\bibnamefont
  {Biroli}}\ and\ \bibinfo {author} {\bibfnamefont {J.-P.}\ \bibnamefont
  {Bouchaud}},\ }\href@noop {} {\bibfield  {journal} {\bibinfo  {journal} {EPL
  (Europhysics Letters)}\ }\textbf {\bibinfo {volume} {67}},\ \bibinfo {pages}
  {21} (\bibinfo {year} {2004})}\BibitemShut {NoStop}%
\bibitem [{\citenamefont {Biroli}\ \emph {et~al.}(2006)\citenamefont {Biroli},
  \citenamefont {Bouchaud}, \citenamefont {Miyazaki},\ and\ \citenamefont
  {Reichman}}]{BBMR}%
  \BibitemOpen
  \bibfield  {author} {\bibinfo {author} {\bibfnamefont {G.}~\bibnamefont
  {Biroli}}, \bibinfo {author} {\bibfnamefont {J.-P.}\ \bibnamefont
  {Bouchaud}}, \bibinfo {author} {\bibfnamefont {K.}~\bibnamefont {Miyazaki}},
  \ and\ \bibinfo {author} {\bibfnamefont {D.~R.}\ \bibnamefont {Reichman}},\
  }\href@noop {} {\bibfield  {journal} {\bibinfo  {journal} {Physical review
  letters}\ }\textbf {\bibinfo {volume} {97}},\ \bibinfo {pages} {195701}
  (\bibinfo {year} {2006})}\BibitemShut {NoStop}%
\bibitem [{\citenamefont {Berthier}\ \emph {et~al.}(2000)\citenamefont
  {Berthier}, \citenamefont {Barrat},\ and\ \citenamefont {Kurchan}}]{BBK}%
  \BibitemOpen
  \bibfield  {author} {\bibinfo {author} {\bibfnamefont {L.}~\bibnamefont
  {Berthier}}, \bibinfo {author} {\bibfnamefont {J.-L.}\ \bibnamefont
  {Barrat}}, \ and\ \bibinfo {author} {\bibfnamefont {J.}~\bibnamefont
  {Kurchan}},\ }\href@noop {} {\bibfield  {journal} {\bibinfo  {journal}
  {Physical Review E}\ }\textbf {\bibinfo {volume} {61}},\ \bibinfo {pages}
  {5464} (\bibinfo {year} {2000})}\BibitemShut {NoStop}%
\bibitem [{\citenamefont {Miyazaki}\ and\ \citenamefont
  {Reichman}(2002)}]{dave}%
  \BibitemOpen
  \bibfield  {author} {\bibinfo {author} {\bibfnamefont {K.}~\bibnamefont
  {Miyazaki}}\ and\ \bibinfo {author} {\bibfnamefont {D.~R.}\ \bibnamefont
  {Reichman}},\ }\href@noop {} {\bibfield  {journal} {\bibinfo  {journal}
  {Physical Review E}\ }\textbf {\bibinfo {volume} {66}},\ \bibinfo {pages}
  {050501} (\bibinfo {year} {2002})}\BibitemShut {NoStop}%
\bibitem [{\citenamefont {Fuchs}\ and\ \citenamefont {Cates}(2002)}]{fuchs}%
  \BibitemOpen
  \bibfield  {author} {\bibinfo {author} {\bibfnamefont {M.}~\bibnamefont
  {Fuchs}}\ and\ \bibinfo {author} {\bibfnamefont {M.~E.}\ \bibnamefont
  {Cates}},\ }\href@noop {} {\bibfield  {journal} {\bibinfo  {journal}
  {Physical review letters}\ }\textbf {\bibinfo {volume} {89}},\ \bibinfo
  {pages} {248304} (\bibinfo {year} {2002})}\BibitemShut {NoStop}%
\bibitem [{\citenamefont {Kirkpatrick}\ \emph {et~al.}(1989)\citenamefont
  {Kirkpatrick}, \citenamefont {Thirumalai},\ and\ \citenamefont
  {Wolynes}}]{KTW2}%
  \BibitemOpen
  \bibfield  {author} {\bibinfo {author} {\bibfnamefont {T.~R.}\ \bibnamefont
  {Kirkpatrick}}, \bibinfo {author} {\bibfnamefont {D.}~\bibnamefont
  {Thirumalai}}, \ and\ \bibinfo {author} {\bibfnamefont {P.~G.}\ \bibnamefont
  {Wolynes}},\ }\href@noop {} {\bibfield  {journal} {\bibinfo  {journal}
  {Physical Review A}\ }\textbf {\bibinfo {volume} {40}},\ \bibinfo {pages}
  {1045} (\bibinfo {year} {1989})}\BibitemShut {NoStop}%
\bibitem [{\citenamefont {Bouchaud}\ and\ \citenamefont {Biroli}(2004)}]{BB2}%
  \BibitemOpen
  \bibfield  {author} {\bibinfo {author} {\bibfnamefont {J.-P.}\ \bibnamefont
  {Bouchaud}}\ and\ \bibinfo {author} {\bibfnamefont {G.}~\bibnamefont
  {Biroli}},\ }\href@noop {} {\bibfield  {journal} {\bibinfo  {journal} {The
  Journal of chemical physics}\ }\textbf {\bibinfo {volume} {121}},\ \bibinfo
  {pages} {7347} (\bibinfo {year} {2004})}\BibitemShut {NoStop}%
\bibitem [{\citenamefont {Dzero}\ \emph {et~al.}(2005)\citenamefont {Dzero},
  \citenamefont {Schmalian},\ and\ \citenamefont {Wolynes}}]{Schmalian}%
  \BibitemOpen
  \bibfield  {author} {\bibinfo {author} {\bibfnamefont {M.}~\bibnamefont
  {Dzero}}, \bibinfo {author} {\bibfnamefont {J.}~\bibnamefont {Schmalian}}, \
  and\ \bibinfo {author} {\bibfnamefont {P.~G.}\ \bibnamefont {Wolynes}},\
  }\href@noop {} {\bibfield  {journal} {\bibinfo  {journal} {Physical Review
  B}\ }\textbf {\bibinfo {volume} {72}},\ \bibinfo {pages} {100201} (\bibinfo
  {year} {2005})}\BibitemShut {NoStop}%
\bibitem [{\citenamefont {Franz}(2006)}]{Franz}%
  \BibitemOpen
  \bibfield  {author} {\bibinfo {author} {\bibfnamefont {S.}~\bibnamefont
  {Franz}},\ }\href@noop {} {\bibfield  {journal} {\bibinfo  {journal} {EPL
  (Europhysics Letters)}\ }\textbf {\bibinfo {volume} {73}},\ \bibinfo {pages}
  {492} (\bibinfo {year} {2006})}\BibitemShut {NoStop}%
\bibitem [{\citenamefont {Biroli}\ and\ \citenamefont
  {Cammarota}(2017)}]{biroli2017fluctuations}%
  \BibitemOpen
  \bibfield  {author} {\bibinfo {author} {\bibfnamefont {G.}~\bibnamefont
  {Biroli}}\ and\ \bibinfo {author} {\bibfnamefont {C.}~\bibnamefont
  {Cammarota}},\ }\href@noop {} {\bibfield  {journal} {\bibinfo  {journal}
  {Physical Review X}\ }\textbf {\bibinfo {volume} {7}},\ \bibinfo {pages}
  {011011} (\bibinfo {year} {2017})}\BibitemShut {NoStop}%
\bibitem [{\citenamefont {Angell}(1997)}]{Angell}%
  \BibitemOpen
  \bibfield  {author} {\bibinfo {author} {\bibfnamefont {C.~A.}\ \bibnamefont
  {Angell}},\ }\href@noop {} {\bibfield  {journal} {\bibinfo  {journal}
  {Journal of research of the National Institute of Standards and Technology}\
  }\textbf {\bibinfo {volume} {102}},\ \bibinfo {pages} {171} (\bibinfo {year}
  {1997})}\BibitemShut {NoStop}%
\bibitem [{\citenamefont {Hodge}(1997)}]{Hodge}%
  \BibitemOpen
  \bibfield  {author} {\bibinfo {author} {\bibfnamefont {I.~M.}\ \bibnamefont
  {Hodge}},\ }\href@noop {} {\bibfield  {journal} {\bibinfo  {journal} {Journal
  of research of the National Institute of Standards and Technology}\ }\textbf
  {\bibinfo {volume} {102}},\ \bibinfo {pages} {195} (\bibinfo {year}
  {1997})}\BibitemShut {NoStop}%
\bibitem [{\citenamefont {Johari}(2000)}]{Johari}%
  \BibitemOpen
  \bibfield  {author} {\bibinfo {author} {\bibfnamefont {G.}~\bibnamefont
  {Johari}},\ }\href@noop {} {\bibfield  {journal} {\bibinfo  {journal} {The
  Journal of Chemical Physics}\ }\textbf {\bibinfo {volume} {112}},\ \bibinfo
  {pages} {7518} (\bibinfo {year} {2000})}\BibitemShut {NoStop}%
\bibitem [{\citenamefont {Ozawa}\ \emph {et~al.}(2019)\citenamefont {Ozawa},
  \citenamefont {Scalliet}, \citenamefont {Ninarello},\ and\ \citenamefont
  {Berthier}}]{ozawa2019does}%
  \BibitemOpen
  \bibfield  {author} {\bibinfo {author} {\bibfnamefont {M.}~\bibnamefont
  {Ozawa}}, \bibinfo {author} {\bibfnamefont {C.}~\bibnamefont {Scalliet}},
  \bibinfo {author} {\bibfnamefont {A.}~\bibnamefont {Ninarello}}, \ and\
  \bibinfo {author} {\bibfnamefont {L.}~\bibnamefont {Berthier}},\ }\href@noop
  {} {\bibfield  {journal} {\bibinfo  {journal} {The Journal of chemical
  physics}\ }\textbf {\bibinfo {volume} {151}},\ \bibinfo {pages} {084504}
  (\bibinfo {year} {2019})}\BibitemShut {NoStop}%
\bibitem [{\citenamefont {M\'ezard}\ and\ \citenamefont
  {Parisi}(1999)}]{MezardParisi}%
  \BibitemOpen
  \bibfield  {author} {\bibinfo {author} {\bibfnamefont {M.}~\bibnamefont
  {M\'ezard}}\ and\ \bibinfo {author} {\bibfnamefont {G.}~\bibnamefont
  {Parisi}},\ }\href {\doibase 10.1103/PhysRevLett.82.747} {\bibfield
  {journal} {\bibinfo  {journal} {Phys. Rev. Lett.}\ }\textbf {\bibinfo
  {volume} {82}},\ \bibinfo {pages} {747} (\bibinfo {year} {1999})}\BibitemShut
  {NoStop}%
\bibitem [{\citenamefont {Ozawa}\ and\ \citenamefont
  {Berthier}(2017)}]{ozawa_does_2017}%
  \BibitemOpen
  \bibfield  {author} {\bibinfo {author} {\bibfnamefont {M.}~\bibnamefont
  {Ozawa}}\ and\ \bibinfo {author} {\bibfnamefont {L.}~\bibnamefont
  {Berthier}},\ }\href {\doibase 10.1063/1.4972525} {\bibfield  {journal}
  {\bibinfo  {journal} {The Journal of Chemical Physics}\ }\textbf {\bibinfo
  {volume} {146}},\ \bibinfo {pages} {014502} (\bibinfo {year}
  {2017})}\BibitemShut {NoStop}%
\bibitem [{\citenamefont {Franz}\ \emph {et~al.}(2011)\citenamefont {Franz},
  \citenamefont {Parisi}, \citenamefont {Ricci-Tersenghi},\ and\ \citenamefont
  {Rizzo}}]{franz2011field}%
  \BibitemOpen
  \bibfield  {author} {\bibinfo {author} {\bibfnamefont {S.}~\bibnamefont
  {Franz}}, \bibinfo {author} {\bibfnamefont {G.}~\bibnamefont {Parisi}},
  \bibinfo {author} {\bibfnamefont {F.}~\bibnamefont {Ricci-Tersenghi}}, \ and\
  \bibinfo {author} {\bibfnamefont {T.}~\bibnamefont {Rizzo}},\ }\href@noop {}
  {\bibfield  {journal} {\bibinfo  {journal} {The European Physical Journal E}\
  }\textbf {\bibinfo {volume} {34}},\ \bibinfo {pages} {1} (\bibinfo {year}
  {2011})}\BibitemShut {NoStop}%
\bibitem [{Note7()}]{Note7}%
  \BibitemOpen
  \bibinfo {note} {This terminology was suggested to us by Jean-Philippe
  Bouchaud.}\BibitemShut {Stop}%
\bibitem [{\citenamefont {Stevenson}\ \emph {et~al.}(2008)\citenamefont
  {Stevenson}, \citenamefont {Walczak}, \citenamefont {Hall},\ and\
  \citenamefont {Wolynes}}]{stevenson2008constructing}%
  \BibitemOpen
  \bibfield  {author} {\bibinfo {author} {\bibfnamefont {J.~D.}\ \bibnamefont
  {Stevenson}}, \bibinfo {author} {\bibfnamefont {A.~M.}\ \bibnamefont
  {Walczak}}, \bibinfo {author} {\bibfnamefont {R.~W.}\ \bibnamefont {Hall}}, \
  and\ \bibinfo {author} {\bibfnamefont {P.~G.}\ \bibnamefont {Wolynes}},\
  }\href@noop {} {\bibfield  {journal} {\bibinfo  {journal} {The Journal of
  chemical physics}\ }\textbf {\bibinfo {volume} {129}},\ \bibinfo {pages}
  {194505} (\bibinfo {year} {2008})}\BibitemShut {NoStop}%
\bibitem [{\citenamefont {Biroli}\ \emph
  {et~al.}(2018{\natexlab{a}})\citenamefont {Biroli}, \citenamefont
  {Cammarota}, \citenamefont {Tarjus},\ and\ \citenamefont
  {Tarzia}}]{biroli2018randoma}%
  \BibitemOpen
  \bibfield  {author} {\bibinfo {author} {\bibfnamefont {G.}~\bibnamefont
  {Biroli}}, \bibinfo {author} {\bibfnamefont {C.}~\bibnamefont {Cammarota}},
  \bibinfo {author} {\bibfnamefont {G.}~\bibnamefont {Tarjus}}, \ and\ \bibinfo
  {author} {\bibfnamefont {M.}~\bibnamefont {Tarzia}},\ }\href@noop {}
  {\bibfield  {journal} {\bibinfo  {journal} {Physical Review B}\ }\textbf
  {\bibinfo {volume} {98}},\ \bibinfo {pages} {174205} (\bibinfo {year}
  {2018}{\natexlab{a}})}\BibitemShut {NoStop}%
\bibitem [{\citenamefont {Biroli}\ \emph
  {et~al.}(2018{\natexlab{b}})\citenamefont {Biroli}, \citenamefont
  {Cammarota}, \citenamefont {Tarjus},\ and\ \citenamefont
  {Tarzia}}]{biroli2018randomb}%
  \BibitemOpen
  \bibfield  {author} {\bibinfo {author} {\bibfnamefont {G.}~\bibnamefont
  {Biroli}}, \bibinfo {author} {\bibfnamefont {C.}~\bibnamefont {Cammarota}},
  \bibinfo {author} {\bibfnamefont {G.}~\bibnamefont {Tarjus}}, \ and\ \bibinfo
  {author} {\bibfnamefont {M.}~\bibnamefont {Tarzia}},\ }\href@noop {}
  {\bibfield  {journal} {\bibinfo  {journal} {Physical Review B}\ }\textbf
  {\bibinfo {volume} {98}},\ \bibinfo {pages} {174206} (\bibinfo {year}
  {2018}{\natexlab{b}})}\BibitemShut {NoStop}%
\bibitem [{\citenamefont {Kirkpatrick}\ and\ \citenamefont
  {Wolynes}(1987{\natexlab{b}})}]{kirkpatrick1987stable}%
  \BibitemOpen
  \bibfield  {author} {\bibinfo {author} {\bibfnamefont {T.}~\bibnamefont
  {Kirkpatrick}}\ and\ \bibinfo {author} {\bibfnamefont {P.}~\bibnamefont
  {Wolynes}},\ }\href@noop {} {\bibfield  {journal} {\bibinfo  {journal}
  {Physical Review B}\ }\textbf {\bibinfo {volume} {36}},\ \bibinfo {pages}
  {8552} (\bibinfo {year} {1987}{\natexlab{b}})}\BibitemShut {NoStop}%
\bibitem [{\citenamefont {Rizzo}(2016)}]{rizzo2016dynamical}%
  \BibitemOpen
  \bibfield  {author} {\bibinfo {author} {\bibfnamefont {T.}~\bibnamefont
  {Rizzo}},\ }\href@noop {} {\bibfield  {journal} {\bibinfo  {journal}
  {Physical Review B}\ }\textbf {\bibinfo {volume} {94}},\ \bibinfo {pages}
  {014202} (\bibinfo {year} {2016})}\BibitemShut {NoStop}%
\bibitem [{\citenamefont {Nandi}\ \emph {et~al.}(2016)\citenamefont {Nandi},
  \citenamefont {Biroli},\ and\ \citenamefont {Tarjus}}]{nandi2016spinodals}%
  \BibitemOpen
  \bibfield  {author} {\bibinfo {author} {\bibfnamefont {S.~K.}\ \bibnamefont
  {Nandi}}, \bibinfo {author} {\bibfnamefont {G.}~\bibnamefont {Biroli}}, \
  and\ \bibinfo {author} {\bibfnamefont {G.}~\bibnamefont {Tarjus}},\
  }\href@noop {} {\bibfield  {journal} {\bibinfo  {journal} {Physical review
  letters}\ }\textbf {\bibinfo {volume} {116}},\ \bibinfo {pages} {145701}
  (\bibinfo {year} {2016})}\BibitemShut {NoStop}%
\bibitem [{\citenamefont {Biroli}\ and\ \citenamefont
  {Bouchaud}(2012)}]{biroli2012random}%
  \BibitemOpen
  \bibfield  {author} {\bibinfo {author} {\bibfnamefont {G.}~\bibnamefont
  {Biroli}}\ and\ \bibinfo {author} {\bibfnamefont {J.-P.}\ \bibnamefont
  {Bouchaud}},\ }\href@noop {} {\bibfield  {journal} {\bibinfo  {journal}
  {Structural Glasses and Supercooled Liquids: Theory, Experiment, and
  Applications}\ ,\ \bibinfo {pages} {31}} (\bibinfo {year}
  {2012})}\BibitemShut {NoStop}%
\bibitem [{\citenamefont {Berthier}\ \emph {et~al.}(2020)\citenamefont
  {Berthier}, \citenamefont {Charbonneau},\ and\ \citenamefont
  {Kundu}}]{berthier2019finite}%
  \BibitemOpen
  \bibfield  {author} {\bibinfo {author} {\bibfnamefont {L.}~\bibnamefont
  {Berthier}}, \bibinfo {author} {\bibfnamefont {P.}~\bibnamefont
  {Charbonneau}}, \ and\ \bibinfo {author} {\bibfnamefont {J.}~\bibnamefont
  {Kundu}},\ }\href@noop {} {\bibfield  {journal} {\bibinfo  {journal}
  {Physical Review Letters}\ }\textbf {\bibinfo {volume} {125}},\ \bibinfo
  {pages} {108001} (\bibinfo {year} {2020})}\BibitemShut {NoStop}%
\bibitem [{\citenamefont {Berthier}\ \emph
  {et~al.}(2019{\natexlab{c}})\citenamefont {Berthier}, \citenamefont {Biroli},
  \citenamefont {Bouchaud},\ and\ \citenamefont {Tarjus}}]{berthier_can_2019}%
  \BibitemOpen
  \bibfield  {author} {\bibinfo {author} {\bibfnamefont {L.}~\bibnamefont
  {Berthier}}, \bibinfo {author} {\bibfnamefont {G.}~\bibnamefont {Biroli}},
  \bibinfo {author} {\bibfnamefont {J.-P.}\ \bibnamefont {Bouchaud}}, \ and\
  \bibinfo {author} {\bibfnamefont {G.}~\bibnamefont {Tarjus}},\ }\href
  {\doibase 10.1063/1.5086509} {\bibfield  {journal} {\bibinfo  {journal} {The
  Journal of Chemical Physics}\ }\textbf {\bibinfo {volume} {150}},\ \bibinfo
  {pages} {094501} (\bibinfo {year} {2019}{\natexlab{c}})}\BibitemShut
  {NoStop}%
\bibitem [{Note8()}]{Note8}%
  \BibitemOpen
  \bibinfo {note} {The difficulty is that the mapping to the RFIM proceeds by
  relating the overlap (for glasses) to the magnetization (for the RFIM). There
  are no natural dynamical equations for the overlap, and in the only cases
  where those have been established--the $\beta $-regime of MCT--these proved
  to be quite complex and different from the corresponding equations for the
  magnetization of the RFIM.}\BibitemShut {Stop}%
\bibitem [{\citenamefont {Castellana}\ \emph {et~al.}(2010)\citenamefont
  {Castellana}, \citenamefont {Decelle}, \citenamefont {Franz}, \citenamefont
  {M{\'e}zard},\ and\ \citenamefont {Parisi}}]{castellana2010hierarchical}%
  \BibitemOpen
  \bibfield  {author} {\bibinfo {author} {\bibfnamefont {M.}~\bibnamefont
  {Castellana}}, \bibinfo {author} {\bibfnamefont {A.}~\bibnamefont {Decelle}},
  \bibinfo {author} {\bibfnamefont {S.}~\bibnamefont {Franz}}, \bibinfo
  {author} {\bibfnamefont {M.}~\bibnamefont {M{\'e}zard}}, \ and\ \bibinfo
  {author} {\bibfnamefont {G.}~\bibnamefont {Parisi}},\ }\href@noop {}
  {\bibfield  {journal} {\bibinfo  {journal} {Physical review letters}\
  }\textbf {\bibinfo {volume} {104}},\ \bibinfo {pages} {127206} (\bibinfo
  {year} {2010})}\BibitemShut {NoStop}%
\bibitem [{\citenamefont {Yeo}\ and\ \citenamefont
  {Moore}(2012)}]{yeo2012origin}%
  \BibitemOpen
  \bibfield  {author} {\bibinfo {author} {\bibfnamefont {J.}~\bibnamefont
  {Yeo}}\ and\ \bibinfo {author} {\bibfnamefont {M.}~\bibnamefont {Moore}},\
  }\href@noop {} {\bibfield  {journal} {\bibinfo  {journal} {Physical Review
  E}\ }\textbf {\bibinfo {volume} {86}},\ \bibinfo {pages} {052501} (\bibinfo
  {year} {2012})}\BibitemShut {NoStop}%
\bibitem [{\citenamefont {Cammarota}\ \emph {et~al.}(2011)\citenamefont
  {Cammarota}, \citenamefont {Biroli}, \citenamefont {Tarzia},\ and\
  \citenamefont {Tarjus}}]{cammarota2011renormalization}%
  \BibitemOpen
  \bibfield  {author} {\bibinfo {author} {\bibfnamefont {C.}~\bibnamefont
  {Cammarota}}, \bibinfo {author} {\bibfnamefont {G.}~\bibnamefont {Biroli}},
  \bibinfo {author} {\bibfnamefont {M.}~\bibnamefont {Tarzia}}, \ and\ \bibinfo
  {author} {\bibfnamefont {G.}~\bibnamefont {Tarjus}},\ }\href@noop {}
  {\bibfield  {journal} {\bibinfo  {journal} {Physical review letters}\
  }\textbf {\bibinfo {volume} {106}},\ \bibinfo {pages} {115705} (\bibinfo
  {year} {2011})}\BibitemShut {NoStop}%
\bibitem [{\citenamefont {Angelini}\ and\ \citenamefont
  {Biroli}(2017)}]{angelini2017real}%
  \BibitemOpen
  \bibfield  {author} {\bibinfo {author} {\bibfnamefont {M.~C.}\ \bibnamefont
  {Angelini}}\ and\ \bibinfo {author} {\bibfnamefont {G.}~\bibnamefont
  {Biroli}},\ }\href@noop {} {\bibfield  {journal} {\bibinfo  {journal}
  {Proceedings of the National Academy of Sciences}\ }\textbf {\bibinfo
  {volume} {114}},\ \bibinfo {pages} {3328} (\bibinfo {year}
  {2017})}\BibitemShut {NoStop}%
\bibitem [{\citenamefont {Fisher}(1986)}]{fisher1986scaling}%
  \BibitemOpen
  \bibfield  {author} {\bibinfo {author} {\bibfnamefont {D.~S.}\ \bibnamefont
  {Fisher}},\ }\href@noop {} {\bibfield  {journal} {\bibinfo  {journal}
  {Physical review letters}\ }\textbf {\bibinfo {volume} {56}},\ \bibinfo
  {pages} {416} (\bibinfo {year} {1986})}\BibitemShut {NoStop}%
\bibitem [{\citenamefont {Kivelson}\ \emph {et~al.}(1995)\citenamefont
  {Kivelson}, \citenamefont {Kivelson}, \citenamefont {Zhao}, \citenamefont
  {Nussinov},\ and\ \citenamefont {Tarjus}}]{gillesphysica}%
  \BibitemOpen
  \bibfield  {author} {\bibinfo {author} {\bibfnamefont {D.}~\bibnamefont
  {Kivelson}}, \bibinfo {author} {\bibfnamefont {S.~A.}\ \bibnamefont
  {Kivelson}}, \bibinfo {author} {\bibfnamefont {X.}~\bibnamefont {Zhao}},
  \bibinfo {author} {\bibfnamefont {Z.}~\bibnamefont {Nussinov}}, \ and\
  \bibinfo {author} {\bibfnamefont {G.}~\bibnamefont {Tarjus}},\ }\href@noop {}
  {\bibfield  {journal} {\bibinfo  {journal} {Physica A: Statistical Mechanics
  and its Applications}\ }\textbf {\bibinfo {volume} {219}},\ \bibinfo {pages}
  {27} (\bibinfo {year} {1995})}\BibitemShut {NoStop}%
\bibitem [{\citenamefont {Berges}\ \emph {et~al.}(2002)\citenamefont {Berges},
  \citenamefont {Tetradis},\ and\ \citenamefont {Wetterich}}]{berges2002non}%
  \BibitemOpen
  \bibfield  {author} {\bibinfo {author} {\bibfnamefont {J.}~\bibnamefont
  {Berges}}, \bibinfo {author} {\bibfnamefont {N.}~\bibnamefont {Tetradis}}, \
  and\ \bibinfo {author} {\bibfnamefont {C.}~\bibnamefont {Wetterich}},\
  }\href@noop {} {\bibfield  {journal} {\bibinfo  {journal} {Physics Reports}\
  }\textbf {\bibinfo {volume} {363}},\ \bibinfo {pages} {223} (\bibinfo {year}
  {2002})}\BibitemShut {NoStop}%
\bibitem [{\citenamefont {Rulquin}\ \emph {et~al.}(2016)\citenamefont
  {Rulquin}, \citenamefont {Urbani}, \citenamefont {Biroli}, \citenamefont
  {Tarjus},\ and\ \citenamefont {Tarzia}}]{rulquin2016nonperturbative}%
  \BibitemOpen
  \bibfield  {author} {\bibinfo {author} {\bibfnamefont {C.}~\bibnamefont
  {Rulquin}}, \bibinfo {author} {\bibfnamefont {P.}~\bibnamefont {Urbani}},
  \bibinfo {author} {\bibfnamefont {G.}~\bibnamefont {Biroli}}, \bibinfo
  {author} {\bibfnamefont {G.}~\bibnamefont {Tarjus}}, \ and\ \bibinfo {author}
  {\bibfnamefont {M.}~\bibnamefont {Tarzia}},\ }\href@noop {} {\bibfield
  {journal} {\bibinfo  {journal} {Journal of Statistical Mechanics: Theory and
  Experiment}\ }\textbf {\bibinfo {volume} {2016}},\ \bibinfo {pages} {023209}
  (\bibinfo {year} {2016})}\BibitemShut {NoStop}%
\bibitem [{\citenamefont {Kob}\ and\ \citenamefont {Andersen}(1993)}]{KAgas}%
  \BibitemOpen
  \bibfield  {author} {\bibinfo {author} {\bibfnamefont {W.}~\bibnamefont
  {Kob}}\ and\ \bibinfo {author} {\bibfnamefont {H.~C.}\ \bibnamefont
  {Andersen}},\ }\href@noop {} {\bibfield  {journal} {\bibinfo  {journal}
  {Physical Review E}\ }\textbf {\bibinfo {volume} {48}},\ \bibinfo {pages}
  {4364} (\bibinfo {year} {1993})}\BibitemShut {NoStop}%
\bibitem [{\citenamefont {Ritort}\ and\ \citenamefont
  {Sollich}(2003)}]{solrit}%
  \BibitemOpen
  \bibfield  {author} {\bibinfo {author} {\bibfnamefont {F.}~\bibnamefont
  {Ritort}}\ and\ \bibinfo {author} {\bibfnamefont {P.}~\bibnamefont
  {Sollich}},\ }\href@noop {} {\bibfield  {journal} {\bibinfo  {journal}
  {Advances in physics}\ }\textbf {\bibinfo {volume} {52}},\ \bibinfo {pages}
  {219} (\bibinfo {year} {2003})}\BibitemShut {NoStop}%
\bibitem [{\citenamefont {Franz}\ \emph {et~al.}(2002)\citenamefont {Franz},
  \citenamefont {Mulet},\ and\ \citenamefont {Parisi}}]{silvioKA}%
  \BibitemOpen
  \bibfield  {author} {\bibinfo {author} {\bibfnamefont {S.}~\bibnamefont
  {Franz}}, \bibinfo {author} {\bibfnamefont {R.}~\bibnamefont {Mulet}}, \ and\
  \bibinfo {author} {\bibfnamefont {G.}~\bibnamefont {Parisi}},\ }\href@noop {}
  {\bibfield  {journal} {\bibinfo  {journal} {Physical Review E}\ }\textbf
  {\bibinfo {volume} {65}},\ \bibinfo {pages} {021506} (\bibinfo {year}
  {2002})}\BibitemShut {NoStop}%
\bibitem [{\citenamefont {Toninelli}\ \emph {et~al.}(2004)\citenamefont
  {Toninelli}, \citenamefont {Biroli},\ and\ \citenamefont {Fisher}}]{TBF}%
  \BibitemOpen
  \bibfield  {author} {\bibinfo {author} {\bibfnamefont {C.}~\bibnamefont
  {Toninelli}}, \bibinfo {author} {\bibfnamefont {G.}~\bibnamefont {Biroli}}, \
  and\ \bibinfo {author} {\bibfnamefont {D.~S.}\ \bibnamefont {Fisher}},\
  }\href@noop {} {\bibfield  {journal} {\bibinfo  {journal} {Physical review
  letters}\ }\textbf {\bibinfo {volume} {92}},\ \bibinfo {pages} {185504}
  (\bibinfo {year} {2004})}\BibitemShut {NoStop}%
\bibitem [{\citenamefont {Pan}\ \emph {et~al.}(2005)\citenamefont {Pan},
  \citenamefont {Garrahan},\ and\ \citenamefont {Chandler}}]{panetal}%
  \BibitemOpen
  \bibfield  {author} {\bibinfo {author} {\bibfnamefont {A.~C.}\ \bibnamefont
  {Pan}}, \bibinfo {author} {\bibfnamefont {J.~P.}\ \bibnamefont {Garrahan}}, \
  and\ \bibinfo {author} {\bibfnamefont {D.}~\bibnamefont {Chandler}},\
  }\href@noop {} {\bibfield  {journal} {\bibinfo  {journal} {Physical Review
  E}\ }\textbf {\bibinfo {volume} {72}},\ \bibinfo {pages} {041106} (\bibinfo
  {year} {2005})}\BibitemShut {NoStop}%
\bibitem [{\citenamefont {Glarum}(1960)}]{glarum}%
  \BibitemOpen
  \bibfield  {author} {\bibinfo {author} {\bibfnamefont {S.~H.}\ \bibnamefont
  {Glarum}},\ }\href@noop {} {\bibfield  {journal} {\bibinfo  {journal} {The
  Journal of Chemical Physics}\ }\textbf {\bibinfo {volume} {33}},\ \bibinfo
  {pages} {639} (\bibinfo {year} {1960})}\BibitemShut {NoStop}%
\bibitem [{\citenamefont {Fredrickson}\ and\ \citenamefont
  {Andersen}(1984)}]{FA}%
  \BibitemOpen
  \bibfield  {author} {\bibinfo {author} {\bibfnamefont {G.~H.}\ \bibnamefont
  {Fredrickson}}\ and\ \bibinfo {author} {\bibfnamefont {H.~C.}\ \bibnamefont
  {Andersen}},\ }\href@noop {} {\bibfield  {journal} {\bibinfo  {journal}
  {Physical review letters}\ }\textbf {\bibinfo {volume} {53}},\ \bibinfo
  {pages} {1244} (\bibinfo {year} {1984})}\BibitemShut {NoStop}%
\bibitem [{\citenamefont {L{\'e}onard}\ \emph {et~al.}(2007)\citenamefont
  {L{\'e}onard}, \citenamefont {Mayer}, \citenamefont {Sollich}, \citenamefont
  {Berthier},\ and\ \citenamefont {Garrahan}}]{leonard}%
  \BibitemOpen
  \bibfield  {author} {\bibinfo {author} {\bibfnamefont {S.}~\bibnamefont
  {L{\'e}onard}}, \bibinfo {author} {\bibfnamefont {P.}~\bibnamefont {Mayer}},
  \bibinfo {author} {\bibfnamefont {P.}~\bibnamefont {Sollich}}, \bibinfo
  {author} {\bibfnamefont {L.}~\bibnamefont {Berthier}}, \ and\ \bibinfo
  {author} {\bibfnamefont {J.~P.}\ \bibnamefont {Garrahan}},\ }\href@noop {}
  {\bibfield  {journal} {\bibinfo  {journal} {Journal of statistical mechanics:
  theory and experiment}\ }\textbf {\bibinfo {volume} {2007}},\ \bibinfo
  {pages} {P07017} (\bibinfo {year} {2007})}\BibitemShut {NoStop}%
\bibitem [{\citenamefont {Cohen}\ and\ \citenamefont {Grest}(1982)}]{freevol}%
  \BibitemOpen
  \bibfield  {author} {\bibinfo {author} {\bibfnamefont {M.~H.}\ \bibnamefont
  {Cohen}}\ and\ \bibinfo {author} {\bibfnamefont {G.}~\bibnamefont {Grest}},\
  }\href@noop {} {\bibfield  {journal} {\bibinfo  {journal} {Physical Review
  B}\ }\textbf {\bibinfo {volume} {26}},\ \bibinfo {pages} {6313} (\bibinfo
  {year} {1982})}\BibitemShut {NoStop}%
\bibitem [{\citenamefont {Fredrickson}\ and\ \citenamefont
  {Brawer}(1986)}]{FA2}%
  \BibitemOpen
  \bibfield  {author} {\bibinfo {author} {\bibfnamefont {G.~H.}\ \bibnamefont
  {Fredrickson}}\ and\ \bibinfo {author} {\bibfnamefont {S.~A.}\ \bibnamefont
  {Brawer}},\ }\href@noop {} {\bibfield  {journal} {\bibinfo  {journal} {The
  Journal of chemical physics}\ }\textbf {\bibinfo {volume} {84}},\ \bibinfo
  {pages} {3351} (\bibinfo {year} {1986})}\BibitemShut {NoStop}%
\bibitem [{\citenamefont {Butler}\ and\ \citenamefont
  {Harrowell}(1991)}]{harro}%
  \BibitemOpen
  \bibfield  {author} {\bibinfo {author} {\bibfnamefont {S.}~\bibnamefont
  {Butler}}\ and\ \bibinfo {author} {\bibfnamefont {P.}~\bibnamefont
  {Harrowell}},\ }\href@noop {} {\bibfield  {journal} {\bibinfo  {journal} {The
  Journal of chemical physics}\ }\textbf {\bibinfo {volume} {95}},\ \bibinfo
  {pages} {4454} (\bibinfo {year} {1991})}\BibitemShut {NoStop}%
\bibitem [{\citenamefont {Toninelli}\ \emph {et~al.}(2006)\citenamefont
  {Toninelli}, \citenamefont {Biroli},\ and\ \citenamefont {Fisher}}]{TBF2}%
  \BibitemOpen
  \bibfield  {author} {\bibinfo {author} {\bibfnamefont {C.}~\bibnamefont
  {Toninelli}}, \bibinfo {author} {\bibfnamefont {G.}~\bibnamefont {Biroli}}, \
  and\ \bibinfo {author} {\bibfnamefont {D.~S.}\ \bibnamefont {Fisher}},\
  }\href@noop {} {\bibfield  {journal} {\bibinfo  {journal} {Physical review
  letters}\ }\textbf {\bibinfo {volume} {96}},\ \bibinfo {pages} {035702}
  (\bibinfo {year} {2006})}\BibitemShut {NoStop}%
\bibitem [{\citenamefont {Elmatad}\ \emph {et~al.}(2009)\citenamefont
  {Elmatad}, \citenamefont {Chandler},\ and\ \citenamefont
  {Garrahan}}]{Elmatad2009}%
  \BibitemOpen
  \bibfield  {author} {\bibinfo {author} {\bibfnamefont {Y.}~\bibnamefont
  {Elmatad}}, \bibinfo {author} {\bibfnamefont {D.}~\bibnamefont {Chandler}}, \
  and\ \bibinfo {author} {\bibfnamefont {J.}~\bibnamefont {Garrahan}},\
  }\href@noop {} {\bibfield  {journal} {\bibinfo  {journal} {The journal of
  physical chemistry. B}\ }\textbf {\bibinfo {volume} {113}},\ \bibinfo {pages}
  {5563} (\bibinfo {year} {2009})}\BibitemShut {NoStop}%
\bibitem [{\citenamefont {Elmatad}\ \emph {et~al.}(2010)\citenamefont
  {Elmatad}, \citenamefont {Jack}, \citenamefont {Chandler},\ and\
  \citenamefont {Garrahan}}]{elmatad_finite-temperature_2010}%
  \BibitemOpen
  \bibfield  {author} {\bibinfo {author} {\bibfnamefont {Y.~S.}\ \bibnamefont
  {Elmatad}}, \bibinfo {author} {\bibfnamefont {R.~L.}\ \bibnamefont {Jack}},
  \bibinfo {author} {\bibfnamefont {D.}~\bibnamefont {Chandler}}, \ and\
  \bibinfo {author} {\bibfnamefont {J.~P.}\ \bibnamefont {Garrahan}},\ }\href
  {\doibase 10.1073/pnas.1006306107} {\bibfield  {journal} {\bibinfo  {journal}
  {Proceedings of the National Academy of Sciences}\ }\textbf {\bibinfo
  {volume} {107}},\ \bibinfo {pages} {12793} (\bibinfo {year}
  {2010})}\BibitemShut {NoStop}%
\bibitem [{\citenamefont {Elmatad}\ and\ \citenamefont
  {Keys}(2012)}]{Elmatad2012}%
  \BibitemOpen
  \bibfield  {author} {\bibinfo {author} {\bibfnamefont {Y.~S.}\ \bibnamefont
  {Elmatad}}\ and\ \bibinfo {author} {\bibfnamefont {A.~S.}\ \bibnamefont
  {Keys}},\ }\href@noop {} {\bibfield  {journal} {\bibinfo  {journal} {Physical
  Review E - Statistical, Nonlinear, and Soft Matter Physics}\ }\textbf
  {\bibinfo {volume} {85}} (\bibinfo {year} {2012})}\BibitemShut {NoStop}%
\bibitem [{\citenamefont {Hartarsky}\ \emph
  {et~al.}(2019{\natexlab{a}})\citenamefont {Hartarsky}, \citenamefont
  {Mar{\^e}ch{\'e}},\ and\ \citenamefont
  {Toninelli}}]{hartarsky2019universality}%
  \BibitemOpen
  \bibfield  {author} {\bibinfo {author} {\bibfnamefont {I.}~\bibnamefont
  {Hartarsky}}, \bibinfo {author} {\bibfnamefont {L.}~\bibnamefont
  {Mar{\^e}ch{\'e}}}, \ and\ \bibinfo {author} {\bibfnamefont {C.}~\bibnamefont
  {Toninelli}},\ }\href@noop {} {\bibfield  {journal} {\bibinfo  {journal}
  {arXiv preprint arXiv:1904.09145}\ } (\bibinfo {year}
  {2019}{\natexlab{a}})}\BibitemShut {NoStop}%
\bibitem [{\citenamefont {Martinelli}\ \emph
  {et~al.}(2019{\natexlab{a}})\citenamefont {Martinelli}, \citenamefont
  {Morris},\ and\ \citenamefont {Toninelli}}]{martinelli2019universality}%
  \BibitemOpen
  \bibfield  {author} {\bibinfo {author} {\bibfnamefont {F.}~\bibnamefont
  {Martinelli}}, \bibinfo {author} {\bibfnamefont {R.}~\bibnamefont {Morris}},
  \ and\ \bibinfo {author} {\bibfnamefont {C.}~\bibnamefont {Toninelli}},\
  }\href@noop {} {\bibfield  {journal} {\bibinfo  {journal} {Communications in
  mathematical physics}\ }\textbf {\bibinfo {volume} {369}},\ \bibinfo {pages}
  {761} (\bibinfo {year} {2019}{\natexlab{a}})}\BibitemShut {NoStop}%
\bibitem [{\citenamefont {Hartarsky}\ \emph
  {et~al.}(2019{\natexlab{b}})\citenamefont {Hartarsky}, \citenamefont
  {Martinelli},\ and\ \citenamefont {Toninelli}}]{hartarsky2019universality2}%
  \BibitemOpen
  \bibfield  {author} {\bibinfo {author} {\bibfnamefont {I.}~\bibnamefont
  {Hartarsky}}, \bibinfo {author} {\bibfnamefont {F.}~\bibnamefont
  {Martinelli}}, \ and\ \bibinfo {author} {\bibfnamefont {C.}~\bibnamefont
  {Toninelli}},\ }\href@noop {} {\bibfield  {journal} {\bibinfo  {journal}
  {arXiv preprint arXiv:1910.06782}\ } (\bibinfo {year}
  {2019}{\natexlab{b}})}\BibitemShut {NoStop}%
\bibitem [{\citenamefont {Martinelli}\ \emph
  {et~al.}(2019{\natexlab{b}})\citenamefont {Martinelli}, \citenamefont
  {Toninelli} \emph {et~al.}}]{martinelli2019towards}%
  \BibitemOpen
  \bibfield  {author} {\bibinfo {author} {\bibfnamefont {F.}~\bibnamefont
  {Martinelli}}, \bibinfo {author} {\bibfnamefont {C.}~\bibnamefont
  {Toninelli}},  \emph {et~al.},\ }\href@noop {} {\bibfield  {journal}
  {\bibinfo  {journal} {The Annals of Probability}\ }\textbf {\bibinfo {volume}
  {47}},\ \bibinfo {pages} {324} (\bibinfo {year}
  {2019}{\natexlab{b}})}\BibitemShut {NoStop}%
\bibitem [{\citenamefont {Keys}\ \emph {et~al.}(2013)\citenamefont {Keys},
  \citenamefont {Garrahan},\ and\ \citenamefont
  {Chandler}}]{keys_calorimetric_2013}%
  \BibitemOpen
  \bibfield  {author} {\bibinfo {author} {\bibfnamefont {A.~S.}\ \bibnamefont
  {Keys}}, \bibinfo {author} {\bibfnamefont {J.~P.}\ \bibnamefont {Garrahan}},
  \ and\ \bibinfo {author} {\bibfnamefont {D.}~\bibnamefont {Chandler}},\
  }\href {\doibase 10.1073/pnas.1302665110} {\bibfield  {journal} {\bibinfo
  {journal} {Proceedings of the National Academy of Sciences}\ }\textbf
  {\bibinfo {volume} {110}},\ \bibinfo {pages} {4482} (\bibinfo {year}
  {2013})}\BibitemShut {NoStop}%
\bibitem [{\citenamefont {Keys}\ \emph {et~al.}(2011)\citenamefont {Keys},
  \citenamefont {Hedges}, \citenamefont {Garrahan}, \citenamefont {Glotzer},\
  and\ \citenamefont {Chandler}}]{Keys2011}%
  \BibitemOpen
  \bibfield  {author} {\bibinfo {author} {\bibfnamefont {A.~S.}\ \bibnamefont
  {Keys}}, \bibinfo {author} {\bibfnamefont {L.~O.}\ \bibnamefont {Hedges}},
  \bibinfo {author} {\bibfnamefont {J.~P.}\ \bibnamefont {Garrahan}}, \bibinfo
  {author} {\bibfnamefont {S.~C.}\ \bibnamefont {Glotzer}}, \ and\ \bibinfo
  {author} {\bibfnamefont {D.}~\bibnamefont {Chandler}},\ }\href {\doibase
  10.1103/PhysRevX.1.021013} {\bibfield  {journal} {\bibinfo  {journal}
  {Physical Review X}\ }\textbf {\bibinfo {volume} {1}},\ \bibinfo {pages} {1}
  (\bibinfo {year} {2011})}\BibitemShut {NoStop}%
\bibitem [{\citenamefont {Garrahan}\ and\ \citenamefont {Chandler}(2002)}]{gc}%
  \BibitemOpen
  \bibfield  {author} {\bibinfo {author} {\bibfnamefont {J.~P.}\ \bibnamefont
  {Garrahan}}\ and\ \bibinfo {author} {\bibfnamefont {D.}~\bibnamefont
  {Chandler}},\ }\href@noop {} {\bibfield  {journal} {\bibinfo  {journal}
  {Physical review letters}\ }\textbf {\bibinfo {volume} {89}},\ \bibinfo
  {pages} {035704} (\bibinfo {year} {2002})}\BibitemShut {NoStop}%
\bibitem [{\citenamefont {Whitelam}\ \emph {et~al.}(2005)\citenamefont
  {Whitelam}, \citenamefont {Berthier},\ and\ \citenamefont
  {Garrahan}}]{steve}%
  \BibitemOpen
  \bibfield  {author} {\bibinfo {author} {\bibfnamefont {S.}~\bibnamefont
  {Whitelam}}, \bibinfo {author} {\bibfnamefont {L.}~\bibnamefont {Berthier}},
  \ and\ \bibinfo {author} {\bibfnamefont {J.~P.}\ \bibnamefont {Garrahan}},\
  }\href@noop {} {\bibfield  {journal} {\bibinfo  {journal} {Physical Review
  E}\ }\textbf {\bibinfo {volume} {71}},\ \bibinfo {pages} {026128} (\bibinfo
  {year} {2005})}\BibitemShut {NoStop}%
\bibitem [{\citenamefont {Berthier}\ and\ \citenamefont
  {Garrahan}(2005)}]{nef}%
  \BibitemOpen
  \bibfield  {author} {\bibinfo {author} {\bibfnamefont {L.}~\bibnamefont
  {Berthier}}\ and\ \bibinfo {author} {\bibfnamefont {J.~P.}\ \bibnamefont
  {Garrahan}},\ }\href@noop {} {\bibfield  {journal} {\bibinfo  {journal} {The
  journal of physical chemistry B}\ }\textbf {\bibinfo {volume} {109}},\
  \bibinfo {pages} {3578} (\bibinfo {year} {2005})}\BibitemShut {NoStop}%
\bibitem [{\citenamefont {Jack}\ \emph
  {et~al.}(2006{\natexlab{a}})\citenamefont {Jack}, \citenamefont {Mayer},\
  and\ \citenamefont {Sollich}}]{mayerjack}%
  \BibitemOpen
  \bibfield  {author} {\bibinfo {author} {\bibfnamefont {R.~L.}\ \bibnamefont
  {Jack}}, \bibinfo {author} {\bibfnamefont {P.}~\bibnamefont {Mayer}}, \ and\
  \bibinfo {author} {\bibfnamefont {P.}~\bibnamefont {Sollich}},\ }\href@noop
  {} {\bibfield  {journal} {\bibinfo  {journal} {Journal of Statistical
  Mechanics: Theory and Experiment}\ }\textbf {\bibinfo {volume} {2006}},\
  \bibinfo {pages} {P03006} (\bibinfo {year} {2006}{\natexlab{a}})}\BibitemShut
  {NoStop}%
\bibitem [{\citenamefont {Berthier}\ \emph {et~al.}(2004)\citenamefont
  {Berthier}, \citenamefont {Chandler},\ and\ \citenamefont
  {Garrahan}}]{berthierepl}%
  \BibitemOpen
  \bibfield  {author} {\bibinfo {author} {\bibfnamefont {L.}~\bibnamefont
  {Berthier}}, \bibinfo {author} {\bibfnamefont {D.}~\bibnamefont {Chandler}},
  \ and\ \bibinfo {author} {\bibfnamefont {J.~P.}\ \bibnamefont {Garrahan}},\
  }\href@noop {} {\bibfield  {journal} {\bibinfo  {journal} {EPL (Europhysics
  Letters)}\ }\textbf {\bibinfo {volume} {69}},\ \bibinfo {pages} {320}
  (\bibinfo {year} {2004})}\BibitemShut {NoStop}%
\bibitem [{\citenamefont {Chandler}\ \emph {et~al.}(2006)\citenamefont
  {Chandler}, \citenamefont {Garrahan}, \citenamefont {Jack}, \citenamefont
  {Maibaum},\ and\ \citenamefont {Pan}}]{chandler}%
  \BibitemOpen
  \bibfield  {author} {\bibinfo {author} {\bibfnamefont {D.}~\bibnamefont
  {Chandler}}, \bibinfo {author} {\bibfnamefont {J.~P.}\ \bibnamefont
  {Garrahan}}, \bibinfo {author} {\bibfnamefont {R.~L.}\ \bibnamefont {Jack}},
  \bibinfo {author} {\bibfnamefont {L.}~\bibnamefont {Maibaum}}, \ and\
  \bibinfo {author} {\bibfnamefont {A.~C.}\ \bibnamefont {Pan}},\ }\href@noop
  {} {\bibfield  {journal} {\bibinfo  {journal} {Physical Review E}\ }\textbf
  {\bibinfo {volume} {74}},\ \bibinfo {pages} {051501} (\bibinfo {year}
  {2006})}\BibitemShut {NoStop}%
\bibitem [{\citenamefont {Whitelam}\ \emph {et~al.}(2004)\citenamefont
  {Whitelam}, \citenamefont {Berthier},\ and\ \citenamefont
  {Garrahan}}]{steve2}%
  \BibitemOpen
  \bibfield  {author} {\bibinfo {author} {\bibfnamefont {S.}~\bibnamefont
  {Whitelam}}, \bibinfo {author} {\bibfnamefont {L.}~\bibnamefont {Berthier}},
  \ and\ \bibinfo {author} {\bibfnamefont {J.~P.}\ \bibnamefont {Garrahan}},\
  }\href@noop {} {\bibfield  {journal} {\bibinfo  {journal} {Physical review
  letters}\ }\textbf {\bibinfo {volume} {92}},\ \bibinfo {pages} {185705}
  (\bibinfo {year} {2004})}\BibitemShut {NoStop}%
\bibitem [{Note9()}]{Note9}%
  \BibitemOpen
  \bibinfo {note} {A critical (different) behaviour is expected and predicted
  for models having a transition \cite {TBF2}.}\BibitemShut {Stop}%
\bibitem [{\citenamefont {Downton}\ and\ \citenamefont
  {Kennett}(2007)}]{kennett}%
  \BibitemOpen
  \bibfield  {author} {\bibinfo {author} {\bibfnamefont {M.~T.}\ \bibnamefont
  {Downton}}\ and\ \bibinfo {author} {\bibfnamefont {M.~P.}\ \bibnamefont
  {Kennett}},\ }\href@noop {} {\bibfield  {journal} {\bibinfo  {journal}
  {Physical Review E}\ }\textbf {\bibinfo {volume} {76}},\ \bibinfo {pages}
  {031502} (\bibinfo {year} {2007})}\BibitemShut {NoStop}%
\bibitem [{\citenamefont {Hedges}\ \emph {et~al.}(2009)\citenamefont {Hedges},
  \citenamefont {Jack}, \citenamefont {Garrahan},\ and\ \citenamefont
  {Chandler}}]{hedges_dynamic_2009}%
  \BibitemOpen
  \bibfield  {author} {\bibinfo {author} {\bibfnamefont {L.~O.}\ \bibnamefont
  {Hedges}}, \bibinfo {author} {\bibfnamefont {R.~L.}\ \bibnamefont {Jack}},
  \bibinfo {author} {\bibfnamefont {J.~P.}\ \bibnamefont {Garrahan}}, \ and\
  \bibinfo {author} {\bibfnamefont {D.}~\bibnamefont {Chandler}},\ }\href
  {\doibase 10.1126/science.1166665} {\bibfield  {journal} {\bibinfo  {journal}
  {Science}\ }\textbf {\bibinfo {volume} {323}},\ \bibinfo {pages} {1309}
  (\bibinfo {year} {2009})}\BibitemShut {NoStop}%
\bibitem [{\citenamefont {Keys}\ \emph {et~al.}(2015)\citenamefont {Keys},
  \citenamefont {Chandler},\ and\ \citenamefont {Garrahan}}]{Keys2015}%
  \BibitemOpen
  \bibfield  {author} {\bibinfo {author} {\bibfnamefont {A.~S.}\ \bibnamefont
  {Keys}}, \bibinfo {author} {\bibfnamefont {D.}~\bibnamefont {Chandler}}, \
  and\ \bibinfo {author} {\bibfnamefont {J.~P.}\ \bibnamefont {Garrahan}},\
  }\href {\doibase 10.1103/PhysRevE.92.022304} {\bibfield  {journal} {\bibinfo
  {journal} {Physical Review E - Statistical, Nonlinear, and Soft Matter
  Physics}\ }\textbf {\bibinfo {volume} {92}},\ \bibinfo {pages} {1} (\bibinfo
  {year} {2015})}\BibitemShut {NoStop}%
\bibitem [{\citenamefont {Isobe}\ \emph {et~al.}(2016)\citenamefont {Isobe},
  \citenamefont {Keys}, \citenamefont {Chandler},\ and\ \citenamefont
  {Garrahan}}]{Isobe2016}%
  \BibitemOpen
  \bibfield  {author} {\bibinfo {author} {\bibfnamefont {M.}~\bibnamefont
  {Isobe}}, \bibinfo {author} {\bibfnamefont {A.~S.}\ \bibnamefont {Keys}},
  \bibinfo {author} {\bibfnamefont {D.}~\bibnamefont {Chandler}}, \ and\
  \bibinfo {author} {\bibfnamefont {J.~P.}\ \bibnamefont {Garrahan}},\ }\href
  {\doibase 10.1103/PhysRevLett.117.145701} {\bibfield  {journal} {\bibinfo
  {journal} {Physical Review Letters}\ }\textbf {\bibinfo {volume} {117}},\
  \bibinfo {pages} {1} (\bibinfo {year} {2016})}\BibitemShut {NoStop}%
\bibitem [{\citenamefont {Garrahan}(2002)}]{juanpe}%
  \BibitemOpen
  \bibfield  {author} {\bibinfo {author} {\bibfnamefont {J.~P.}\ \bibnamefont
  {Garrahan}},\ }\href@noop {} {\bibfield  {journal} {\bibinfo  {journal}
  {Journal of Physics: Condensed Matter}\ }\textbf {\bibinfo {volume} {14}},\
  \bibinfo {pages} {1571} (\bibinfo {year} {2002})}\BibitemShut {NoStop}%
\bibitem [{\citenamefont {Turner}\ \emph {et~al.}(2015)\citenamefont {Turner},
  \citenamefont {Jack},\ and\ \citenamefont {Garrahan}}]{turner_overlap_2015}%
  \BibitemOpen
  \bibfield  {author} {\bibinfo {author} {\bibfnamefont {R.~M.}\ \bibnamefont
  {Turner}}, \bibinfo {author} {\bibfnamefont {R.~L.}\ \bibnamefont {Jack}}, \
  and\ \bibinfo {author} {\bibfnamefont {J.~P.}\ \bibnamefont {Garrahan}},\
  }\href {\doibase 10.1103/PhysRevE.92.022115} {\bibfield  {journal} {\bibinfo
  {journal} {Physical Review E}\ }\textbf {\bibinfo {volume} {92}} (\bibinfo
  {year} {2015}),\ 10.1103/PhysRevE.92.022115}\BibitemShut {NoStop}%
\bibitem [{\citenamefont {Jack}\ \emph {et~al.}(2005)\citenamefont {Jack},
  \citenamefont {Berthier},\ and\ \citenamefont {Garrahan}}]{spm}%
  \BibitemOpen
  \bibfield  {author} {\bibinfo {author} {\bibfnamefont {R.~L.}\ \bibnamefont
  {Jack}}, \bibinfo {author} {\bibfnamefont {L.}~\bibnamefont {Berthier}}, \
  and\ \bibinfo {author} {\bibfnamefont {J.~P.}\ \bibnamefont {Garrahan}},\
  }\href@noop {} {\bibfield  {journal} {\bibinfo  {journal} {Physical Review
  E}\ }\textbf {\bibinfo {volume} {72}},\ \bibinfo {pages} {016103} (\bibinfo
  {year} {2005})}\BibitemShut {NoStop}%
\bibitem [{Note10()}]{Note10}%
  \BibitemOpen
  \bibinfo {note} {This type of plaquette models, and other spin models, were
  introduced originally~\cite {Lipowsky,Sethna} to show how ultra-slow glassy
  dynamics can emerge because of growing free energy barriers.}\BibitemShut
  {Stop}%
\bibitem [{\citenamefont {Biroli}\ \emph {et~al.}(2005)\citenamefont {Biroli},
  \citenamefont {Bouchaud},\ and\ \citenamefont {Tarjus}}]{BBT}%
  \BibitemOpen
  \bibfield  {author} {\bibinfo {author} {\bibfnamefont {G.}~\bibnamefont
  {Biroli}}, \bibinfo {author} {\bibfnamefont {J.-P.}\ \bibnamefont
  {Bouchaud}}, \ and\ \bibinfo {author} {\bibfnamefont {G.}~\bibnamefont
  {Tarjus}},\ }\href@noop {} {\bibfield  {journal} {\bibinfo  {journal} {The
  Journal of chemical physics}\ }\textbf {\bibinfo {volume} {123}},\ \bibinfo
  {pages} {044510} (\bibinfo {year} {2005})}\BibitemShut {NoStop}%
\bibitem [{\citenamefont {Berthier}\ and\ \citenamefont
  {Garrahan}(2003)}]{nontopo}%
  \BibitemOpen
  \bibfield  {author} {\bibinfo {author} {\bibfnamefont {L.}~\bibnamefont
  {Berthier}}\ and\ \bibinfo {author} {\bibfnamefont {J.~P.}\ \bibnamefont
  {Garrahan}},\ }\href@noop {} {\bibfield  {journal} {\bibinfo  {journal} {The
  Journal of chemical physics}\ }\textbf {\bibinfo {volume} {119}},\ \bibinfo
  {pages} {4367} (\bibinfo {year} {2003})}\BibitemShut {NoStop}%
\bibitem [{\citenamefont {Whitelam}\ and\ \citenamefont
  {Garrahan}(2004)}]{steve3}%
  \BibitemOpen
  \bibfield  {author} {\bibinfo {author} {\bibfnamefont {S.}~\bibnamefont
  {Whitelam}}\ and\ \bibinfo {author} {\bibfnamefont {J.~P.}\ \bibnamefont
  {Garrahan}},\ }\href@noop {} {\bibfield  {journal} {\bibinfo  {journal} {The
  Journal of Physical Chemistry B}\ }\textbf {\bibinfo {volume} {108}},\
  \bibinfo {pages} {6611} (\bibinfo {year} {2004})}\BibitemShut {NoStop}%
\bibitem [{\citenamefont {Jack}\ and\ \citenamefont {Garrahan}(2005)}]{rob2}%
  \BibitemOpen
  \bibfield  {author} {\bibinfo {author} {\bibfnamefont {R.~L.}\ \bibnamefont
  {Jack}}\ and\ \bibinfo {author} {\bibfnamefont {J.~P.}\ \bibnamefont
  {Garrahan}},\ }\href@noop {} {\bibfield  {journal} {\bibinfo  {journal} {The
  Journal of chemical physics}\ }\textbf {\bibinfo {volume} {123}},\ \bibinfo
  {pages} {164508} (\bibinfo {year} {2005})}\BibitemShut {NoStop}%
\bibitem [{Note11()}]{Note11}%
  \BibitemOpen
  \bibinfo {note} {Most KCMs do not have a finite temperature dynamical
  transition and the ones displaying a transition have critical properties
  different from MCT.}\BibitemShut {Stop}%
\bibitem [{\citenamefont {Berthier}\ \emph {et~al.}(2012)\citenamefont
  {Berthier}, \citenamefont {Biroli}, \citenamefont {Coslovich}, \citenamefont
  {Kob},\ and\ \citenamefont {Toninelli}}]{berthier2012finite}%
  \BibitemOpen
  \bibfield  {author} {\bibinfo {author} {\bibfnamefont {L.}~\bibnamefont
  {Berthier}}, \bibinfo {author} {\bibfnamefont {G.}~\bibnamefont {Biroli}},
  \bibinfo {author} {\bibfnamefont {D.}~\bibnamefont {Coslovich}}, \bibinfo
  {author} {\bibfnamefont {W.}~\bibnamefont {Kob}}, \ and\ \bibinfo {author}
  {\bibfnamefont {C.}~\bibnamefont {Toninelli}},\ }\href@noop {} {\bibfield
  {journal} {\bibinfo  {journal} {Physical Review E}\ }\textbf {\bibinfo
  {volume} {86}},\ \bibinfo {pages} {031502} (\bibinfo {year}
  {2012})}\BibitemShut {NoStop}%
\bibitem [{\citenamefont {Frank}(1952)}]{frank}%
  \BibitemOpen
  \bibfield  {author} {\bibinfo {author} {\bibfnamefont {F.~C.}\ \bibnamefont
  {Frank}},\ }\href@noop {} {\bibfield  {journal} {\bibinfo  {journal}
  {Proceedings of the Royal Society of London. Series A. Mathematical and
  Physical Sciences}\ }\textbf {\bibinfo {volume} {215}},\ \bibinfo {pages}
  {43} (\bibinfo {year} {1952})}\BibitemShut {NoStop}%
\bibitem [{\citenamefont {Sausset}\ \emph {et~al.}(2009)\citenamefont
  {Sausset}, \citenamefont {Tarjus},\ and\ \citenamefont
  {Viot}}]{sausset_thermodynamics_2009}%
  \BibitemOpen
  \bibfield  {author} {\bibinfo {author} {\bibfnamefont {F.}~\bibnamefont
  {Sausset}}, \bibinfo {author} {\bibfnamefont {G.}~\bibnamefont {Tarjus}}, \
  and\ \bibinfo {author} {\bibfnamefont {P.}~\bibnamefont {Viot}},\ }\href
  {\doibase 10.1088/1742-5468/2009/04/P04022} {\bibfield  {journal} {\bibinfo
  {journal} {Journal of Statistical Mechanics: Theory and Experiment}\ }\textbf
  {\bibinfo {volume} {2009}},\ \bibinfo {pages} {P04022} (\bibinfo {year}
  {2009})}\BibitemShut {NoStop}%
\bibitem [{\citenamefont {Vest}\ \emph {et~al.}(2014)\citenamefont {Vest},
  \citenamefont {Tarjus},\ and\ \citenamefont {Viot}}]{vest_dynamics_2014}%
  \BibitemOpen
  \bibfield  {author} {\bibinfo {author} {\bibfnamefont {J.-P.}\ \bibnamefont
  {Vest}}, \bibinfo {author} {\bibfnamefont {G.}~\bibnamefont {Tarjus}}, \ and\
  \bibinfo {author} {\bibfnamefont {P.}~\bibnamefont {Viot}},\ }\href {\doibase
  10.1080/00268976.2014.901568} {\bibfield  {journal} {\bibinfo  {journal}
  {Molecular Physics}\ }\textbf {\bibinfo {volume} {112}},\ \bibinfo {pages}
  {1330} (\bibinfo {year} {2014})}\BibitemShut {NoStop}%
\bibitem [{\citenamefont {Vest}\ \emph {et~al.}(2015)\citenamefont {Vest},
  \citenamefont {Tarjus},\ and\ \citenamefont
  {Viot}}]{vest_mode-coupling_2015}%
  \BibitemOpen
  \bibfield  {author} {\bibinfo {author} {\bibfnamefont {J.-P.}\ \bibnamefont
  {Vest}}, \bibinfo {author} {\bibfnamefont {G.}~\bibnamefont {Tarjus}}, \ and\
  \bibinfo {author} {\bibfnamefont {P.}~\bibnamefont {Viot}},\ }\href {\doibase
  10.1063/1.4928513} {\bibfield  {journal} {\bibinfo  {journal} {The Journal of
  Chemical Physics}\ }\textbf {\bibinfo {volume} {143}},\ \bibinfo {pages}
  {084505} (\bibinfo {year} {2015})}\BibitemShut {NoStop}%
\bibitem [{\citenamefont {Turci}\ \emph
  {et~al.}(2017{\natexlab{a}})\citenamefont {Turci}, \citenamefont {Tarjus},\
  and\ \citenamefont {Royall}}]{Turci2017}%
  \BibitemOpen
  \bibfield  {author} {\bibinfo {author} {\bibfnamefont {F.}~\bibnamefont
  {Turci}}, \bibinfo {author} {\bibfnamefont {G.}~\bibnamefont {Tarjus}}, \
  and\ \bibinfo {author} {\bibfnamefont {C.~P.}\ \bibnamefont {Royall}},\
  }\href {\doibase 10.1103/PhysRevLett.118.215501} {\bibfield  {journal}
  {\bibinfo  {journal} {Physical Review Letters}\ }\textbf {\bibinfo {volume}
  {118}},\ \bibinfo {pages} {1} (\bibinfo {year}
  {2017}{\natexlab{a}})}\BibitemShut {NoStop}%
\bibitem [{\citenamefont {Coslovich}\ and\ \citenamefont
  {Pastore}(2007)}]{coslovich}%
  \BibitemOpen
  \bibfield  {author} {\bibinfo {author} {\bibfnamefont {D.}~\bibnamefont
  {Coslovich}}\ and\ \bibinfo {author} {\bibfnamefont {G.}~\bibnamefont
  {Pastore}},\ }\href@noop {} {\bibfield  {journal} {\bibinfo  {journal} {The
  Journal of chemical physics}\ }\textbf {\bibinfo {volume} {127}},\ \bibinfo
  {pages} {124504} (\bibinfo {year} {2007})}\BibitemShut {NoStop}%
\bibitem [{\citenamefont {Coslovich}(2011)}]{Coslovich2011}%
  \BibitemOpen
  \bibfield  {author} {\bibinfo {author} {\bibfnamefont {D.}~\bibnamefont
  {Coslovich}},\ }\href {\doibase 10.1103/PhysRevE.83.051505} {\bibfield
  {journal} {\bibinfo  {journal} {Physical Review E - Statistical, Nonlinear,
  and Soft Matter Physics}\ }\textbf {\bibinfo {volume} {83}},\ \bibinfo
  {pages} {1} (\bibinfo {year} {2011})}\BibitemShut {NoStop}%
\bibitem [{\citenamefont {Royall}\ and\ \citenamefont
  {Williams}(2015)}]{royall2015role}%
  \BibitemOpen
  \bibfield  {author} {\bibinfo {author} {\bibfnamefont {C.~P.}\ \bibnamefont
  {Royall}}\ and\ \bibinfo {author} {\bibfnamefont {S.~R.}\ \bibnamefont
  {Williams}},\ }\href@noop {} {\bibfield  {journal} {\bibinfo  {journal}
  {Physics Reports}\ }\textbf {\bibinfo {volume} {560}},\ \bibinfo {pages} {1}
  (\bibinfo {year} {2015})}\BibitemShut {NoStop}%
\bibitem [{\citenamefont {Malins}\ \emph
  {et~al.}(2013{\natexlab{a}})\citenamefont {Malins}, \citenamefont {Eggers},
  \citenamefont {Royall}, \citenamefont {Williams},\ and\ \citenamefont
  {Tanaka}}]{malins2013identification}%
  \BibitemOpen
  \bibfield  {author} {\bibinfo {author} {\bibfnamefont {A.}~\bibnamefont
  {Malins}}, \bibinfo {author} {\bibfnamefont {J.}~\bibnamefont {Eggers}},
  \bibinfo {author} {\bibfnamefont {C.~P.}\ \bibnamefont {Royall}}, \bibinfo
  {author} {\bibfnamefont {S.~R.}\ \bibnamefont {Williams}}, \ and\ \bibinfo
  {author} {\bibfnamefont {H.}~\bibnamefont {Tanaka}},\ }\href@noop {}
  {\bibfield  {journal} {\bibinfo  {journal} {The Journal of chemical physics}\
  }\textbf {\bibinfo {volume} {138}},\ \bibinfo {pages} {12A535} (\bibinfo
  {year} {2013}{\natexlab{a}})}\BibitemShut {NoStop}%
\bibitem [{\citenamefont {Malins}\ \emph
  {et~al.}(2013{\natexlab{b}})\citenamefont {Malins}, \citenamefont {Williams},
  \citenamefont {Eggers},\ and\ \citenamefont
  {Royall}}]{malins2013identificationb}%
  \BibitemOpen
  \bibfield  {author} {\bibinfo {author} {\bibfnamefont {A.}~\bibnamefont
  {Malins}}, \bibinfo {author} {\bibfnamefont {S.~R.}\ \bibnamefont
  {Williams}}, \bibinfo {author} {\bibfnamefont {J.}~\bibnamefont {Eggers}}, \
  and\ \bibinfo {author} {\bibfnamefont {C.~P.}\ \bibnamefont {Royall}},\
  }\href@noop {} {\bibfield  {journal} {\bibinfo  {journal} {The Journal of
  chemical physics}\ }\textbf {\bibinfo {volume} {139}},\ \bibinfo {pages}
  {234506} (\bibinfo {year} {2013}{\natexlab{b}})}\BibitemShut {NoStop}%
\bibitem [{\citenamefont {Royall}\ and\ \citenamefont
  {Kob}(2017)}]{Royall2017}%
  \BibitemOpen
  \bibfield  {author} {\bibinfo {author} {\bibfnamefont {C.~P.}\ \bibnamefont
  {Royall}}\ and\ \bibinfo {author} {\bibfnamefont {W.}~\bibnamefont {Kob}},\
  }\href {\doibase 10.1088/1742-5468/aa4e92} {\bibfield  {journal} {\bibinfo
  {journal} {Journal of Statistical Mechanics: Theory and Experiment}\ }\textbf
  {\bibinfo {volume} {2017}} (\bibinfo {year} {2017}),\
  10.1088/1742-5468/aa4e92}\BibitemShut {NoStop}%
\bibitem [{\citenamefont {Turci}\ \emph
  {et~al.}(2017{\natexlab{b}})\citenamefont {Turci}, \citenamefont {Royall},\
  and\ \citenamefont {Speck}}]{turci_nonequilibrium_2017}%
  \BibitemOpen
  \bibfield  {author} {\bibinfo {author} {\bibfnamefont {F.}~\bibnamefont
  {Turci}}, \bibinfo {author} {\bibfnamefont {C.~P.}\ \bibnamefont {Royall}}, \
  and\ \bibinfo {author} {\bibfnamefont {T.}~\bibnamefont {Speck}},\ }\href
  {\doibase 10.1103/PhysRevX.7.031028} {\bibfield  {journal} {\bibinfo
  {journal} {Physical Review X}\ }\textbf {\bibinfo {volume} {7}} (\bibinfo
  {year} {2017}{\natexlab{b}}),\ 10.1103/PhysRevX.7.031028}\BibitemShut
  {NoStop}%
\bibitem [{\citenamefont {Turci}\ \emph {et~al.}(2018)\citenamefont {Turci},
  \citenamefont {Speck},\ and\ \citenamefont
  {Royall}}]{turci_structural-dynamical_2018}%
  \BibitemOpen
  \bibfield  {author} {\bibinfo {author} {\bibfnamefont {F.}~\bibnamefont
  {Turci}}, \bibinfo {author} {\bibfnamefont {T.}~\bibnamefont {Speck}}, \ and\
  \bibinfo {author} {\bibfnamefont {C.~P.}\ \bibnamefont {Royall}},\ }\href
  {\doibase 10.1140/epje/i2018-11662-3} {\bibfield  {journal} {\bibinfo
  {journal} {The European Physical Journal E}\ }\textbf {\bibinfo {volume}
  {41}} (\bibinfo {year} {2018}),\ 10.1140/epje/i2018-11662-3}\BibitemShut
  {NoStop}%
\bibitem [{\citenamefont {Turci}\ \emph {et~al.}(2019)\citenamefont {Turci},
  \citenamefont {Patrick~Royall},\ and\ \citenamefont
  {Speck}}]{turci_devitrification_2019}%
  \BibitemOpen
  \bibfield  {author} {\bibinfo {author} {\bibfnamefont {F.}~\bibnamefont
  {Turci}}, \bibinfo {author} {\bibfnamefont {C.}~\bibnamefont
  {Patrick~Royall}}, \ and\ \bibinfo {author} {\bibfnamefont {T.}~\bibnamefont
  {Speck}},\ }\href {\doibase 10.1088/1742-6596/1252/1/012012} {\bibfield
  {journal} {\bibinfo  {journal} {Journal of Physics: Conference Series}\
  }\textbf {\bibinfo {volume} {1252}},\ \bibinfo {pages} {012012} (\bibinfo
  {year} {2019})}\BibitemShut {NoStop}%
\bibitem [{\citenamefont {Tong}\ and\ \citenamefont
  {Tanaka}(2018)}]{tong_revealing_2018}%
  \BibitemOpen
  \bibfield  {author} {\bibinfo {author} {\bibfnamefont {H.}~\bibnamefont
  {Tong}}\ and\ \bibinfo {author} {\bibfnamefont {H.}~\bibnamefont {Tanaka}},\
  }\href {\doibase 10.1103/PhysRevX.8.011041} {\bibfield  {journal} {\bibinfo
  {journal} {Physical Review X}\ }\textbf {\bibinfo {volume} {8}} (\bibinfo
  {year} {2018}),\ 10.1103/PhysRevX.8.011041}\BibitemShut {NoStop}%
\bibitem [{\citenamefont {Shi}\ and\ \citenamefont
  {Tanaka}(2019)}]{shi_distinct_2019}%
  \BibitemOpen
  \bibfield  {author} {\bibinfo {author} {\bibfnamefont {R.}~\bibnamefont
  {Shi}}\ and\ \bibinfo {author} {\bibfnamefont {H.}~\bibnamefont {Tanaka}},\
  }\href {\doibase 10.1126/sciadv.aav3194} {\bibfield  {journal} {\bibinfo
  {journal} {Science Advances}\ }\textbf {\bibinfo {volume} {5}},\ \bibinfo
  {pages} {eaav3194} (\bibinfo {year} {2019})}\BibitemShut {NoStop}%
\bibitem [{\citenamefont {Pinchaipat}\ \emph {et~al.}(2017)\citenamefont
  {Pinchaipat}, \citenamefont {Campo}, \citenamefont {Turci}, \citenamefont
  {Hallett}, \citenamefont {Speck},\ and\ \citenamefont
  {Royall}}]{pinchaipat_experimental_2017}%
  \BibitemOpen
  \bibfield  {author} {\bibinfo {author} {\bibfnamefont {R.}~\bibnamefont
  {Pinchaipat}}, \bibinfo {author} {\bibfnamefont {M.}~\bibnamefont {Campo}},
  \bibinfo {author} {\bibfnamefont {F.}~\bibnamefont {Turci}}, \bibinfo
  {author} {\bibfnamefont {J.~E.}\ \bibnamefont {Hallett}}, \bibinfo {author}
  {\bibfnamefont {T.}~\bibnamefont {Speck}}, \ and\ \bibinfo {author}
  {\bibfnamefont {C.~P.}\ \bibnamefont {Royall}},\ }\href {\doibase
  10.1103/PhysRevLett.119.028004} {\bibfield  {journal} {\bibinfo  {journal}
  {Physical Review Letters}\ }\textbf {\bibinfo {volume} {119}} (\bibinfo
  {year} {2017}),\ 10.1103/PhysRevLett.119.028004}\BibitemShut {NoStop}%
\bibitem [{\citenamefont {Mossa}\ and\ \citenamefont
  {Tarjus}(2006)}]{mossa_operational_2006}%
  \BibitemOpen
  \bibfield  {author} {\bibinfo {author} {\bibfnamefont {S.}~\bibnamefont
  {Mossa}}\ and\ \bibinfo {author} {\bibfnamefont {G.}~\bibnamefont {Tarjus}},\
  }\href {\doibase 10.1016/j.jnoncrysol.2005.12.060} {\bibfield  {journal}
  {\bibinfo  {journal} {Journal of Non-Crystalline Solids}\ }\textbf {\bibinfo
  {volume} {352}},\ \bibinfo {pages} {4847} (\bibinfo {year}
  {2006})}\BibitemShut {NoStop}%
\bibitem [{\citenamefont {Ronhovde}\ \emph {et~al.}(2011)\citenamefont
  {Ronhovde}, \citenamefont {Chakrabarty}, \citenamefont {Hu}, \citenamefont
  {Sahu}, \citenamefont {Sahu}, \citenamefont {Kelton}, \citenamefont {Mauro},\
  and\ \citenamefont {Nussinov}}]{ronhovde2011detecting}%
  \BibitemOpen
  \bibfield  {author} {\bibinfo {author} {\bibfnamefont {P.}~\bibnamefont
  {Ronhovde}}, \bibinfo {author} {\bibfnamefont {S.}~\bibnamefont
  {Chakrabarty}}, \bibinfo {author} {\bibfnamefont {D.}~\bibnamefont {Hu}},
  \bibinfo {author} {\bibfnamefont {M.}~\bibnamefont {Sahu}}, \bibinfo {author}
  {\bibfnamefont {K.}~\bibnamefont {Sahu}}, \bibinfo {author} {\bibfnamefont
  {K.}~\bibnamefont {Kelton}}, \bibinfo {author} {\bibfnamefont
  {N.}~\bibnamefont {Mauro}}, \ and\ \bibinfo {author} {\bibfnamefont
  {Z.}~\bibnamefont {Nussinov}},\ }\href@noop {} {\bibfield  {journal}
  {\bibinfo  {journal} {The European Physical Journal E}\ }\textbf {\bibinfo
  {volume} {34}},\ \bibinfo {pages} {105} (\bibinfo {year} {2011})}\BibitemShut
  {NoStop}%
\bibitem [{\citenamefont {Ronhovde}\ \emph {et~al.}(2012)\citenamefont
  {Ronhovde}, \citenamefont {Chakrabarty}, \citenamefont {Hu}, \citenamefont
  {Sahu}, \citenamefont {Sahu}, \citenamefont {Kelton}, \citenamefont {Mauro},\
  and\ \citenamefont {Nussinov}}]{ronhovde2012detection}%
  \BibitemOpen
  \bibfield  {author} {\bibinfo {author} {\bibfnamefont {P.}~\bibnamefont
  {Ronhovde}}, \bibinfo {author} {\bibfnamefont {S.}~\bibnamefont
  {Chakrabarty}}, \bibinfo {author} {\bibfnamefont {D.}~\bibnamefont {Hu}},
  \bibinfo {author} {\bibfnamefont {M.}~\bibnamefont {Sahu}}, \bibinfo {author}
  {\bibfnamefont {K.~K.}\ \bibnamefont {Sahu}}, \bibinfo {author}
  {\bibfnamefont {K.~F.}\ \bibnamefont {Kelton}}, \bibinfo {author}
  {\bibfnamefont {N.~A.}\ \bibnamefont {Mauro}}, \ and\ \bibinfo {author}
  {\bibfnamefont {Z.}~\bibnamefont {Nussinov}},\ }\href@noop {} {\bibfield
  {journal} {\bibinfo  {journal} {Scientific reports}\ }\textbf {\bibinfo
  {volume} {2}},\ \bibinfo {pages} {1} (\bibinfo {year} {2012})}\BibitemShut
  {NoStop}%
\bibitem [{\citenamefont {Paret}\ \emph {et~al.}(2020)\citenamefont {Paret},
  \citenamefont {Jack},\ and\ \citenamefont {Coslovich}}]{paret2020assessing}%
  \BibitemOpen
  \bibfield  {author} {\bibinfo {author} {\bibfnamefont {J.}~\bibnamefont
  {Paret}}, \bibinfo {author} {\bibfnamefont {R.~L.}\ \bibnamefont {Jack}}, \
  and\ \bibinfo {author} {\bibfnamefont {D.}~\bibnamefont {Coslovich}},\
  }\href@noop {} {\bibfield  {journal} {\bibinfo  {journal} {The Journal of
  chemical physics}\ }\textbf {\bibinfo {volume} {152}},\ \bibinfo {pages}
  {144502} (\bibinfo {year} {2020})}\BibitemShut {NoStop}%
\bibitem [{\citenamefont {Boattini}\ \emph {et~al.}(2020)\citenamefont
  {Boattini}, \citenamefont {Mar{\'\i}n-Aguilar}, \citenamefont {Mitra},
  \citenamefont {Foffi}, \citenamefont {Smallenburg},\ and\ \citenamefont
  {Filion}}]{boattini2020autonomously}%
  \BibitemOpen
  \bibfield  {author} {\bibinfo {author} {\bibfnamefont {E.}~\bibnamefont
  {Boattini}}, \bibinfo {author} {\bibfnamefont {S.}~\bibnamefont
  {Mar{\'\i}n-Aguilar}}, \bibinfo {author} {\bibfnamefont {S.}~\bibnamefont
  {Mitra}}, \bibinfo {author} {\bibfnamefont {G.}~\bibnamefont {Foffi}},
  \bibinfo {author} {\bibfnamefont {F.}~\bibnamefont {Smallenburg}}, \ and\
  \bibinfo {author} {\bibfnamefont {L.}~\bibnamefont {Filion}},\ }\href@noop {}
  {\bibfield  {journal} {\bibinfo  {journal} {arXiv preprint arXiv:2003.00586}\
  } (\bibinfo {year} {2020})}\BibitemShut {NoStop}%
\bibitem [{\citenamefont {Biroli}\ and\ \citenamefont
  {Urbani}(2018)}]{BiroliUrbani}%
  \BibitemOpen
  \bibfield  {author} {\bibinfo {author} {\bibfnamefont {G.}~\bibnamefont
  {Biroli}}\ and\ \bibinfo {author} {\bibfnamefont {P.}~\bibnamefont
  {Urbani}},\ }\href@noop {} {\bibfield  {journal} {\bibinfo  {journal}
  {SciPost Physics}\ }\textbf {\bibinfo {volume} {4}},\ \bibinfo {pages} {020}
  (\bibinfo {year} {2018})}\BibitemShut {NoStop}%
\bibitem [{\citenamefont {Rainone}\ and\ \citenamefont
  {Urbani}(2016)}]{RainoneUrbani}%
  \BibitemOpen
  \bibfield  {author} {\bibinfo {author} {\bibfnamefont {C.}~\bibnamefont
  {Rainone}}\ and\ \bibinfo {author} {\bibfnamefont {P.}~\bibnamefont
  {Urbani}},\ }\href@noop {} {\bibfield  {journal} {\bibinfo  {journal}
  {Journal of Statistical Mechanics: Theory and Experiment}\ }\textbf {\bibinfo
  {volume} {2016}},\ \bibinfo {pages} {053302} (\bibinfo {year}
  {2016})}\BibitemShut {NoStop}%
\bibitem [{\citenamefont {Rainone}\ \emph {et~al.}(2015)\citenamefont
  {Rainone}, \citenamefont {Urbani}, \citenamefont {Yoshino},\ and\
  \citenamefont {Zamponi}}]{RainoneUrbaniYoshinoZamponi}%
  \BibitemOpen
  \bibfield  {author} {\bibinfo {author} {\bibfnamefont {C.}~\bibnamefont
  {Rainone}}, \bibinfo {author} {\bibfnamefont {P.}~\bibnamefont {Urbani}},
  \bibinfo {author} {\bibfnamefont {H.}~\bibnamefont {Yoshino}}, \ and\
  \bibinfo {author} {\bibfnamefont {F.}~\bibnamefont {Zamponi}},\ }\href@noop
  {} {\bibfield  {journal} {\bibinfo  {journal} {Physical review letters}\
  }\textbf {\bibinfo {volume} {114}},\ \bibinfo {pages} {015701} (\bibinfo
  {year} {2015})}\BibitemShut {NoStop}%
\bibitem [{\citenamefont {Scalliet}\ \emph
  {et~al.}(2019{\natexlab{a}})\citenamefont {Scalliet}, \citenamefont
  {Berthier},\ and\ \citenamefont {Zamponi}}]{softmeanfield}%
  \BibitemOpen
  \bibfield  {author} {\bibinfo {author} {\bibfnamefont {C.}~\bibnamefont
  {Scalliet}}, \bibinfo {author} {\bibfnamefont {L.}~\bibnamefont {Berthier}},
  \ and\ \bibinfo {author} {\bibfnamefont {F.}~\bibnamefont {Zamponi}},\
  }\href@noop {} {\bibfield  {journal} {\bibinfo  {journal} {Physical Review
  E}\ }\textbf {\bibinfo {volume} {99}},\ \bibinfo {pages} {012107} (\bibinfo
  {year} {2019}{\natexlab{a}})}\BibitemShut {NoStop}%
\bibitem [{\citenamefont {Parisi}\ and\ \citenamefont
  {Slanina}(2000)}]{ParisiSlanina}%
  \BibitemOpen
  \bibfield  {author} {\bibinfo {author} {\bibfnamefont {G.}~\bibnamefont
  {Parisi}}\ and\ \bibinfo {author} {\bibfnamefont {F.}~\bibnamefont
  {Slanina}},\ }\href@noop {} {\bibfield  {journal} {\bibinfo  {journal}
  {Physical Review E}\ }\textbf {\bibinfo {volume} {62}},\ \bibinfo {pages}
  {6554} (\bibinfo {year} {2000})}\BibitemShut {NoStop}%
\bibitem [{\citenamefont {Parisi}\ and\ \citenamefont
  {Zamponi}(2010)}]{parisi_mean-field_2010}%
  \BibitemOpen
  \bibfield  {author} {\bibinfo {author} {\bibfnamefont {G.}~\bibnamefont
  {Parisi}}\ and\ \bibinfo {author} {\bibfnamefont {F.}~\bibnamefont
  {Zamponi}},\ }\href {\doibase 10.1103/RevModPhys.82.789} {\bibfield
  {journal} {\bibinfo  {journal} {Rev. Mod. Phys.}\ }\textbf {\bibinfo {volume}
  {82}},\ \bibinfo {pages} {789} (\bibinfo {year} {2010})}\BibitemShut
  {NoStop}%
\bibitem [{\citenamefont {Gardner}(1985)}]{Gardner}%
  \BibitemOpen
  \bibfield  {author} {\bibinfo {author} {\bibfnamefont {E.}~\bibnamefont
  {Gardner}},\ }\href@noop {} {\bibfield  {journal} {\bibinfo  {journal}
  {Nuclear Physics B}\ }\textbf {\bibinfo {volume} {257}},\ \bibinfo {pages}
  {747} (\bibinfo {year} {1985})}\BibitemShut {NoStop}%
\bibitem [{\citenamefont {Berthier}\ \emph
  {et~al.}(2019{\natexlab{d}})\citenamefont {Berthier}, \citenamefont {Biroli},
  \citenamefont {Charbonneau}, \citenamefont {Corwin}, \citenamefont {Franz},\
  and\ \citenamefont {Zamponi}}]{berthier2019gardner}%
  \BibitemOpen
  \bibfield  {author} {\bibinfo {author} {\bibfnamefont {L.}~\bibnamefont
  {Berthier}}, \bibinfo {author} {\bibfnamefont {G.}~\bibnamefont {Biroli}},
  \bibinfo {author} {\bibfnamefont {P.}~\bibnamefont {Charbonneau}}, \bibinfo
  {author} {\bibfnamefont {E.~I.}\ \bibnamefont {Corwin}}, \bibinfo {author}
  {\bibfnamefont {S.}~\bibnamefont {Franz}}, \ and\ \bibinfo {author}
  {\bibfnamefont {F.}~\bibnamefont {Zamponi}},\ }\href@noop {} {\bibfield
  {journal} {\bibinfo  {journal} {The Journal of chemical physics}\ }\textbf
  {\bibinfo {volume} {151}},\ \bibinfo {pages} {010901} (\bibinfo {year}
  {2019}{\natexlab{d}})}\BibitemShut {NoStop}%
\bibitem [{\citenamefont {Franz}\ \emph {et~al.}(2017)\citenamefont {Franz},
  \citenamefont {Parisi}, \citenamefont {Sevelev}, \citenamefont {Urbani},\
  and\ \citenamefont {Zamponi}}]{universality_franz_2017}%
  \BibitemOpen
  \bibfield  {author} {\bibinfo {author} {\bibfnamefont {S.}~\bibnamefont
  {Franz}}, \bibinfo {author} {\bibfnamefont {G.}~\bibnamefont {Parisi}},
  \bibinfo {author} {\bibfnamefont {M.}~\bibnamefont {Sevelev}}, \bibinfo
  {author} {\bibfnamefont {P.}~\bibnamefont {Urbani}}, \ and\ \bibinfo {author}
  {\bibfnamefont {F.}~\bibnamefont {Zamponi}},\ }\href@noop {} {\bibfield
  {journal} {\bibinfo  {journal} {SciPost Phys}\ }\textbf {\bibinfo {volume}
  {2}},\ \bibinfo {pages} {019} (\bibinfo {year} {2017})}\BibitemShut {NoStop}%
\bibitem [{\citenamefont {Scalliet}\ and\ \citenamefont
  {Berthier}(2019)}]{CamilleLudo}%
  \BibitemOpen
  \bibfield  {author} {\bibinfo {author} {\bibfnamefont {C.}~\bibnamefont
  {Scalliet}}\ and\ \bibinfo {author} {\bibfnamefont {L.}~\bibnamefont
  {Berthier}},\ }\href@noop {} {\bibfield  {journal} {\bibinfo  {journal}
  {Physical review letters}\ }\textbf {\bibinfo {volume} {122}},\ \bibinfo
  {pages} {255502} (\bibinfo {year} {2019})}\BibitemShut {NoStop}%
\bibitem [{\citenamefont {Liao}\ and\ \citenamefont
  {Berthier}(2019)}]{liao2019hierarchical}%
  \BibitemOpen
  \bibfield  {author} {\bibinfo {author} {\bibfnamefont {Q.}~\bibnamefont
  {Liao}}\ and\ \bibinfo {author} {\bibfnamefont {L.}~\bibnamefont
  {Berthier}},\ }\href@noop {} {\bibfield  {journal} {\bibinfo  {journal}
  {Physical Review X}\ }\textbf {\bibinfo {volume} {9}},\ \bibinfo {pages}
  {011049} (\bibinfo {year} {2019})}\BibitemShut {NoStop}%
\bibitem [{\citenamefont {Scalliet}\ \emph
  {et~al.}(2019{\natexlab{b}})\citenamefont {Scalliet}, \citenamefont
  {Berthier},\ and\ \citenamefont {Zamponi}}]{scalliet2019nature}%
  \BibitemOpen
  \bibfield  {author} {\bibinfo {author} {\bibfnamefont {C.}~\bibnamefont
  {Scalliet}}, \bibinfo {author} {\bibfnamefont {L.}~\bibnamefont {Berthier}},
  \ and\ \bibinfo {author} {\bibfnamefont {F.}~\bibnamefont {Zamponi}},\
  }\href@noop {} {\bibfield  {journal} {\bibinfo  {journal} {Nature
  communications}\ }\textbf {\bibinfo {volume} {10}},\ \bibinfo {pages} {1}
  (\bibinfo {year} {2019}{\natexlab{b}})}\BibitemShut {NoStop}%
\bibitem [{\citenamefont {Berthier}\ \emph
  {et~al.}(2016{\natexlab{a}})\citenamefont {Berthier}, \citenamefont
  {Charbonneau}, \citenamefont {Jin}, \citenamefont {Parisi}, \citenamefont
  {Seoane},\ and\ \citenamefont {Zamponi}}]{PNASgardner}%
  \BibitemOpen
  \bibfield  {author} {\bibinfo {author} {\bibfnamefont {L.}~\bibnamefont
  {Berthier}}, \bibinfo {author} {\bibfnamefont {P.}~\bibnamefont
  {Charbonneau}}, \bibinfo {author} {\bibfnamefont {Y.}~\bibnamefont {Jin}},
  \bibinfo {author} {\bibfnamefont {G.}~\bibnamefont {Parisi}}, \bibinfo
  {author} {\bibfnamefont {B.}~\bibnamefont {Seoane}}, \ and\ \bibinfo {author}
  {\bibfnamefont {F.}~\bibnamefont {Zamponi}},\ }\href@noop {} {\bibfield
  {journal} {\bibinfo  {journal} {Proceedings of the National Academy of
  Sciences}\ }\textbf {\bibinfo {volume} {113}},\ \bibinfo {pages} {8397}
  (\bibinfo {year} {2016}{\natexlab{a}})}\BibitemShut {NoStop}%
\bibitem [{\citenamefont {Biroli}\ and\ \citenamefont
  {Urbani}(2016)}]{biroli2016breakdown}%
  \BibitemOpen
  \bibfield  {author} {\bibinfo {author} {\bibfnamefont {G.}~\bibnamefont
  {Biroli}}\ and\ \bibinfo {author} {\bibfnamefont {P.}~\bibnamefont
  {Urbani}},\ }\href@noop {} {\bibfield  {journal} {\bibinfo  {journal} {Nature
  physics}\ }\textbf {\bibinfo {volume} {12}},\ \bibinfo {pages} {1130}
  (\bibinfo {year} {2016})}\BibitemShut {NoStop}%
\bibitem [{\citenamefont {Scalliet}\ \emph {et~al.}(2017)\citenamefont
  {Scalliet}, \citenamefont {Berthier},\ and\ \citenamefont
  {Zamponi}}]{scalliet2017absence}%
  \BibitemOpen
  \bibfield  {author} {\bibinfo {author} {\bibfnamefont {C.}~\bibnamefont
  {Scalliet}}, \bibinfo {author} {\bibfnamefont {L.}~\bibnamefont {Berthier}},
  \ and\ \bibinfo {author} {\bibfnamefont {F.}~\bibnamefont {Zamponi}},\
  }\href@noop {} {\bibfield  {journal} {\bibinfo  {journal} {Physical review
  letters}\ }\textbf {\bibinfo {volume} {119}},\ \bibinfo {pages} {205501}
  (\bibinfo {year} {2017})}\BibitemShut {NoStop}%
\bibitem [{\citenamefont {Chaudhuri}\ \emph {et~al.}(2010)\citenamefont
  {Chaudhuri}, \citenamefont {Berthier},\ and\ \citenamefont
  {Sastry}}]{chaudhuri2010jamming}%
  \BibitemOpen
  \bibfield  {author} {\bibinfo {author} {\bibfnamefont {P.}~\bibnamefont
  {Chaudhuri}}, \bibinfo {author} {\bibfnamefont {L.}~\bibnamefont {Berthier}},
  \ and\ \bibinfo {author} {\bibfnamefont {S.}~\bibnamefont {Sastry}},\
  }\href@noop {} {\bibfield  {journal} {\bibinfo  {journal} {Physical review
  letters}\ }\textbf {\bibinfo {volume} {104}},\ \bibinfo {pages} {165701}
  (\bibinfo {year} {2010})}\BibitemShut {NoStop}%
\bibitem [{\citenamefont {Liu}\ and\ \citenamefont
  {Nagel}(2010)}]{liu2010jamming}%
  \BibitemOpen
  \bibfield  {author} {\bibinfo {author} {\bibfnamefont {A.~J.}\ \bibnamefont
  {Liu}}\ and\ \bibinfo {author} {\bibfnamefont {S.~R.}\ \bibnamefont
  {Nagel}},\ }\href@noop {} {\bibfield  {journal} {\bibinfo  {journal} {Annu.
  Rev. Condens. Matter Phys.}\ }\textbf {\bibinfo {volume} {1}},\ \bibinfo
  {pages} {347} (\bibinfo {year} {2010})}\BibitemShut {NoStop}%
\bibitem [{\citenamefont {O'Hern}\ \emph {et~al.}(2002)\citenamefont {O'Hern},
  \citenamefont {Langer}, \citenamefont {Liu},\ and\ \citenamefont
  {Nagel}}]{ohern_random_2002}%
  \BibitemOpen
  \bibfield  {author} {\bibinfo {author} {\bibfnamefont {C.~S.}\ \bibnamefont
  {O'Hern}}, \bibinfo {author} {\bibfnamefont {S.~A.}\ \bibnamefont {Langer}},
  \bibinfo {author} {\bibfnamefont {A.~J.}\ \bibnamefont {Liu}}, \ and\
  \bibinfo {author} {\bibfnamefont {S.~R.}\ \bibnamefont {Nagel}},\ }\href@noop
  {} {\bibfield  {journal} {\bibinfo  {journal} {Physical Review Letters}\
  }\textbf {\bibinfo {volume} {88}},\ \bibinfo {pages} {075507} (\bibinfo
  {year} {2002})}\BibitemShut {NoStop}%
\bibitem [{\citenamefont {O’hern}\ \emph {et~al.}(2003)\citenamefont
  {O’hern}, \citenamefont {Silbert}, \citenamefont {Liu},\ and\ \citenamefont
  {Nagel}}]{ohern_random_2003}%
  \BibitemOpen
  \bibfield  {author} {\bibinfo {author} {\bibfnamefont {C.~S.}\ \bibnamefont
  {O’hern}}, \bibinfo {author} {\bibfnamefont {L.~E.}\ \bibnamefont
  {Silbert}}, \bibinfo {author} {\bibfnamefont {A.~J.}\ \bibnamefont {Liu}}, \
  and\ \bibinfo {author} {\bibfnamefont {S.~R.}\ \bibnamefont {Nagel}},\
  }\href@noop {} {\bibfield  {journal} {\bibinfo  {journal} {Physical Review
  E}\ }\textbf {\bibinfo {volume} {68}},\ \bibinfo {pages} {011306} (\bibinfo
  {year} {2003})}\BibitemShut {NoStop}%
\bibitem [{\citenamefont {Durian}(1995)}]{durian95}%
  \BibitemOpen
  \bibfield  {author} {\bibinfo {author} {\bibfnamefont {D.~J.}\ \bibnamefont
  {Durian}},\ }\href@noop {} {\bibfield  {journal} {\bibinfo  {journal}
  {Physical review letters}\ }\textbf {\bibinfo {volume} {75}},\ \bibinfo
  {pages} {4780} (\bibinfo {year} {1995})}\BibitemShut {NoStop}%
\bibitem [{\citenamefont {Wyart}(2012)}]{wyart2012marginal}%
  \BibitemOpen
  \bibfield  {author} {\bibinfo {author} {\bibfnamefont {M.}~\bibnamefont
  {Wyart}},\ }\href@noop {} {\bibfield  {journal} {\bibinfo  {journal}
  {Physical review letters}\ }\textbf {\bibinfo {volume} {109}},\ \bibinfo
  {pages} {125502} (\bibinfo {year} {2012})}\BibitemShut {NoStop}%
\bibitem [{\citenamefont {Wyart}\ \emph {et~al.}(2005)\citenamefont {Wyart},
  \citenamefont {Nagel},\ and\ \citenamefont {Witten}}]{cutting-argument}%
  \BibitemOpen
  \bibfield  {author} {\bibinfo {author} {\bibfnamefont {M.}~\bibnamefont
  {Wyart}}, \bibinfo {author} {\bibfnamefont {S.~R.}\ \bibnamefont {Nagel}}, \
  and\ \bibinfo {author} {\bibfnamefont {T.~A.}\ \bibnamefont {Witten}},\
  }\href@noop {} {\bibfield  {journal} {\bibinfo  {journal} {EPL (Europhysics
  Letters)}\ }\textbf {\bibinfo {volume} {72}},\ \bibinfo {pages} {486}
  (\bibinfo {year} {2005})}\BibitemShut {NoStop}%
\bibitem [{\citenamefont {Wyart}(2010)}]{wyart_scaling_2010}%
  \BibitemOpen
  \bibfield  {author} {\bibinfo {author} {\bibfnamefont {M.}~\bibnamefont
  {Wyart}},\ }\href@noop {} {\bibfield  {journal} {\bibinfo  {journal} {EPL
  (Europhysics Letters)}\ }\textbf {\bibinfo {volume} {89}},\ \bibinfo {pages}
  {64001} (\bibinfo {year} {2010})}\BibitemShut {NoStop}%
\bibitem [{\citenamefont {Degiuli}\ \emph {et~al.}(2015)\citenamefont
  {Degiuli}, \citenamefont {Lerner},\ and\ \citenamefont
  {Wyart}}]{degiuli_lerner_wyart_theory_2015}%
  \BibitemOpen
  \bibfield  {author} {\bibinfo {author} {\bibfnamefont {E.}~\bibnamefont
  {Degiuli}}, \bibinfo {author} {\bibfnamefont {E.}~\bibnamefont {Lerner}}, \
  and\ \bibinfo {author} {\bibfnamefont {M.}~\bibnamefont {Wyart}},\
  }\href@noop {} {\bibfield  {journal} {\bibinfo  {journal} {The Journal of
  chemical physics}\ }\textbf {\bibinfo {volume} {142}},\ \bibinfo {pages}
  {164503} (\bibinfo {year} {2015})}\BibitemShut {NoStop}%
\bibitem [{\citenamefont {Charbonneau}\ \emph {et~al.}(2012)\citenamefont
  {Charbonneau}, \citenamefont {Corwin}, \citenamefont {Parisi},\ and\
  \citenamefont {Zamponi}}]{charbonneau_universal_2012}%
  \BibitemOpen
  \bibfield  {author} {\bibinfo {author} {\bibfnamefont {P.}~\bibnamefont
  {Charbonneau}}, \bibinfo {author} {\bibfnamefont {E.~I.}\ \bibnamefont
  {Corwin}}, \bibinfo {author} {\bibfnamefont {G.}~\bibnamefont {Parisi}}, \
  and\ \bibinfo {author} {\bibfnamefont {F.}~\bibnamefont {Zamponi}},\
  }\href@noop {} {\bibfield  {journal} {\bibinfo  {journal} {Physical review
  letters}\ }\textbf {\bibinfo {volume} {109}},\ \bibinfo {pages} {205501}
  (\bibinfo {year} {2012})}\BibitemShut {NoStop}%
\bibitem [{\citenamefont {Charbonneau}\ \emph {et~al.}(2015)\citenamefont
  {Charbonneau}, \citenamefont {Corwin}, \citenamefont {Parisi},\ and\
  \citenamefont {Zamponi}}]{charbonneau_jamming_2015}%
  \BibitemOpen
  \bibfield  {author} {\bibinfo {author} {\bibfnamefont {P.}~\bibnamefont
  {Charbonneau}}, \bibinfo {author} {\bibfnamefont {E.~I.}\ \bibnamefont
  {Corwin}}, \bibinfo {author} {\bibfnamefont {G.}~\bibnamefont {Parisi}}, \
  and\ \bibinfo {author} {\bibfnamefont {F.}~\bibnamefont {Zamponi}},\
  }\href@noop {} {\bibfield  {journal} {\bibinfo  {journal} {Physical review
  letters}\ }\textbf {\bibinfo {volume} {114}},\ \bibinfo {pages} {125504}
  (\bibinfo {year} {2015})}\BibitemShut {NoStop}%
\bibitem [{\citenamefont {DeGiuli}\ \emph {et~al.}(2014)\citenamefont
  {DeGiuli}, \citenamefont {Lerner}, \citenamefont {Brito},\ and\ \citenamefont
  {Wyart}}]{degiuli_force_2014}%
  \BibitemOpen
  \bibfield  {author} {\bibinfo {author} {\bibfnamefont {E.}~\bibnamefont
  {DeGiuli}}, \bibinfo {author} {\bibfnamefont {E.}~\bibnamefont {Lerner}},
  \bibinfo {author} {\bibfnamefont {C.}~\bibnamefont {Brito}}, \ and\ \bibinfo
  {author} {\bibfnamefont {M.}~\bibnamefont {Wyart}},\ }\href@noop {}
  {\bibfield  {journal} {\bibinfo  {journal} {Proceedings of the National
  Academy of Sciences}\ }\textbf {\bibinfo {volume} {111}},\ \bibinfo {pages}
  {17054} (\bibinfo {year} {2014})}\BibitemShut {NoStop}%
\bibitem [{\citenamefont {Lerner}\ \emph {et~al.}(2013)\citenamefont {Lerner},
  \citenamefont {D{\"u}ring},\ and\ \citenamefont {Wyart}}]{lerner13}%
  \BibitemOpen
  \bibfield  {author} {\bibinfo {author} {\bibfnamefont {E.}~\bibnamefont
  {Lerner}}, \bibinfo {author} {\bibfnamefont {G.}~\bibnamefont {D{\"u}ring}},
  \ and\ \bibinfo {author} {\bibfnamefont {M.}~\bibnamefont {Wyart}},\
  }\href@noop {} {\bibfield  {journal} {\bibinfo  {journal} {Soft Matter}\
  }\textbf {\bibinfo {volume} {9}},\ \bibinfo {pages} {8252} (\bibinfo {year}
  {2013})}\BibitemShut {NoStop}%
\bibitem [{\citenamefont {Ikeda}\ \emph {et~al.}(2013)\citenamefont {Ikeda},
  \citenamefont {Berthier},\ and\ \citenamefont
  {Biroli}}]{ikeda_berthier_biroli}%
  \BibitemOpen
  \bibfield  {author} {\bibinfo {author} {\bibfnamefont {A.}~\bibnamefont
  {Ikeda}}, \bibinfo {author} {\bibfnamefont {L.}~\bibnamefont {Berthier}}, \
  and\ \bibinfo {author} {\bibfnamefont {G.}~\bibnamefont {Biroli}},\
  }\href@noop {} {\bibfield  {journal} {\bibinfo  {journal} {The Journal of
  chemical physics}\ }\textbf {\bibinfo {volume} {138}},\ \bibinfo {pages}
  {12A507} (\bibinfo {year} {2013})}\BibitemShut {NoStop}%
\bibitem [{\citenamefont {Brito}\ and\ \citenamefont
  {Wyart}(2009)}]{BritoWyart}%
  \BibitemOpen
  \bibfield  {author} {\bibinfo {author} {\bibfnamefont {C.}~\bibnamefont
  {Brito}}\ and\ \bibinfo {author} {\bibfnamefont {M.}~\bibnamefont {Wyart}},\
  }\href@noop {} {\bibfield  {journal} {\bibinfo  {journal} {The Journal of
  chemical physics}\ }\textbf {\bibinfo {volume} {131}},\ \bibinfo {pages}
  {149} (\bibinfo {year} {2009})}\BibitemShut {NoStop}%
\bibitem [{\citenamefont {Schreck}\ \emph {et~al.}(2011)\citenamefont
  {Schreck}, \citenamefont {Bertrand}, \citenamefont {O’Hern},\ and\
  \citenamefont {Shattuck}}]{schreck2011repulsive}%
  \BibitemOpen
  \bibfield  {author} {\bibinfo {author} {\bibfnamefont {C.~F.}\ \bibnamefont
  {Schreck}}, \bibinfo {author} {\bibfnamefont {T.}~\bibnamefont {Bertrand}},
  \bibinfo {author} {\bibfnamefont {C.~S.}\ \bibnamefont {O’Hern}}, \ and\
  \bibinfo {author} {\bibfnamefont {M.}~\bibnamefont {Shattuck}},\ }\href@noop
  {} {\bibfield  {journal} {\bibinfo  {journal} {Physical review letters}\
  }\textbf {\bibinfo {volume} {107}},\ \bibinfo {pages} {078301} (\bibinfo
  {year} {2011})}\BibitemShut {NoStop}%
\bibitem [{\citenamefont {Mizuno}\ \emph {et~al.}(2017)\citenamefont {Mizuno},
  \citenamefont {Shiba},\ and\ \citenamefont {Ikeda}}]{mizuno_continuum_2017}%
  \BibitemOpen
  \bibfield  {author} {\bibinfo {author} {\bibfnamefont {H.}~\bibnamefont
  {Mizuno}}, \bibinfo {author} {\bibfnamefont {H.}~\bibnamefont {Shiba}}, \
  and\ \bibinfo {author} {\bibfnamefont {A.}~\bibnamefont {Ikeda}},\
  }\href@noop {} {\bibfield  {journal} {\bibinfo  {journal} {Proceedings of the
  National Academy of Sciences}\ }\textbf {\bibinfo {volume} {114}},\ \bibinfo
  {pages} {E9767} (\bibinfo {year} {2017})}\BibitemShut {NoStop}%
\bibitem [{\citenamefont {Franz}\ \emph {et~al.}(2015)\citenamefont {Franz},
  \citenamefont {Parisi}, \citenamefont {Urbani},\ and\ \citenamefont
  {Zamponi}}]{franz_universal_2015}%
  \BibitemOpen
  \bibfield  {author} {\bibinfo {author} {\bibfnamefont {S.}~\bibnamefont
  {Franz}}, \bibinfo {author} {\bibfnamefont {G.}~\bibnamefont {Parisi}},
  \bibinfo {author} {\bibfnamefont {P.}~\bibnamefont {Urbani}}, \ and\ \bibinfo
  {author} {\bibfnamefont {F.}~\bibnamefont {Zamponi}},\ }\href@noop {}
  {\bibfield  {journal} {\bibinfo  {journal} {Proceedings of the National
  Academy of Sciences}\ }\textbf {\bibinfo {volume} {112}},\ \bibinfo {pages}
  {14539} (\bibinfo {year} {2015})}\BibitemShut {NoStop}%
\bibitem [{\citenamefont {Berthier}\ \emph
  {et~al.}(2011{\natexlab{b}})\citenamefont {Berthier}, \citenamefont
  {Jacquin},\ and\ \citenamefont {Zamponi}}]{jacquin_microscopic_2011}%
  \BibitemOpen
  \bibfield  {author} {\bibinfo {author} {\bibfnamefont {L.}~\bibnamefont
  {Berthier}}, \bibinfo {author} {\bibfnamefont {H.}~\bibnamefont {Jacquin}}, \
  and\ \bibinfo {author} {\bibfnamefont {F.}~\bibnamefont {Zamponi}},\
  }\href@noop {} {\bibfield  {journal} {\bibinfo  {journal} {Physical Review
  E}\ }\textbf {\bibinfo {volume} {84}},\ \bibinfo {pages} {051103} (\bibinfo
  {year} {2011}{\natexlab{b}})}\BibitemShut {NoStop}%
\bibitem [{\citenamefont {Charbonneau}\ \emph
  {et~al.}(2016{\natexlab{a}})\citenamefont {Charbonneau}, \citenamefont
  {Corwin}, \citenamefont {Parisi}, \citenamefont {Poncet},\ and\ \citenamefont
  {Zamponi}}]{charbonneau_universal_2016}%
  \BibitemOpen
  \bibfield  {author} {\bibinfo {author} {\bibfnamefont {P.}~\bibnamefont
  {Charbonneau}}, \bibinfo {author} {\bibfnamefont {E.~I.}\ \bibnamefont
  {Corwin}}, \bibinfo {author} {\bibfnamefont {G.}~\bibnamefont {Parisi}},
  \bibinfo {author} {\bibfnamefont {A.}~\bibnamefont {Poncet}}, \ and\ \bibinfo
  {author} {\bibfnamefont {F.}~\bibnamefont {Zamponi}},\ }\href@noop {}
  {\bibfield  {journal} {\bibinfo  {journal} {Physical review letters}\
  }\textbf {\bibinfo {volume} {117}},\ \bibinfo {pages} {045503} (\bibinfo
  {year} {2016}{\natexlab{a}})}\BibitemShut {NoStop}%
\bibitem [{\citenamefont {Lerner}\ \emph {et~al.}(2016)\citenamefont {Lerner},
  \citenamefont {D{\"u}ring},\ and\ \citenamefont
  {Bouchbinder}}]{lerner_statistics_2016}%
  \BibitemOpen
  \bibfield  {author} {\bibinfo {author} {\bibfnamefont {E.}~\bibnamefont
  {Lerner}}, \bibinfo {author} {\bibfnamefont {G.}~\bibnamefont {D{\"u}ring}},
  \ and\ \bibinfo {author} {\bibfnamefont {E.}~\bibnamefont {Bouchbinder}},\
  }\href@noop {} {\bibfield  {journal} {\bibinfo  {journal} {Physical review
  letters}\ }\textbf {\bibinfo {volume} {117}},\ \bibinfo {pages} {035501}
  (\bibinfo {year} {2016})}\BibitemShut {NoStop}%
\bibitem [{\citenamefont {Yoshino}\ and\ \citenamefont
  {Zamponi}(2014)}]{yoshino_shear_2014}%
  \BibitemOpen
  \bibfield  {author} {\bibinfo {author} {\bibfnamefont {H.}~\bibnamefont
  {Yoshino}}\ and\ \bibinfo {author} {\bibfnamefont {F.}~\bibnamefont
  {Zamponi}},\ }\href@noop {} {\bibfield  {journal} {\bibinfo  {journal}
  {Physical Review E}\ }\textbf {\bibinfo {volume} {90}},\ \bibinfo {pages}
  {022302} (\bibinfo {year} {2014})}\BibitemShut {NoStop}%
\bibitem [{\citenamefont {Urbani}\ and\ \citenamefont
  {Zamponi}(2017)}]{urbani2017shear}%
  \BibitemOpen
  \bibfield  {author} {\bibinfo {author} {\bibfnamefont {P.}~\bibnamefont
  {Urbani}}\ and\ \bibinfo {author} {\bibfnamefont {F.}~\bibnamefont
  {Zamponi}},\ }\href@noop {} {\bibfield  {journal} {\bibinfo  {journal}
  {Physical review letters}\ }\textbf {\bibinfo {volume} {118}},\ \bibinfo
  {pages} {038001} (\bibinfo {year} {2017})}\BibitemShut {NoStop}%
\bibitem [{\citenamefont {Peters}\ \emph {et~al.}(2016)\citenamefont {Peters},
  \citenamefont {Majumdar},\ and\ \citenamefont {Jaeger}}]{peters2016direct}%
  \BibitemOpen
  \bibfield  {author} {\bibinfo {author} {\bibfnamefont {I.~R.}\ \bibnamefont
  {Peters}}, \bibinfo {author} {\bibfnamefont {S.}~\bibnamefont {Majumdar}}, \
  and\ \bibinfo {author} {\bibfnamefont {H.~M.}\ \bibnamefont {Jaeger}},\
  }\href@noop {} {\bibfield  {journal} {\bibinfo  {journal} {Nature}\ }\textbf
  {\bibinfo {volume} {532}},\ \bibinfo {pages} {214} (\bibinfo {year}
  {2016})}\BibitemShut {NoStop}%
\bibitem [{\citenamefont {Parisi}\ \emph {et~al.}(2017)\citenamefont {Parisi},
  \citenamefont {Procaccia}, \citenamefont {Rainone},\ and\ \citenamefont
  {Singh}}]{parisi_shear_2017}%
  \BibitemOpen
  \bibfield  {author} {\bibinfo {author} {\bibfnamefont {G.}~\bibnamefont
  {Parisi}}, \bibinfo {author} {\bibfnamefont {I.}~\bibnamefont {Procaccia}},
  \bibinfo {author} {\bibfnamefont {C.}~\bibnamefont {Rainone}}, \ and\
  \bibinfo {author} {\bibfnamefont {M.}~\bibnamefont {Singh}},\ }\href@noop {}
  {\bibfield  {journal} {\bibinfo  {journal} {Proceedings of the National
  Academy of Sciences}\ }\textbf {\bibinfo {volume} {114}},\ \bibinfo {pages}
  {5577} (\bibinfo {year} {2017})}\BibitemShut {NoStop}%
\bibitem [{\citenamefont {Lin}\ \emph {et~al.}(2014)\citenamefont {Lin},
  \citenamefont {Lerner}, \citenamefont {Rosso},\ and\ \citenamefont
  {Wyart}}]{lin_scaling_2014}%
  \BibitemOpen
  \bibfield  {author} {\bibinfo {author} {\bibfnamefont {J.}~\bibnamefont
  {Lin}}, \bibinfo {author} {\bibfnamefont {E.}~\bibnamefont {Lerner}},
  \bibinfo {author} {\bibfnamefont {A.}~\bibnamefont {Rosso}}, \ and\ \bibinfo
  {author} {\bibfnamefont {M.}~\bibnamefont {Wyart}},\ }\href@noop {}
  {\bibfield  {journal} {\bibinfo  {journal} {Proceedings of the National
  Academy of Sciences}\ }\textbf {\bibinfo {volume} {111}},\ \bibinfo {pages}
  {14382} (\bibinfo {year} {2014})}\BibitemShut {NoStop}%
\bibitem [{\citenamefont {Ozawa}\ \emph
  {et~al.}(2018{\natexlab{b}})\citenamefont {Ozawa}, \citenamefont {Berthier},
  \citenamefont {Biroli}, \citenamefont {Rosso},\ and\ \citenamefont
  {Tarjus}}]{ozawa_random_2018}%
  \BibitemOpen
  \bibfield  {author} {\bibinfo {author} {\bibfnamefont {M.}~\bibnamefont
  {Ozawa}}, \bibinfo {author} {\bibfnamefont {L.}~\bibnamefont {Berthier}},
  \bibinfo {author} {\bibfnamefont {G.}~\bibnamefont {Biroli}}, \bibinfo
  {author} {\bibfnamefont {A.}~\bibnamefont {Rosso}}, \ and\ \bibinfo {author}
  {\bibfnamefont {G.}~\bibnamefont {Tarjus}},\ }\href@noop {} {\bibfield
  {journal} {\bibinfo  {journal} {Proceedings of the National Academy of
  Sciences}\ }\textbf {\bibinfo {volume} {115}},\ \bibinfo {pages} {6656}
  (\bibinfo {year} {2018}{\natexlab{b}})}\BibitemShut {NoStop}%
\bibitem [{\citenamefont {Berthier}\ \emph
  {et~al.}(2016{\natexlab{b}})\citenamefont {Berthier}, \citenamefont
  {Coslovich}, \citenamefont {Ninarello},\ and\ \citenamefont
  {Ozawa}}]{berthier_equilibrium_2016}%
  \BibitemOpen
  \bibfield  {author} {\bibinfo {author} {\bibfnamefont {L.}~\bibnamefont
  {Berthier}}, \bibinfo {author} {\bibfnamefont {D.}~\bibnamefont {Coslovich}},
  \bibinfo {author} {\bibfnamefont {A.}~\bibnamefont {Ninarello}}, \ and\
  \bibinfo {author} {\bibfnamefont {M.}~\bibnamefont {Ozawa}},\ }\href
  {\doibase 10.1103/PhysRevLett.116.238002} {\bibfield  {journal} {\bibinfo
  {journal} {Physical Review Letters}\ }\textbf {\bibinfo {volume} {116}}
  (\bibinfo {year} {2016}{\natexlab{b}}),\
  10.1103/PhysRevLett.116.238002}\BibitemShut {NoStop}%
\bibitem [{\citenamefont {Ninarello}\ \emph {et~al.}(2017)\citenamefont
  {Ninarello}, \citenamefont {Berthier},\ and\ \citenamefont
  {Coslovich}}]{ninarello_models_2017}%
  \BibitemOpen
  \bibfield  {author} {\bibinfo {author} {\bibfnamefont {A.}~\bibnamefont
  {Ninarello}}, \bibinfo {author} {\bibfnamefont {L.}~\bibnamefont {Berthier}},
  \ and\ \bibinfo {author} {\bibfnamefont {D.}~\bibnamefont {Coslovich}},\
  }\href {\doibase 10.1103/PhysRevX.7.021039} {\bibfield  {journal} {\bibinfo
  {journal} {Physical Review X}\ }\textbf {\bibinfo {volume} {7}} (\bibinfo
  {year} {2017}),\ 10.1103/PhysRevX.7.021039}\BibitemShut {NoStop}%
\bibitem [{\citenamefont {Gazzillo}\ and\ \citenamefont
  {Pastore}(1989)}]{gazzillo_equation_1989}%
  \BibitemOpen
  \bibfield  {author} {\bibinfo {author} {\bibfnamefont {D.}~\bibnamefont
  {Gazzillo}}\ and\ \bibinfo {author} {\bibfnamefont {G.}~\bibnamefont
  {Pastore}},\ }\href {\doibase 10.1016/0009-2614(89)87505-0} {\bibfield
  {journal} {\bibinfo  {journal} {Chemical Physics Letters}\ }\textbf {\bibinfo
  {volume} {159}},\ \bibinfo {pages} {388} (\bibinfo {year}
  {1989})}\BibitemShut {NoStop}%
\bibitem [{\citenamefont {Grigera}\ and\ \citenamefont
  {Parisi}(2001)}]{grigera_fast_2001}%
  \BibitemOpen
  \bibfield  {author} {\bibinfo {author} {\bibfnamefont {T.~S.}\ \bibnamefont
  {Grigera}}\ and\ \bibinfo {author} {\bibfnamefont {G.}~\bibnamefont
  {Parisi}},\ }\href {\doibase 10.1103/PhysRevE.63.045102} {\bibfield
  {journal} {\bibinfo  {journal} {Physical Review E}\ }\textbf {\bibinfo
  {volume} {63}} (\bibinfo {year} {2001}),\
  10.1103/PhysRevE.63.045102}\BibitemShut {NoStop}%
\bibitem [{\citenamefont {Ninarello}(2017)}]{ninarello_computer_nodate}%
  \BibitemOpen
  \bibfield  {author} {\bibinfo {author} {\bibfnamefont {A.~S.}\ \bibnamefont
  {Ninarello}},\ }\emph {\bibinfo {title} {Computer simulations of supercooled
  liquids near the experimental glass transition}},\ \href@noop {} {Ph.D.
  thesis},\ \bibinfo  {school} {Montpellier} (\bibinfo {year}
  {2017})\BibitemShut {NoStop}%
\bibitem [{\citenamefont {Brito}\ \emph {et~al.}(2018)\citenamefont {Brito},
  \citenamefont {Lerner},\ and\ \citenamefont {Wyart}}]{brito_theory_2018}%
  \BibitemOpen
  \bibfield  {author} {\bibinfo {author} {\bibfnamefont {C.}~\bibnamefont
  {Brito}}, \bibinfo {author} {\bibfnamefont {E.}~\bibnamefont {Lerner}}, \
  and\ \bibinfo {author} {\bibfnamefont {M.}~\bibnamefont {Wyart}},\ }\href
  {\doibase 10.1103/PhysRevX.8.031050} {\bibfield  {journal} {\bibinfo
  {journal} {Physical Review X}\ }\textbf {\bibinfo {volume} {8}} (\bibinfo
  {year} {2018}),\ 10.1103/PhysRevX.8.031050}\BibitemShut {NoStop}%
\bibitem [{\citenamefont {Kapteijns}\ \emph {et~al.}(2019)\citenamefont
  {Kapteijns}, \citenamefont {Ji}, \citenamefont {Brito}, \citenamefont
  {Wyart},\ and\ \citenamefont {Lerner}}]{kapteijns_fast_2019}%
  \BibitemOpen
  \bibfield  {author} {\bibinfo {author} {\bibfnamefont {G.}~\bibnamefont
  {Kapteijns}}, \bibinfo {author} {\bibfnamefont {W.}~\bibnamefont {Ji}},
  \bibinfo {author} {\bibfnamefont {C.}~\bibnamefont {Brito}}, \bibinfo
  {author} {\bibfnamefont {M.}~\bibnamefont {Wyart}}, \ and\ \bibinfo {author}
  {\bibfnamefont {E.}~\bibnamefont {Lerner}},\ }\href {\doibase
  10.1103/PhysRevE.99.012106} {\bibfield  {journal} {\bibinfo  {journal}
  {Physical Review E}\ }\textbf {\bibinfo {volume} {99}} (\bibinfo {year}
  {2019}),\ 10.1103/PhysRevE.99.012106}\BibitemShut {NoStop}%
\bibitem [{\citenamefont {Berthier}\ \emph
  {et~al.}(2019{\natexlab{e}})\citenamefont {Berthier}, \citenamefont
  {Flenner}, \citenamefont {Fullerton}, \citenamefont {Scalliet},\ and\
  \citenamefont {Singh}}]{berthier2019efficient}%
  \BibitemOpen
  \bibfield  {author} {\bibinfo {author} {\bibfnamefont {L.}~\bibnamefont
  {Berthier}}, \bibinfo {author} {\bibfnamefont {E.}~\bibnamefont {Flenner}},
  \bibinfo {author} {\bibfnamefont {C.~J.}\ \bibnamefont {Fullerton}}, \bibinfo
  {author} {\bibfnamefont {C.}~\bibnamefont {Scalliet}}, \ and\ \bibinfo
  {author} {\bibfnamefont {M.}~\bibnamefont {Singh}},\ }\href@noop {}
  {\bibfield  {journal} {\bibinfo  {journal} {Journal of Statistical Mechanics:
  Theory and Experiment}\ }\textbf {\bibinfo {volume} {2019}},\ \bibinfo
  {pages} {064004} (\bibinfo {year} {2019}{\natexlab{e}})}\BibitemShut
  {NoStop}%
\bibitem [{\citenamefont {Berthier}\ \emph
  {et~al.}(2019{\natexlab{f}})\citenamefont {Berthier}, \citenamefont
  {Charbonneau},\ and\ \citenamefont {Kundu}}]{berthier_bypassing_2019}%
  \BibitemOpen
  \bibfield  {author} {\bibinfo {author} {\bibfnamefont {L.}~\bibnamefont
  {Berthier}}, \bibinfo {author} {\bibfnamefont {P.}~\bibnamefont
  {Charbonneau}}, \ and\ \bibinfo {author} {\bibfnamefont {J.}~\bibnamefont
  {Kundu}},\ }\href {\doibase 10.1103/PhysRevE.99.031301} {\bibfield  {journal}
  {\bibinfo  {journal} {Physical Review E}\ }\textbf {\bibinfo {volume} {99}}
  (\bibinfo {year} {2019}{\natexlab{f}}),\
  10.1103/PhysRevE.99.031301}\BibitemShut {NoStop}%
\bibitem [{\citenamefont {Parmar}\ \emph {et~al.}(2020)\citenamefont {Parmar},
  \citenamefont {Ozawa},\ and\ \citenamefont
  {Berthier}}]{parmar2020ultrastable}%
  \BibitemOpen
  \bibfield  {author} {\bibinfo {author} {\bibfnamefont {A.~D.~S.}\
  \bibnamefont {Parmar}}, \bibinfo {author} {\bibfnamefont {M.}~\bibnamefont
  {Ozawa}}, \ and\ \bibinfo {author} {\bibfnamefont {L.}~\bibnamefont
  {Berthier}},\ }\href {\doibase 10.1103/PhysRevLett.125.085505} {\bibfield
  {journal} {\bibinfo  {journal} {Phys. Rev. Lett.}\ }\textbf {\bibinfo
  {volume} {125}},\ \bibinfo {pages} {085505} (\bibinfo {year}
  {2020})}\BibitemShut {NoStop}%
\bibitem [{\citenamefont {Coslovich}\ \emph {et~al.}(2018)\citenamefont
  {Coslovich}, \citenamefont {Ozawa},\ and\ \citenamefont
  {Berthier}}]{coslovich2018local}%
  \BibitemOpen
  \bibfield  {author} {\bibinfo {author} {\bibfnamefont {D.}~\bibnamefont
  {Coslovich}}, \bibinfo {author} {\bibfnamefont {M.}~\bibnamefont {Ozawa}}, \
  and\ \bibinfo {author} {\bibfnamefont {L.}~\bibnamefont {Berthier}},\
  }\href@noop {} {\bibfield  {journal} {\bibinfo  {journal} {Journal of
  Physics: Condensed Matter}\ }\textbf {\bibinfo {volume} {30}},\ \bibinfo
  {pages} {144004} (\bibinfo {year} {2018})}\BibitemShut {NoStop}%
\bibitem [{\citenamefont {Ozawa}\ \emph
  {et~al.}(2018{\natexlab{c}})\citenamefont {Ozawa}, \citenamefont {Parisi},\
  and\ \citenamefont {Berthier}}]{ozawa_configurational_2018}%
  \BibitemOpen
  \bibfield  {author} {\bibinfo {author} {\bibfnamefont {M.}~\bibnamefont
  {Ozawa}}, \bibinfo {author} {\bibfnamefont {G.}~\bibnamefont {Parisi}}, \
  and\ \bibinfo {author} {\bibfnamefont {L.}~\bibnamefont {Berthier}},\ }\href
  {\doibase 10.1063/1.5040975} {\bibfield  {journal} {\bibinfo  {journal} {The
  Journal of Chemical Physics}\ }\textbf {\bibinfo {volume} {149}},\ \bibinfo
  {pages} {154501} (\bibinfo {year} {2018}{\natexlab{c}})}\BibitemShut
  {NoStop}%
\bibitem [{\citenamefont {Berthier}\ \emph
  {et~al.}(2019{\natexlab{g}})\citenamefont {Berthier}, \citenamefont
  {Charbonneau}, \citenamefont {Ninarello}, \citenamefont {Ozawa},\ and\
  \citenamefont {Yaida}}]{berthier_zero-temperature_2019}%
  \BibitemOpen
  \bibfield  {author} {\bibinfo {author} {\bibfnamefont {L.}~\bibnamefont
  {Berthier}}, \bibinfo {author} {\bibfnamefont {P.}~\bibnamefont
  {Charbonneau}}, \bibinfo {author} {\bibfnamefont {A.}~\bibnamefont
  {Ninarello}}, \bibinfo {author} {\bibfnamefont {M.}~\bibnamefont {Ozawa}}, \
  and\ \bibinfo {author} {\bibfnamefont {S.}~\bibnamefont {Yaida}},\ }\href
  {\doibase 10.1038/s41467-019-09512-3} {\bibfield  {journal} {\bibinfo
  {journal} {Nature Communications}\ }\textbf {\bibinfo {volume} {10}}
  (\bibinfo {year} {2019}{\natexlab{g}}),\
  10.1038/s41467-019-09512-3}\BibitemShut {NoStop}%
\bibitem [{\citenamefont {Berthier}\ \emph
  {et~al.}(2017{\natexlab{c}})\citenamefont {Berthier}, \citenamefont
  {Charbonneau}, \citenamefont {Coslovich}, \citenamefont {Ninarello},
  \citenamefont {Ozawa},\ and\ \citenamefont
  {Yaida}}]{berthier_configurational_2017}%
  \BibitemOpen
  \bibfield  {author} {\bibinfo {author} {\bibfnamefont {L.}~\bibnamefont
  {Berthier}}, \bibinfo {author} {\bibfnamefont {P.}~\bibnamefont
  {Charbonneau}}, \bibinfo {author} {\bibfnamefont {D.}~\bibnamefont
  {Coslovich}}, \bibinfo {author} {\bibfnamefont {A.}~\bibnamefont
  {Ninarello}}, \bibinfo {author} {\bibfnamefont {M.}~\bibnamefont {Ozawa}}, \
  and\ \bibinfo {author} {\bibfnamefont {S.}~\bibnamefont {Yaida}},\ }\href
  {\doibase 10.1073/pnas.1706860114} {\bibfield  {journal} {\bibinfo  {journal}
  {Proceedings of the National Academy of Sciences}\ }\textbf {\bibinfo
  {volume} {114}},\ \bibinfo {pages} {11356} (\bibinfo {year}
  {2017}{\natexlab{c}})}\BibitemShut {NoStop}%
\bibitem [{\citenamefont {Yaida}\ \emph {et~al.}(2016)\citenamefont {Yaida},
  \citenamefont {Berthier}, \citenamefont {Charbonneau},\ and\ \citenamefont
  {Tarjus}}]{yaida_point--set_2016}%
  \BibitemOpen
  \bibfield  {author} {\bibinfo {author} {\bibfnamefont {S.}~\bibnamefont
  {Yaida}}, \bibinfo {author} {\bibfnamefont {L.}~\bibnamefont {Berthier}},
  \bibinfo {author} {\bibfnamefont {P.}~\bibnamefont {Charbonneau}}, \ and\
  \bibinfo {author} {\bibfnamefont {G.}~\bibnamefont {Tarjus}},\ }\href
  {\doibase 10.1103/PhysRevE.94.032605} {\bibfield  {journal} {\bibinfo
  {journal} {Physical Review E}\ }\textbf {\bibinfo {volume} {94}} (\bibinfo
  {year} {2016}),\ 10.1103/PhysRevE.94.032605}\BibitemShut {NoStop}%
\bibitem [{\citenamefont {Guiselin}\ \emph {et~al.}(2020)\citenamefont
  {Guiselin}, \citenamefont {Berthier},\ and\ \citenamefont
  {Tarjus}}]{guiselin2020random}%
  \BibitemOpen
  \bibfield  {author} {\bibinfo {author} {\bibfnamefont {B.}~\bibnamefont
  {Guiselin}}, \bibinfo {author} {\bibfnamefont {L.}~\bibnamefont {Berthier}},
  \ and\ \bibinfo {author} {\bibfnamefont {G.}~\bibnamefont {Tarjus}},\
  }\href@noop {} {\bibfield  {journal} {\bibinfo  {journal} {arXiv preprint
  arXiv:2004.10555}\ } (\bibinfo {year} {2020})}\BibitemShut {NoStop}%
\bibitem [{\citenamefont {Wang}\ \emph
  {et~al.}(2019{\natexlab{a}})\citenamefont {Wang}, \citenamefont {Ninarello},
  \citenamefont {Guan}, \citenamefont {Berthier}, \citenamefont {Szamel},\ and\
  \citenamefont {Flenner}}]{wang2019low}%
  \BibitemOpen
  \bibfield  {author} {\bibinfo {author} {\bibfnamefont {L.}~\bibnamefont
  {Wang}}, \bibinfo {author} {\bibfnamefont {A.}~\bibnamefont {Ninarello}},
  \bibinfo {author} {\bibfnamefont {P.}~\bibnamefont {Guan}}, \bibinfo {author}
  {\bibfnamefont {L.}~\bibnamefont {Berthier}}, \bibinfo {author}
  {\bibfnamefont {G.}~\bibnamefont {Szamel}}, \ and\ \bibinfo {author}
  {\bibfnamefont {E.}~\bibnamefont {Flenner}},\ }\href@noop {} {\bibfield
  {journal} {\bibinfo  {journal} {Nature communications}\ }\textbf {\bibinfo
  {volume} {10}},\ \bibinfo {pages} {1} (\bibinfo {year}
  {2019}{\natexlab{a}})}\BibitemShut {NoStop}%
\bibitem [{\citenamefont {Wang}\ \emph
  {et~al.}(2019{\natexlab{b}})\citenamefont {Wang}, \citenamefont {Berthier},
  \citenamefont {Flenner}, \citenamefont {Guan},\ and\ \citenamefont
  {Szamel}}]{wang2019sound}%
  \BibitemOpen
  \bibfield  {author} {\bibinfo {author} {\bibfnamefont {L.}~\bibnamefont
  {Wang}}, \bibinfo {author} {\bibfnamefont {L.}~\bibnamefont {Berthier}},
  \bibinfo {author} {\bibfnamefont {E.}~\bibnamefont {Flenner}}, \bibinfo
  {author} {\bibfnamefont {P.}~\bibnamefont {Guan}}, \ and\ \bibinfo {author}
  {\bibfnamefont {G.}~\bibnamefont {Szamel}},\ }\href@noop {} {\bibfield
  {journal} {\bibinfo  {journal} {Soft matter}\ }\textbf {\bibinfo {volume}
  {15}},\ \bibinfo {pages} {7018} (\bibinfo {year}
  {2019}{\natexlab{b}})}\BibitemShut {NoStop}%
\bibitem [{\citenamefont {Wyart}\ and\ \citenamefont
  {Cates}(2017)}]{wyart_does_2017}%
  \BibitemOpen
  \bibfield  {author} {\bibinfo {author} {\bibfnamefont {M.}~\bibnamefont
  {Wyart}}\ and\ \bibinfo {author} {\bibfnamefont {M.~E.}\ \bibnamefont
  {Cates}},\ }\href {\doibase 10.1103/PhysRevLett.119.195501} {\bibfield
  {journal} {\bibinfo  {journal} {Physical Review Letters}\ }\textbf {\bibinfo
  {volume} {119}} (\bibinfo {year} {2017}),\
  10.1103/PhysRevLett.119.195501}\BibitemShut {NoStop}%
\bibitem [{\citenamefont {Guti{\'e}rrez}\ \emph {et~al.}(2019)\citenamefont
  {Guti{\'e}rrez}, \citenamefont {Garrahan},\ and\ \citenamefont
  {Jack}}]{gutierrez2019accelerated}%
  \BibitemOpen
  \bibfield  {author} {\bibinfo {author} {\bibfnamefont {R.}~\bibnamefont
  {Guti{\'e}rrez}}, \bibinfo {author} {\bibfnamefont {J.~P.}\ \bibnamefont
  {Garrahan}}, \ and\ \bibinfo {author} {\bibfnamefont {R.~L.}\ \bibnamefont
  {Jack}},\ }\href@noop {} {\bibfield  {journal} {\bibinfo  {journal} {Journal
  of Statistical Mechanics: Theory and Experiment}\ }\textbf {\bibinfo {volume}
  {2019}},\ \bibinfo {pages} {094006} (\bibinfo {year} {2019})}\BibitemShut
  {NoStop}%
\bibitem [{\citenamefont {Ikeda}\ \emph
  {et~al.}(2017{\natexlab{b}})\citenamefont {Ikeda}, \citenamefont {Zamponi},\
  and\ \citenamefont {Ikeda}}]{ikeda_mean_2017}%
  \BibitemOpen
  \bibfield  {author} {\bibinfo {author} {\bibfnamefont {H.}~\bibnamefont
  {Ikeda}}, \bibinfo {author} {\bibfnamefont {F.}~\bibnamefont {Zamponi}}, \
  and\ \bibinfo {author} {\bibfnamefont {A.}~\bibnamefont {Ikeda}},\ }\href
  {\doibase 10.1063/1.5009116} {\bibfield  {journal} {\bibinfo  {journal} {The
  Journal of Chemical Physics}\ }\textbf {\bibinfo {volume} {147}},\ \bibinfo
  {pages} {234506} (\bibinfo {year} {2017}{\natexlab{b}})}\BibitemShut
  {NoStop}%
\bibitem [{\citenamefont {Szamel}(2018)}]{szamel_theory_2018}%
  \BibitemOpen
  \bibfield  {author} {\bibinfo {author} {\bibfnamefont {G.}~\bibnamefont
  {Szamel}},\ }\href {\doibase 10.1103/PhysRevE.98.050601} {\bibfield
  {journal} {\bibinfo  {journal} {Physical Review E}\ }\textbf {\bibinfo
  {volume} {98}} (\bibinfo {year} {2018}),\
  10.1103/PhysRevE.98.050601}\BibitemShut {NoStop}%
\bibitem [{\citenamefont {Franz}\ and\ \citenamefont
  {Parisi}(1997)}]{franz_phase_1997}%
  \BibitemOpen
  \bibfield  {author} {\bibinfo {author} {\bibfnamefont {S.}~\bibnamefont
  {Franz}}\ and\ \bibinfo {author} {\bibfnamefont {G.}~\bibnamefont {Parisi}},\
  }\href {\doibase 10.1103/PhysRevLett.79.2486} {\bibfield  {journal} {\bibinfo
   {journal} {Physical Review Letters}\ }\textbf {\bibinfo {volume} {79}},\
  \bibinfo {pages} {2486} (\bibinfo {year} {1997})}\BibitemShut {NoStop}%
\bibitem [{\citenamefont {Franz}\ and\ \citenamefont
  {Parisi}(1998)}]{franz_e_1998}%
  \BibitemOpen
  \bibfield  {author} {\bibinfo {author} {\bibfnamefont {S.}~\bibnamefont
  {Franz}}\ and\ \bibinfo {author} {\bibfnamefont {G.}~\bibnamefont {Parisi}},\
  }\href@noop {} {\bibfield  {journal} {\bibinfo  {journal} {Physica A}\ ,\
  \bibinfo {pages} {23}} (\bibinfo {year} {1998})}\BibitemShut {NoStop}%
\bibitem [{\citenamefont {Cardenas}\ \emph {et~al.}(1999)\citenamefont
  {Cardenas}, \citenamefont {Franz},\ and\ \citenamefont
  {Parisi}}]{cardenas_constrained_1999}%
  \BibitemOpen
  \bibfield  {author} {\bibinfo {author} {\bibfnamefont {M.}~\bibnamefont
  {Cardenas}}, \bibinfo {author} {\bibfnamefont {S.}~\bibnamefont {Franz}}, \
  and\ \bibinfo {author} {\bibfnamefont {G.}~\bibnamefont {Parisi}},\ }\href
  {\doibase 10.1063/1.478028} {\bibfield  {journal} {\bibinfo  {journal} {The
  Journal of Chemical Physics}\ }\textbf {\bibinfo {volume} {110}},\ \bibinfo
  {pages} {1726} (\bibinfo {year} {1999})}\BibitemShut {NoStop}%
\bibitem [{\citenamefont {Donati}\ \emph {et~al.}(2002)\citenamefont {Donati},
  \citenamefont {Franz}, \citenamefont {Glotzer},\ and\ \citenamefont
  {Parisi}}]{donati_theory_2002}%
  \BibitemOpen
  \bibfield  {author} {\bibinfo {author} {\bibfnamefont {C.}~\bibnamefont
  {Donati}}, \bibinfo {author} {\bibfnamefont {S.}~\bibnamefont {Franz}},
  \bibinfo {author} {\bibfnamefont {S.~C.}\ \bibnamefont {Glotzer}}, \ and\
  \bibinfo {author} {\bibfnamefont {G.}~\bibnamefont {Parisi}},\ }\href
  {\doibase 10.1016/S0022-3093(02)01461-8} {\bibfield  {journal} {\bibinfo
  {journal} {Journal of Non-Crystalline Solids}\ }\textbf {\bibinfo {volume}
  {307-310}},\ \bibinfo {pages} {215} (\bibinfo {year} {2002})}\BibitemShut
  {NoStop}%
\bibitem [{\citenamefont {Biroli}\ \emph {et~al.}(2014)\citenamefont {Biroli},
  \citenamefont {Cammarota}, \citenamefont {Tarjus},\ and\ \citenamefont
  {Tarzia}}]{biroli2014random}%
  \BibitemOpen
  \bibfield  {author} {\bibinfo {author} {\bibfnamefont {G.}~\bibnamefont
  {Biroli}}, \bibinfo {author} {\bibfnamefont {C.}~\bibnamefont {Cammarota}},
  \bibinfo {author} {\bibfnamefont {G.}~\bibnamefont {Tarjus}}, \ and\ \bibinfo
  {author} {\bibfnamefont {M.}~\bibnamefont {Tarzia}},\ }\href@noop {}
  {\bibfield  {journal} {\bibinfo  {journal} {Physical review letters}\
  }\textbf {\bibinfo {volume} {112}},\ \bibinfo {pages} {175701} (\bibinfo
  {year} {2014})}\BibitemShut {NoStop}%
\bibitem [{\citenamefont {Berthier}(2013)}]{berthier_overlap_2013}%
  \BibitemOpen
  \bibfield  {author} {\bibinfo {author} {\bibfnamefont {L.}~\bibnamefont
  {Berthier}},\ }\href {\doibase 10.1103/PhysRevE.88.022313} {\bibfield
  {journal} {\bibinfo  {journal} {Physical Review E}\ }\textbf {\bibinfo
  {volume} {88}} (\bibinfo {year} {2013}),\
  10.1103/PhysRevE.88.022313}\BibitemShut {NoStop}%
\bibitem [{\citenamefont {Berthier}\ and\ \citenamefont
  {Jack}(2015)}]{berthier_evidence_2015}%
  \BibitemOpen
  \bibfield  {author} {\bibinfo {author} {\bibfnamefont {L.}~\bibnamefont
  {Berthier}}\ and\ \bibinfo {author} {\bibfnamefont {R.~L.}\ \bibnamefont
  {Jack}},\ }\href {\doibase 10.1103/PhysRevLett.114.205701} {\bibfield
  {journal} {\bibinfo  {journal} {Physical Review Letters}\ }\textbf {\bibinfo
  {volume} {114}} (\bibinfo {year} {2015}),\
  10.1103/PhysRevLett.114.205701}\BibitemShut {NoStop}%
\bibitem [{\citenamefont {Jack}\ and\ \citenamefont
  {Garrahan}(2016)}]{jack_phase_2016}%
  \BibitemOpen
  \bibfield  {author} {\bibinfo {author} {\bibfnamefont {R.~L.}\ \bibnamefont
  {Jack}}\ and\ \bibinfo {author} {\bibfnamefont {J.~P.}\ \bibnamefont
  {Garrahan}},\ }\href {\doibase 10.1103/PhysRevLett.116.055702} {\bibfield
  {journal} {\bibinfo  {journal} {Physical Review Letters}\ }\textbf {\bibinfo
  {volume} {116}} (\bibinfo {year} {2016}),\
  10.1103/PhysRevLett.116.055702}\BibitemShut {NoStop}%
\bibitem [{\citenamefont {Parisi}\ and\ \citenamefont
  {Seoane}(2014)}]{parisi2014liquid}%
  \BibitemOpen
  \bibfield  {author} {\bibinfo {author} {\bibfnamefont {G.}~\bibnamefont
  {Parisi}}\ and\ \bibinfo {author} {\bibfnamefont {B.}~\bibnamefont
  {Seoane}},\ }\href@noop {} {\bibfield  {journal} {\bibinfo  {journal}
  {Physical Review E}\ }\textbf {\bibinfo {volume} {89}},\ \bibinfo {pages}
  {022309} (\bibinfo {year} {2014})}\BibitemShut {NoStop}%
\bibitem [{\citenamefont {Biroli}\ \emph {et~al.}(2016)\citenamefont {Biroli},
  \citenamefont {Rulquin}, \citenamefont {Tarjus},\ and\ \citenamefont
  {Tarzia}}]{biroli_role_2016}%
  \BibitemOpen
  \bibfield  {author} {\bibinfo {author} {\bibfnamefont {G.}~\bibnamefont
  {Biroli}}, \bibinfo {author} {\bibfnamefont {C.}~\bibnamefont {Rulquin}},
  \bibinfo {author} {\bibfnamefont {G.}~\bibnamefont {Tarjus}}, \ and\ \bibinfo
  {author} {\bibfnamefont {M.}~\bibnamefont {Tarzia}},\ }\href {\doibase
  10.21468/SciPostPhys.1.1.007} {\bibfield  {journal} {\bibinfo  {journal}
  {SciPost Physics}\ }\textbf {\bibinfo {volume} {1}} (\bibinfo {year}
  {2016}),\ 10.21468/SciPostPhys.1.1.007}\BibitemShut {NoStop}%
\bibitem [{\citenamefont {Berthier}\ and\ \citenamefont
  {Coslovich}(2014)}]{berthier_novel_2014}%
  \BibitemOpen
  \bibfield  {author} {\bibinfo {author} {\bibfnamefont {L.}~\bibnamefont
  {Berthier}}\ and\ \bibinfo {author} {\bibfnamefont {D.}~\bibnamefont
  {Coslovich}},\ }\href {\doibase 10.1073/pnas.1407934111} {\bibfield
  {journal} {\bibinfo  {journal} {Proceedings of the National Academy of
  Sciences}\ }\textbf {\bibinfo {volume} {111}},\ \bibinfo {pages} {11668}
  (\bibinfo {year} {2014})}\BibitemShut {NoStop}%
\bibitem [{\citenamefont {Montanari}\ and\ \citenamefont
  {Semerjian}(2006)}]{montanari2006rigorous}%
  \BibitemOpen
  \bibfield  {author} {\bibinfo {author} {\bibfnamefont {A.}~\bibnamefont
  {Montanari}}\ and\ \bibinfo {author} {\bibfnamefont {G.}~\bibnamefont
  {Semerjian}},\ }\href@noop {} {\bibfield  {journal} {\bibinfo  {journal}
  {Journal of statistical physics}\ }\textbf {\bibinfo {volume} {125}},\
  \bibinfo {pages} {23} (\bibinfo {year} {2006})}\BibitemShut {NoStop}%
\bibitem [{\citenamefont {M{\'e}zard}\ and\ \citenamefont
  {Montanari}(2006)}]{mezard2006reconstruction}%
  \BibitemOpen
  \bibfield  {author} {\bibinfo {author} {\bibfnamefont {M.}~\bibnamefont
  {M{\'e}zard}}\ and\ \bibinfo {author} {\bibfnamefont {A.}~\bibnamefont
  {Montanari}},\ }\href@noop {} {\bibfield  {journal} {\bibinfo  {journal}
  {Journal of statistical physics}\ }\textbf {\bibinfo {volume} {124}},\
  \bibinfo {pages} {1317} (\bibinfo {year} {2006})}\BibitemShut {NoStop}%
\bibitem [{\citenamefont {Biroli}\ \emph {et~al.}(2008)\citenamefont {Biroli},
  \citenamefont {Bouchaud}, \citenamefont {Cavagna}, \citenamefont {Grigera},\
  and\ \citenamefont {Verrocchio}}]{biroli_thermodynamic_2008}%
  \BibitemOpen
  \bibfield  {author} {\bibinfo {author} {\bibfnamefont {G.}~\bibnamefont
  {Biroli}}, \bibinfo {author} {\bibfnamefont {J.-P.}\ \bibnamefont
  {Bouchaud}}, \bibinfo {author} {\bibfnamefont {A.}~\bibnamefont {Cavagna}},
  \bibinfo {author} {\bibfnamefont {T.~S.}\ \bibnamefont {Grigera}}, \ and\
  \bibinfo {author} {\bibfnamefont {P.}~\bibnamefont {Verrocchio}},\ }\href
  {\doibase 10.1038/nphys1050} {\bibfield  {journal} {\bibinfo  {journal}
  {Nature Physics}\ }\textbf {\bibinfo {volume} {4}},\ \bibinfo {pages} {771}
  (\bibinfo {year} {2008})}\BibitemShut {NoStop}%
\bibitem [{\citenamefont {Cavagna}\ \emph {et~al.}(2012)\citenamefont
  {Cavagna}, \citenamefont {Grigera},\ and\ \citenamefont
  {Verrocchio}}]{cavagna_dynamic_2012}%
  \BibitemOpen
  \bibfield  {author} {\bibinfo {author} {\bibfnamefont {A.}~\bibnamefont
  {Cavagna}}, \bibinfo {author} {\bibfnamefont {T.~S.}\ \bibnamefont
  {Grigera}}, \ and\ \bibinfo {author} {\bibfnamefont {P.}~\bibnamefont
  {Verrocchio}},\ }\href {\doibase 10.1063/1.4720477} {\bibfield  {journal}
  {\bibinfo  {journal} {The Journal of Chemical Physics}\ }\textbf {\bibinfo
  {volume} {136}},\ \bibinfo {pages} {204502} (\bibinfo {year}
  {2012})}\BibitemShut {NoStop}%
\bibitem [{\citenamefont {Berthier}\ \emph
  {et~al.}(2016{\natexlab{c}})\citenamefont {Berthier}, \citenamefont
  {Charbonneau},\ and\ \citenamefont {Yaida}}]{berthier_efficient_2016}%
  \BibitemOpen
  \bibfield  {author} {\bibinfo {author} {\bibfnamefont {L.}~\bibnamefont
  {Berthier}}, \bibinfo {author} {\bibfnamefont {P.}~\bibnamefont
  {Charbonneau}}, \ and\ \bibinfo {author} {\bibfnamefont {S.}~\bibnamefont
  {Yaida}},\ }\href {\doibase 10.1063/1.4939640} {\bibfield  {journal}
  {\bibinfo  {journal} {The Journal of Chemical Physics}\ }\textbf {\bibinfo
  {volume} {144}},\ \bibinfo {pages} {024501} (\bibinfo {year}
  {2016}{\natexlab{c}})}\BibitemShut {NoStop}%
\bibitem [{\citenamefont {Charbonneau}\ \emph
  {et~al.}(2016{\natexlab{b}})\citenamefont {Charbonneau}, \citenamefont
  {Dyer}, \citenamefont {Lee},\ and\ \citenamefont
  {Yaida}}]{charbonneau_linking_2016}%
  \BibitemOpen
  \bibfield  {author} {\bibinfo {author} {\bibfnamefont {P.}~\bibnamefont
  {Charbonneau}}, \bibinfo {author} {\bibfnamefont {E.}~\bibnamefont {Dyer}},
  \bibinfo {author} {\bibfnamefont {J.}~\bibnamefont {Lee}}, \ and\ \bibinfo
  {author} {\bibfnamefont {S.}~\bibnamefont {Yaida}},\ }\href {\doibase
  10.1088/1742-5468/2016/07/074004} {\bibfield  {journal} {\bibinfo  {journal}
  {Journal of Statistical Mechanics: Theory and Experiment}\ }\textbf {\bibinfo
  {volume} {2016}},\ \bibinfo {pages} {074004} (\bibinfo {year}
  {2016}{\natexlab{b}})}\BibitemShut {NoStop}%
\bibitem [{\citenamefont {Russo}\ and\ \citenamefont
  {Tanaka}(2015)}]{russo2015assessing}%
  \BibitemOpen
  \bibfield  {author} {\bibinfo {author} {\bibfnamefont {J.}~\bibnamefont
  {Russo}}\ and\ \bibinfo {author} {\bibfnamefont {H.}~\bibnamefont {Tanaka}},\
  }\href@noop {} {\bibfield  {journal} {\bibinfo  {journal} {Proceedings of the
  National Academy of Sciences}\ }\textbf {\bibinfo {volume} {112}},\ \bibinfo
  {pages} {6920} (\bibinfo {year} {2015})}\BibitemShut {NoStop}%
\bibitem [{\citenamefont {Li}\ \emph {et~al.}(2014)\citenamefont {Li},
  \citenamefont {Xu},\ and\ \citenamefont {Sun}}]{li_growing_2014}%
  \BibitemOpen
  \bibfield  {author} {\bibinfo {author} {\bibfnamefont {Y.-W.}\ \bibnamefont
  {Li}}, \bibinfo {author} {\bibfnamefont {W.-S.}\ \bibnamefont {Xu}}, \ and\
  \bibinfo {author} {\bibfnamefont {Z.-Y.}\ \bibnamefont {Sun}},\ }\href
  {\doibase 10.1063/1.4868987} {\bibfield  {journal} {\bibinfo  {journal} {The
  Journal of Chemical Physics}\ }\textbf {\bibinfo {volume} {140}},\ \bibinfo
  {pages} {124502} (\bibinfo {year} {2014})}\BibitemShut {NoStop}%
\bibitem [{\citenamefont {Garrahan}\ \emph {et~al.}(2007)\citenamefont
  {Garrahan}, \citenamefont {Jack}, \citenamefont {Lecomte}, \citenamefont
  {Pitard}, \citenamefont {van Duijvendijk},\ and\ \citenamefont {van
  Wijland}}]{garrahan_dynamical_2007}%
  \BibitemOpen
  \bibfield  {author} {\bibinfo {author} {\bibfnamefont {J.~P.}\ \bibnamefont
  {Garrahan}}, \bibinfo {author} {\bibfnamefont {R.~L.}\ \bibnamefont {Jack}},
  \bibinfo {author} {\bibfnamefont {V.}~\bibnamefont {Lecomte}}, \bibinfo
  {author} {\bibfnamefont {E.}~\bibnamefont {Pitard}}, \bibinfo {author}
  {\bibfnamefont {K.}~\bibnamefont {van Duijvendijk}}, \ and\ \bibinfo {author}
  {\bibfnamefont {F.}~\bibnamefont {van Wijland}},\ }\href {\doibase
  10.1103/PhysRevLett.98.195702} {\bibfield  {journal} {\bibinfo  {journal}
  {Physical Review Letters}\ }\textbf {\bibinfo {volume} {98}} (\bibinfo {year}
  {2007}),\ 10.1103/PhysRevLett.98.195702}\BibitemShut {NoStop}%
\bibitem [{\citenamefont {Garrahan}\ \emph {et~al.}(2009)\citenamefont
  {Garrahan}, \citenamefont {Jack}, \citenamefont {Lecomte}, \citenamefont
  {Pitard}, \citenamefont {van Duijvendijk},\ and\ \citenamefont {van
  Wijland}}]{garrahan_first-order_2009}%
  \BibitemOpen
  \bibfield  {author} {\bibinfo {author} {\bibfnamefont {J.~P.}\ \bibnamefont
  {Garrahan}}, \bibinfo {author} {\bibfnamefont {R.~L.}\ \bibnamefont {Jack}},
  \bibinfo {author} {\bibfnamefont {V.}~\bibnamefont {Lecomte}}, \bibinfo
  {author} {\bibfnamefont {E.}~\bibnamefont {Pitard}}, \bibinfo {author}
  {\bibfnamefont {K.}~\bibnamefont {van Duijvendijk}}, \ and\ \bibinfo {author}
  {\bibfnamefont {F.}~\bibnamefont {van Wijland}},\ }\href {\doibase
  10.1088/1751-8113/42/7/075007} {\bibfield  {journal} {\bibinfo  {journal}
  {Journal of Physics A: Mathematical and Theoretical}\ }\textbf {\bibinfo
  {volume} {42}},\ \bibinfo {pages} {075007} (\bibinfo {year}
  {2009})}\BibitemShut {NoStop}%
\bibitem [{\citenamefont {Noe}\ \emph {et~al.}(2009)\citenamefont {Noe},
  \citenamefont {Schutte}, \citenamefont {Vanden-Eijnden}, \citenamefont
  {Reich},\ and\ \citenamefont {Weikl}}]{Noe2009}%
  \BibitemOpen
  \bibfield  {author} {\bibinfo {author} {\bibfnamefont {F.}~\bibnamefont
  {Noe}}, \bibinfo {author} {\bibfnamefont {C.}~\bibnamefont {Schutte}},
  \bibinfo {author} {\bibfnamefont {E.}~\bibnamefont {Vanden-Eijnden}},
  \bibinfo {author} {\bibfnamefont {L.}~\bibnamefont {Reich}}, \ and\ \bibinfo
  {author} {\bibfnamefont {T.~R.}\ \bibnamefont {Weikl}},\ }\href {\doibase
  10.1073/pnas.0905466106} {\bibfield  {journal} {\bibinfo  {journal}
  {Proceedings of the National Academy of Sciences}\ }\textbf {\bibinfo
  {volume} {106}},\ \bibinfo {pages} {19011} (\bibinfo {year} {2009})},\
  \bibinfo {note} {iSBN: 0027-8424}\BibitemShut {NoStop}%
\bibitem [{\citenamefont {Jack}\ \emph {et~al.}(2011)\citenamefont {Jack},
  \citenamefont {Hedges}, \citenamefont {Garrahan},\ and\ \citenamefont
  {Chandler}}]{jack_preparation_2011}%
  \BibitemOpen
  \bibfield  {author} {\bibinfo {author} {\bibfnamefont {R.~L.}\ \bibnamefont
  {Jack}}, \bibinfo {author} {\bibfnamefont {L.~O.}\ \bibnamefont {Hedges}},
  \bibinfo {author} {\bibfnamefont {J.~P.}\ \bibnamefont {Garrahan}}, \ and\
  \bibinfo {author} {\bibfnamefont {D.}~\bibnamefont {Chandler}},\ }\href
  {\doibase 10.1103/PhysRevLett.107.275702} {\bibfield  {journal} {\bibinfo
  {journal} {Physical Review Letters}\ }\textbf {\bibinfo {volume} {107}}
  (\bibinfo {year} {2011}),\ 10.1103/PhysRevLett.107.275702}\BibitemShut
  {NoStop}%
\bibitem [{\citenamefont {Jack}\ and\ \citenamefont
  {Garrahan}(2010)}]{jack2010metastable}%
  \BibitemOpen
  \bibfield  {author} {\bibinfo {author} {\bibfnamefont {R.~L.}\ \bibnamefont
  {Jack}}\ and\ \bibinfo {author} {\bibfnamefont {J.~P.}\ \bibnamefont
  {Garrahan}},\ }\href@noop {} {\bibfield  {journal} {\bibinfo  {journal}
  {Physical Review E}\ }\textbf {\bibinfo {volume} {81}},\ \bibinfo {pages}
  {011111} (\bibinfo {year} {2010})}\BibitemShut {NoStop}%
\bibitem [{\citenamefont {Schoenholz}(2018)}]{Schoenholz2017}%
  \BibitemOpen
  \bibfield  {author} {\bibinfo {author} {\bibfnamefont {S.~S.}\ \bibnamefont
  {Schoenholz}},\ }in\ \href@noop {} {\emph {\bibinfo {booktitle} {Journal of
  Physics: Conference Series}}},\ Vol.\ \bibinfo {volume} {1036}\ (\bibinfo
  {organization} {IOP Publishing},\ \bibinfo {year} {2018})\ p.\ \bibinfo
  {pages} {012021}\BibitemShut {NoStop}%
\bibitem [{\citenamefont {Candelier}\ \emph {et~al.}(2009)\citenamefont
  {Candelier}, \citenamefont {Dauchot},\ and\ \citenamefont
  {Biroli}}]{Candelier2009}%
  \BibitemOpen
  \bibfield  {author} {\bibinfo {author} {\bibfnamefont {R.}~\bibnamefont
  {Candelier}}, \bibinfo {author} {\bibfnamefont {O.}~\bibnamefont {Dauchot}},
  \ and\ \bibinfo {author} {\bibfnamefont {G.}~\bibnamefont {Biroli}},\ }\href
  {\doibase 10.1103/PhysRevLett.102.088001} {\bibfield  {journal} {\bibinfo
  {journal} {Physical Review Letters}\ }\textbf {\bibinfo {volume} {102}},\
  \bibinfo {pages} {1} (\bibinfo {year} {2009})}\BibitemShut {NoStop}%
\bibitem [{\citenamefont {Candelier}\ \emph
  {et~al.}(2010{\natexlab{a}})\citenamefont {Candelier}, \citenamefont
  {Widmer-Cooper}, \citenamefont {Kummerfeld}, \citenamefont {Dauchot},
  \citenamefont {Biroli}, \citenamefont {Harrowell},\ and\ \citenamefont
  {Reichman}}]{Candelier2010a}%
  \BibitemOpen
  \bibfield  {author} {\bibinfo {author} {\bibfnamefont {R.}~\bibnamefont
  {Candelier}}, \bibinfo {author} {\bibfnamefont {A.}~\bibnamefont
  {Widmer-Cooper}}, \bibinfo {author} {\bibfnamefont {J.~K.}\ \bibnamefont
  {Kummerfeld}}, \bibinfo {author} {\bibfnamefont {O.}~\bibnamefont {Dauchot}},
  \bibinfo {author} {\bibfnamefont {G.}~\bibnamefont {Biroli}}, \bibinfo
  {author} {\bibfnamefont {P.}~\bibnamefont {Harrowell}}, \ and\ \bibinfo
  {author} {\bibfnamefont {D.~R.}\ \bibnamefont {Reichman}},\ }\href {\doibase
  10.1103/PhysRevLett.105.135702} {\bibfield  {journal} {\bibinfo  {journal}
  {Physical Review Letters}\ }\textbf {\bibinfo {volume} {105}},\ \bibinfo
  {pages} {135702} (\bibinfo {year} {2010}{\natexlab{a}})}\BibitemShut
  {NoStop}%
\bibitem [{\citenamefont {Candelier}\ \emph
  {et~al.}(2010{\natexlab{b}})\citenamefont {Candelier}, \citenamefont
  {Dauchot},\ and\ \citenamefont {Biroli}}]{Candelier2010c}%
  \BibitemOpen
  \bibfield  {author} {\bibinfo {author} {\bibfnamefont {R.}~\bibnamefont
  {Candelier}}, \bibinfo {author} {\bibfnamefont {O.}~\bibnamefont {Dauchot}},
  \ and\ \bibinfo {author} {\bibfnamefont {G.}~\bibnamefont {Biroli}},\ }\href
  {\doibase 10.1209/0295-5075/92/24003} {\bibfield  {journal} {\bibinfo
  {journal} {EPL (Europhysics Letters)}\ }\textbf {\bibinfo {volume} {92}},\
  \bibinfo {pages} {24003} (\bibinfo {year} {2010}{\natexlab{b}})}\BibitemShut
  {NoStop}%
\bibitem [{\citenamefont {Candelier}\ \emph
  {et~al.}(2010{\natexlab{c}})\citenamefont {Candelier}, \citenamefont
  {Widmer-Cooper}, \citenamefont {Kummerfeld}, \citenamefont {Dauchot},
  \citenamefont {Biroli}, \citenamefont {Harrowell},\ and\ \citenamefont
  {Reichman}}]{candelier_avalanches_2010}%
  \BibitemOpen
  \bibfield  {author} {\bibinfo {author} {\bibfnamefont {R.}~\bibnamefont
  {Candelier}}, \bibinfo {author} {\bibfnamefont {A.}~\bibnamefont
  {Widmer-Cooper}}, \bibinfo {author} {\bibfnamefont {J.~K.}\ \bibnamefont
  {Kummerfeld}}, \bibinfo {author} {\bibfnamefont {O.}~\bibnamefont {Dauchot}},
  \bibinfo {author} {\bibfnamefont {G.}~\bibnamefont {Biroli}}, \bibinfo
  {author} {\bibfnamefont {P.}~\bibnamefont {Harrowell}}, \ and\ \bibinfo
  {author} {\bibfnamefont {D.~R.}\ \bibnamefont {Reichman}},\ }\href {\doibase
  10.1103/PhysRevLett.105.135702} {\bibfield  {journal} {\bibinfo  {journal}
  {Physical Review Letters}\ }\textbf {\bibinfo {volume} {105}} (\bibinfo
  {year} {2010}{\natexlab{c}}),\ 10.1103/PhysRevLett.105.135702}\BibitemShut
  {NoStop}%
\bibitem [{\citenamefont {Cubuk}\ \emph {et~al.}(2016)\citenamefont {Cubuk},
  \citenamefont {Schoenholz}, \citenamefont {Kaxiras},\ and\ \citenamefont
  {Liu}}]{Cubuk2016}%
  \BibitemOpen
  \bibfield  {author} {\bibinfo {author} {\bibfnamefont {E.~D.}\ \bibnamefont
  {Cubuk}}, \bibinfo {author} {\bibfnamefont {S.~S.}\ \bibnamefont
  {Schoenholz}}, \bibinfo {author} {\bibfnamefont {E.}~\bibnamefont {Kaxiras}},
  \ and\ \bibinfo {author} {\bibfnamefont {A.~J.}\ \bibnamefont {Liu}},\ }\href
  {\doibase 10.1021/acs.jpcb.6b02144} {\bibfield  {journal} {\bibinfo
  {journal} {Journal of Physical Chemistry B}\ }\textbf {\bibinfo {volume}
  {120}},\ \bibinfo {pages} {6139} (\bibinfo {year} {2016})},\ \bibinfo {note}
  {iSBN: 1520-5207 (Electronic) 1520-5207 (Linking)}\BibitemShut {NoStop}%
\bibitem [{\citenamefont {Schoenholz}\ \emph {et~al.}(2016)\citenamefont
  {Schoenholz}, \citenamefont {Cubuk}, \citenamefont {Kaxiras},\ and\
  \citenamefont {Liu}}]{Schoenholz2016}%
  \BibitemOpen
  \bibfield  {author} {\bibinfo {author} {\bibfnamefont {S.~S.}\ \bibnamefont
  {Schoenholz}}, \bibinfo {author} {\bibfnamefont {E.~D.}\ \bibnamefont
  {Cubuk}}, \bibinfo {author} {\bibfnamefont {E.}~\bibnamefont {Kaxiras}}, \
  and\ \bibinfo {author} {\bibfnamefont {A.~J.}\ \bibnamefont {Liu}},\ }\href
  {\doibase 10.1073/pnas.1610204114} {\bibfield  {journal} {\bibinfo  {journal}
  {Proceedings of the National Academy of Sciences}\ }\textbf {\bibinfo
  {volume} {114}},\ \bibinfo {pages} {263} (\bibinfo {year}
  {2016})}\BibitemShut {NoStop}%
\bibitem [{\citenamefont {Landes}\ \emph {et~al.}(2020)\citenamefont {Landes},
  \citenamefont {Biroli}, \citenamefont {Dauchot}, \citenamefont {Liu},\ and\
  \citenamefont {Reichman}}]{landes_attractive_2019}%
  \BibitemOpen
  \bibfield  {author} {\bibinfo {author} {\bibfnamefont {F.~P.}\ \bibnamefont
  {Landes}}, \bibinfo {author} {\bibfnamefont {G.}~\bibnamefont {Biroli}},
  \bibinfo {author} {\bibfnamefont {O.}~\bibnamefont {Dauchot}}, \bibinfo
  {author} {\bibfnamefont {A.~J.}\ \bibnamefont {Liu}}, \ and\ \bibinfo
  {author} {\bibfnamefont {D.~R.}\ \bibnamefont {Reichman}},\ }\href@noop {}
  {\bibfield  {journal} {\bibinfo  {journal} {Physical Review E}\ }\textbf
  {\bibinfo {volume} {101}},\ \bibinfo {pages} {010602} (\bibinfo {year}
  {2020})}\BibitemShut {NoStop}%
\bibitem [{\citenamefont {Cubuk}\ \emph {et~al.}(2015)\citenamefont {Cubuk},
  \citenamefont {Schoenholz}, \citenamefont {Rieser}, \citenamefont {Malone},
  \citenamefont {Rottler}, \citenamefont {Durian}, \citenamefont {Kaxiras},\
  and\ \citenamefont {Liu}}]{Cubuk2015}%
  \BibitemOpen
  \bibfield  {author} {\bibinfo {author} {\bibfnamefont {E.~D.}\ \bibnamefont
  {Cubuk}}, \bibinfo {author} {\bibfnamefont {S.~S.}\ \bibnamefont
  {Schoenholz}}, \bibinfo {author} {\bibfnamefont {J.~M.}\ \bibnamefont
  {Rieser}}, \bibinfo {author} {\bibfnamefont {B.~D.}\ \bibnamefont {Malone}},
  \bibinfo {author} {\bibfnamefont {J.}~\bibnamefont {Rottler}}, \bibinfo
  {author} {\bibfnamefont {D.~J.}\ \bibnamefont {Durian}}, \bibinfo {author}
  {\bibfnamefont {E.}~\bibnamefont {Kaxiras}}, \ and\ \bibinfo {author}
  {\bibfnamefont {A.~J.}\ \bibnamefont {Liu}},\ }\href {\doibase
  10.1103/PhysRevLett.114.108001} {\bibfield  {journal} {\bibinfo  {journal}
  {Physical Review Letters}\ }\textbf {\bibinfo {volume} {114}},\ \bibinfo
  {pages} {1} (\bibinfo {year} {2015})}\BibitemShut {NoStop}%
\bibitem [{\citenamefont {Schoenholz}\ \emph {et~al.}(2015)\citenamefont
  {Schoenholz}, \citenamefont {Cubuk}, \citenamefont {Sussman}, \citenamefont
  {Kaxiras},\ and\ \citenamefont {Liu}}]{Schoenholz2016a}%
  \BibitemOpen
  \bibfield  {author} {\bibinfo {author} {\bibfnamefont {S.~S.}\ \bibnamefont
  {Schoenholz}}, \bibinfo {author} {\bibfnamefont {E.~D.}\ \bibnamefont
  {Cubuk}}, \bibinfo {author} {\bibfnamefont {D.~M.}\ \bibnamefont {Sussman}},
  \bibinfo {author} {\bibfnamefont {E.}~\bibnamefont {Kaxiras}}, \ and\
  \bibinfo {author} {\bibfnamefont {A.~J.}\ \bibnamefont {Liu}},\ }\href
  {\doibase 10.1038/nphys3644} {\bibfield  {journal} {\bibinfo  {journal}
  {Nature Physics}\ }\textbf {\bibinfo {volume} {12}},\ \bibinfo {pages} {469}
  (\bibinfo {year} {2015})}\BibitemShut {NoStop}%
\bibitem [{\citenamefont {Ivancic}\ \emph {et~al.}(2017)\citenamefont
  {Ivancic}, \citenamefont {Cubuk}, \citenamefont {Schoenholz}, \citenamefont
  {Strickland}, \citenamefont {Gianola},\ and\ \citenamefont
  {Liu}}]{Mechanics}%
  \BibitemOpen
  \bibfield  {author} {\bibinfo {author} {\bibfnamefont {R.}~\bibnamefont
  {Ivancic}}, \bibinfo {author} {\bibfnamefont {E.}~\bibnamefont {Cubuk}},
  \bibinfo {author} {\bibfnamefont {S.}~\bibnamefont {Schoenholz}}, \bibinfo
  {author} {\bibfnamefont {D.}~\bibnamefont {Strickland}}, \bibinfo {author}
  {\bibfnamefont {D.}~\bibnamefont {Gianola}}, \ and\ \bibinfo {author}
  {\bibfnamefont {A.}~\bibnamefont {Liu}},\ }\href@noop {} {\bibfield
  {journal} {\bibinfo  {journal} {Bulletin of the American Physical Society}\
  }\textbf {\bibinfo {volume} {62}} (\bibinfo {year} {2017})}\BibitemShut
  {NoStop}%
\bibitem [{\citenamefont {Cubuk}\ \emph {et~al.}(2017)\citenamefont {Cubuk},
  \citenamefont {Ivancic}, \citenamefont {Schoenholz}, \citenamefont
  {Strickland}, \citenamefont {Basu}, \citenamefont {Davidson}, \citenamefont
  {Fontaine}, \citenamefont {Hor}, \citenamefont {Huang}, \citenamefont
  {Jiang}, \citenamefont {Keim}, \citenamefont {Koshigan}, \citenamefont
  {Lefever}, \citenamefont {Liu}, \citenamefont {Ma}, \citenamefont
  {Magagnosc}, \citenamefont {Morrow}, \citenamefont {Ortiz}, \citenamefont
  {Rieser}, \citenamefont {Shavit}, \citenamefont {Still}, \citenamefont {Xu},
  \citenamefont {Zhang}, \citenamefont {Nordstrom}, \citenamefont {Arratia},
  \citenamefont {Carpick}, \citenamefont {Durian}, \citenamefont {Fakhraai},
  \citenamefont {Jerolmack}, \citenamefont {Lee}, \citenamefont {Li},
  \citenamefont {Riggleman}, \citenamefont {Turner}, \citenamefont {Yodh},
  \citenamefont {Gianola},\ and\ \citenamefont {Liu}}]{Definiujemy2012}%
  \BibitemOpen
  \bibfield  {author} {\bibinfo {author} {\bibfnamefont {E.~D.}\ \bibnamefont
  {Cubuk}}, \bibinfo {author} {\bibfnamefont {R.~J.~S.}\ \bibnamefont
  {Ivancic}}, \bibinfo {author} {\bibfnamefont {S.~S.}\ \bibnamefont
  {Schoenholz}}, \bibinfo {author} {\bibfnamefont {D.~J.}\ \bibnamefont
  {Strickland}}, \bibinfo {author} {\bibfnamefont {A.}~\bibnamefont {Basu}},
  \bibinfo {author} {\bibfnamefont {Z.~S.}\ \bibnamefont {Davidson}}, \bibinfo
  {author} {\bibfnamefont {J.}~\bibnamefont {Fontaine}}, \bibinfo {author}
  {\bibfnamefont {J.~L.}\ \bibnamefont {Hor}}, \bibinfo {author} {\bibfnamefont
  {Y.-R.}\ \bibnamefont {Huang}}, \bibinfo {author} {\bibfnamefont
  {Y.}~\bibnamefont {Jiang}}, \bibinfo {author} {\bibfnamefont {N.~C.}\
  \bibnamefont {Keim}}, \bibinfo {author} {\bibfnamefont {K.~D.}\ \bibnamefont
  {Koshigan}}, \bibinfo {author} {\bibfnamefont {J.~A.}\ \bibnamefont
  {Lefever}}, \bibinfo {author} {\bibfnamefont {T.}~\bibnamefont {Liu}},
  \bibinfo {author} {\bibfnamefont {X.-G.}\ \bibnamefont {Ma}}, \bibinfo
  {author} {\bibfnamefont {D.~J.}\ \bibnamefont {Magagnosc}}, \bibinfo {author}
  {\bibfnamefont {E.}~\bibnamefont {Morrow}}, \bibinfo {author} {\bibfnamefont
  {C.~P.}\ \bibnamefont {Ortiz}}, \bibinfo {author} {\bibfnamefont {J.~M.}\
  \bibnamefont {Rieser}}, \bibinfo {author} {\bibfnamefont {A.}~\bibnamefont
  {Shavit}}, \bibinfo {author} {\bibfnamefont {T.}~\bibnamefont {Still}},
  \bibinfo {author} {\bibfnamefont {Y.}~\bibnamefont {Xu}}, \bibinfo {author}
  {\bibfnamefont {Y.}~\bibnamefont {Zhang}}, \bibinfo {author} {\bibfnamefont
  {K.~N.}\ \bibnamefont {Nordstrom}}, \bibinfo {author} {\bibfnamefont {P.~E.}\
  \bibnamefont {Arratia}}, \bibinfo {author} {\bibfnamefont {R.~W.}\
  \bibnamefont {Carpick}}, \bibinfo {author} {\bibfnamefont {D.~J.}\
  \bibnamefont {Durian}}, \bibinfo {author} {\bibfnamefont {Z.}~\bibnamefont
  {Fakhraai}}, \bibinfo {author} {\bibfnamefont {D.~J.}\ \bibnamefont
  {Jerolmack}}, \bibinfo {author} {\bibfnamefont {D.}~\bibnamefont {Lee}},
  \bibinfo {author} {\bibfnamefont {J.}~\bibnamefont {Li}}, \bibinfo {author}
  {\bibfnamefont {R.}~\bibnamefont {Riggleman}}, \bibinfo {author}
  {\bibfnamefont {K.~T.}\ \bibnamefont {Turner}}, \bibinfo {author}
  {\bibfnamefont {A.~G.}\ \bibnamefont {Yodh}}, \bibinfo {author}
  {\bibfnamefont {D.~S.}\ \bibnamefont {Gianola}}, \ and\ \bibinfo {author}
  {\bibfnamefont {A.~J.}\ \bibnamefont {Liu}},\ }\href {\doibase
  10.1126/science.aai8830} {\bibfield  {journal} {\bibinfo  {journal}
  {Science}\ }\textbf {\bibinfo {volume} {358}},\ \bibinfo {pages} {1033}
  (\bibinfo {year} {2017})},\ \bibinfo {note} {iSBN: 3487716170192}\BibitemShut
  {NoStop}%
\bibitem [{\citenamefont {Sharp}\ \emph {et~al.}(2018)\citenamefont {Sharp},
  \citenamefont {Thomas}, \citenamefont {Cubuk}, \citenamefont {Schoenholz},
  \citenamefont {Srolovitz},\ and\ \citenamefont {Liu}}]{sharp2018machine}%
  \BibitemOpen
  \bibfield  {author} {\bibinfo {author} {\bibfnamefont {T.~A.}\ \bibnamefont
  {Sharp}}, \bibinfo {author} {\bibfnamefont {S.~L.}\ \bibnamefont {Thomas}},
  \bibinfo {author} {\bibfnamefont {E.~D.}\ \bibnamefont {Cubuk}}, \bibinfo
  {author} {\bibfnamefont {S.~S.}\ \bibnamefont {Schoenholz}}, \bibinfo
  {author} {\bibfnamefont {D.~J.}\ \bibnamefont {Srolovitz}}, \ and\ \bibinfo
  {author} {\bibfnamefont {A.~J.}\ \bibnamefont {Liu}},\ }\href@noop {}
  {\bibfield  {journal} {\bibinfo  {journal} {Proceedings of the National
  Academy of Sciences}\ }\textbf {\bibinfo {volume} {115}},\ \bibinfo {pages}
  {10943} (\bibinfo {year} {2018})}\BibitemShut {NoStop}%
\bibitem [{\citenamefont {Bapst}\ \emph {et~al.}(2020)\citenamefont {Bapst},
  \citenamefont {Keck}, \citenamefont {Grabska-Barwi{\'n}ska}, \citenamefont
  {Donner}, \citenamefont {Cubuk}, \citenamefont {Schoenholz}, \citenamefont
  {Obika}, \citenamefont {Nelson}, \citenamefont {Back}, \citenamefont
  {Hassabis} \emph {et~al.}}]{bapst2020unveiling}%
  \BibitemOpen
  \bibfield  {author} {\bibinfo {author} {\bibfnamefont {V.}~\bibnamefont
  {Bapst}}, \bibinfo {author} {\bibfnamefont {T.}~\bibnamefont {Keck}},
  \bibinfo {author} {\bibfnamefont {A.}~\bibnamefont {Grabska-Barwi{\'n}ska}},
  \bibinfo {author} {\bibfnamefont {C.}~\bibnamefont {Donner}}, \bibinfo
  {author} {\bibfnamefont {E.~D.}\ \bibnamefont {Cubuk}}, \bibinfo {author}
  {\bibfnamefont {S.}~\bibnamefont {Schoenholz}}, \bibinfo {author}
  {\bibfnamefont {A.}~\bibnamefont {Obika}}, \bibinfo {author} {\bibfnamefont
  {A.}~\bibnamefont {Nelson}}, \bibinfo {author} {\bibfnamefont
  {T.}~\bibnamefont {Back}}, \bibinfo {author} {\bibfnamefont {D.}~\bibnamefont
  {Hassabis}},  \emph {et~al.},\ }\href@noop {} {\bibfield  {journal} {\bibinfo
   {journal} {Nature Physics}\ }\textbf {\bibinfo {volume} {16}},\ \bibinfo
  {pages} {448} (\bibinfo {year} {2020})}\BibitemShut {NoStop}%
\bibitem [{\citenamefont {Widmer-Cooper}\ and\ \citenamefont
  {Harrowell}(2006)}]{widmer2006predicting}%
  \BibitemOpen
  \bibfield  {author} {\bibinfo {author} {\bibfnamefont {A.}~\bibnamefont
  {Widmer-Cooper}}\ and\ \bibinfo {author} {\bibfnamefont {P.}~\bibnamefont
  {Harrowell}},\ }\href@noop {} {\bibfield  {journal} {\bibinfo  {journal}
  {Physical review letters}\ }\textbf {\bibinfo {volume} {96}},\ \bibinfo
  {pages} {185701} (\bibinfo {year} {2006})}\BibitemShut {NoStop}%
\bibitem [{\citenamefont {Bouchaud}(1992)}]{bouchaud1992weak}%
  \BibitemOpen
  \bibfield  {author} {\bibinfo {author} {\bibfnamefont {J.-P.}\ \bibnamefont
  {Bouchaud}},\ }\href@noop {} {\bibfield  {journal} {\bibinfo  {journal}
  {Journal de Physique I}\ }\textbf {\bibinfo {volume} {2}},\ \bibinfo {pages}
  {1705} (\bibinfo {year} {1992})}\BibitemShut {NoStop}%
\bibitem [{\citenamefont {Bray}\ and\ \citenamefont {Moore}(1984)}]{braymoore}%
  \BibitemOpen
  \bibfield  {author} {\bibinfo {author} {\bibfnamefont {A.}~\bibnamefont
  {Bray}}\ and\ \bibinfo {author} {\bibfnamefont {M.}~\bibnamefont {Moore}},\
  }\href@noop {} {\bibfield  {journal} {\bibinfo  {journal} {Journal of Physics
  C: Solid State Physics}\ }\textbf {\bibinfo {volume} {17}},\ \bibinfo {pages}
  {L463} (\bibinfo {year} {1984})}\BibitemShut {NoStop}%
\bibitem [{\citenamefont {van Hemmen}\ and\ \citenamefont
  {Morgenstern}(1987)}]{braymoore2}%
  \BibitemOpen
  \bibfield  {author} {\bibinfo {author} {\bibfnamefont {J.}~\bibnamefont {van
  Hemmen}}\ and\ \bibinfo {author} {\bibfnamefont {I.}~\bibnamefont
  {Morgenstern}},\ }in\ \href@noop {} {\emph {\bibinfo {booktitle} {Lecture
  Notes in Physics, Berlin Springer Verlag}}},\ Vol.\ \bibinfo {volume} {275}\
  (\bibinfo {year} {1987})\BibitemShut {NoStop}%
\bibitem [{\citenamefont {Fisher}\ and\ \citenamefont {Huse}(1986)}]{FH}%
  \BibitemOpen
  \bibfield  {author} {\bibinfo {author} {\bibfnamefont {D.~S.}\ \bibnamefont
  {Fisher}}\ and\ \bibinfo {author} {\bibfnamefont {D.~A.}\ \bibnamefont
  {Huse}},\ }\href@noop {} {\bibfield  {journal} {\bibinfo  {journal} {Physical
  review letters}\ }\textbf {\bibinfo {volume} {56}},\ \bibinfo {pages} {1601}
  (\bibinfo {year} {1986})}\BibitemShut {NoStop}%
\bibitem [{\citenamefont {Kisker}\ \emph {et~al.}(1996)\citenamefont {Kisker},
  \citenamefont {Santen}, \citenamefont {Schreckenberg},\ and\ \citenamefont
  {Rieger}}]{heiko}%
  \BibitemOpen
  \bibfield  {author} {\bibinfo {author} {\bibfnamefont {J.}~\bibnamefont
  {Kisker}}, \bibinfo {author} {\bibfnamefont {L.}~\bibnamefont {Santen}},
  \bibinfo {author} {\bibfnamefont {M.}~\bibnamefont {Schreckenberg}}, \ and\
  \bibinfo {author} {\bibfnamefont {H.}~\bibnamefont {Rieger}},\ }\href@noop {}
  {\bibfield  {journal} {\bibinfo  {journal} {Physical Review B}\ }\textbf
  {\bibinfo {volume} {53}},\ \bibinfo {pages} {6418} (\bibinfo {year}
  {1996})}\BibitemShut {NoStop}%
\bibitem [{\citenamefont {Refregier}\ \emph {et~al.}(1987)\citenamefont
  {Refregier}, \citenamefont {Vincent}, \citenamefont {Hammann},\ and\
  \citenamefont {Ocio}}]{refregier}%
  \BibitemOpen
  \bibfield  {author} {\bibinfo {author} {\bibfnamefont {P.}~\bibnamefont
  {Refregier}}, \bibinfo {author} {\bibfnamefont {E.}~\bibnamefont {Vincent}},
  \bibinfo {author} {\bibfnamefont {J.}~\bibnamefont {Hammann}}, \ and\
  \bibinfo {author} {\bibfnamefont {M.}~\bibnamefont {Ocio}},\ }\href@noop {}
  {\bibfield  {journal} {\bibinfo  {journal} {Journal De Physique}\ }\textbf
  {\bibinfo {volume} {48}},\ \bibinfo {pages} {1533} (\bibinfo {year}
  {1987})}\BibitemShut {NoStop}%
\bibitem [{\citenamefont {Berthier}\ and\ \citenamefont
  {Young}(2005)}]{berthieryoung}%
  \BibitemOpen
  \bibfield  {author} {\bibinfo {author} {\bibfnamefont {L.}~\bibnamefont
  {Berthier}}\ and\ \bibinfo {author} {\bibfnamefont {A.}~\bibnamefont
  {Young}},\ }\href@noop {} {\bibfield  {journal} {\bibinfo  {journal}
  {Physical Review B}\ }\textbf {\bibinfo {volume} {71}},\ \bibinfo {pages}
  {214429} (\bibinfo {year} {2005})}\BibitemShut {NoStop}%
\bibitem [{\citenamefont {Berthier}\ and\ \citenamefont
  {Bouchaud}(2002)}]{berthierbouchaud}%
  \BibitemOpen
  \bibfield  {author} {\bibinfo {author} {\bibfnamefont {L.}~\bibnamefont
  {Berthier}}\ and\ \bibinfo {author} {\bibfnamefont {J.-P.}\ \bibnamefont
  {Bouchaud}},\ }\href@noop {} {\bibfield  {journal} {\bibinfo  {journal}
  {Physical Review B}\ }\textbf {\bibinfo {volume} {66}},\ \bibinfo {pages}
  {054404} (\bibinfo {year} {2002})}\BibitemShut {NoStop}%
\bibitem [{\citenamefont {Bouchaud}\ \emph {et~al.}(2001)\citenamefont
  {Bouchaud}, \citenamefont {Dupuis}, \citenamefont {Hammann},\ and\
  \citenamefont {Vincent}}]{microT}%
  \BibitemOpen
  \bibfield  {author} {\bibinfo {author} {\bibfnamefont {J.-P.}\ \bibnamefont
  {Bouchaud}}, \bibinfo {author} {\bibfnamefont {V.}~\bibnamefont {Dupuis}},
  \bibinfo {author} {\bibfnamefont {J.}~\bibnamefont {Hammann}}, \ and\
  \bibinfo {author} {\bibfnamefont {E.}~\bibnamefont {Vincent}},\ }\href@noop
  {} {\bibfield  {journal} {\bibinfo  {journal} {Physical review B}\ }\textbf
  {\bibinfo {volume} {65}},\ \bibinfo {pages} {024439} (\bibinfo {year}
  {2001})}\BibitemShut {NoStop}%
\bibitem [{\citenamefont {Bert}\ \emph {et~al.}(2004)\citenamefont {Bert},
  \citenamefont {Dupuis}, \citenamefont {Vincent}, \citenamefont {Hammann},\
  and\ \citenamefont {Bouchaud}}]{bert}%
  \BibitemOpen
  \bibfield  {author} {\bibinfo {author} {\bibfnamefont {F.}~\bibnamefont
  {Bert}}, \bibinfo {author} {\bibfnamefont {V.}~\bibnamefont {Dupuis}},
  \bibinfo {author} {\bibfnamefont {E.}~\bibnamefont {Vincent}}, \bibinfo
  {author} {\bibfnamefont {J.}~\bibnamefont {Hammann}}, \ and\ \bibinfo
  {author} {\bibfnamefont {J.-P.}\ \bibnamefont {Bouchaud}},\ }\href@noop {}
  {\bibfield  {journal} {\bibinfo  {journal} {Physical review letters}\
  }\textbf {\bibinfo {volume} {92}},\ \bibinfo {pages} {167203} (\bibinfo
  {year} {2004})}\BibitemShut {NoStop}%
\bibitem [{\citenamefont {Bray}\ and\ \citenamefont {Moore}(1987)}]{chaos1}%
  \BibitemOpen
  \bibfield  {author} {\bibinfo {author} {\bibfnamefont {A.}~\bibnamefont
  {Bray}}\ and\ \bibinfo {author} {\bibfnamefont {M.}~\bibnamefont {Moore}},\
  }\href@noop {} {\bibfield  {journal} {\bibinfo  {journal} {Physical review
  letters}\ }\textbf {\bibinfo {volume} {58}},\ \bibinfo {pages} {57} (\bibinfo
  {year} {1987})}\BibitemShut {NoStop}%
\bibitem [{\citenamefont {J{\"o}nsson}\ \emph {et~al.}(2004)\citenamefont
  {J{\"o}nsson}, \citenamefont {Mathieu}, \citenamefont {Nordblad},
  \citenamefont {Yoshino}, \citenamefont {Katori},\ and\ \citenamefont
  {Ito}}]{chaos2}%
  \BibitemOpen
  \bibfield  {author} {\bibinfo {author} {\bibfnamefont {P.}~\bibnamefont
  {J{\"o}nsson}}, \bibinfo {author} {\bibfnamefont {R.}~\bibnamefont
  {Mathieu}}, \bibinfo {author} {\bibfnamefont {P.}~\bibnamefont {Nordblad}},
  \bibinfo {author} {\bibfnamefont {H.}~\bibnamefont {Yoshino}}, \bibinfo
  {author} {\bibfnamefont {H.~A.}\ \bibnamefont {Katori}}, \ and\ \bibinfo
  {author} {\bibfnamefont {A.}~\bibnamefont {Ito}},\ }\href@noop {} {\bibfield
  {journal} {\bibinfo  {journal} {Physical Review B}\ }\textbf {\bibinfo
  {volume} {70}},\ \bibinfo {pages} {174402} (\bibinfo {year}
  {2004})}\BibitemShut {NoStop}%
\bibitem [{\citenamefont {Cugliandolo}\ and\ \citenamefont
  {Kurchan}(1993)}]{CugKur1}%
  \BibitemOpen
  \bibfield  {author} {\bibinfo {author} {\bibfnamefont {L.~F.}\ \bibnamefont
  {Cugliandolo}}\ and\ \bibinfo {author} {\bibfnamefont {J.}~\bibnamefont
  {Kurchan}},\ }\href@noop {} {\bibfield  {journal} {\bibinfo  {journal}
  {Physical Review Letters}\ }\textbf {\bibinfo {volume} {71}},\ \bibinfo
  {pages} {173} (\bibinfo {year} {1993})}\BibitemShut {NoStop}%
\bibitem [{\citenamefont {Cugliandolo}\ and\ \citenamefont
  {Kurchan}(1994)}]{CugKur2}%
  \BibitemOpen
  \bibfield  {author} {\bibinfo {author} {\bibfnamefont {L.~F.}\ \bibnamefont
  {Cugliandolo}}\ and\ \bibinfo {author} {\bibfnamefont {J.}~\bibnamefont
  {Kurchan}},\ }\href@noop {} {\bibfield  {journal} {\bibinfo  {journal}
  {Journal of Physics A: Mathematical and General}\ }\textbf {\bibinfo {volume}
  {27}},\ \bibinfo {pages} {5749} (\bibinfo {year} {1994})}\BibitemShut
  {NoStop}%
\bibitem [{\citenamefont {Crisanti}\ and\ \citenamefont
  {Ritort}(2003)}]{crisrit}%
  \BibitemOpen
  \bibfield  {author} {\bibinfo {author} {\bibfnamefont {A.}~\bibnamefont
  {Crisanti}}\ and\ \bibinfo {author} {\bibfnamefont {F.}~\bibnamefont
  {Ritort}},\ }\href@noop {} {\bibfield  {journal} {\bibinfo  {journal}
  {Journal of Physics A: Mathematical and General}\ }\textbf {\bibinfo {volume}
  {36}},\ \bibinfo {pages} {R181} (\bibinfo {year} {2003})}\BibitemShut
  {NoStop}%
\bibitem [{\citenamefont {Kurchan}\ \emph {et~al.}(2016)\citenamefont
  {Kurchan}, \citenamefont {Maimbourg},\ and\ \citenamefont
  {Zamponi}}]{kurchan2016statics}%
  \BibitemOpen
  \bibfield  {author} {\bibinfo {author} {\bibfnamefont {J.}~\bibnamefont
  {Kurchan}}, \bibinfo {author} {\bibfnamefont {T.}~\bibnamefont {Maimbourg}},
  \ and\ \bibinfo {author} {\bibfnamefont {F.}~\bibnamefont {Zamponi}},\
  }\href@noop {} {\bibfield  {journal} {\bibinfo  {journal} {Journal of
  Statistical Mechanics: Theory and Experiment}\ }\textbf {\bibinfo {volume}
  {2016}},\ \bibinfo {pages} {033210} (\bibinfo {year} {2016})}\BibitemShut
  {NoStop}%
\bibitem [{\citenamefont {Manacorda}\ \emph {et~al.}(2020)\citenamefont
  {Manacorda}, \citenamefont {Schehr},\ and\ \citenamefont
  {Zamponi}}]{manacorda2020numerical}%
  \BibitemOpen
  \bibfield  {author} {\bibinfo {author} {\bibfnamefont {A.}~\bibnamefont
  {Manacorda}}, \bibinfo {author} {\bibfnamefont {G.}~\bibnamefont {Schehr}}, \
  and\ \bibinfo {author} {\bibfnamefont {F.}~\bibnamefont {Zamponi}},\
  }\href@noop {} {\bibfield  {journal} {\bibinfo  {journal} {The Journal of
  Chemical Physics}\ }\textbf {\bibinfo {volume} {152}},\ \bibinfo {pages}
  {164506} (\bibinfo {year} {2020})}\BibitemShut {NoStop}%
\bibitem [{\citenamefont {Agoritsas}\ \emph {et~al.}(2018)\citenamefont
  {Agoritsas}, \citenamefont {Biroli}, \citenamefont {Urbani},\ and\
  \citenamefont {Zamponi}}]{agoritsas2018out}%
  \BibitemOpen
  \bibfield  {author} {\bibinfo {author} {\bibfnamefont {E.}~\bibnamefont
  {Agoritsas}}, \bibinfo {author} {\bibfnamefont {G.}~\bibnamefont {Biroli}},
  \bibinfo {author} {\bibfnamefont {P.}~\bibnamefont {Urbani}}, \ and\ \bibinfo
  {author} {\bibfnamefont {F.}~\bibnamefont {Zamponi}},\ }\href@noop {}
  {\bibfield  {journal} {\bibinfo  {journal} {Journal of Physics A:
  Mathematical and Theoretical}\ }\textbf {\bibinfo {volume} {51}},\ \bibinfo
  {pages} {085002} (\bibinfo {year} {2018})}\BibitemShut {NoStop}%
\bibitem [{\citenamefont {Agoritsas}\ \emph {et~al.}(2019)\citenamefont
  {Agoritsas}, \citenamefont {Maimbourg},\ and\ \citenamefont
  {Zamponi}}]{agoritsas2019out}%
  \BibitemOpen
  \bibfield  {author} {\bibinfo {author} {\bibfnamefont {E.}~\bibnamefont
  {Agoritsas}}, \bibinfo {author} {\bibfnamefont {T.}~\bibnamefont
  {Maimbourg}}, \ and\ \bibinfo {author} {\bibfnamefont {F.}~\bibnamefont
  {Zamponi}},\ }\href@noop {} {\bibfield  {journal} {\bibinfo  {journal}
  {Journal of Physics A: Mathematical and Theoretical}\ }\textbf {\bibinfo
  {volume} {52}},\ \bibinfo {pages} {144002} (\bibinfo {year}
  {2019})}\BibitemShut {NoStop}%
\bibitem [{\citenamefont {Altieri}\ \emph {et~al.}(2020)\citenamefont
  {Altieri}, \citenamefont {Biroli},\ and\ \citenamefont
  {Cammarota}}]{altieri2020dynamical}%
  \BibitemOpen
  \bibfield  {author} {\bibinfo {author} {\bibfnamefont {A.}~\bibnamefont
  {Altieri}}, \bibinfo {author} {\bibfnamefont {G.}~\bibnamefont {Biroli}}, \
  and\ \bibinfo {author} {\bibfnamefont {C.}~\bibnamefont {Cammarota}},\
  }\href@noop {} {\bibfield  {journal} {\bibinfo  {journal} {arXiv preprint
  arXiv:2005.05118}\ } (\bibinfo {year} {2020})}\BibitemShut {NoStop}%
\bibitem [{\citenamefont {Kurchan}\ and\ \citenamefont
  {Laloux}(1996)}]{laloux}%
  \BibitemOpen
  \bibfield  {author} {\bibinfo {author} {\bibfnamefont {J.}~\bibnamefont
  {Kurchan}}\ and\ \bibinfo {author} {\bibfnamefont {L.}~\bibnamefont
  {Laloux}},\ }\href@noop {} {\bibfield  {journal} {\bibinfo  {journal}
  {Journal of Physics A: Mathematical and General}\ }\textbf {\bibinfo {volume}
  {29}},\ \bibinfo {pages} {1929} (\bibinfo {year} {1996})}\BibitemShut
  {NoStop}%
\bibitem [{\citenamefont {Cugliandolo}\ \emph {et~al.}(1997)\citenamefont
  {Cugliandolo}, \citenamefont {Kurchan},\ and\ \citenamefont
  {Peliti}}]{CugKurPel97}%
  \BibitemOpen
  \bibfield  {author} {\bibinfo {author} {\bibfnamefont {L.~F.}\ \bibnamefont
  {Cugliandolo}}, \bibinfo {author} {\bibfnamefont {J.}~\bibnamefont
  {Kurchan}}, \ and\ \bibinfo {author} {\bibfnamefont {L.}~\bibnamefont
  {Peliti}},\ }\href@noop {} {\bibfield  {journal} {\bibinfo  {journal}
  {Physical Review E}\ }\textbf {\bibinfo {volume} {55}},\ \bibinfo {pages}
  {3898} (\bibinfo {year} {1997})}\BibitemShut {NoStop}%
\bibitem [{\citenamefont {Kurchan}(2005)}]{Kurchan}%
  \BibitemOpen
  \bibfield  {author} {\bibinfo {author} {\bibfnamefont {J.}~\bibnamefont
  {Kurchan}},\ }\href@noop {} {\bibfield  {journal} {\bibinfo  {journal}
  {Nature}\ }\textbf {\bibinfo {volume} {433}},\ \bibinfo {pages} {222}
  (\bibinfo {year} {2005})}\BibitemShut {NoStop}%
\bibitem [{\citenamefont {Berthier}\ and\ \citenamefont
  {Barrat}(2002{\natexlab{a}})}]{jl}%
  \BibitemOpen
  \bibfield  {author} {\bibinfo {author} {\bibfnamefont {L.}~\bibnamefont
  {Berthier}}\ and\ \bibinfo {author} {\bibfnamefont {J.-L.}\ \bibnamefont
  {Barrat}},\ }\href@noop {} {\bibfield  {journal} {\bibinfo  {journal}
  {Physical review letters}\ }\textbf {\bibinfo {volume} {89}},\ \bibinfo
  {pages} {095702} (\bibinfo {year} {2002}{\natexlab{a}})}\BibitemShut
  {NoStop}%
\bibitem [{\citenamefont {Franz}\ \emph {et~al.}(1998)\citenamefont {Franz},
  \citenamefont {M{\'e}zard}, \citenamefont {Parisi},\ and\ \citenamefont
  {Peliti}}]{FraMezParPel98}%
  \BibitemOpen
  \bibfield  {author} {\bibinfo {author} {\bibfnamefont {S.}~\bibnamefont
  {Franz}}, \bibinfo {author} {\bibfnamefont {M.}~\bibnamefont {M{\'e}zard}},
  \bibinfo {author} {\bibfnamefont {G.}~\bibnamefont {Parisi}}, \ and\ \bibinfo
  {author} {\bibfnamefont {L.}~\bibnamefont {Peliti}},\ }\href@noop {}
  {\bibfield  {journal} {\bibinfo  {journal} {Physical Review Letters}\
  }\textbf {\bibinfo {volume} {81}},\ \bibinfo {pages} {1758} (\bibinfo {year}
  {1998})}\BibitemShut {NoStop}%
\bibitem [{\citenamefont {Berthier}(2007{\natexlab{b}})}]{prlsilica}%
  \BibitemOpen
  \bibfield  {author} {\bibinfo {author} {\bibfnamefont {L.}~\bibnamefont
  {Berthier}},\ }\href@noop {} {\bibfield  {journal} {\bibinfo  {journal}
  {Physical review letters}\ }\textbf {\bibinfo {volume} {98}},\ \bibinfo
  {pages} {220601} (\bibinfo {year} {2007}{\natexlab{b}})}\BibitemShut
  {NoStop}%
\bibitem [{\citenamefont {Grigera}\ and\ \citenamefont
  {Israeloff}(1999)}]{Grigera99}%
  \BibitemOpen
  \bibfield  {author} {\bibinfo {author} {\bibfnamefont {T.~S.}\ \bibnamefont
  {Grigera}}\ and\ \bibinfo {author} {\bibfnamefont {N.}~\bibnamefont
  {Israeloff}},\ }\href@noop {} {\bibfield  {journal} {\bibinfo  {journal}
  {Physical Review Letters}\ }\textbf {\bibinfo {volume} {83}},\ \bibinfo
  {pages} {5038} (\bibinfo {year} {1999})}\BibitemShut {NoStop}%
\bibitem [{\citenamefont {Abou}\ and\ \citenamefont {Gallet}(2004)}]{Abou04}%
  \BibitemOpen
  \bibfield  {author} {\bibinfo {author} {\bibfnamefont {B.}~\bibnamefont
  {Abou}}\ and\ \bibinfo {author} {\bibfnamefont {F.}~\bibnamefont {Gallet}},\
  }\href@noop {} {\bibfield  {journal} {\bibinfo  {journal} {Physical Review
  Letters}\ }\textbf {\bibinfo {volume} {93}},\ \bibinfo {pages} {160603}
  (\bibinfo {year} {2004})}\BibitemShut {NoStop}%
\bibitem [{\citenamefont {Wang}\ \emph {et~al.}(2006)\citenamefont {Wang},
  \citenamefont {Song},\ and\ \citenamefont {Makse}}]{Wang06}%
  \BibitemOpen
  \bibfield  {author} {\bibinfo {author} {\bibfnamefont {P.}~\bibnamefont
  {Wang}}, \bibinfo {author} {\bibfnamefont {C.}~\bibnamefont {Song}}, \ and\
  \bibinfo {author} {\bibfnamefont {H.~A.}\ \bibnamefont {Makse}},\ }\href@noop
  {} {\bibfield  {journal} {\bibinfo  {journal} {Nature Physics}\ }\textbf
  {\bibinfo {volume} {2}},\ \bibinfo {pages} {526} (\bibinfo {year}
  {2006})}\BibitemShut {NoStop}%
\bibitem [{\citenamefont {Bellon}\ \emph {et~al.}(2001)\citenamefont {Bellon},
  \citenamefont {Ciliberto},\ and\ \citenamefont {Laroche}}]{Bellon1}%
  \BibitemOpen
  \bibfield  {author} {\bibinfo {author} {\bibfnamefont {L.}~\bibnamefont
  {Bellon}}, \bibinfo {author} {\bibfnamefont {S.}~\bibnamefont {Ciliberto}}, \
  and\ \bibinfo {author} {\bibfnamefont {C.}~\bibnamefont {Laroche}},\
  }\href@noop {} {\bibfield  {journal} {\bibinfo  {journal} {EPL (Europhysics
  Letters)}\ }\textbf {\bibinfo {volume} {53}},\ \bibinfo {pages} {511}
  (\bibinfo {year} {2001})}\BibitemShut {NoStop}%
\bibitem [{\citenamefont {Bellon}\ and\ \citenamefont
  {Ciliberto}(2002)}]{Bellon2}%
  \BibitemOpen
  \bibfield  {author} {\bibinfo {author} {\bibfnamefont {L.}~\bibnamefont
  {Bellon}}\ and\ \bibinfo {author} {\bibfnamefont {S.}~\bibnamefont
  {Ciliberto}},\ }\href@noop {} {\bibfield  {journal} {\bibinfo  {journal}
  {Physica D: Nonlinear Phenomena}\ }\textbf {\bibinfo {volume} {168}},\
  \bibinfo {pages} {325} (\bibinfo {year} {2002})}\BibitemShut {NoStop}%
\bibitem [{\citenamefont {Buisson}\ \emph
  {et~al.}(2003{\natexlab{a}})\citenamefont {Buisson}, \citenamefont {Bellon},\
  and\ \citenamefont {Ciliberto}}]{Buisson1}%
  \BibitemOpen
  \bibfield  {author} {\bibinfo {author} {\bibfnamefont {L.}~\bibnamefont
  {Buisson}}, \bibinfo {author} {\bibfnamefont {L.}~\bibnamefont {Bellon}}, \
  and\ \bibinfo {author} {\bibfnamefont {S.}~\bibnamefont {Ciliberto}},\
  }\href@noop {} {\bibfield  {journal} {\bibinfo  {journal} {Journal of
  Physics: Condensed Matter}\ }\textbf {\bibinfo {volume} {15}},\ \bibinfo
  {pages} {S1163} (\bibinfo {year} {2003}{\natexlab{a}})}\BibitemShut {NoStop}%
\bibitem [{\citenamefont {Buisson}\ \emph
  {et~al.}(2003{\natexlab{b}})\citenamefont {Buisson}, \citenamefont
  {Ciliberto},\ and\ \citenamefont {Garcimartin}}]{Buisson2}%
  \BibitemOpen
  \bibfield  {author} {\bibinfo {author} {\bibfnamefont {L.}~\bibnamefont
  {Buisson}}, \bibinfo {author} {\bibfnamefont {S.}~\bibnamefont {Ciliberto}},
  \ and\ \bibinfo {author} {\bibfnamefont {A.}~\bibnamefont {Garcimartin}},\
  }\href@noop {} {\bibfield  {journal} {\bibinfo  {journal} {EPL (Europhysics
  Letters)}\ }\textbf {\bibinfo {volume} {63}},\ \bibinfo {pages} {603}
  (\bibinfo {year} {2003}{\natexlab{b}})}\BibitemShut {NoStop}%
\bibitem [{\citenamefont {Talbot}\ \emph {et~al.}(2003)\citenamefont {Talbot},
  \citenamefont {Tarjus},\ and\ \citenamefont {Viot}}]{Viot03}%
  \BibitemOpen
  \bibfield  {author} {\bibinfo {author} {\bibfnamefont {J.}~\bibnamefont
  {Talbot}}, \bibinfo {author} {\bibfnamefont {G.}~\bibnamefont {Tarjus}}, \
  and\ \bibinfo {author} {\bibfnamefont {P.}~\bibnamefont {Viot}},\ }\href@noop
  {} {\bibfield  {journal} {\bibinfo  {journal} {Journal of Physics A:
  Mathematical and General}\ }\textbf {\bibinfo {volume} {36}},\ \bibinfo
  {pages} {9009} (\bibinfo {year} {2003})}\BibitemShut {NoStop}%
\bibitem [{\citenamefont {Nicodemi}(1999)}]{nicodemi}%
  \BibitemOpen
  \bibfield  {author} {\bibinfo {author} {\bibfnamefont {M.}~\bibnamefont
  {Nicodemi}},\ }\href@noop {} {\bibfield  {journal} {\bibinfo  {journal}
  {Physical review letters}\ }\textbf {\bibinfo {volume} {82}},\ \bibinfo
  {pages} {3734} (\bibinfo {year} {1999})}\BibitemShut {NoStop}%
\bibitem [{\citenamefont {Krzkala}(2005)}]{kr}%
  \BibitemOpen
  \bibfield  {author} {\bibinfo {author} {\bibfnamefont {F.}~\bibnamefont
  {Krzkala}},\ }\href@noop {} {\bibfield  {journal} {\bibinfo  {journal}
  {Physical review letters}\ }\textbf {\bibinfo {volume} {94}},\ \bibinfo
  {pages} {077204} (\bibinfo {year} {2005})}\BibitemShut {NoStop}%
\bibitem [{\citenamefont {Depken}\ and\ \citenamefont
  {Stinchcombe}(2005)}]{DepSti}%
  \BibitemOpen
  \bibfield  {author} {\bibinfo {author} {\bibfnamefont {M.}~\bibnamefont
  {Depken}}\ and\ \bibinfo {author} {\bibfnamefont {R.}~\bibnamefont
  {Stinchcombe}},\ }\href@noop {} {\bibfield  {journal} {\bibinfo  {journal}
  {Physical Review E}\ }\textbf {\bibinfo {volume} {71}},\ \bibinfo {pages}
  {065102} (\bibinfo {year} {2005})}\BibitemShut {NoStop}%
\bibitem [{\citenamefont {Fielding}\ and\ \citenamefont
  {Sollich}(2002)}]{FieSol02}%
  \BibitemOpen
  \bibfield  {author} {\bibinfo {author} {\bibfnamefont {S.}~\bibnamefont
  {Fielding}}\ and\ \bibinfo {author} {\bibfnamefont {P.}~\bibnamefont
  {Sollich}},\ }\href@noop {} {\bibfield  {journal} {\bibinfo  {journal}
  {Physical review letters}\ }\textbf {\bibinfo {volume} {88}},\ \bibinfo
  {pages} {050603} (\bibinfo {year} {2002})}\BibitemShut {NoStop}%
\bibitem [{\citenamefont {Bray}(1994)}]{reviewbray}%
  \BibitemOpen
  \bibfield  {author} {\bibinfo {author} {\bibfnamefont {A.}~\bibnamefont
  {Bray}},\ }\href@noop {} {\bibfield  {journal} {\bibinfo  {journal} {Advances
  in Physics}\ }\textbf {\bibinfo {volume} {43}},\ \bibinfo {pages} {357}
  (\bibinfo {year} {1994})}\BibitemShut {NoStop}%
\bibitem [{\citenamefont {Barrat}(1998)}]{barrat}%
  \BibitemOpen
  \bibfield  {author} {\bibinfo {author} {\bibfnamefont {A.}~\bibnamefont
  {Barrat}},\ }\href@noop {} {\bibfield  {journal} {\bibinfo  {journal}
  {Physical Review E}\ }\textbf {\bibinfo {volume} {57}},\ \bibinfo {pages}
  {3629} (\bibinfo {year} {1998})}\BibitemShut {NoStop}%
\bibitem [{\citenamefont {Berthier}\ \emph {et~al.}(1999)\citenamefont
  {Berthier}, \citenamefont {Barrat},\ and\ \citenamefont {Kurchan}}]{BBK2}%
  \BibitemOpen
  \bibfield  {author} {\bibinfo {author} {\bibfnamefont {L.}~\bibnamefont
  {Berthier}}, \bibinfo {author} {\bibfnamefont {J.-L.}\ \bibnamefont
  {Barrat}}, \ and\ \bibinfo {author} {\bibfnamefont {J.}~\bibnamefont
  {Kurchan}},\ }\href@noop {} {\bibfield  {journal} {\bibinfo  {journal} {The
  European Physical Journal B-Condensed Matter and Complex Systems}\ }\textbf
  {\bibinfo {volume} {11}},\ \bibinfo {pages} {635} (\bibinfo {year}
  {1999})}\BibitemShut {NoStop}%
\bibitem [{\citenamefont {Godr{\`e}che}\ and\ \citenamefont
  {Luck}(2000)}]{ising1d}%
  \BibitemOpen
  \bibfield  {author} {\bibinfo {author} {\bibfnamefont {C.}~\bibnamefont
  {Godr{\`e}che}}\ and\ \bibinfo {author} {\bibfnamefont {J.}~\bibnamefont
  {Luck}},\ }\href@noop {} {\bibfield  {journal} {\bibinfo  {journal} {Journal
  of Physics A: Mathematical and General}\ }\textbf {\bibinfo {volume} {33}},\
  \bibinfo {pages} {1151} (\bibinfo {year} {2000})}\BibitemShut {NoStop}%
\bibitem [{\citenamefont {Lippiello}\ and\ \citenamefont
  {Zannetti}(2000)}]{ising1dLippiello}%
  \BibitemOpen
  \bibfield  {author} {\bibinfo {author} {\bibfnamefont {E.}~\bibnamefont
  {Lippiello}}\ and\ \bibinfo {author} {\bibfnamefont {M.}~\bibnamefont
  {Zannetti}},\ }\href@noop {} {\bibfield  {journal} {\bibinfo  {journal}
  {Physical Review E}\ }\textbf {\bibinfo {volume} {61}},\ \bibinfo {pages}
  {3369} (\bibinfo {year} {2000})}\BibitemShut {NoStop}%
\bibitem [{\citenamefont {Godreche}\ and\ \citenamefont
  {Luck}(2000)}]{godluck}%
  \BibitemOpen
  \bibfield  {author} {\bibinfo {author} {\bibfnamefont {C.}~\bibnamefont
  {Godreche}}\ and\ \bibinfo {author} {\bibfnamefont {J.}~\bibnamefont
  {Luck}},\ }\href@noop {} {\bibfield  {journal} {\bibinfo  {journal} {Journal
  of Physics A: Mathematical and General}\ }\textbf {\bibinfo {volume} {33}},\
  \bibinfo {pages} {9141} (\bibinfo {year} {2000})}\BibitemShut {NoStop}%
\bibitem [{\citenamefont {Mayer}\ \emph {et~al.}(2003)\citenamefont {Mayer},
  \citenamefont {Berthier}, \citenamefont {Garrahan},\ and\ \citenamefont
  {Sollich}}]{pre}%
  \BibitemOpen
  \bibfield  {author} {\bibinfo {author} {\bibfnamefont {P.}~\bibnamefont
  {Mayer}}, \bibinfo {author} {\bibfnamefont {L.}~\bibnamefont {Berthier}},
  \bibinfo {author} {\bibfnamefont {J.~P.}\ \bibnamefont {Garrahan}}, \ and\
  \bibinfo {author} {\bibfnamefont {P.}~\bibnamefont {Sollich}},\ }\href@noop
  {} {\bibfield  {journal} {\bibinfo  {journal} {Physical Review E}\ }\textbf
  {\bibinfo {volume} {68}},\ \bibinfo {pages} {016116} (\bibinfo {year}
  {2003})}\BibitemShut {NoStop}%
\bibitem [{\citenamefont {Calabrese}\ and\ \citenamefont
  {Gambassi}(2005)}]{pasquale}%
  \BibitemOpen
  \bibfield  {author} {\bibinfo {author} {\bibfnamefont {P.}~\bibnamefont
  {Calabrese}}\ and\ \bibinfo {author} {\bibfnamefont {A.}~\bibnamefont
  {Gambassi}},\ }\href@noop {} {\bibfield  {journal} {\bibinfo  {journal}
  {Journal of Physics A: Mathematical and General}\ }\textbf {\bibinfo {volume}
  {38}},\ \bibinfo {pages} {R133} (\bibinfo {year} {2005})}\BibitemShut
  {NoStop}%
\bibitem [{\citenamefont {Mayer}\ and\ \citenamefont {Sollich}(2007)}]{peters}%
  \BibitemOpen
  \bibfield  {author} {\bibinfo {author} {\bibfnamefont {P.}~\bibnamefont
  {Mayer}}\ and\ \bibinfo {author} {\bibfnamefont {P.}~\bibnamefont
  {Sollich}},\ }\href@noop {} {\bibfield  {journal} {\bibinfo  {journal}
  {Journal of Physics A: Mathematical and Theoretical}\ }\textbf {\bibinfo
  {volume} {40}},\ \bibinfo {pages} {5823} (\bibinfo {year}
  {2007})}\BibitemShut {NoStop}%
\bibitem [{\citenamefont {Mayer}\ \emph {et~al.}(2006)\citenamefont {Mayer},
  \citenamefont {L{\'e}onard}, \citenamefont {Berthier}, \citenamefont
  {Garrahan},\ and\ \citenamefont {Sollich}}]{nveX}%
  \BibitemOpen
  \bibfield  {author} {\bibinfo {author} {\bibfnamefont {P.}~\bibnamefont
  {Mayer}}, \bibinfo {author} {\bibfnamefont {S.}~\bibnamefont {L{\'e}onard}},
  \bibinfo {author} {\bibfnamefont {L.}~\bibnamefont {Berthier}}, \bibinfo
  {author} {\bibfnamefont {J.~P.}\ \bibnamefont {Garrahan}}, \ and\ \bibinfo
  {author} {\bibfnamefont {P.}~\bibnamefont {Sollich}},\ }\href@noop {}
  {\bibfield  {journal} {\bibinfo  {journal} {Physical review letters}\
  }\textbf {\bibinfo {volume} {96}},\ \bibinfo {pages} {030602} (\bibinfo
  {year} {2006})}\BibitemShut {NoStop}%
\bibitem [{\citenamefont {Jack}\ \emph
  {et~al.}(2006{\natexlab{b}})\citenamefont {Jack}, \citenamefont {Berthier},\
  and\ \citenamefont {Garrahan}}]{robfdt}%
  \BibitemOpen
  \bibfield  {author} {\bibinfo {author} {\bibfnamefont {R.~L.}\ \bibnamefont
  {Jack}}, \bibinfo {author} {\bibfnamefont {L.}~\bibnamefont {Berthier}}, \
  and\ \bibinfo {author} {\bibfnamefont {J.~P.}\ \bibnamefont {Garrahan}},\
  }\href@noop {} {\bibfield  {journal} {\bibinfo  {journal} {Journal of
  Statistical Mechanics: Theory and Experiment}\ }\textbf {\bibinfo {volume}
  {2006}},\ \bibinfo {pages} {P12005} (\bibinfo {year}
  {2006}{\natexlab{b}})}\BibitemShut {NoStop}%
\bibitem [{\citenamefont {Crisanti}\ and\ \citenamefont
  {Ritort}(2000{\natexlab{a}})}]{crisanti2000potential}%
  \BibitemOpen
  \bibfield  {author} {\bibinfo {author} {\bibfnamefont {A.}~\bibnamefont
  {Crisanti}}\ and\ \bibinfo {author} {\bibfnamefont {F.}~\bibnamefont
  {Ritort}},\ }\href@noop {} {\bibfield  {journal} {\bibinfo  {journal} {EPL
  (Europhysics Letters)}\ }\textbf {\bibinfo {volume} {51}},\ \bibinfo {pages}
  {147} (\bibinfo {year} {2000}{\natexlab{a}})}\BibitemShut {NoStop}%
\bibitem [{\citenamefont {Crisanti}\ and\ \citenamefont
  {Ritort}(2000{\natexlab{b}})}]{crisanti2000activated}%
  \BibitemOpen
  \bibfield  {author} {\bibinfo {author} {\bibfnamefont {A.}~\bibnamefont
  {Crisanti}}\ and\ \bibinfo {author} {\bibfnamefont {F.}~\bibnamefont
  {Ritort}},\ }\href@noop {} {\bibfield  {journal} {\bibinfo  {journal} {EPL
  (Europhysics Letters)}\ }\textbf {\bibinfo {volume} {52}},\ \bibinfo {pages}
  {640} (\bibinfo {year} {2000}{\natexlab{b}})}\BibitemShut {NoStop}%
\bibitem [{\citenamefont {Baity-Jesi}\ \emph
  {et~al.}(2018{\natexlab{a}})\citenamefont {Baity-Jesi}, \citenamefont
  {Achard-de Lustrac},\ and\ \citenamefont {Biroli}}]{baity2018activated}%
  \BibitemOpen
  \bibfield  {author} {\bibinfo {author} {\bibfnamefont {M.}~\bibnamefont
  {Baity-Jesi}}, \bibinfo {author} {\bibfnamefont {A.}~\bibnamefont {Achard-de
  Lustrac}}, \ and\ \bibinfo {author} {\bibfnamefont {G.}~\bibnamefont
  {Biroli}},\ }\href@noop {} {\bibfield  {journal} {\bibinfo  {journal}
  {Physical Review E}\ }\textbf {\bibinfo {volume} {98}},\ \bibinfo {pages}
  {012133} (\bibinfo {year} {2018}{\natexlab{a}})}\BibitemShut {NoStop}%
\bibitem [{\citenamefont {Baity-Jesi}\ \emph
  {et~al.}(2018{\natexlab{b}})\citenamefont {Baity-Jesi}, \citenamefont
  {Biroli},\ and\ \citenamefont {Cammarota}}]{baity2018activated2}%
  \BibitemOpen
  \bibfield  {author} {\bibinfo {author} {\bibfnamefont {M.}~\bibnamefont
  {Baity-Jesi}}, \bibinfo {author} {\bibfnamefont {G.}~\bibnamefont {Biroli}},
  \ and\ \bibinfo {author} {\bibfnamefont {C.}~\bibnamefont {Cammarota}},\
  }\href@noop {} {\bibfield  {journal} {\bibinfo  {journal} {Journal of
  Statistical Mechanics: Theory and Experiment}\ }\textbf {\bibinfo {volume}
  {2018}},\ \bibinfo {pages} {013301} (\bibinfo {year}
  {2018}{\natexlab{b}})}\BibitemShut {NoStop}%
\bibitem [{\citenamefont {Billoire}\ \emph {et~al.}(2005)\citenamefont
  {Billoire}, \citenamefont {Giomi},\ and\ \citenamefont
  {Marinari}}]{billoire2005mean}%
  \BibitemOpen
  \bibfield  {author} {\bibinfo {author} {\bibfnamefont {A.}~\bibnamefont
  {Billoire}}, \bibinfo {author} {\bibfnamefont {L.}~\bibnamefont {Giomi}}, \
  and\ \bibinfo {author} {\bibfnamefont {E.}~\bibnamefont {Marinari}},\
  }\href@noop {} {\bibfield  {journal} {\bibinfo  {journal} {EPL (Europhysics
  Letters)}\ }\textbf {\bibinfo {volume} {71}},\ \bibinfo {pages} {824}
  (\bibinfo {year} {2005})}\BibitemShut {NoStop}%
\bibitem [{\citenamefont {Stariolo}\ and\ \citenamefont
  {Cugliandolo}(2019)}]{stariolo2019activated}%
  \BibitemOpen
  \bibfield  {author} {\bibinfo {author} {\bibfnamefont {D.~A.}\ \bibnamefont
  {Stariolo}}\ and\ \bibinfo {author} {\bibfnamefont {L.~F.}\ \bibnamefont
  {Cugliandolo}},\ }\href@noop {} {\bibfield  {journal} {\bibinfo  {journal}
  {EPL (Europhysics Letters)}\ }\textbf {\bibinfo {volume} {127}},\ \bibinfo
  {pages} {16002} (\bibinfo {year} {2019})}\BibitemShut {NoStop}%
\bibitem [{\citenamefont {Arous}\ \emph {et~al.}(2002)\citenamefont {Arous},
  \citenamefont {Bovier},\ and\ \citenamefont {Gayrard}}]{arous2002aging}%
  \BibitemOpen
  \bibfield  {author} {\bibinfo {author} {\bibfnamefont {G.~B.}\ \bibnamefont
  {Arous}}, \bibinfo {author} {\bibfnamefont {A.}~\bibnamefont {Bovier}}, \
  and\ \bibinfo {author} {\bibfnamefont {V.}~\bibnamefont {Gayrard}},\
  }\href@noop {} {\bibfield  {journal} {\bibinfo  {journal} {Physical review
  letters}\ }\textbf {\bibinfo {volume} {88}},\ \bibinfo {pages} {087201}
  (\bibinfo {year} {2002})}\BibitemShut {NoStop}%
\bibitem [{\citenamefont {Gayrard}(2019)}]{gayrard2019aging}%
  \BibitemOpen
  \bibfield  {author} {\bibinfo {author} {\bibfnamefont {V.}~\bibnamefont
  {Gayrard}},\ }\href@noop {} {\bibfield  {journal} {\bibinfo  {journal}
  {Probability Theory and Related Fields}\ }\textbf {\bibinfo {volume} {174}},\
  \bibinfo {pages} {501} (\bibinfo {year} {2019})}\BibitemShut {NoStop}%
\bibitem [{\citenamefont {Dyre}(1987)}]{dyre1987master}%
  \BibitemOpen
  \bibfield  {author} {\bibinfo {author} {\bibfnamefont {J.~C.}\ \bibnamefont
  {Dyre}},\ }\href@noop {} {\bibfield  {journal} {\bibinfo  {journal} {Physical
  review letters}\ }\textbf {\bibinfo {volume} {58}},\ \bibinfo {pages} {792}
  (\bibinfo {year} {1987})}\BibitemShut {NoStop}%
\bibitem [{\citenamefont {Barrat}\ and\ \citenamefont
  {M{\'e}zard}(1995)}]{barrat1995phase}%
  \BibitemOpen
  \bibfield  {author} {\bibinfo {author} {\bibfnamefont {A.}~\bibnamefont
  {Barrat}}\ and\ \bibinfo {author} {\bibfnamefont {M.}~\bibnamefont
  {M{\'e}zard}},\ }\href@noop {} {\bibfield  {journal} {\bibinfo  {journal}
  {Journal de Physique I}\ }\textbf {\bibinfo {volume} {5}},\ \bibinfo {pages}
  {941} (\bibinfo {year} {1995})}\BibitemShut {NoStop}%
\bibitem [{\citenamefont {Cammarota}\ and\ \citenamefont
  {Marinari}(2018)}]{cammarota2018numerical}%
  \BibitemOpen
  \bibfield  {author} {\bibinfo {author} {\bibfnamefont {C.}~\bibnamefont
  {Cammarota}}\ and\ \bibinfo {author} {\bibfnamefont {E.}~\bibnamefont
  {Marinari}},\ }\href@noop {} {\bibfield  {journal} {\bibinfo  {journal}
  {Journal of Statistical Mechanics: Theory and Experiment}\ }\textbf {\bibinfo
  {volume} {2018}},\ \bibinfo {pages} {043303} (\bibinfo {year}
  {2018})}\BibitemShut {NoStop}%
\bibitem [{\citenamefont {Cammarota}\ and\ \citenamefont
  {Marinari}(2015)}]{cammarota2015spontaneous}%
  \BibitemOpen
  \bibfield  {author} {\bibinfo {author} {\bibfnamefont {C.}~\bibnamefont
  {Cammarota}}\ and\ \bibinfo {author} {\bibfnamefont {E.}~\bibnamefont
  {Marinari}},\ }\href@noop {} {\bibfield  {journal} {\bibinfo  {journal}
  {Physical Review E}\ }\textbf {\bibinfo {volume} {92}},\ \bibinfo {pages}
  {010301} (\bibinfo {year} {2015})}\BibitemShut {NoStop}%
\bibitem [{\citenamefont {Ros}\ \emph {et~al.}(2019)\citenamefont {Ros},
  \citenamefont {Biroli},\ and\ \citenamefont {Cammarota}}]{ros2019complexity}%
  \BibitemOpen
  \bibfield  {author} {\bibinfo {author} {\bibfnamefont {V.}~\bibnamefont
  {Ros}}, \bibinfo {author} {\bibfnamefont {G.}~\bibnamefont {Biroli}}, \ and\
  \bibinfo {author} {\bibfnamefont {C.}~\bibnamefont {Cammarota}},\ }\href@noop
  {} {\bibfield  {journal} {\bibinfo  {journal} {EPL (Europhysics Letters)}\
  }\textbf {\bibinfo {volume} {126}},\ \bibinfo {pages} {20003} (\bibinfo
  {year} {2019})}\BibitemShut {NoStop}%
\bibitem [{\citenamefont {B{\"u}chner}\ and\ \citenamefont
  {Heuer}(1999)}]{buchner1999potential}%
  \BibitemOpen
  \bibfield  {author} {\bibinfo {author} {\bibfnamefont {S.}~\bibnamefont
  {B{\"u}chner}}\ and\ \bibinfo {author} {\bibfnamefont {A.}~\bibnamefont
  {Heuer}},\ }\href@noop {} {\bibfield  {journal} {\bibinfo  {journal}
  {Physical Review E}\ }\textbf {\bibinfo {volume} {60}},\ \bibinfo {pages}
  {6507} (\bibinfo {year} {1999})}\BibitemShut {NoStop}%
\bibitem [{\citenamefont {Doliwa}\ and\ \citenamefont
  {Heuer}(2003)}]{doliwa2003hopping}%
  \BibitemOpen
  \bibfield  {author} {\bibinfo {author} {\bibfnamefont {B.}~\bibnamefont
  {Doliwa}}\ and\ \bibinfo {author} {\bibfnamefont {A.}~\bibnamefont {Heuer}},\
  }\href@noop {} {\bibfield  {journal} {\bibinfo  {journal} {Physical Review
  E}\ }\textbf {\bibinfo {volume} {67}},\ \bibinfo {pages} {030501} (\bibinfo
  {year} {2003})}\BibitemShut {NoStop}%
\bibitem [{\citenamefont {Heuer}(2008)}]{heuer2008exploring}%
  \BibitemOpen
  \bibfield  {author} {\bibinfo {author} {\bibfnamefont {A.}~\bibnamefont
  {Heuer}},\ }\href@noop {} {\bibfield  {journal} {\bibinfo  {journal} {Journal
  of Physics: Condensed Matter}\ }\textbf {\bibinfo {volume} {20}},\ \bibinfo
  {pages} {373101} (\bibinfo {year} {2008})}\BibitemShut {NoStop}%
\bibitem [{\citenamefont {Denny}\ \emph {et~al.}(2003)\citenamefont {Denny},
  \citenamefont {Reichman},\ and\ \citenamefont {Bouchaud}}]{denny2003trap}%
  \BibitemOpen
  \bibfield  {author} {\bibinfo {author} {\bibfnamefont {R.~A.}\ \bibnamefont
  {Denny}}, \bibinfo {author} {\bibfnamefont {D.~R.}\ \bibnamefont {Reichman}},
  \ and\ \bibinfo {author} {\bibfnamefont {J.-P.}\ \bibnamefont {Bouchaud}},\
  }\href@noop {} {\bibfield  {journal} {\bibinfo  {journal} {Physical review
  letters}\ }\textbf {\bibinfo {volume} {90}},\ \bibinfo {pages} {025503}
  (\bibinfo {year} {2003})}\BibitemShut {NoStop}%
\bibitem [{\citenamefont {McKenna}\ and\ \citenamefont
  {Kovacs}(1984)}]{mckenna}%
  \BibitemOpen
  \bibfield  {author} {\bibinfo {author} {\bibfnamefont {G.}~\bibnamefont
  {McKenna}}\ and\ \bibinfo {author} {\bibfnamefont {A.}~\bibnamefont
  {Kovacs}},\ }\href@noop {} {\bibfield  {journal} {\bibinfo  {journal}
  {Polymer Engineering \& Science}\ }\textbf {\bibinfo {volume} {24}},\
  \bibinfo {pages} {1138} (\bibinfo {year} {1984})}\BibitemShut {NoStop}%
\bibitem [{\citenamefont {Sollich}\ \emph {et~al.}(1997)\citenamefont
  {Sollich}, \citenamefont {Lequeux}, \citenamefont {H{\'e}braud},\ and\
  \citenamefont {Cates}}]{prlsollich}%
  \BibitemOpen
  \bibfield  {author} {\bibinfo {author} {\bibfnamefont {P.}~\bibnamefont
  {Sollich}}, \bibinfo {author} {\bibfnamefont {F.}~\bibnamefont {Lequeux}},
  \bibinfo {author} {\bibfnamefont {P.}~\bibnamefont {H{\'e}braud}}, \ and\
  \bibinfo {author} {\bibfnamefont {M.~E.}\ \bibnamefont {Cates}},\ }\href@noop
  {} {\bibfield  {journal} {\bibinfo  {journal} {Physical review letters}\
  }\textbf {\bibinfo {volume} {78}},\ \bibinfo {pages} {2020} (\bibinfo {year}
  {1997})}\BibitemShut {NoStop}%
\bibitem [{\citenamefont {Ikeda}\ \emph {et~al.}(2012)\citenamefont {Ikeda},
  \citenamefont {Berthier},\ and\ \citenamefont {Sollich}}]{ikeda2012unified}%
  \BibitemOpen
  \bibfield  {author} {\bibinfo {author} {\bibfnamefont {A.}~\bibnamefont
  {Ikeda}}, \bibinfo {author} {\bibfnamefont {L.}~\bibnamefont {Berthier}}, \
  and\ \bibinfo {author} {\bibfnamefont {P.}~\bibnamefont {Sollich}},\
  }\href@noop {} {\bibfield  {journal} {\bibinfo  {journal} {Physical review
  letters}\ }\textbf {\bibinfo {volume} {109}},\ \bibinfo {pages} {018301}
  (\bibinfo {year} {2012})}\BibitemShut {NoStop}%
\bibitem [{\citenamefont {Sollich}(1998)}]{presollich}%
  \BibitemOpen
  \bibfield  {author} {\bibinfo {author} {\bibfnamefont {P.}~\bibnamefont
  {Sollich}},\ }\href@noop {} {\bibfield  {journal} {\bibinfo  {journal}
  {Physical Review E}\ }\textbf {\bibinfo {volume} {58}},\ \bibinfo {pages}
  {738} (\bibinfo {year} {1998})}\BibitemShut {NoStop}%
\bibitem [{\citenamefont {Berthier}\ and\ \citenamefont
  {Barrat}(2002{\natexlab{b}})}]{jllong}%
  \BibitemOpen
  \bibfield  {author} {\bibinfo {author} {\bibfnamefont {L.}~\bibnamefont
  {Berthier}}\ and\ \bibinfo {author} {\bibfnamefont {J.-L.}\ \bibnamefont
  {Barrat}},\ }\href@noop {} {\bibfield  {journal} {\bibinfo  {journal} {The
  Journal of Chemical Physics}\ }\textbf {\bibinfo {volume} {116}},\ \bibinfo
  {pages} {6228} (\bibinfo {year} {2002}{\natexlab{b}})}\BibitemShut {NoStop}%
\bibitem [{\citenamefont {Nandkishore}\ and\ \citenamefont
  {Huse}(2015)}]{nandkishore2015many}%
  \BibitemOpen
  \bibfield  {author} {\bibinfo {author} {\bibfnamefont {R.}~\bibnamefont
  {Nandkishore}}\ and\ \bibinfo {author} {\bibfnamefont {D.~A.}\ \bibnamefont
  {Huse}},\ }\href@noop {} {\bibfield  {journal} {\bibinfo  {journal} {Annu.
  Rev. Condens. Matter Phys.}\ }\textbf {\bibinfo {volume} {6}},\ \bibinfo
  {pages} {15} (\bibinfo {year} {2015})}\BibitemShut {NoStop}%
\bibitem [{\citenamefont {Turner}\ \emph {et~al.}(2018)\citenamefont {Turner},
  \citenamefont {Michailidis}, \citenamefont {Abanin}, \citenamefont {Serbyn},\
  and\ \citenamefont {Papi{\'c}}}]{turner2018weak}%
  \BibitemOpen
  \bibfield  {author} {\bibinfo {author} {\bibfnamefont {C.~J.}\ \bibnamefont
  {Turner}}, \bibinfo {author} {\bibfnamefont {A.~A.}\ \bibnamefont
  {Michailidis}}, \bibinfo {author} {\bibfnamefont {D.~A.}\ \bibnamefont
  {Abanin}}, \bibinfo {author} {\bibfnamefont {M.}~\bibnamefont {Serbyn}}, \
  and\ \bibinfo {author} {\bibfnamefont {Z.}~\bibnamefont {Papi{\'c}}},\
  }\href@noop {} {\bibfield  {journal} {\bibinfo  {journal} {Nature Physics}\
  }\textbf {\bibinfo {volume} {14}},\ \bibinfo {pages} {745} (\bibinfo {year}
  {2018})}\BibitemShut {NoStop}%
\bibitem [{\citenamefont {Pancotti}\ \emph {et~al.}(2020)\citenamefont
  {Pancotti}, \citenamefont {Giudice}, \citenamefont {Cirac}, \citenamefont
  {Garrahan},\ and\ \citenamefont {Ba{\~n}uls}}]{pancotti2020quantum}%
  \BibitemOpen
  \bibfield  {author} {\bibinfo {author} {\bibfnamefont {N.}~\bibnamefont
  {Pancotti}}, \bibinfo {author} {\bibfnamefont {G.}~\bibnamefont {Giudice}},
  \bibinfo {author} {\bibfnamefont {J.~I.}\ \bibnamefont {Cirac}}, \bibinfo
  {author} {\bibfnamefont {J.~P.}\ \bibnamefont {Garrahan}}, \ and\ \bibinfo
  {author} {\bibfnamefont {M.~C.}\ \bibnamefont {Ba{\~n}uls}},\ }\href@noop {}
  {\bibfield  {journal} {\bibinfo  {journal} {Physical Review X}\ }\textbf
  {\bibinfo {volume} {10}},\ \bibinfo {pages} {021051} (\bibinfo {year}
  {2020})}\BibitemShut {NoStop}%
\bibitem [{\citenamefont {Facoetti}\ \emph {et~al.}(2019)\citenamefont
  {Facoetti}, \citenamefont {Biroli}, \citenamefont {Kurchan},\ and\
  \citenamefont {Reichman}}]{facoetti2019classical}%
  \BibitemOpen
  \bibfield  {author} {\bibinfo {author} {\bibfnamefont {D.}~\bibnamefont
  {Facoetti}}, \bibinfo {author} {\bibfnamefont {G.}~\bibnamefont {Biroli}},
  \bibinfo {author} {\bibfnamefont {J.}~\bibnamefont {Kurchan}}, \ and\
  \bibinfo {author} {\bibfnamefont {D.~R.}\ \bibnamefont {Reichman}},\
  }\href@noop {} {\bibfield  {journal} {\bibinfo  {journal} {Physical Review
  B}\ }\textbf {\bibinfo {volume} {100}},\ \bibinfo {pages} {205108} (\bibinfo
  {year} {2019})}\BibitemShut {NoStop}%
\bibitem [{\citenamefont {Skoge}\ \emph {et~al.}(2006)\citenamefont {Skoge},
  \citenamefont {Donev}, \citenamefont {Stillinger},\ and\ \citenamefont
  {Torquato}}]{Skoge}%
  \BibitemOpen
  \bibfield  {author} {\bibinfo {author} {\bibfnamefont {M.}~\bibnamefont
  {Skoge}}, \bibinfo {author} {\bibfnamefont {A.}~\bibnamefont {Donev}},
  \bibinfo {author} {\bibfnamefont {F.~H.}\ \bibnamefont {Stillinger}}, \ and\
  \bibinfo {author} {\bibfnamefont {S.}~\bibnamefont {Torquato}},\ }\href@noop
  {} {\bibfield  {journal} {\bibinfo  {journal} {Physical Review E}\ }\textbf
  {\bibinfo {volume} {74}},\ \bibinfo {pages} {041127} (\bibinfo {year}
  {2006})}\BibitemShut {NoStop}%
\bibitem [{\citenamefont {van Meel}\ \emph {et~al.}(2009)\citenamefont {van
  Meel}, \citenamefont {Charbonneau}, \citenamefont {Fortini},\ and\
  \citenamefont {Charbonneau}}]{vanmeelhardsphere2009}%
  \BibitemOpen
  \bibfield  {author} {\bibinfo {author} {\bibfnamefont {J.~A.}\ \bibnamefont
  {van Meel}}, \bibinfo {author} {\bibfnamefont {B.}~\bibnamefont
  {Charbonneau}}, \bibinfo {author} {\bibfnamefont {A.}~\bibnamefont
  {Fortini}}, \ and\ \bibinfo {author} {\bibfnamefont {P.}~\bibnamefont
  {Charbonneau}},\ }\href@noop {} {\bibfield  {journal} {\bibinfo  {journal}
  {Physical Review E}\ }\textbf {\bibinfo {volume} {80}},\ \bibinfo {pages}
  {061110} (\bibinfo {year} {2009})}\BibitemShut {NoStop}%
\bibitem [{\citenamefont {Lipowski}\ \emph {et~al.}(2000)\citenamefont
  {Lipowski}, \citenamefont {Johnston},\ and\ \citenamefont
  {Espriu}}]{Lipowsky}%
  \BibitemOpen
  \bibfield  {author} {\bibinfo {author} {\bibfnamefont {A.}~\bibnamefont
  {Lipowski}}, \bibinfo {author} {\bibfnamefont {D.}~\bibnamefont {Johnston}},
  \ and\ \bibinfo {author} {\bibfnamefont {D.}~\bibnamefont {Espriu}},\
  }\href@noop {} {\bibfield  {journal} {\bibinfo  {journal} {Physical Review
  E}\ }\textbf {\bibinfo {volume} {62}},\ \bibinfo {pages} {3404} (\bibinfo
  {year} {2000})}\BibitemShut {NoStop}%
\bibitem [{\citenamefont {Sethna}\ \emph {et~al.}(1991)\citenamefont {Sethna},
  \citenamefont {Shore},\ and\ \citenamefont {Huang}}]{Sethna}%
  \BibitemOpen
  \bibfield  {author} {\bibinfo {author} {\bibfnamefont {J.~P.}\ \bibnamefont
  {Sethna}}, \bibinfo {author} {\bibfnamefont {J.~D.}\ \bibnamefont {Shore}}, \
  and\ \bibinfo {author} {\bibfnamefont {M.}~\bibnamefont {Huang}},\
  }\href@noop {} {\bibfield  {journal} {\bibinfo  {journal} {Physical Review
  B}\ }\textbf {\bibinfo {volume} {44}},\ \bibinfo {pages} {4943} (\bibinfo
  {year} {1991})}\BibitemShut {NoStop}%
\end{thebibliography}%

\end{document}